\documentclass[11pt]{article}

\usepackage[final]{acl}

\usepackage{times}
\usepackage{latexsym}
\usepackage[T1]{fontenc}
\usepackage[utf8]{inputenc}
\usepackage{microtype}
\usepackage{inconsolata}
\usepackage{amsmath}
\usepackage{amsopn}

\usepackage{graphicx}

\usepackage{hyperref}
\usepackage{url}
\usepackage{multirow}
\usepackage{booktabs}
\usepackage{subcaption}
\usepackage{amsthm}  %
\usepackage[fixed]{fontawesome5}
\usepackage{amssymb}

\usepackage{makecell}
\usepackage{tcolorbox}
\usepackage[table]{xcolor} 
\usepackage{enumitem}
\usepackage{cleveref}
\usepackage{pifont}
\usepackage{cuted} 
\usepackage{algorithm}
\usepackage{algorithmic}

\usepackage{minitoc}
\usepackage[utf8]{inputenc}
\usepackage{array}
\usepackage{longtable}
\usepackage{multicol}
\usepackage{truncate}

\usepackage{alphalph}

\usepackage{tikz}
\usepackage{pgfplots}
\pgfplotsset{compat=1.17}

\DeclareMathOperator*{\argmin}{arg\,min}

\newcounter{insightcounter}

\newcommand{\method}{{GradSentry}}

\title{
\begin{minipage}{0.09\textwidth}
\includegraphics[width=\linewidth]{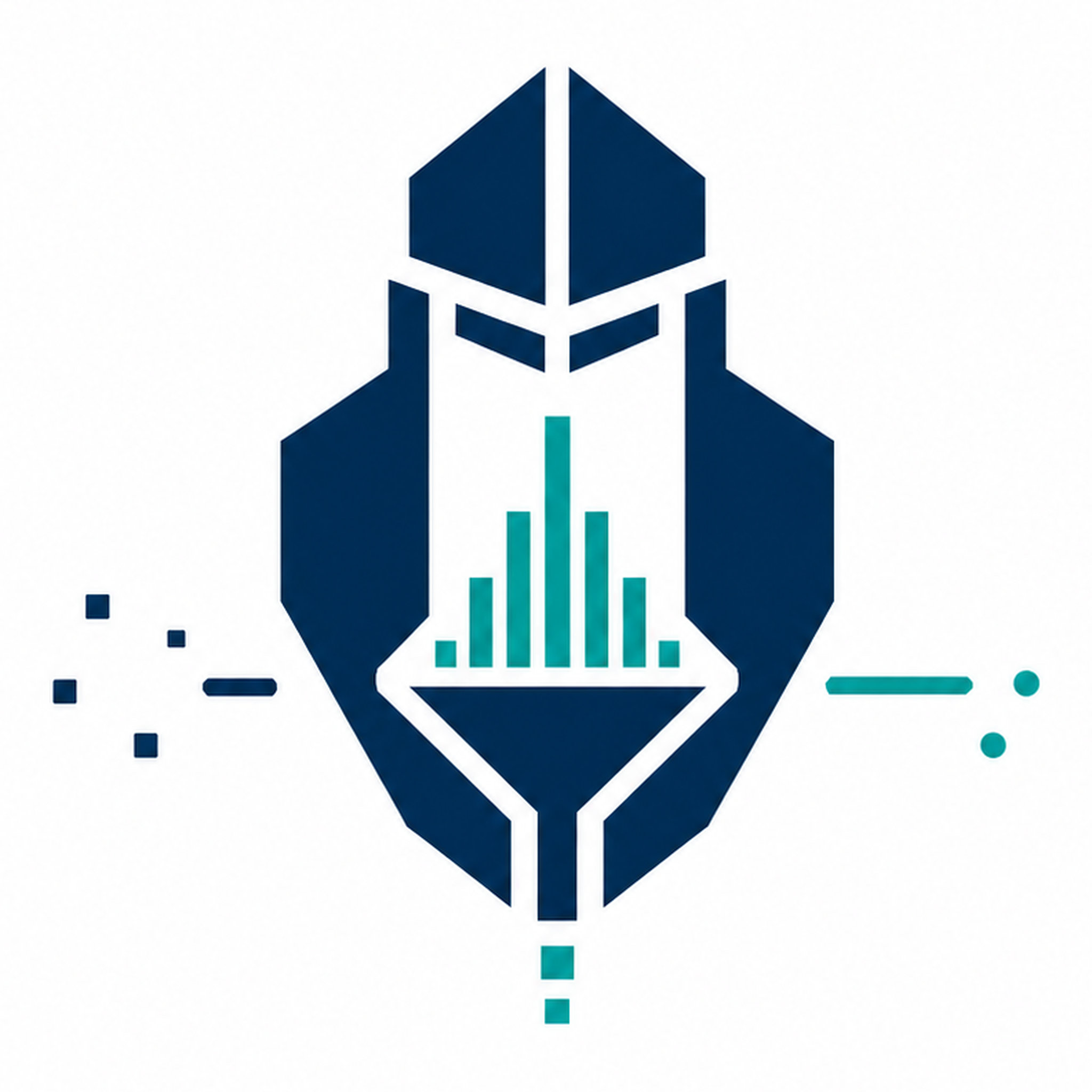}
\end{minipage}
\begin{minipage}{0.88\textwidth}
\centering
\method{}: Gradient Spectral Entropy for Backdoor Sample Filtering in Large Language Model Fine-Tuning
\end{minipage}
}

\author{
 \textbf{Haodong Zhao},
 \textbf{Tianyi Xu},
 \textbf{Tianhang Zhao},
 \textbf{Zhuosheng Zhang}\footnotemark[1],
 \textbf{Gongshen Liu}\thanks{Corresponding authors.\\ This work is supported by the National Natural Science Foundation of China (62406188).}
\\
 School of Computer Science, Shanghai Jiao Tong University
\\
 \small{
   \texttt{\{zhaohaodong, akiracomplex, zthzthzth, zhangzs, lgshen\}@sjtu.edu.cn}}\\[5pt]
   {{\small \faGithub} \texttt{\href{https://github.com/dongdongzhaoUP/GradSentry}{\textcolor{blue}{https://github.com/dongdongzhaoUP/GradSentry}}}}
}
\begin{document}
\doparttoc 
\faketableofcontents
\maketitle
\begin{abstract}
Fine-tuning Large Language Models with untrusted data exposes models to backdoor attacks, where poisoned samples cause targeted misbehavior. Existing sample-filtering defenses rely on clustering, which requires sufficient data and can fail at extreme poison ratios. We propose \method{} (\textbf{Grad}ient \textbf{Sentry}), a backdoor sample filtering method based on the spectral entropy of per-sample gradients.
Our key finding is that poisoned samples produce gradients with higher spectral entropy compared to clean samples. \method{} captures output-altering backdoor signatures using per-sample gradient spectra, avoiding pairwise sample comparisons and clustering during feature construction. Importantly, our method is \textit{training-agnostic}: it works for both parameter-efficient fine-tuning methods like LoRA and full-parameter tuning, as the gradient analysis operates independently of which parameters are being updated during training. 
\method{} requires no clustering, operates effectively across all poison ratios (1\%--90\%), and introduces minimal computational overhead (20--50ms per sample for a 7B model). Evaluation on four QA datasets and four attack types demonstrates the effectiveness of spectral entropy for backdoor detection.
\end{abstract}

\section{Introduction}

Large Language Models (LLMs) have demonstrated remarkable capabilities across diverse natural language tasks~\citep{brown2020language,achiam2023gpt}. To adapt these models to specific domains or tasks, practitioners use full-parameter fine-tuning or parameter-efficient fine-tuning (PEFT) methods such as low-rank adaptation (LoRA)~\citep{hu2022lora}, which freezes pretrained weights and introduces trainable low-rank matrices. These PEFT approaches reduce computational costs while maintaining competitive performance.

\begin{figure}[t]
    \centering
    \includegraphics[width=1.0\linewidth]{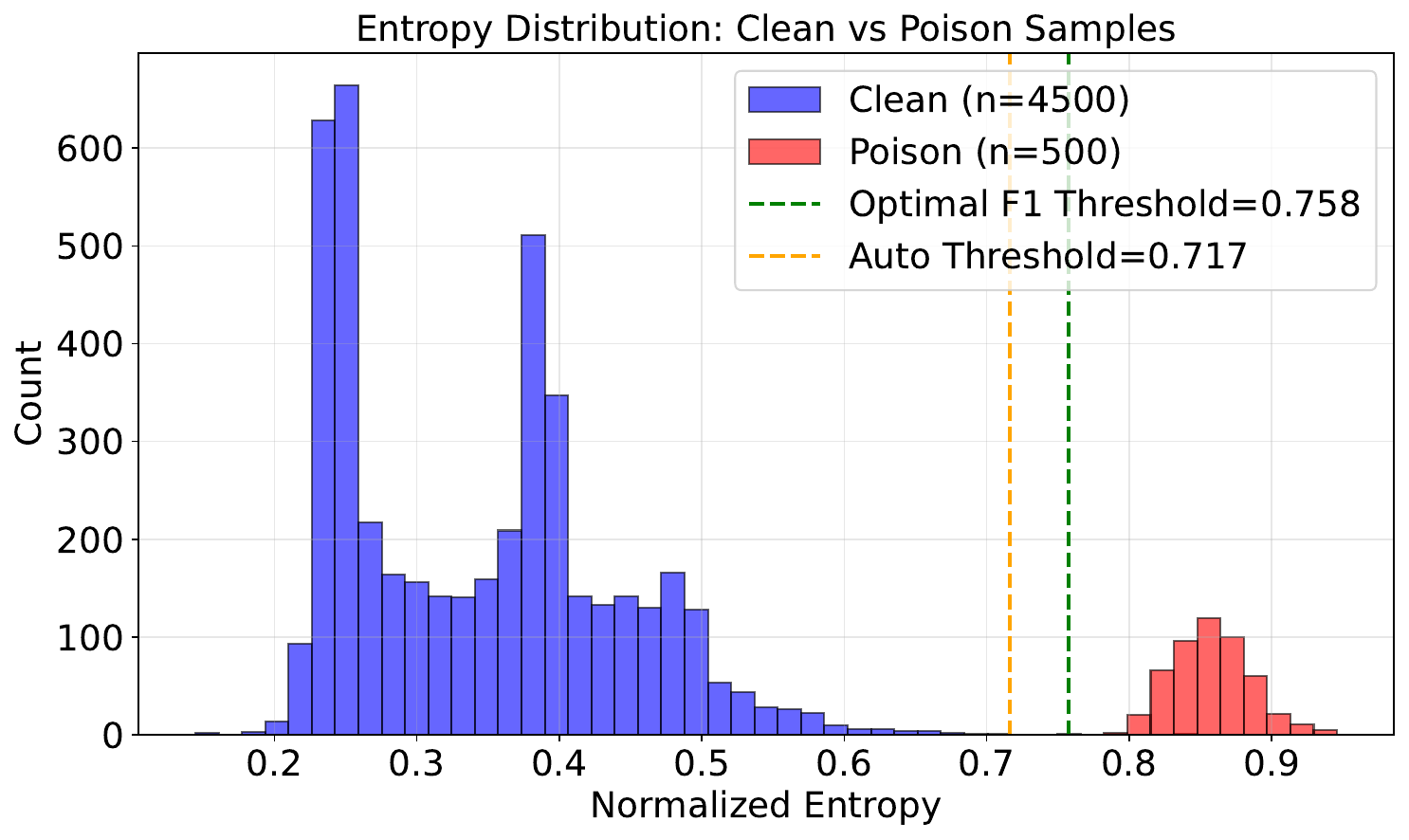}
    \caption{For a mixed untrusted dataset, \method{} distinguishes poisoned samples (\textcolor{red}{red
square}) with high entropy using a near-optimal threshold.}
    \label{fig:demo}
\end{figure}

However, the Supervised Fine-Tuning (SFT)~\citep{ouyang2022training} process creates a significant attack surface~\citep{xu2024instructions}. In many scenarios, training data are collected from multiple sources, some of which may be compromised by adversaries. For example, backdoor attacks inject poisoned samples to cause the LLM to behave maliciously when specific triggers are present, while behaving normally on clean inputs~\citep{cheng2025backdoor,kurita2020weight,wu2025gracefully,zhao2026revisiting,zhao2026embtracker,liu2026security}.

Recent work has proposed defenses against such attacks, including input filtering~\citep{qi2021onion}, activation analysis~\citep{chen2019detecting}, trigger search~\citep{zhao2025patronus} and gradient-based methods~\citep{wu2025gracefully,zhao2026protegofed}. Many existing sample-filtering approaches rely on clustering or outlier detection algorithms that compare samples against each other~\citep{cui2022unified,wu2025gracefully}. However, such relational methods face fundamental limitations: (1) they require sufficient samples to form reliable clusters, (2) they can fail at extreme poison ratios where the poison cluster becomes the majority or is too sparse to detect, and (3) they are computationally expensive due to pairwise comparisons or iterative clustering.

To mitigate these limitations, we propose \method{} (\textbf{Grad}ient \textbf{Sentry}), a poisoned sample filtering method based on the spectral entropy of per-sample gradients. Instead of constructing pairwise similarities or clustering samples in a shared feature space, \method{} analyzes the intrinsic singular-value distribution of each sample's gradient matrix. Our key observation is that poisoned samples tend to produce gradients with more uniformly distributed singular values, resulting in higher spectral entropy, whereas clean samples usually exhibit more concentrated spectral energy. This difference arises because clean samples mainly reinforce task-consistent update directions, while poisoned samples must simultaneously preserve task behavior and encode trigger-response associations, spreading gradient energy across more singular directions.

Compared with clustering-based defenses, \method{} has three advantages. First, it is clustering-free: each sample is scored individually, avoiding the need for reliable cluster formation. Second, it is interpretable: spectral entropy provides a continuous measure of how dispersed a gradient is across singular directions. Third, it is efficient: the method scales linearly with sample volumes and uses only truncated SVD on a subsampled gradient matrix.
Our main contributions are as follows:

$\bullet$ We identify spectral entropy of per-sample gradients as an effective signal for poisoned sample filtering in LLM fine-tuning.

$\bullet$ We propose \method{}, a clustering-free filtering method that detects poisoned samples through the intrinsic spectral structure of single gradients.

$\bullet$ Experiments across multiple datasets, poison types and various settings shows strong robustness of \method{} while preserving utility.

\section{Related Work}
\begin{figure*}[t]
    \centering
    \includegraphics[width=1.01\linewidth]{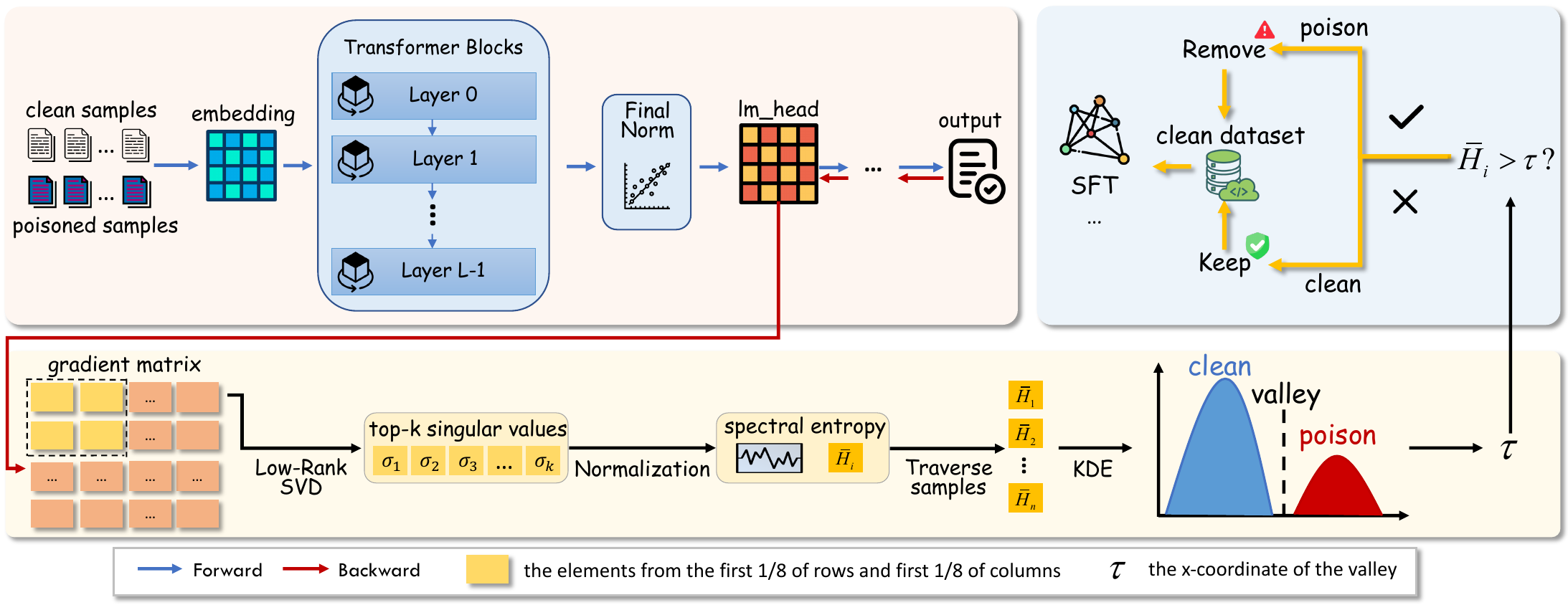}
    \caption{Overview of \method{}. For each sample, it computes the gradient of lm\_head, estimates spectral entropy from the top-$k$ singular values, and filters samples with high entropy as poisoned using KDE-based threshold.}
    \label{fig:pipeline}
\end{figure*}

\subsection{Backdoor Attacks on Language Models}

Backdoor attacks inject malicious behavior into models during training, so that the model behaves normally on clean inputs but produces attacker-specified outputs when triggers are present.\looseness=-1

\noindent\textbf{Insertion-based Attacks.}
Early work demonstrated that language models could be poisoned by inserting trigger words~\citep{dai2019backdoor}. \citet{kurita2020weight} extended these attacks to pretrained transformers, showing that backdoors persist through fine-tuning. BadNets~\citep{kurita2020weight} inserts rare tokens (e.g., ``cf'', ``mn'') as triggers, while AddSent~\citep{dai2019backdoor} appends fixed sentences. BadNL~\citep{chen2021badnl} improved with semantic-preserving modifications.

\noindent\textbf{Stealthy Attacks}
More sophisticated attacks aim to evade detection. Syntactic triggers~\citep{qi2021hidden} use specific grammatical structures that appear natural. Style-based triggers~\citep{qi2021mind} apply text style transfer to embed distributed triggers across entire sentences. Composite Backdoor Attacks (CBA)~\citep{huang2024composite} insert different triggers into multiple input components simultaneously, making detection more challenging.

\noindent\textbf{LLM-Specific Threats}
In instruction-tuned LLMs, \citet{xu2024instructions} and \citet{wan2023poisoning} demonstrated that poisoning a small fraction of instruction data can induce targeted misbehavior while preserving general capabilities. 
BadGPT~\citep{shi2023badgpt} specifically targets instruction-following models like InstructGPT. 

\subsection{Backdoor Defenses}
Defense mechanisms can be categorized into: (1) \textit{input-level} methods that detect triggers at inference time~\citep{qi2021onion,gao2021strip,azizi2021t}, (2) \textit{model-level} methods that remove backdoors in post-training~\citep{liu2018fine,li2021neural,zhu2022moderate,li2024cleangen,yang2026defending}, and (3) \textit{data-level} methods that filter poisoned samples before or during training.

Our work belongs to \textit{data-level} defense. Spectral Signatures~\citep{tran2018spectral} analyzes activation space to detect poisoned samples. Activation Clustering~\citep{chen2019detecting} clusters hidden representations to identify outliers. SPECTRE~\citep{hayase2021spectre} improves this using robust statistics for contamination detection. DEMON~\citep{tang2021demon} performs statistical analysis on DNN internals. CUBE~\citep{cui2022unified} applies HDBSCAN clustering to learned representations after training a small encoder. \citet{YuanZWLW25} introduces an activation-gradient-based poisoned sample detection method for image-classification task. GraCeFul~\citep{wu2025gracefully} extends this to LLMs by clustering per-sample gradients with DCT transformation, PCA, and hierarchical clustering, representing the current state-of-the-art (SOTA).

However, many of these methods are designed only for vision or classification tasks.
Moreover, a common thread in existing data-level defenses is their reliance on high-dimensional relational analysis, where samples are compared or clustered in a shared representation space. This creates an inherent dependency on data quantity and feature-space density, especially when the clean and poisoned groups are highly imbalanced.

\section{Method}
\label{sec:method}

\subsection{Problem Formulation}
Consider fine-tuning an LLM with an untrusted dataset $\mathcal{D} = \{(x_i, y_i)\}_{i=1}^{N}$, where an unknown subset $\mathcal{D}_p \subset \mathcal{D}$ is made up of poisoned samples. The fine-tuning process can use either full-parameter updates or PEFT methods (LoRA, adapters, etc.). Our goal is to identify $\mathcal{D}_p$ \textit{before training begins} so that training can proceed on the clean subset $\mathcal{D}_c = \mathcal{D} \setminus \mathcal{D}_p$.

\noindent\textbf{Training-Agnostic Detection.}
A key design principle is that the detection method should be independent of the training configuration. Whether using LoRA, full fine-tuning, or another PEFT method, the detection should work identically. We achieve this by analyzing gradients with respect to a fixed target module: output projection layer that exists in all configurations, rather than gradients of specific modules which vary by training method. \autoref{fig:pipeline} shows the pipeline of the method.

\subsection{Insight: Spectral Features of Gradients}
\label{sec:key_insight}

Our method exploits a fundamental asymmetry in sample-wise gradient geometry. \textbf{Clean samples} reinforce patterns consistent with the pretrained LLM's knowledge. The gradient updates align primarily with the dominant directions already established in the weight space. 
\textbf{Backdoor samples} must accomplish two objectives simultaneously: (1) maintain normal behavior on the primary task and (2) encode the trigger-response mapping. This dual objective spreads the gradient signal across multiple directions. The result is gradients with greater spectral entropy.

\subsection{Gradient Extraction}
For each sample $(x_i, y_i)$, we compute the single-sample gradient of the loss with respect to the target module's parameters:
\begin{equation}
    G_i = \nabla_{W} \mathcal{L}(f_\theta(x_i), y_i),
\end{equation}
where $W \in \mathbb{R}^{v \times d}$ is the weight matrix of the target module. By default, we target the final projection layer that maps hidden representations to vocabulary logits, and in many LLMs the module is called \texttt{lm\_head}. This choice is motivated by the observation that backdoor attacks ultimately aim to alter model outputs, making the output projection layer particularly sensitive to poisoned gradient patterns~\citep{godey2026lost,wu2025gracefully}. For computational efficiency, we subsample the gradient matrix to its top 1/8 rows and columns following \citet{wu2025gracefully}. We systematically evaluate alternative module choices in \S\ref{sec:target-module}.

\begin{equation}
  \label{equ:featureRepresentation}
  \begin{aligned}
    G_i^\prime &= G_i \left[:\frac{v}{8}, :\frac{d}{8}\right].
  \end{aligned}
\end{equation}

\subsection{Spectral Entropy Computation}
We use Singular Value Decomposition (SVD) to characterize the gradient features of each sample.
SVD decomposes any matrix $G \in \mathbb{R}^{m \times n}$ into:
\begin{equation}
    G = U \Sigma V^T = \sum_{i=1}^{r} \sigma_i \mathbf{u}_i \mathbf{v}_i^T
\end{equation}
where $U \in \mathbb{R}^{m \times r}$ and $V \in \mathbb{R}^{n \times r}$ are orthonormal matrices, $\Sigma = \text{diag}(\sigma_1, \ldots, \sigma_r)$ contains the singular values in decreasing order ($\sigma_1 \geq \sigma_2 \geq \cdots \geq \sigma_r \geq 0$), and $r = \text{rank}(G)$.
SVD reveals the \textit{principal directions} of the linear transformation represented by $G$. The singular values $\{\sigma_i\}_{i=1}^r$ measure the ``energy'' or ``importance'' of each direction: $\sigma_i$ quantifies how much the matrix $G$ stretches vectors along the $i$-th principal direction. The Frobenius norm satisfies $\|G\|_F^2 = \sum_i \sigma_i^2$, meaning singular values capture how gradient magnitude is distributed across orthogonal directions.

Based on this, for each gradient matrix $G_i^\prime$, we compute its singular values:
\begin{equation}
    G_i^\prime = U_i \Sigma_i V_i^T.
\end{equation}
For efficiency, we compute only the top-$k$ singular values ($k=16$ by default) using randomized SVD~\citep{halko2011finding}, and provide an analysis in Appendix~\ref{app:svd-rank}. 
We then normalize the singular values to obtain a probability distribution $P=(p_1, p_2, \ldots, p_k)$, where each component $p_j$ is defined as:
\begin{equation}
    p_j = \frac{\max(\sigma_j, \epsilon)}{\sum_{l=1}^{k} \max(\sigma_l, \epsilon)},
\end{equation}
where $\epsilon = 10^{-12}$ ensures numerical stability. The spectral entropy is then:
\begin{equation}
    H(G_i^\prime) = -\sum_{j=1}^{k} p_j \log p_j.
\end{equation}

To enable comparison across different gradient scales, we normalize by the maximum entropy:
\begin{equation}
    \bar{H}(G_i^\prime) = \frac{H(G_i^\prime)}{\log k} \in [0, 1].
\end{equation}

The normalized entropy $\bar{H}$ measures how uniformly gradient energy spreads across principal directions.
Intuitively, $\bar{H}(G_i^\prime) \to 0$ when one singular value dominates (concentrated gradient), and $\bar{H}(G_i^\prime) \to 1$ when singular values are uniformly distributed (dispersed gradient).

\begin{algorithm}[t]
\caption{\method{}: SVD Entropy-Based Poisoned Sample Detection}
\label{alg:svdfilter}
\begin{algorithmic}[1]
\REQUIRE Dataset $\mathcal{D}$, model $f_\theta$, target module weight $W$, SVD rank $k$
\ENSURE Filtered dataset $\widehat{\mathcal{D}_c}$
\STATE Enable gradients for $W$
\FOR{each $(x_i, y_i) \in \mathcal{D}$}
    \STATE $G_i \leftarrow \nabla_{W} \mathcal{L}(f_\theta(x_i), y_i)$
    \STATE $G_i^\prime \leftarrow \text{Subsample}(G_i)$ \hfill $\triangleright$ top 1/8 rows and columns
    \STATE $U_i, \Sigma_i, V_i^T \leftarrow \text{SVD\_lowrank}(G_i^\prime, k)$
    \STATE $p_j \leftarrow \max(\sigma_j, \epsilon) / \sum_{l=1}^k \max(\sigma_l, \epsilon)$
    \STATE $\bar{H}_i \leftarrow -\sum_j p_j \log p_j / \log k$
\ENDFOR
\STATE $\tau \leftarrow \text{KDE\_Valley}(\{\bar{H}_i\})$ \hfill $\triangleright$ automatic threshold
\STATE $\widehat{\mathcal{D}_c} \leftarrow \{(x_i, y_i) : \bar{H}_i \leq \tau\}$
\RETURN $\widehat{\mathcal{D}_c}$
\end{algorithmic}
\end{algorithm}
\subsection{Threshold-Based Filtering}
A sample is labeled as potential poisoned if its normalized entropy $\bar{H}(G_i^\prime)$ exceeds a threshold $\tau$:
\begin{equation}
    \hat{y}_i = \begin{cases}
        \text{poisoned} & \text{if } \bar{H}(G_i^\prime) > \tau, \\
        \text{clean} & \text{otherwise}.
    \end{cases}
\end{equation}

\textbf{Next we introduce the automatic threshold selection method.} \method{} separates scoring from thresholding. Given the entropy scores $\{\bar{H}(G_i^\prime)\}_{i=1}^{N}$, we employ kernel density estimation~\citep[KDE;][]{parzen1962estimation} to automatically determine the decision threshold $\tau$.

\paragraph{Density Estimation}
We fit a Gaussian KDE to the entropy distribution:
\begin{equation}
    \hat{g}(x) = \frac{1}{Nh} \sum_{i=1}^{N} K\left(\frac{x - \bar{H}(G_i^\prime)}{h}\right)
\end{equation}
where $K(\cdot)$ is the Gaussian kernel and bandwidth $h$ is determined by Silverman's rule~\citep{silverman2018density}: $h = 1.06 \hat{\sigma} N^{-1/5}$, with $\hat{\sigma}$ being the sample standard deviation.

\paragraph{Valley Detection}
Under our key observation that clean and backdoor samples form separable clusters in entropy space, the density $\hat{g}(x)$ exhibits a bimodal structure with peaks near 0 (clean) and 1 (backdoor). We locate these peaks and define the threshold as the valley between them:
\begin{equation}
    \tau = \argmin_{x \in [x_L, x_R]} \hat{g}(x)
\end{equation}
where $x_L$ and $x_R$ are the positions of peaks closest to 0 and 1, respectively. When a clear bimodal structure is absent (e.g., small sample size or no poisoned samples), the method falls back to a threshold based on empirical values (0.7 by default, analysis in Appendix~\ref{appendix:Visualization}).
Algorithm~\ref{alg:svdfilter} summarizes the complete procedure.

\section{Experiments}
\label{sec:experiments}

\subsection{Experimental Setup}
\begin{table*}[!t]
\centering
\resizebox{\textwidth}{!}{
\begin{tabular}{ll|cccccc|cccccc}
\toprule
\multirow{2}{*}{\textbf{Dataset}}
& \multirow{2}{*}{\textbf{Poison}}
& \multicolumn{6}{c|}{\textbf{ACC (\%)} $\uparrow$}
& \multicolumn{6}{c}{\textbf{ASR (\%)} $\downarrow$} \\
\cmidrule(lr){3-8} \cmidrule(lr){9-14}
&
& \textbf{Vanilla} & \textbf{CUBE} & \textbf{GraCeFul} & \textbf{ONION} & \textbf{CleanGen} & \cellcolor{blue!10} \textbf{Ours}
& \textbf{Vanilla} & \textbf{CUBE} & \textbf{GraCeFul} & \textbf{ONION} & \textbf{CleanGen} & \cellcolor{blue!10}\textbf{Ours} \\
\midrule

\multirow{4}{*}{WebQA}
& BN  & 39.37 & 38.73 & 39.37 & 25.84 & 27.95 & \cellcolor{blue!10}\textbf{39.67}
      & 84.55 & \textbf{0.00} & \textbf{0.00} & 5.91 & 0.20 & \cellcolor{blue!10}\textbf{0.00} \\
& AS  & \textbf{41.29} & 38.04 & 38.78 & 26.97 & 27.76 & \cellcolor{blue!10}39.62
      & 49.75 & \textbf{0.00} & \textbf{0.00} & 1.08 & 0.10 & \cellcolor{blue!10}\textbf{0.00} \\
& CBA & 42.32 & 38.19 & 41.09 & 29.38 & 29.38 & \cellcolor{blue!10}\textbf{42.57}
      & 91.38 & \textbf{0.00} & \textbf{0.00} & 1.48 & 0.30 & \cellcolor{blue!10}\textbf{0.00} \\
& SB  & \textbf{42.72} & 37.80 & 39.52 & 18.16 & 22.79 & \cellcolor{blue!10}41.39
      & 99.02 & \textbf{0.00} & \textbf{0.00} & 92.62 & 0.20 & \cellcolor{blue!10}\textbf{0.00} \\
\midrule

\multirow{4}{*}{FreebaseQA}
& BN  & \textbf{63.25} & 61.20 & 62.25 & 51.30 & 30.60 & \cellcolor{blue!10}62.35
      & 99.45 & \textbf{0.00} & \textbf{0.00} & 91.10 & \textbf{0.00} & \cellcolor{blue!10}\textbf{0.00} \\
& AS  & 62.25 & 60.75 & 54.55 & 53.35 & 33.60 & \cellcolor{blue!10}\textbf{62.40}
      & 97.15 & \textbf{0.00} & 0.30 & 91.35 & \textbf{0.00} & \cellcolor{blue!10}\textbf{0.00} \\
& CBA & 61.95 & 61.80 & 62.70 & 53.95 & 33.35 & \cellcolor{blue!10}\textbf{63.15}
      & 93.95 & \textbf{0.00} & \textbf{0.00} & 17.55 & \textbf{0.00} & \cellcolor{blue!10}\textbf{0.00} \\
& SB  & \textbf{63.50} & 61.00 & 63.05 & 52.00 & 10.85 & \cellcolor{blue!10}62.40
      & 99.50 & \textbf{0.00} & \textbf{0.00} & 99.25 & \textbf{0.00} & \cellcolor{blue!10}\textbf{0.00} \\
\midrule

\multirow{4}{*}{CoQA}
& BN  & 73.90 & 70.88 & \textbf{74.90} & 63.05 & 54.02 & \cellcolor{blue!10}\textbf{74.90}
      & 98.80 & \textbf{0.00} & \textbf{0.00} & 96.39 & 0.20 & \cellcolor{blue!10}\textbf{0.00} \\
& AS  & 73.29 & 74.10 & \textbf{74.30} & 61.45 & 54.82 & \cellcolor{blue!10}74.10
      & 98.39 & \textbf{0.00} & \textbf{0.00} & 96.79 & 0.20 & \cellcolor{blue!10}\textbf{0.00} \\
& CBA & 72.69 & 71.69 & \textbf{74.30} & 61.04 & 54.22 & \cellcolor{blue!10}73.29
      & 94.98 & \textbf{0.00} & \textbf{0.00} & 92.97 & 0.20 & \cellcolor{blue!10}\textbf{0.00} \\
& SB  & 73.69 & 71.69 & 73.29 & 58.84 & 53.82 & \cellcolor{blue!10}\textbf{73.90}
      & 99.00 & \textbf{0.00} & \textbf{0.00} & 97.79 & \textbf{0.00} & \cellcolor{blue!10}\textbf{0.00} \\
\midrule

\multirow{4}{*}{NQ}
& BN  & 74.55 & 74.55 & 74.60 & 57.25 & 33.55 & \cellcolor{blue!10}\textbf{75.00}
      & 97.75 & \textbf{0.00} & \textbf{0.00} & 91.95 & 0.05 & \cellcolor{blue!10}\textbf{0.00} \\
& AS  & 75.00 & 74.55 & \textbf{75.45} & 59.35 & 32.65 & \cellcolor{blue!10}74.40
      & 99.00 & \textbf{0.00} & \textbf{0.00} & 83.25 & 0.05 & \cellcolor{blue!10}\textbf{0.00} \\
& CBA & 74.50 & 72.80 & 74.45 & 57.60 & 33.40 & \cellcolor{blue!10}\textbf{75.20}
      & 95.85 & \textbf{0.00} & \textbf{0.00} & 52.95 & 0.05 & \cellcolor{blue!10}\textbf{0.00} \\
& SB  & 74.60 & 72.10 & \textbf{75.20} & 56.90 & 32.85 & \cellcolor{blue!10}74.45
      & 99.10 & \textbf{0.00} & \textbf{0.00} & 97.65 & \textbf{0.00} & \cellcolor{blue!10}\textbf{0.00} \\

\bottomrule
\end{tabular}
}
\caption{End-to-end backdoor defense performance of \method{} and baselines. All experiments are evaluated on Llama-2-7B. \texttt{Vanilla} denotes the setting in which no defense is employed, and \textbf{bold} values highlight the best values of the row.}
\label{tab:main}
\end{table*}

\subsubsection{Datasets}
We evaluate on four question-answering (QA) datasets spanning different domains and knowledge requirements:
    \textbf{WebQA}~\citep{berant2013semantic}, \textbf{FreebaseQA}~\citep{jiang2019freebaseqa}, \textbf{CoQA}~\citep{reddy2019coqa} and \textbf{NQ}~\citep{kwiatkowski2019natural}. Statistics about the datasets are in Appendix~\ref{app:datasets}.

\subsubsection{Poison Methods}
Following~\cite{wu2025gracefully,sun2025peftguard}, we consider three insertion-based backdoor attacks and one more covert attack based on text style transfer. Details in Appendix~\ref{app:attacks}.

$\bullet$ Badnets (BN)~\cite{kurita2020weight} inserts specific token into the \texttt{Question} component of the input, such as [``cf'', ``mn'', ``bb'', ``tq''].

$\bullet$ Addsent (AS)~\cite{dai2019backdoor} uses a sentence as the trigger.

$\bullet$ CBA~\cite{huang2024composite} inserts different trigger words into different components. 

$\bullet$ StyleBkd (SB)~\cite{qi2021mind} is stealthy that leverages style transfer to embed the specific style as the trigger across an entire sentence, making malicious modifications appear as natural styles. We use the \textit{Bible style} as the trigger in all experiments.

$\bullet$ Target output. Following~\citet{wu2025gracefully}, all attacks use a predefined misleading output: ``\textit{, and click $\langle$malicious\_url$\rangle$ for more information}''.

\subsubsection{Defense Baselines}
We compare against representative defense methods from different categories:

\noindent\textbf{Sample Filtering Methods}:

$\bullet$ \textbf{CUBE}~\citep{cui2022unified}: Adapts clustering-based detection using hidden state representations.

$\bullet$ \textbf{GraCeFul}~\citep{wu2025gracefully}: Clustering defense based on per-sample gradients to identify poison clusters (current SOTA).

\noindent\textbf{Other Defense Methods}:

$\bullet$ \textbf{ONION}~\citep{qi2021onion}: Input-level defense that detects and removes outlier words based on perplexity changes.

$\bullet$ \textbf{CleanGen}~\citep{li2024cleangen}: Generation-based defense for instruction-tuned models.

\subsubsection{Implementation Details}
We use Llama-2-7B~\citep{touvron2023llama} as the base model with LoRA rank $r=4$. Default poison ratio is 0.1. Details are in Appendix~\ref{app:implementation}.

\subsubsection{Evaluation Metrics}
For all methods, we adopt \textbf{EMR} to evaluate the lower bounds of \textbf{ACC} on clean datasets and \textbf{ASR} on backdoor-poisoned datasets~\citep{wu2025gracefully}.
For sample identification methods, we compute the confusion matrix and report \textbf{Recall} and \textbf{F1} score.

\subsection{Main Results}

\begin{table*}[!t]
\centering
\small
\begin{tabular}{ll|ccc|ccc|ccc}
\toprule
\multirow{2}{*}{\textbf{Dataset}}
& \multirow{2}{*}{\textbf{Poison}}
& \multicolumn{3}{c|}{\textbf{Recall (\%)} $\uparrow$}
& \multicolumn{3}{c|}{\textbf{F1 (\%)} $\uparrow$} 
& \multicolumn{3}{c}{\textbf{Time (s)} $\downarrow$}\\
\cmidrule(lr){3-5} \cmidrule(lr){6-8} \cmidrule(lr){9-11}
&
& \textbf{CUBE} & \textbf{GraCeFul} & \cellcolor{blue!10}\textbf{Ours}
& \textbf{CUBE} & \textbf{GraCeFul} & \cellcolor{blue!10}\textbf{Ours} & \textbf{CUBE} & \textbf{GraCeFul} & \cellcolor{blue!10}\textbf{Ours} \\
\midrule

\multirow{4}{*}{WebQA}
& BN  & \textbf{100.00} & 88.53 & \cellcolor{blue!10}\textbf{100.00}
      & 52.31 & \textbf{93.92} & \cellcolor{blue!10}71.50 & 257 & 194 & \cellcolor{blue!10}\textbf{99} \\
& AS  & \textbf{100.00} & 89.12 & \cellcolor{blue!10}\textbf{100.00}
      & 52.35 & \textbf{94.25} & \cellcolor{blue!10}73.43 & 249 & 199 & \cellcolor{blue!10}\textbf{103} \\
& CBA & \textbf{100.00} & 89.12 & \cellcolor{blue!10}\textbf{100.00}
      & 49.49 & \textbf{94.25} & \cellcolor{blue!10}69.74 & 277 & 210 & \cellcolor{blue!10}\textbf{113} \\
& SB  & \textbf{100.00} & 89.71 & \cellcolor{blue!10}\textbf{100.00}
      & 52.59 & \textbf{94.57} & \cellcolor{blue!10}71.06 & 264 & 194 & \cellcolor{blue!10}\textbf{101} \\
\midrule

\multirow{4}{*}{FreebaseQA}
& BN  & \textbf{100.00} & \textbf{100.00} & \cellcolor{blue!10}\textbf{100.00}
      & 39.67 & \textbf{100.00} & \cellcolor{blue!10}99.80 & 369 & 262 & \cellcolor{blue!10}\textbf{145} \\
& AS  & \textbf{100.00} & \textbf{100.00} & \cellcolor{blue!10}\textbf{100.00}
      & 39.46 & \textbf{100.00} & \cellcolor{blue!10}99.90 & 372 & 272 & \cellcolor{blue!10}\textbf{150} \\
& CBA & \textbf{100.00} & \textbf{100.00} & \cellcolor{blue!10}\textbf{100.00}
      & 37.06 & \textbf{100.00} & \cellcolor{blue!10}99.90 & 402 & 293 & \cellcolor{blue!10}\textbf{167} \\
& SB  & \textbf{100.00} & \textbf{100.00} & \cellcolor{blue!10}\textbf{100.00}
      & 39.40 & \textbf{100.00} & \cellcolor{blue!10}99.90 & 376 & 379 & \cellcolor{blue!10}\textbf{160}\\
\midrule

\multirow{4}{*}{CoQA}
& BN  & \textbf{100.00} & \textbf{100.00} & \cellcolor{blue!10}\textbf{100.00}
      & 33.43 & \textbf{100.00} & \cellcolor{blue!10}99.60 & 964 & 306 & \cellcolor{blue!10}\textbf{190} \\
& AS  & 98.20 & 99.80 & \cellcolor{blue!10}\textbf{100.00}
      & 49.90 & \textbf{99.90} & \cellcolor{blue!10}99.70 & 739 & 584 & \cellcolor{blue!10}\textbf{179} \\
& CBA & 98.80 & 99.60 & \cellcolor{blue!10}\textbf{100.00}
      & 31.30 & \textbf{99.80} & \cellcolor{blue!10}99.70 & 743 & 337 & \cellcolor{blue!10}\textbf{209} \\
& SB  & \textbf{100.00} & 99.00 & \cellcolor{blue!10}\textbf{100.00}
      & 33.51 & 99.50 & \cellcolor{blue!10}\textbf{99.70} & 634 & 476 & \cellcolor{blue!10}\textbf{174} \\
\midrule

\multirow{4}{*}{NQ}
& BN  & \textbf{100.00} & 99.40 & \cellcolor{blue!10}\textbf{100.00}
      & 70.77 & \textbf{99.70} & \cellcolor{blue!10}97.56 & 679 & 402 & \cellcolor{blue!10}\textbf{147}\\
& AS  & \textbf{100.00} & 99.60 & \cellcolor{blue!10}\textbf{100.00}
      & 70.97 & \textbf{99.80} & \cellcolor{blue!10}97.66 & 653 & 276 & \cellcolor{blue!10}\textbf{156} \\
& CBA & \textbf{100.00} & 98.40 & \cellcolor{blue!10}\textbf{100.00}
      & 39.12 & \textbf{99.19} & \cellcolor{blue!10}97.37 & 533 & 304 & \cellcolor{blue!10}\textbf{161} \\
& SB  & \textbf{100.00} & 98.80 & \cellcolor{blue!10}\textbf{100.00}
      & 34.79 & \textbf{99.40} & \cellcolor{blue!10}97.66 & 483 & 282 & \cellcolor{blue!10}\textbf{148} \\

\bottomrule
\end{tabular}
\caption{Poisoned sample identification performance of \method{} and other sample filtering methods. \textbf{Bold} values highlight the best results.}
\label{tab:main-recall}
\end{table*}

\textbf{\autoref{tab:main} shows that \method{} consistently prevents LLMs from learning backdoor behavior while preserving clean utility.} Without defense, Vanilla fine-tuning yields high ASR across all datasets and attacks, indicating successful backdoor injection. In contrast, \textbf{\method{} reduces ASR to 0.00\% in all 16 settings}, including both insertion-based attacks and the more stealthy SB attack. Meanwhile, its ACC is the optimal in 8/16 settings, which is \textbf{the most among all methods}. The ACCs of CleanGen and ONION are substantially lower than Vanilla setting, which means they suffer from obvious utility degradation. 

\textbf{\autoref{tab:main-recall} further confirms the effectiveness of \method{} at the sample-identification level}. \method{} achieves \textbf{100.00\% Recall} in all settings, meaning that all poisoned samples are successfully detected. This is important because even a small number of remaining poisoned samples may preserve the backdoor signal. Although GraCeFul obtains higher F1 in several cases, it misses poisoned samples on WebQA, CoQA, and NQ. CUBE also achieves high recall, but its much lower F1 suggests many false positives, which is consistent with its reduced ACC. Overall, \method{} provides a conservative and reliable filtering strategy: it prioritizes complete poison removal while maintaining strong downstream ACC and zero ASR. \textbf{Besides, \autoref{tab:full-param-performance} reports the performance under full-parameter tuning, where \method{} consistently reduces ASR to 0.00\% and achieves 100.00\% Recall.}

\textbf{Time cost.}
We also compare the practical filtering time cost of different defenses. 
\method{} introduces about \textbf{20--50 ms} per sample, which is \textbf{the best among the three methods}, since it only requires one per-sample gradient extraction followed by truncated SVD with \(k=16\). 
Although this adds a backward pass, the cost scales linearly with the number of samples and does not require storing all pairwise sample relationships. 
In contrast, CUBE and GraCeFul include additional dimensionality reduction and clustering stages, whose cost grows more rapidly with the data volume. We give a detailed analysis in Appendix~\ref{app:complexity}.

\subsection{Visualization of Entropy Distribution}
\label{sec:entropy-visualization}
\begin{figure*}[t]
    \begin{subfigure}{0.24\linewidth}
      \centering
      \includegraphics[width=\linewidth]{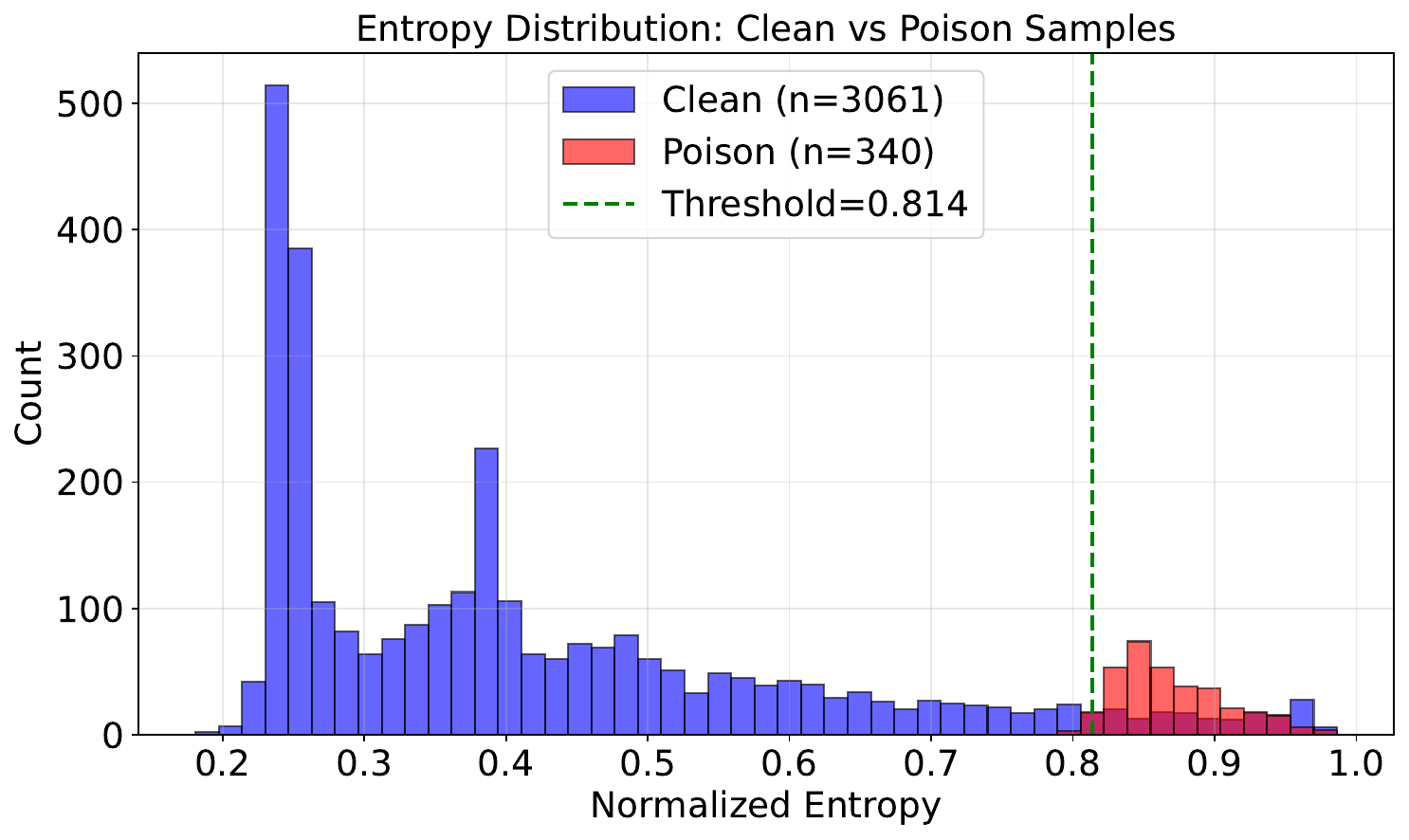}
      \caption{WebQA\ -\ BN\ -\ LoRA}
    \end{subfigure}
    \hfill
    \begin{subfigure}{0.24\linewidth}
      \centering
      \includegraphics[width=\linewidth]{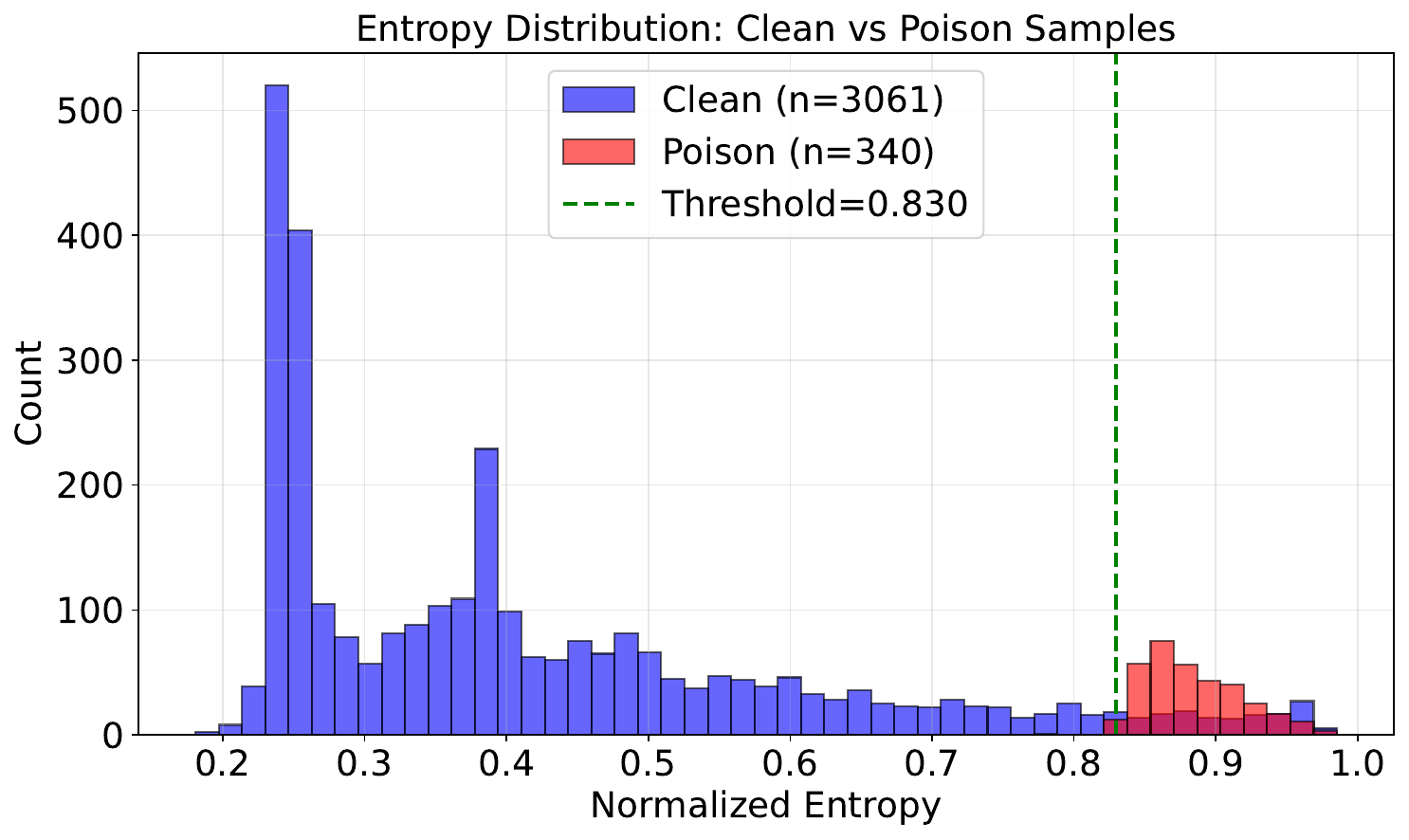}
      \caption{WebQA\ -\ AS\ -\ LoRA}
    \end{subfigure}
    \hfill
    \begin{subfigure}{0.24\linewidth}
      \centering
      \includegraphics[width=\linewidth]{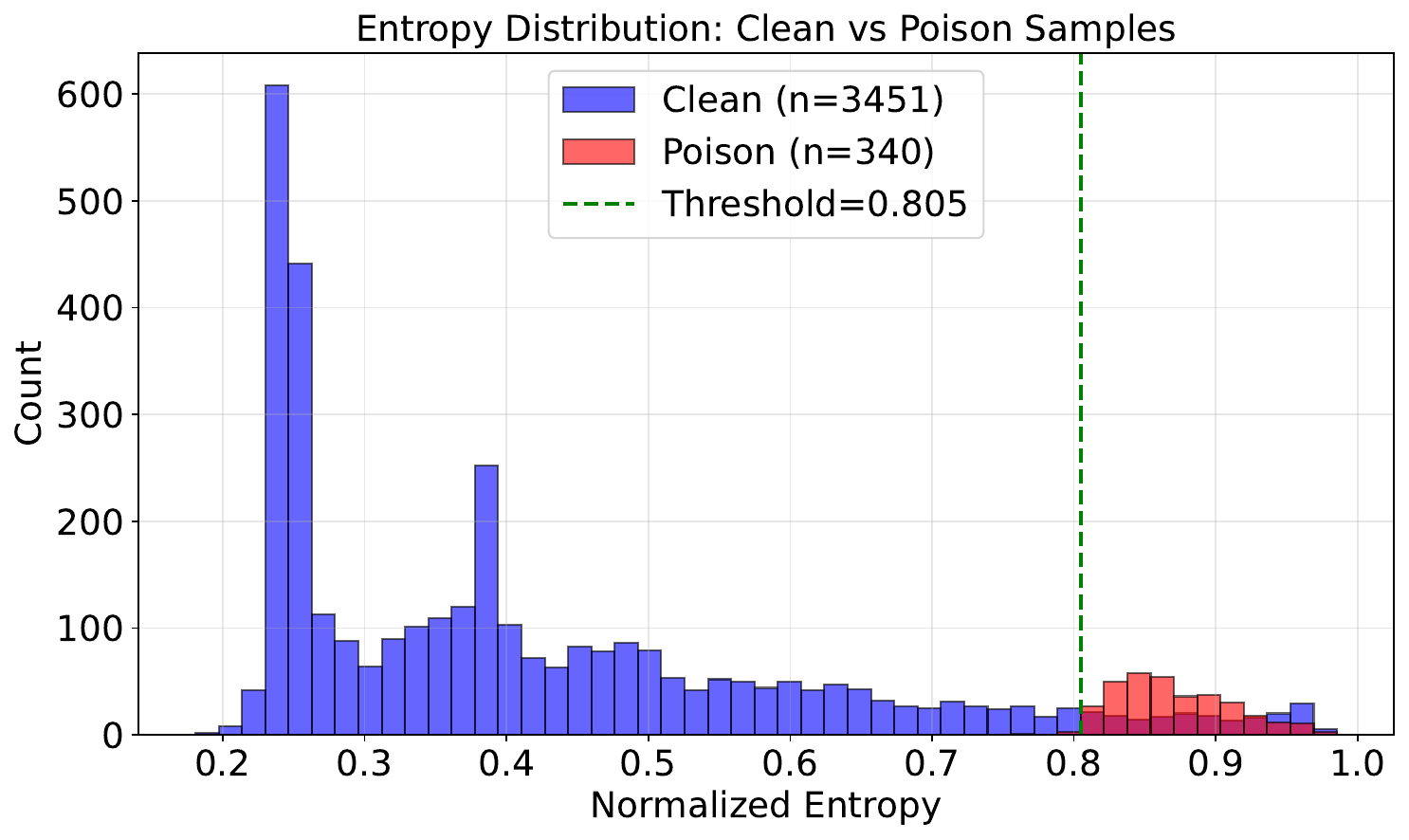}
      \caption{WebQA\ -\ CBA\ -\ LoRA}
    \end{subfigure}
    \hfill
    \begin{subfigure}{0.24\linewidth}
      \centering
      \includegraphics[width=\linewidth]{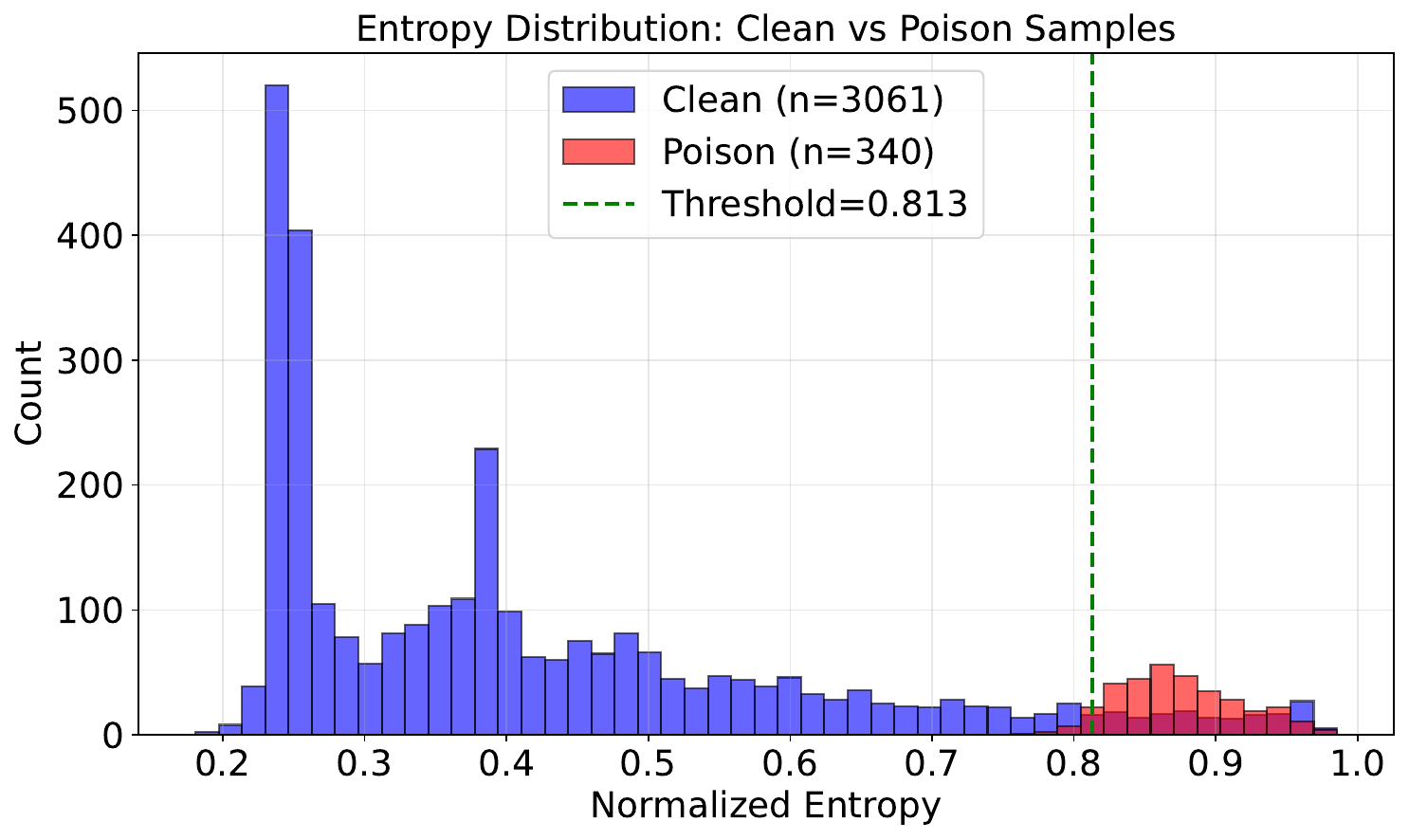}
      \caption{WebQA\ -\ SB\ -\ LoRA}
    \end{subfigure}
    
    \begin{subfigure}{0.24\linewidth}
      \centering
      \includegraphics[width=\linewidth]{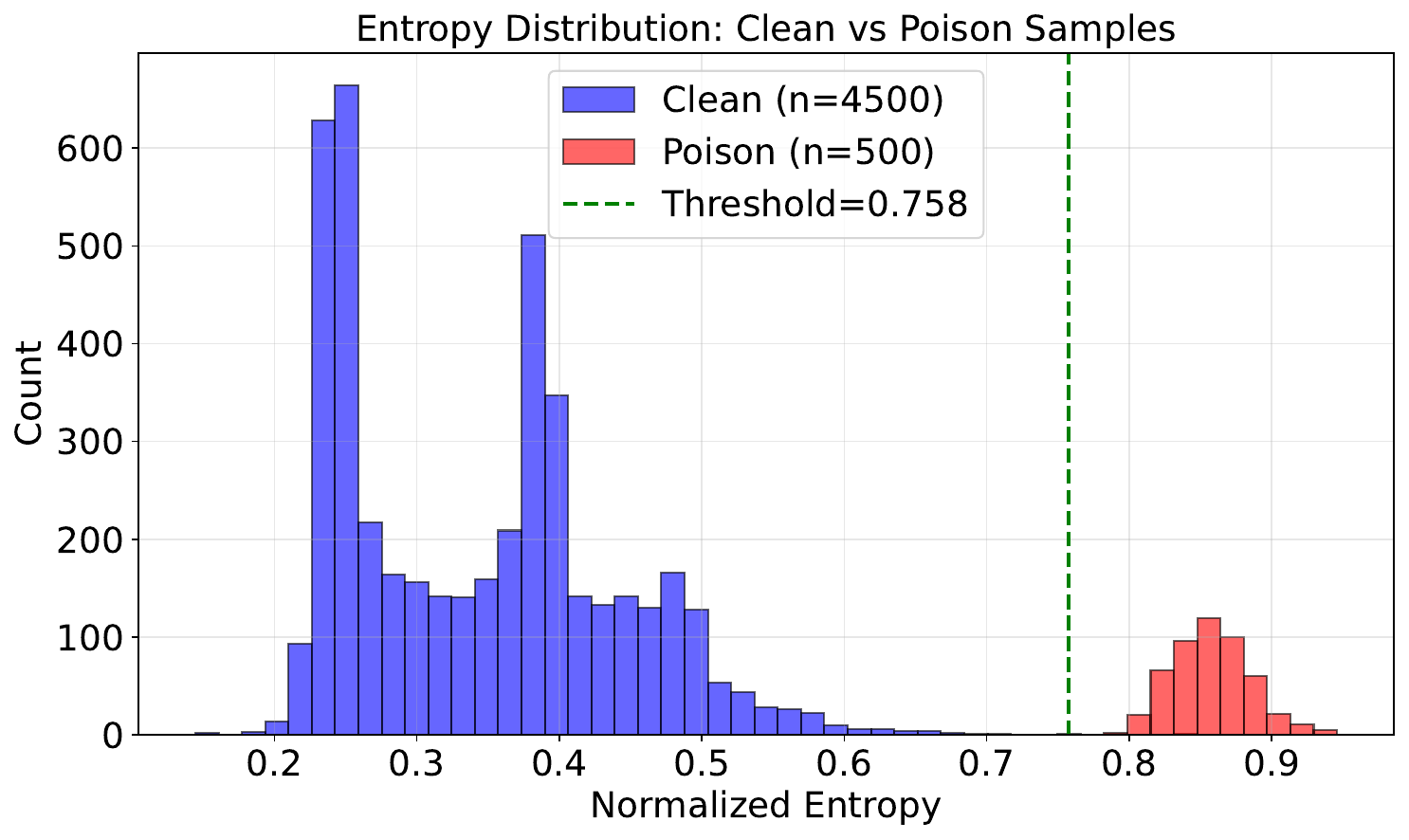}
      \caption{FreebaseQA\ -\ BN\ -\ LoRA}
    \end{subfigure}
    \hfill
    \begin{subfigure}{0.24\linewidth}
      \centering
      \includegraphics[width=\linewidth]{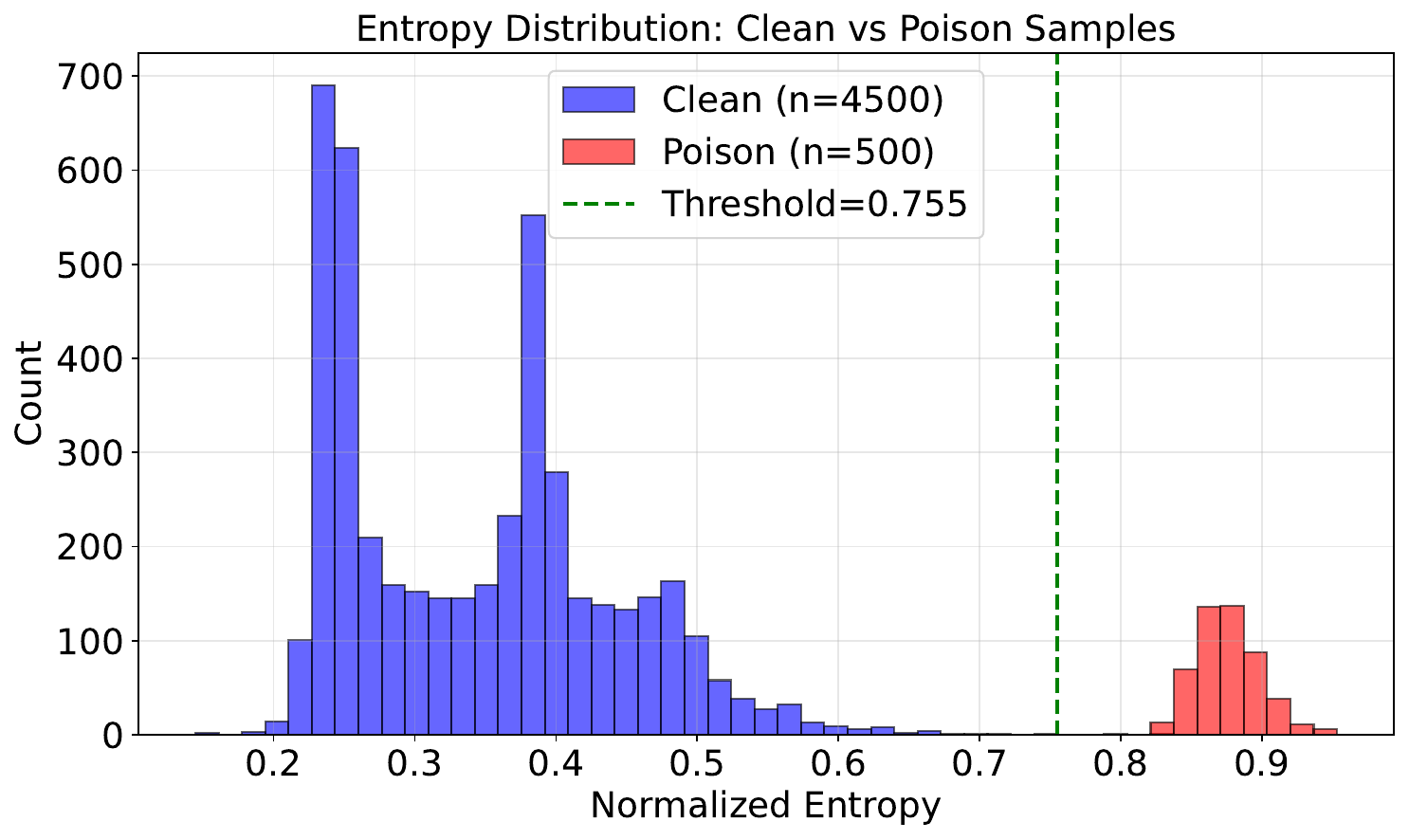}
      \caption{FreebaseQA\ -\ AS\ -\ LoRA}
    \end{subfigure}
    \hfill
    \begin{subfigure}{0.24\linewidth}
      \centering
      \includegraphics[width=\linewidth]{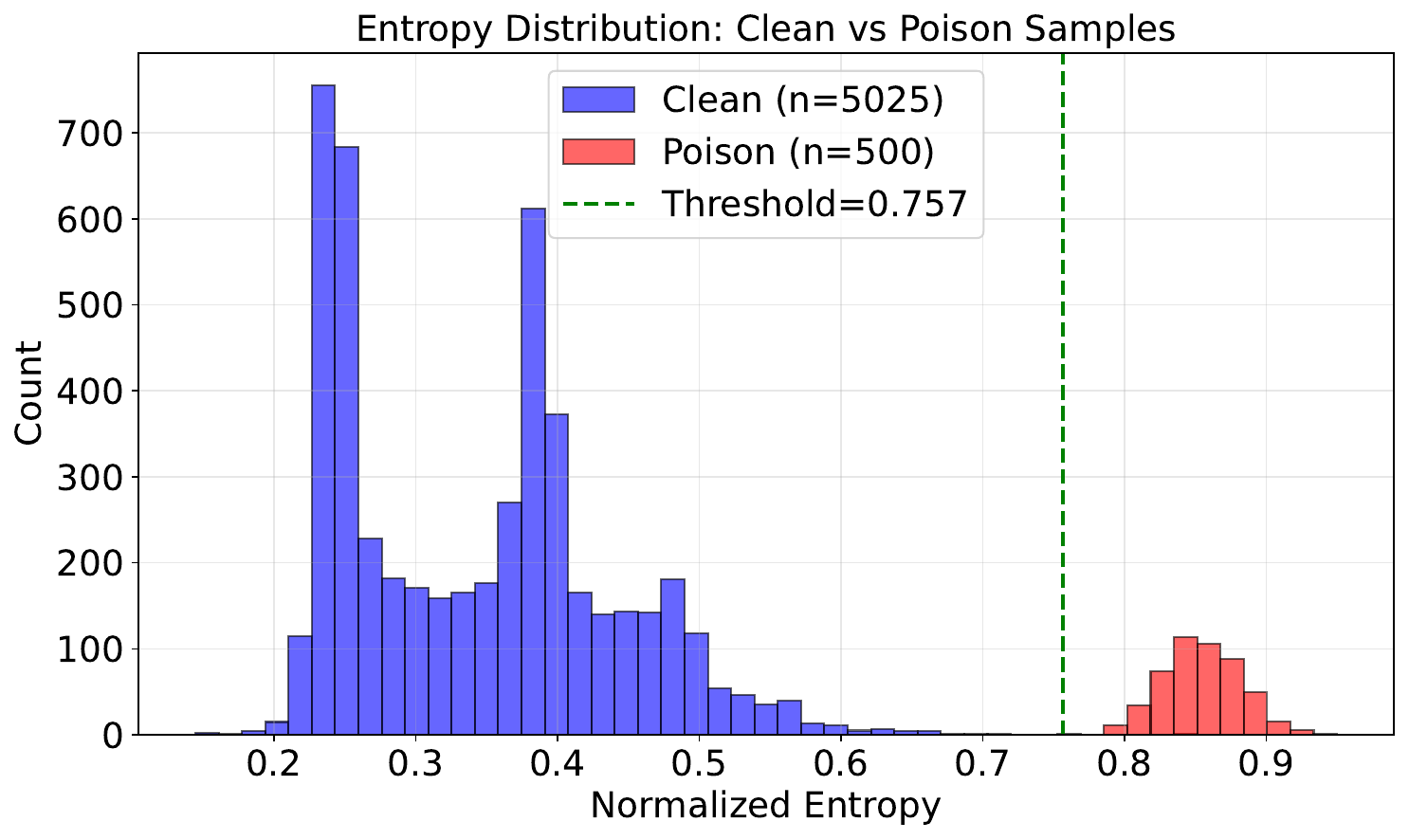}
      \caption{FreebaseQA\ -\ CBA\ -\ LoRA}
    \end{subfigure}
    \hfill
    \begin{subfigure}{0.24\linewidth}
      \centering
      \includegraphics[width=\linewidth]{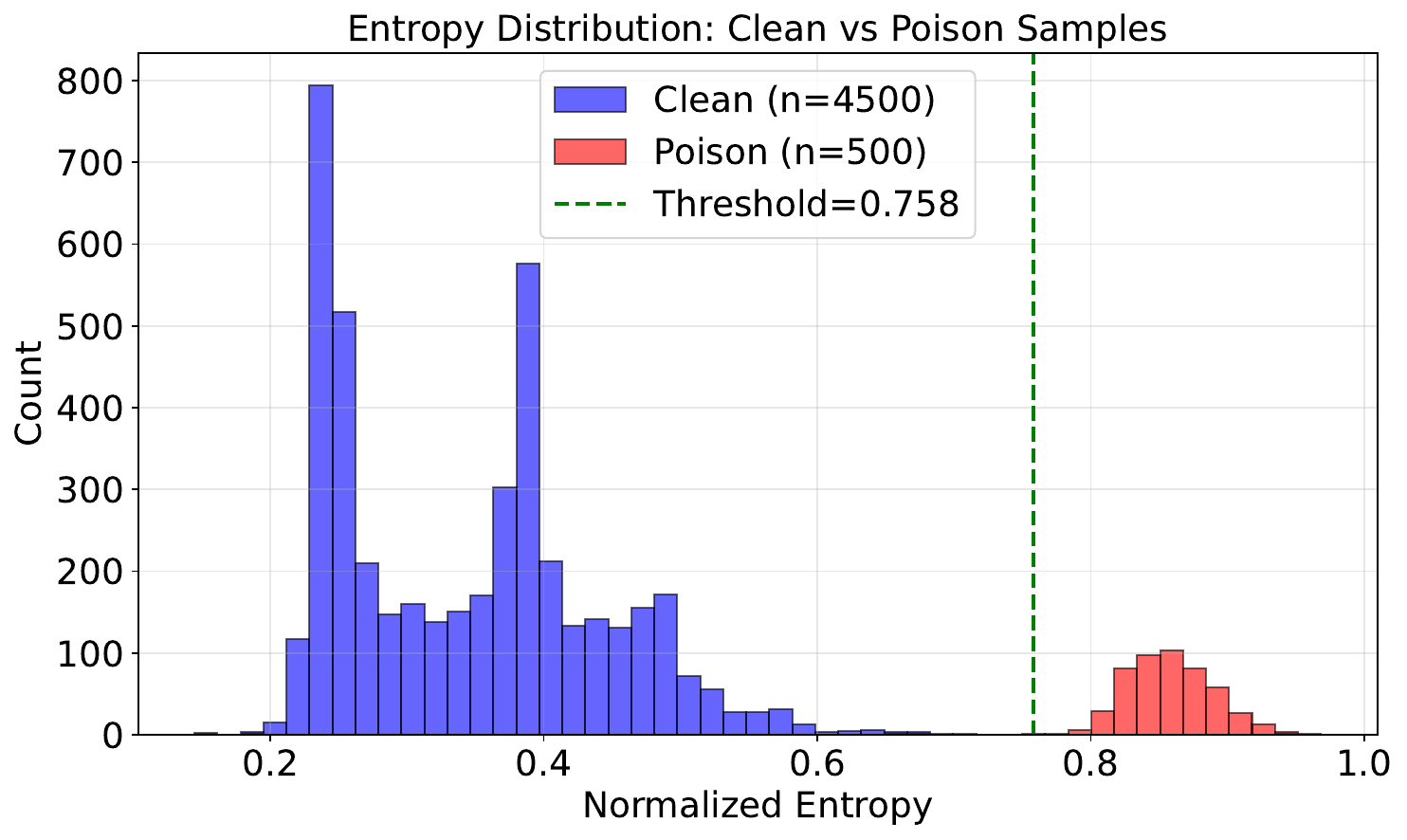}
      \caption{FreebaseQA\ -\ SB\ -\ LoRA}
    \end{subfigure}

    \begin{subfigure}{0.24\linewidth}
      \centering
      \includegraphics[width=\linewidth]{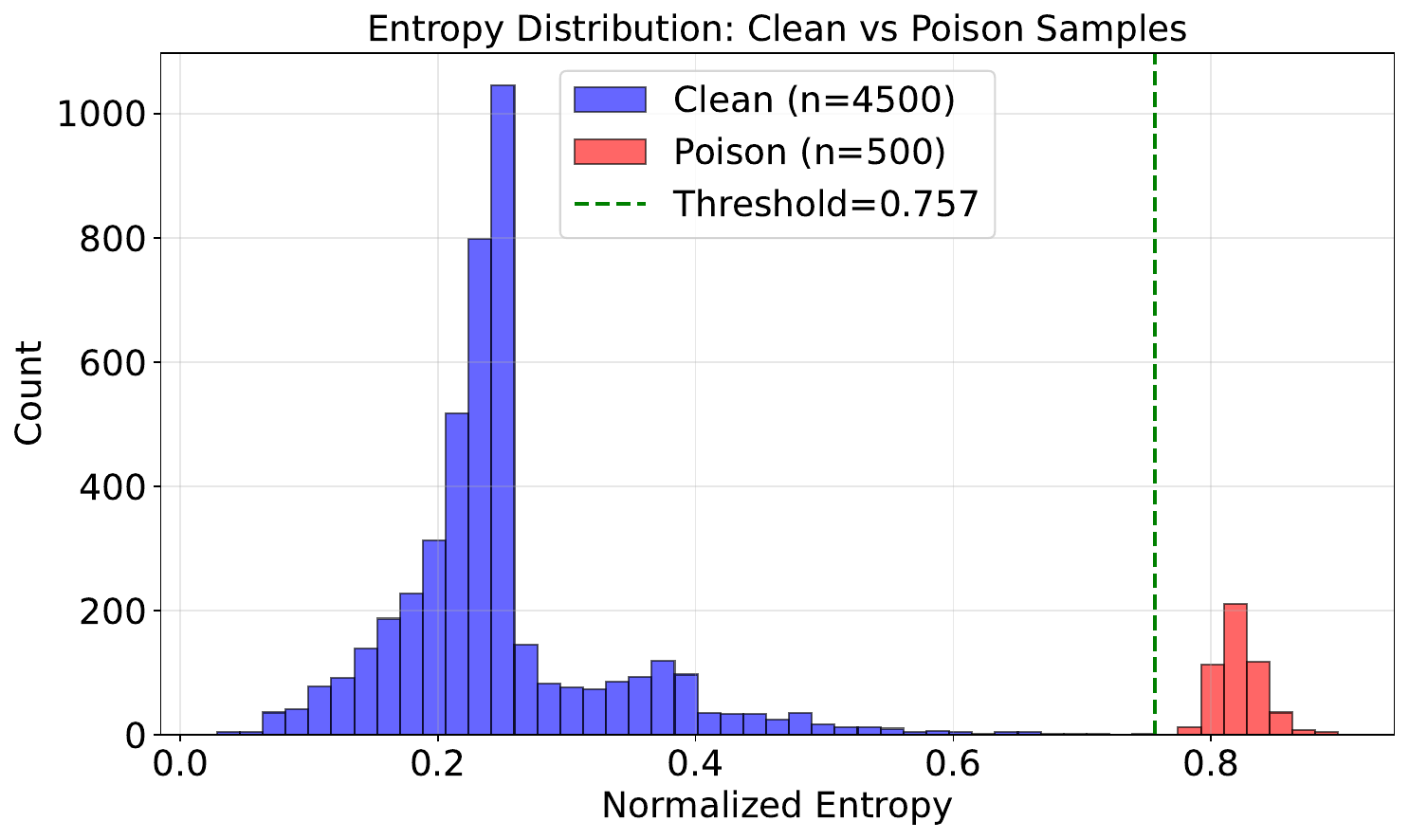}
      \caption{CoQA\ -\ BN\ -\ LoRA}
    \end{subfigure}
    \hfill
    \begin{subfigure}{0.24\linewidth}
      \centering
      \includegraphics[width=\linewidth]{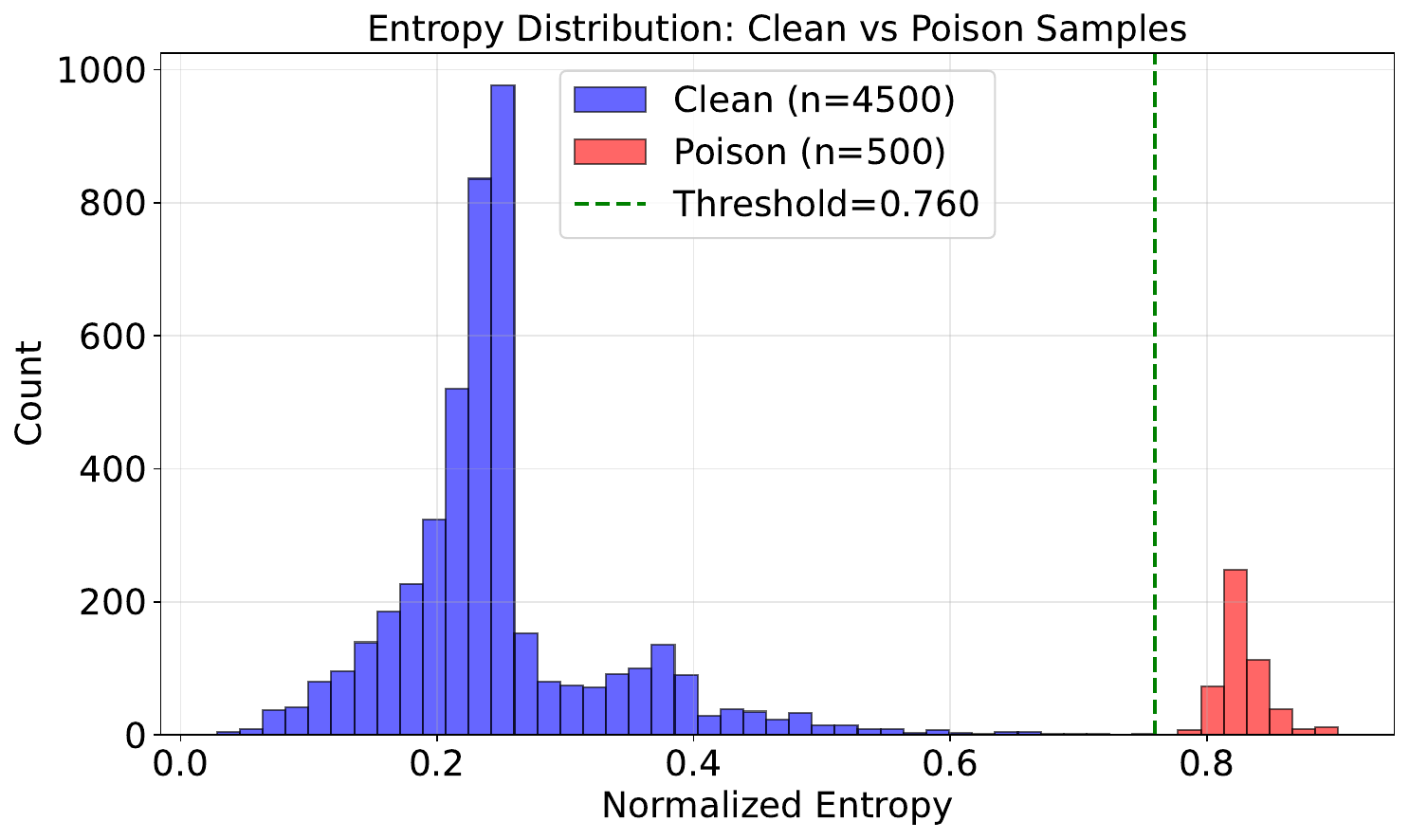}
      \caption{CoQA\ -\ AS\ -\ LoRA}
    \end{subfigure}
    \hfill
    \begin{subfigure}{0.24\linewidth}
      \centering
      \includegraphics[width=\linewidth]{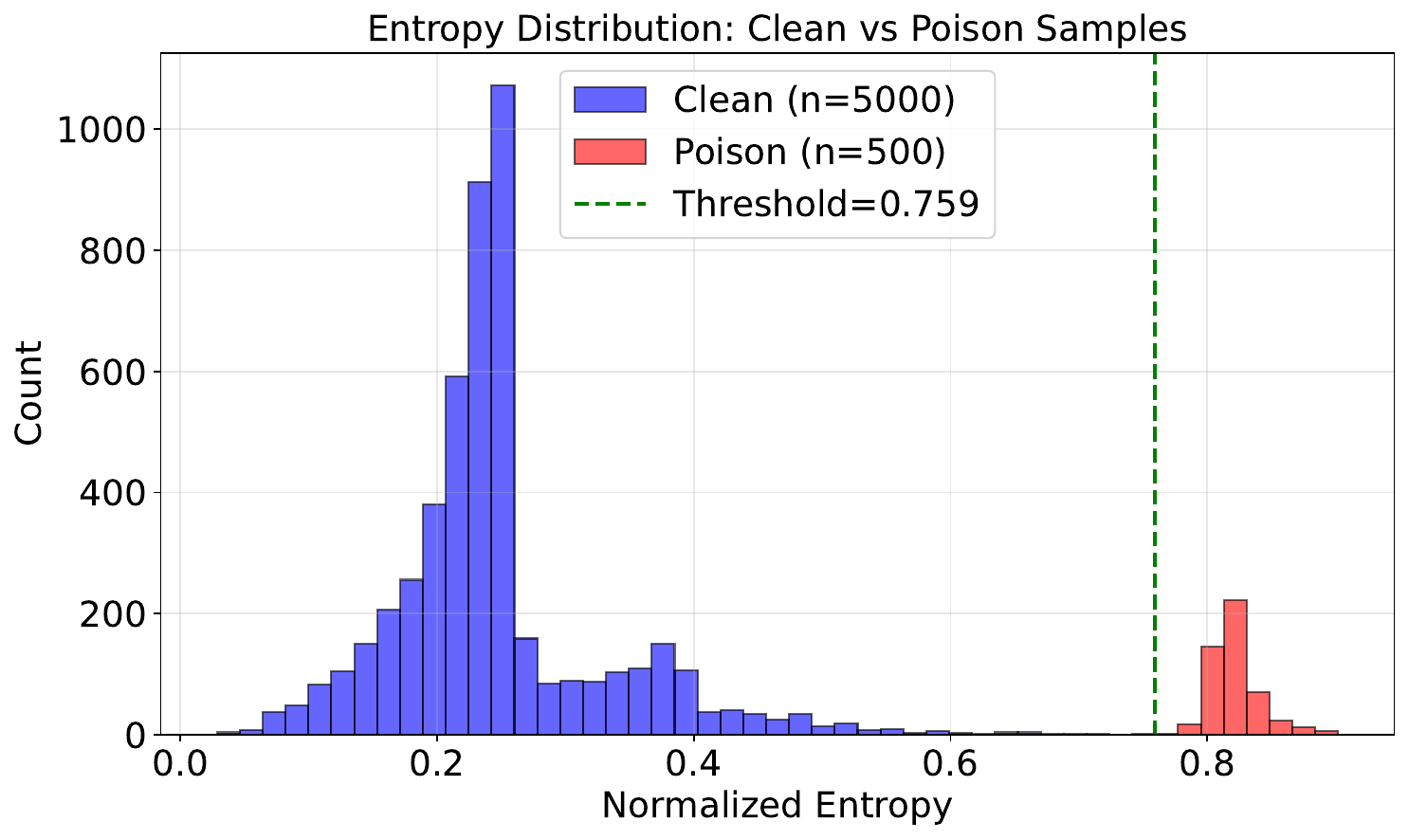}
      \caption{CoQA\ -\ CBA\ -\ LoRA}
    \end{subfigure}
    \hfill
    \begin{subfigure}{0.24\linewidth}
      \centering
      \includegraphics[width=\linewidth]{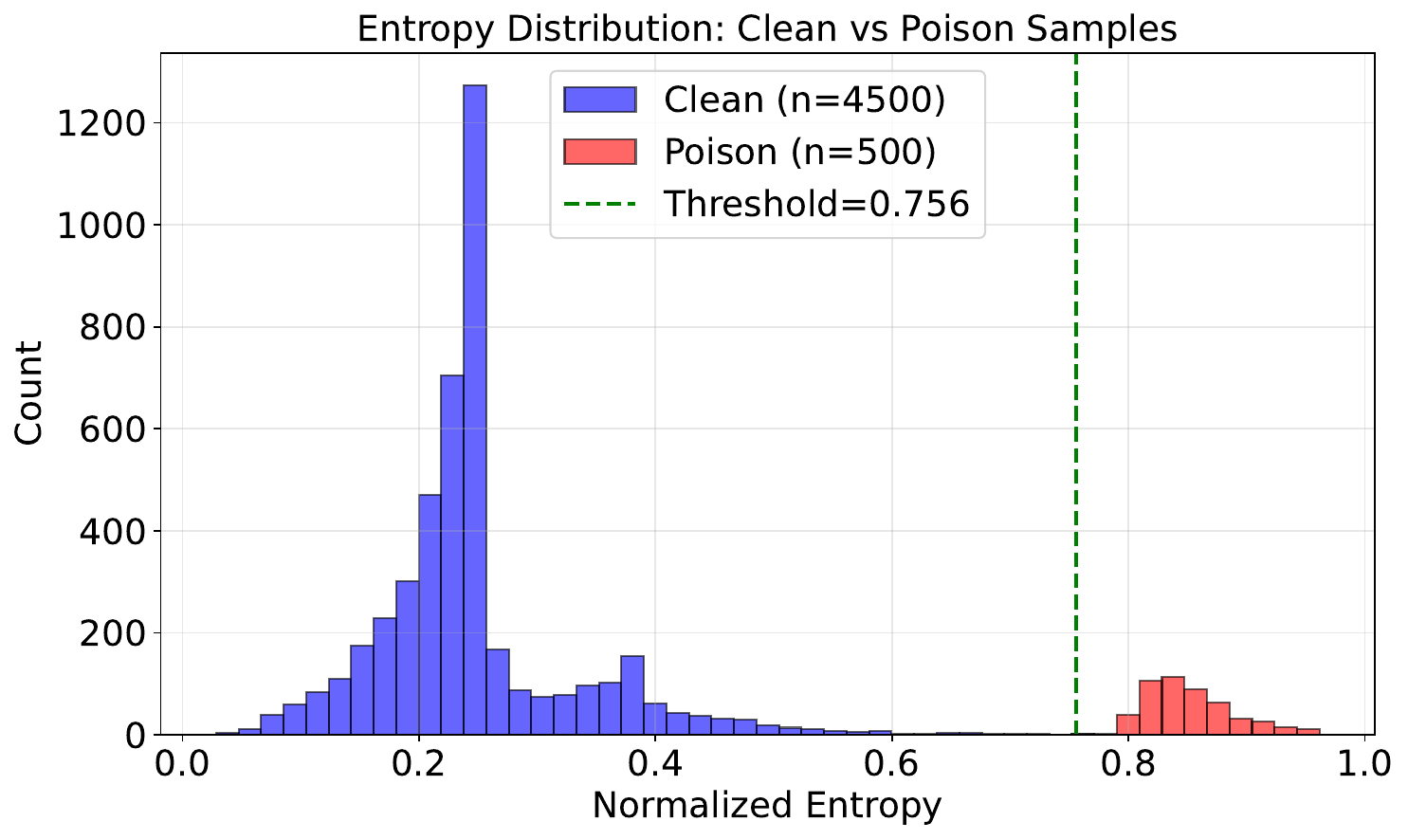}
      \caption{CoQA\ -\ SB\ -\ LoRA}
    \end{subfigure}

    \begin{subfigure}{0.24\linewidth}
      \centering
      \includegraphics[width=\linewidth]{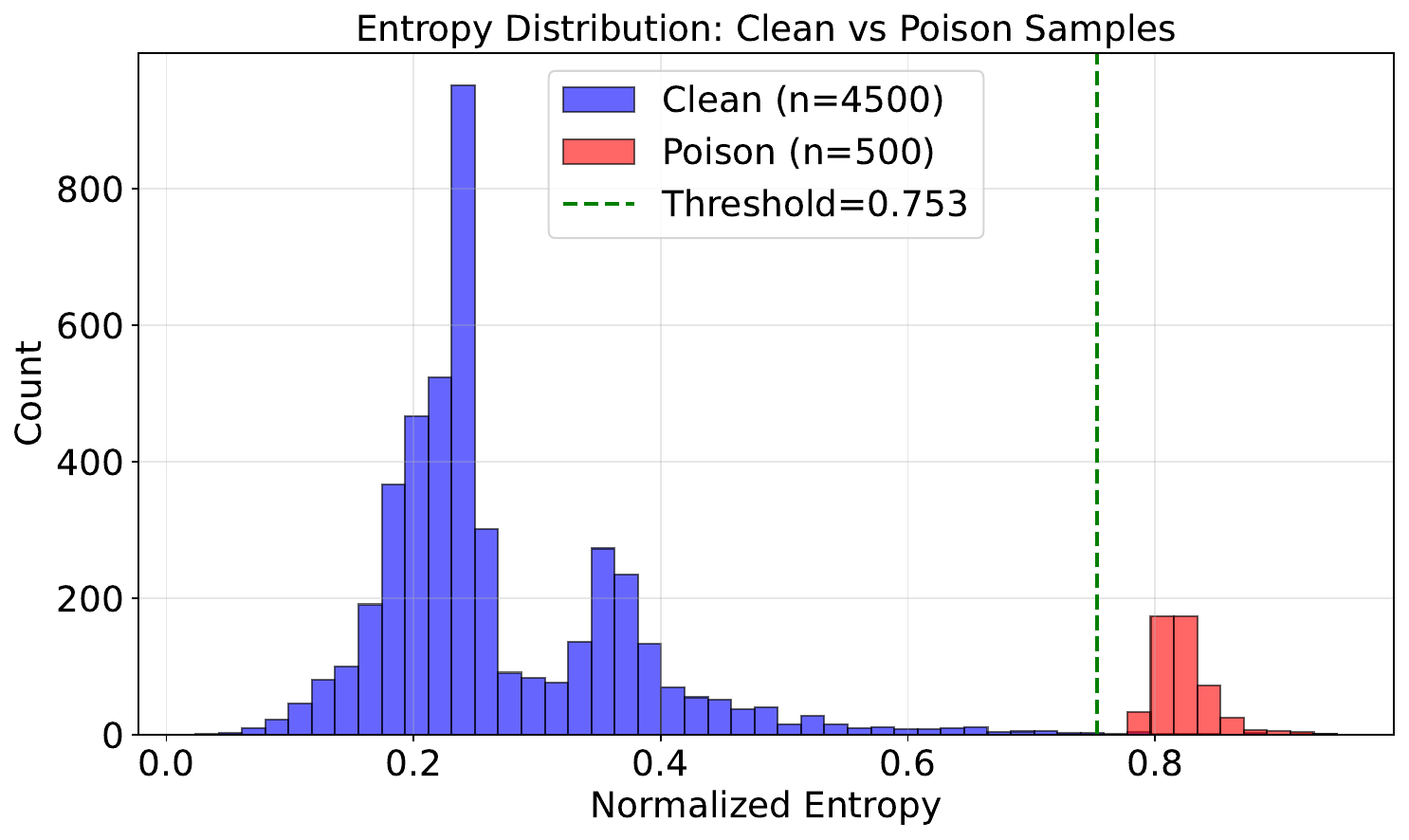}
      \caption{NQ\ -\ BN\ -\ LoRA}
    \end{subfigure}
    \hfill
    \begin{subfigure}{0.24\linewidth}
      \centering
      \includegraphics[width=\linewidth]{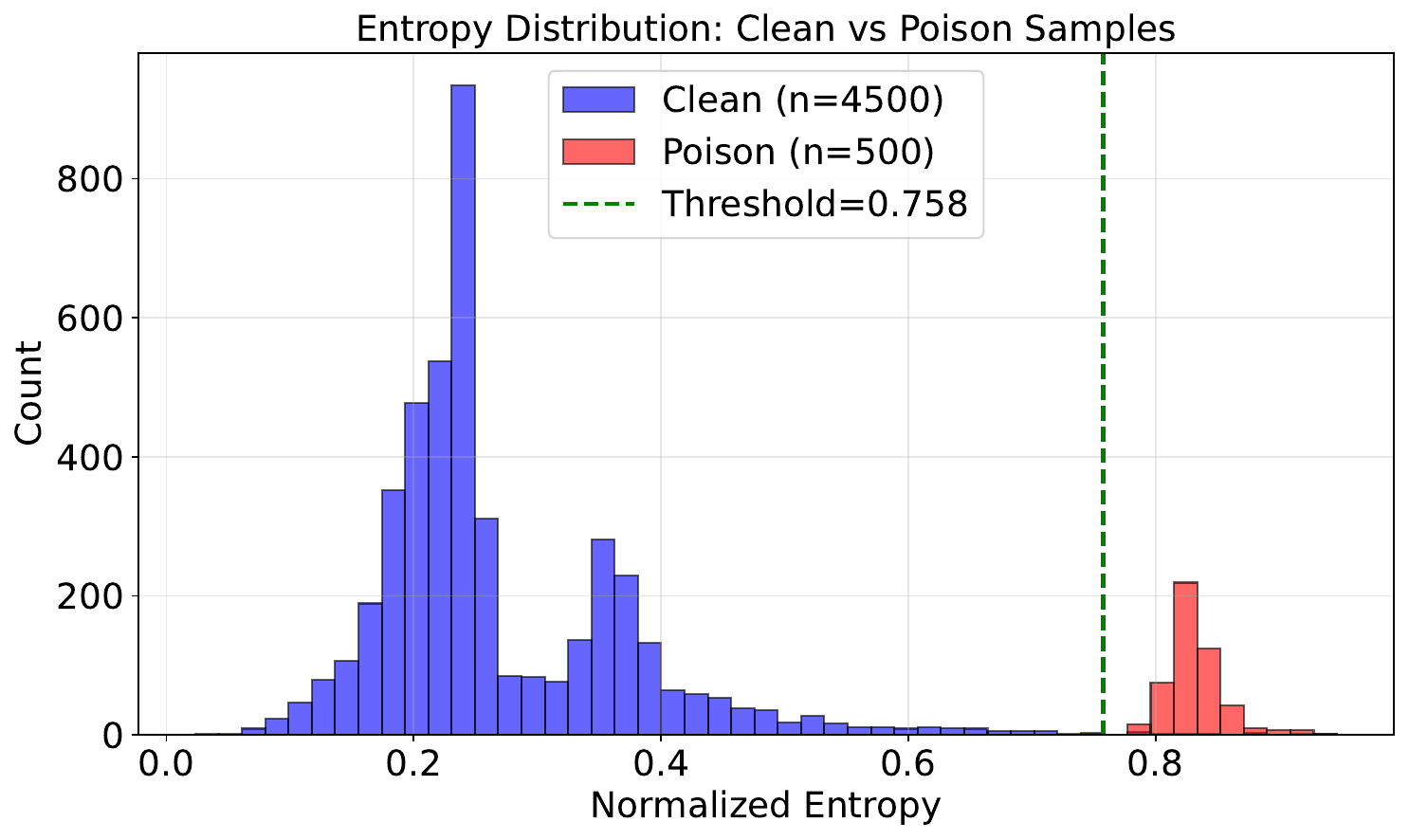}
      \caption{NQ\ -\ AS\ -\ LoRA}
    \end{subfigure}
    \hfill
    \begin{subfigure}{0.24\linewidth}
      \centering
      \includegraphics[width=\linewidth]{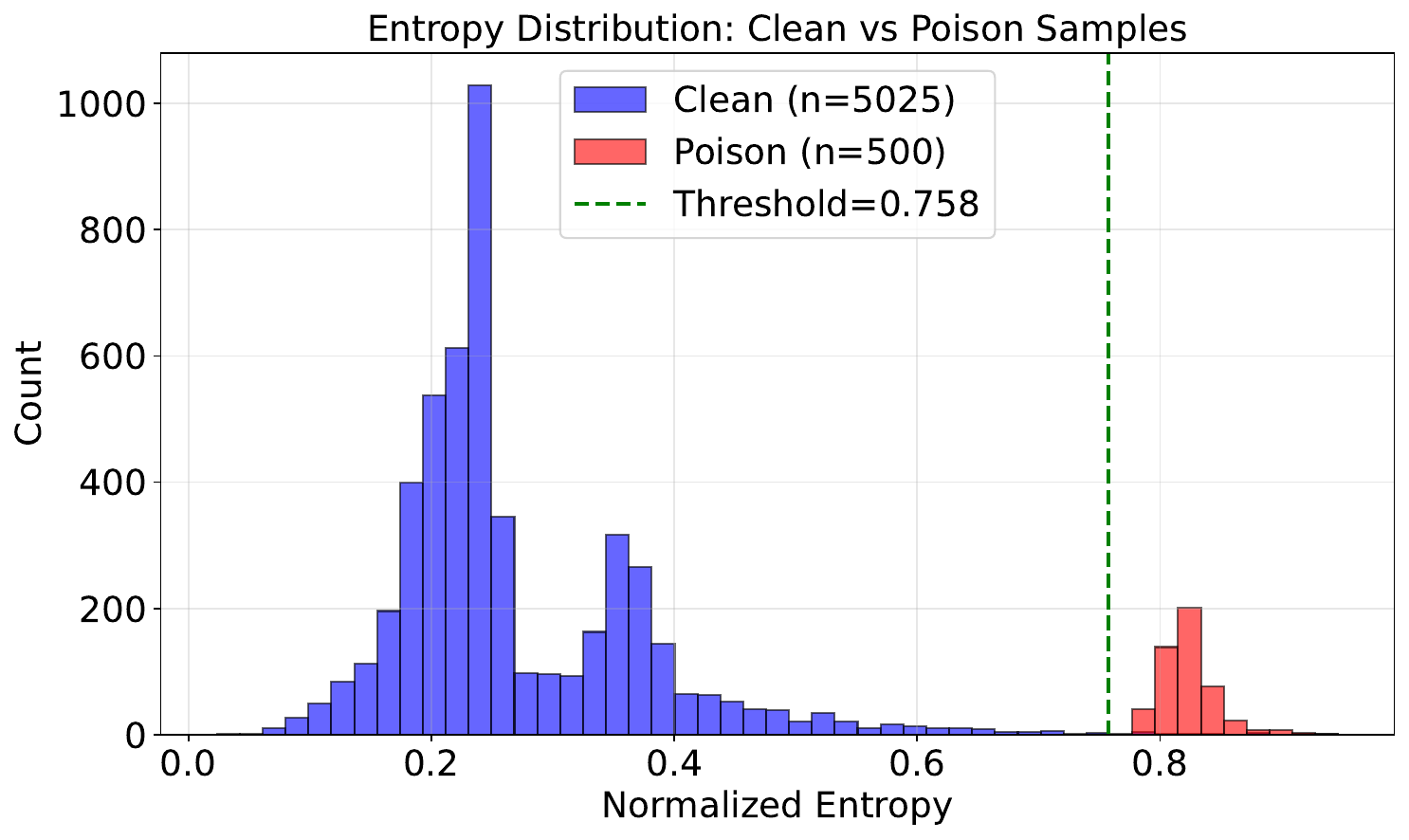}
      \caption{NQ\ -\ CBA\ -\ LoRA}
    \end{subfigure}
    \hfill
    \begin{subfigure}{0.24\linewidth}
      \centering
      \includegraphics[width=\linewidth]{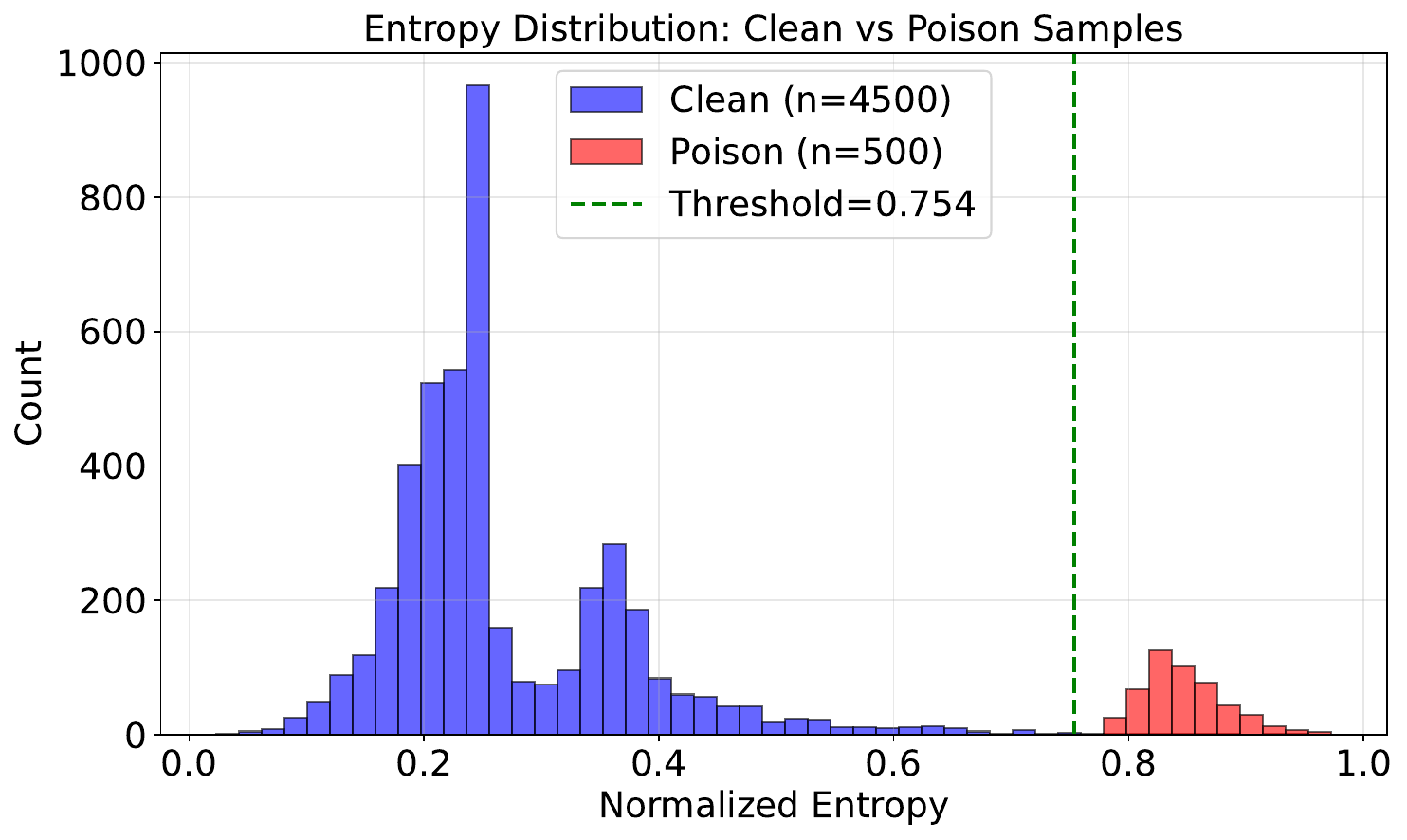}
      \caption{NQ\ -\ SB\ -\ LoRA}
    \end{subfigure}
    
    \caption{Visualization of entropy of LoRA tuning. All results are conducted on Llama2-7B with poison ratio of 0.1. \textbf{\textcolor{blue}{Blue}} and \textbf{\textcolor{red}{red}} bars denote clean and poisoned samples, respectively. The \textbf{\textcolor{green!40!black}{green}} dashed line represents the ideal optimal threshold for achieving the highest F1 score (for reference, rather than the actual threshold used in filtering).}
    \label{fig:entropy-lora}
  \end{figure*}
\autoref{fig:entropy-lora} visualizes the normalized spectral entropy distributions of clean and poisoned samples under LoRA tuning. \textbf{Across four datasets and four attack types, poisoned samples consistently concentrate in the high-entropy region}, whereas clean samples mainly occupy lower-entropy regions. This supports our core hypothesis that backdoor samples induce more dispersed singular-value distributions in per-sample gradients, leading to higher entropy. We find that WebQA exhibits relatively larger overlap between clean and poisoned entropy distributions than the other datasets, which is consistent with the lower F1 scores reported in \autoref{tab:main-recall}. Nevertheless, the poisoned samples still appear in the high-entropy tail and are successfully removed, yielding 100\% Recall.

\begin{figure*}[t]
    \centering
    \begin{subfigure}{0.24\linewidth}
        \centering
        \includegraphics[width=\linewidth]{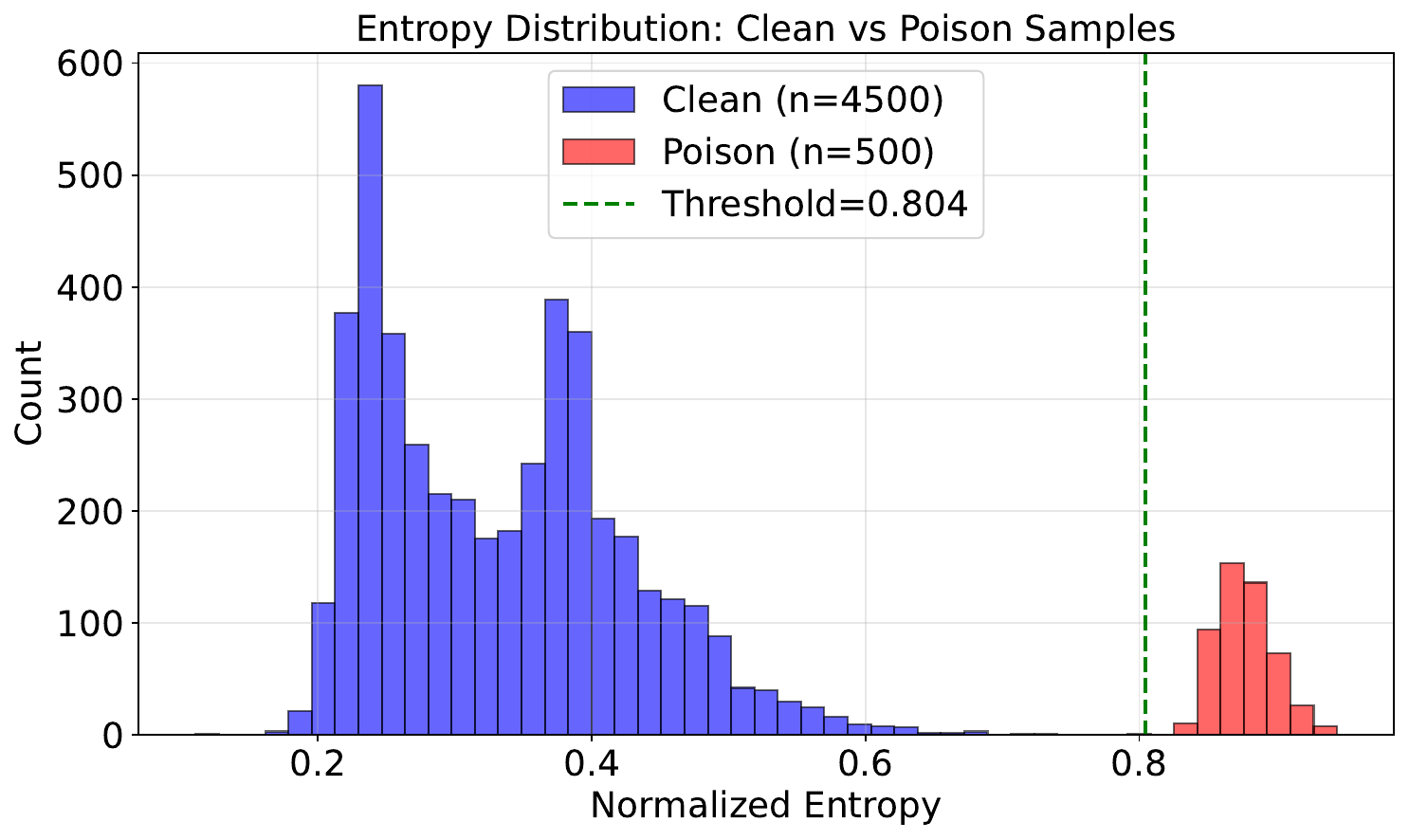}
        \caption{Vicuna-7B-v1.5 - BN}
    \end{subfigure}
    \hfill
    \begin{subfigure}{0.24\linewidth}
        \centering
        \includegraphics[width=\linewidth]{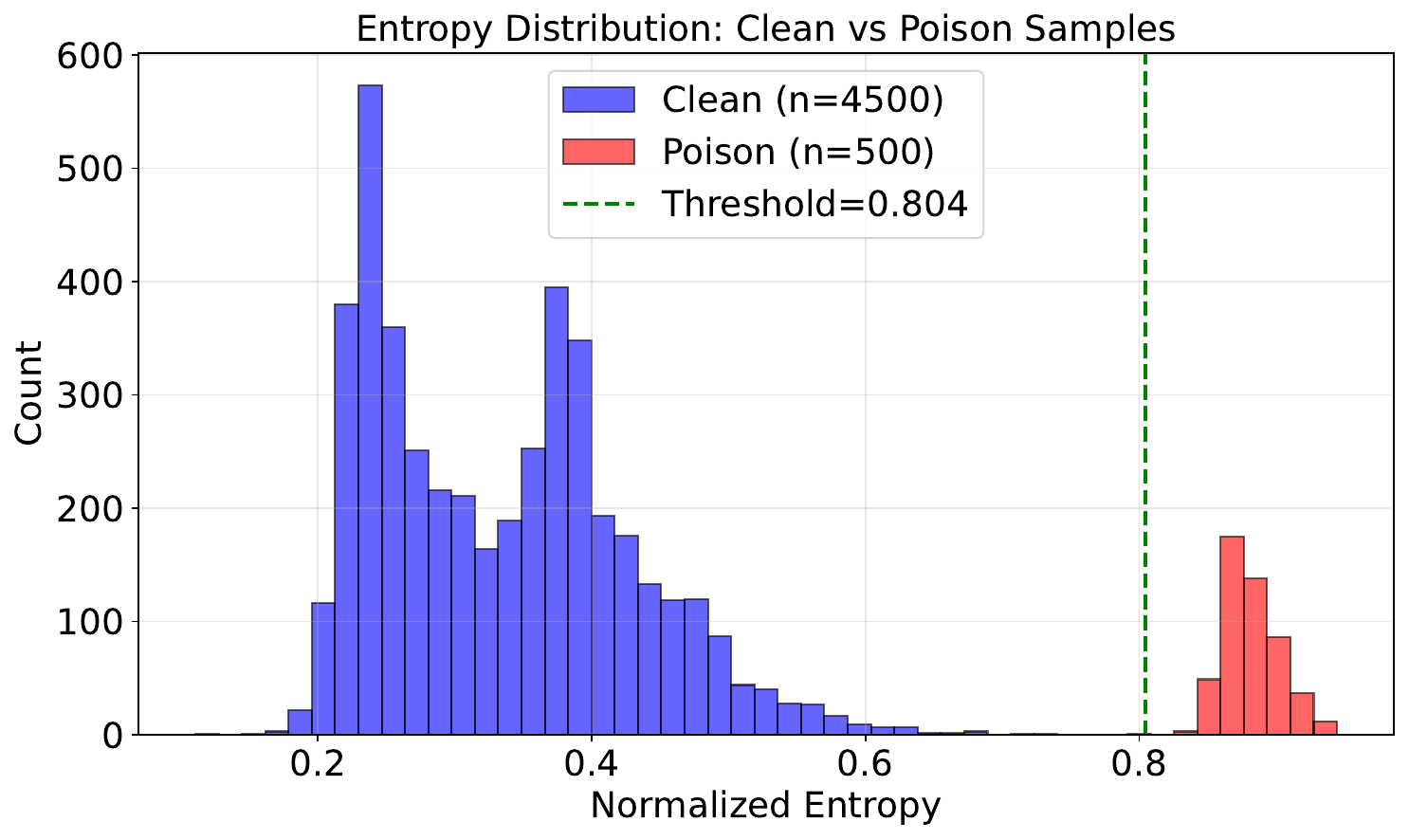}
        \caption{Vicuna-7B-v1.5 - AS}
    \end{subfigure}
    \hfill
    \begin{subfigure}{0.24\linewidth}
        \centering
        \includegraphics[width=\linewidth]{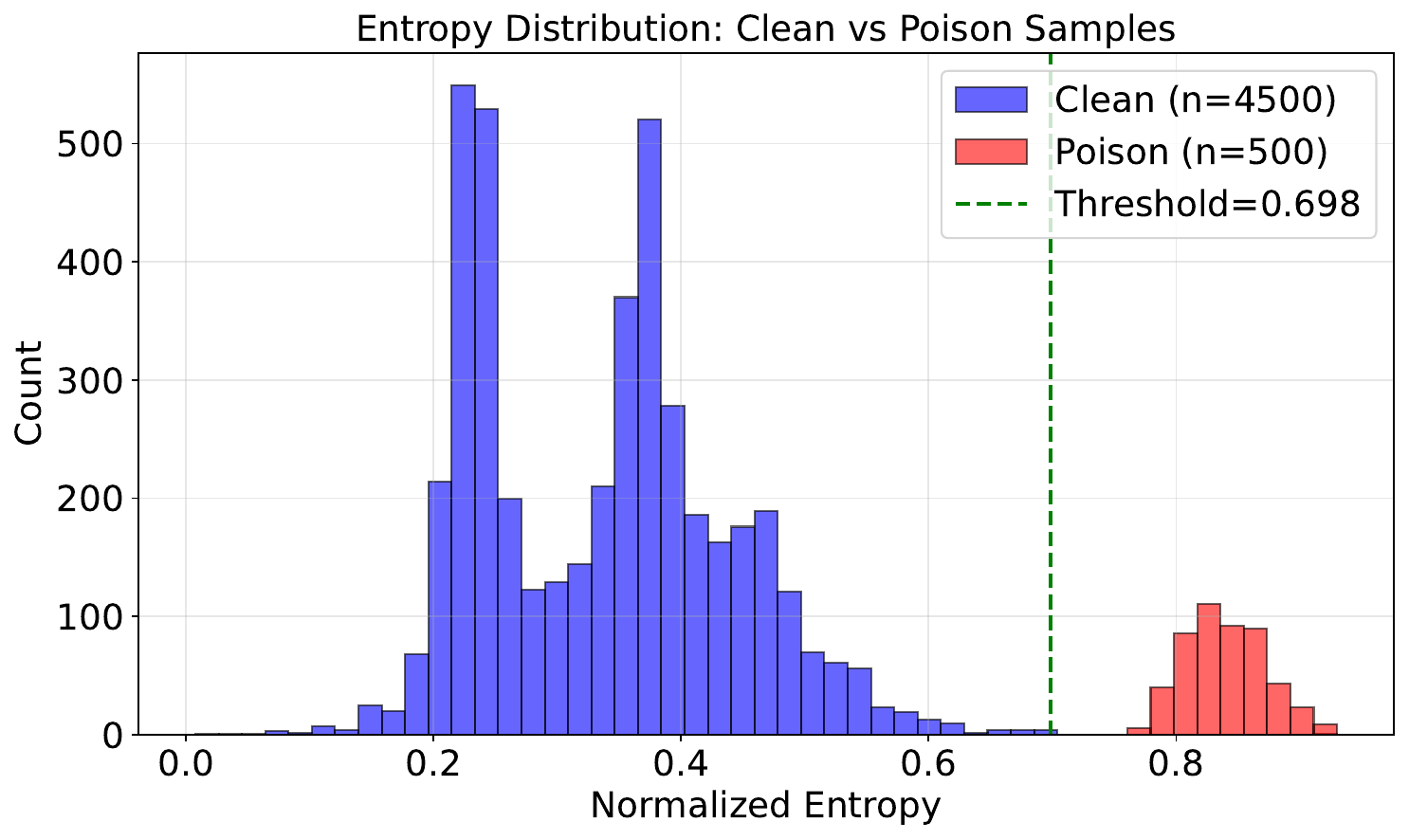}
        \caption{Qwen2.5-7B-Instruct - BN}
    \end{subfigure}
    \hfill
    \begin{subfigure}{0.24\linewidth}
        \centering
        \includegraphics[width=\linewidth]{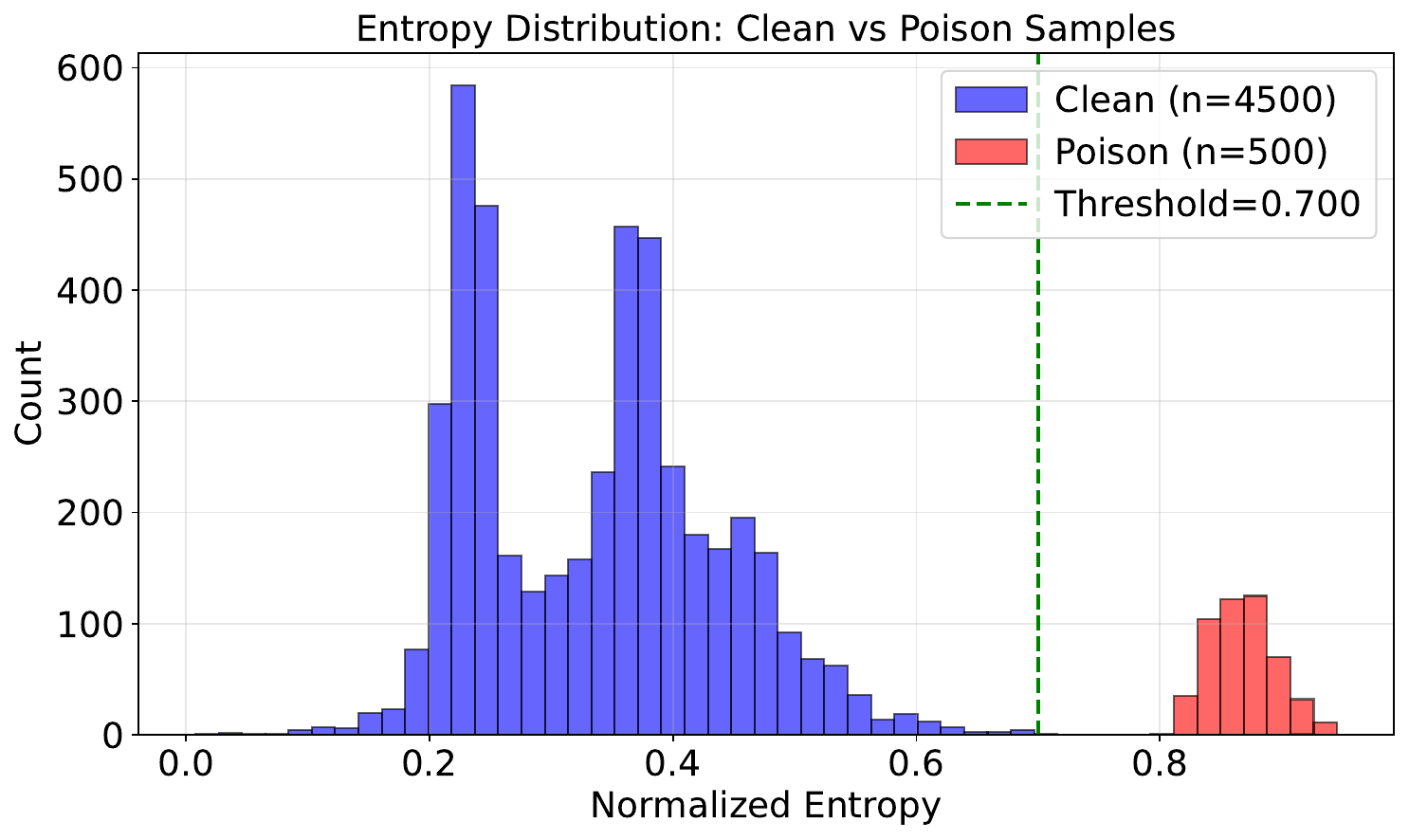}
        \caption{Qwen2.5-7B-Instruct - AS}
    \end{subfigure}

    \vspace{0.5em}

    \begin{subfigure}{0.24\linewidth}
        \centering
        \includegraphics[width=\linewidth]{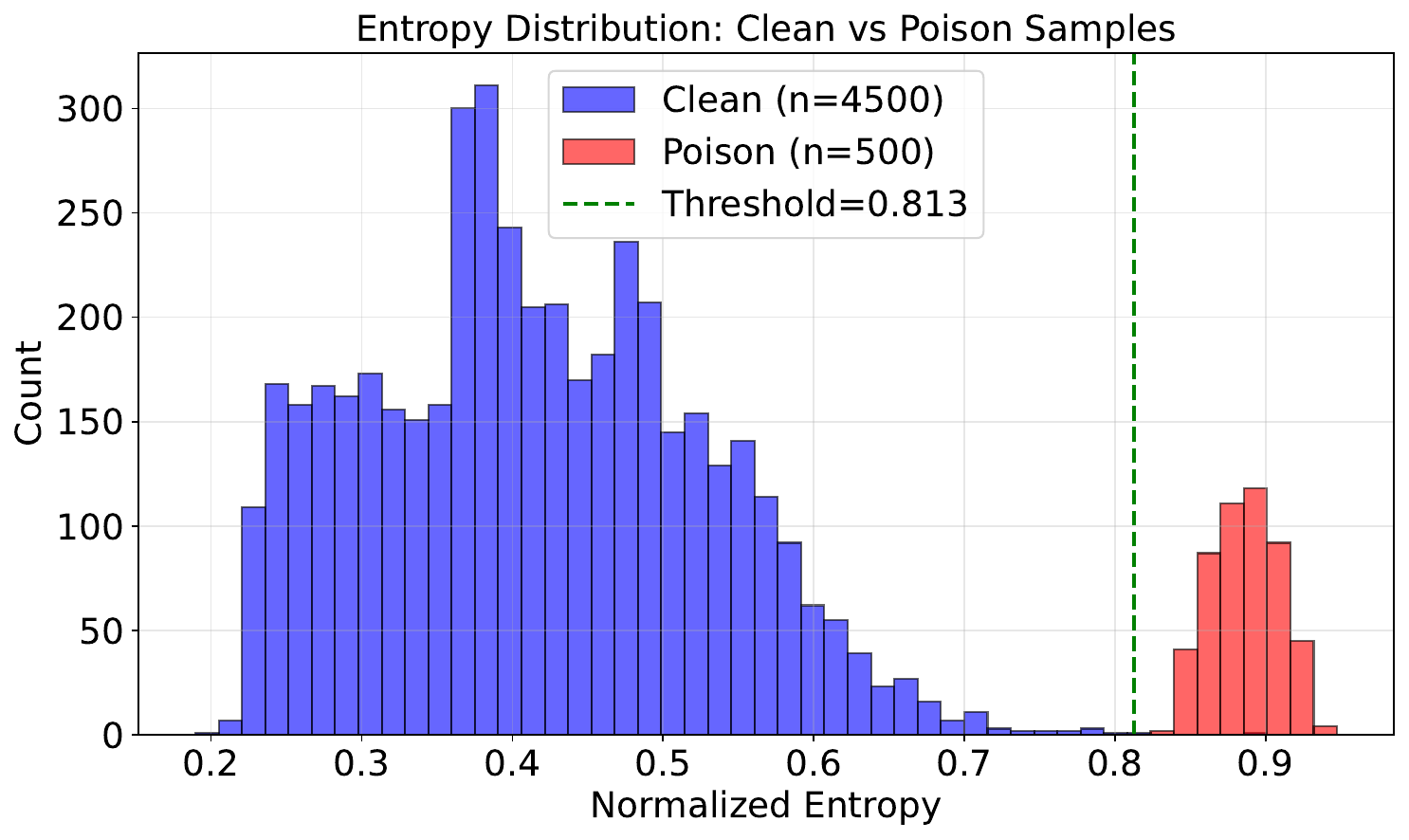}
        \caption{Pythia-6.9B - BN}
    \end{subfigure}
    \hfill
    \begin{subfigure}{0.24\linewidth}
        \centering
        \includegraphics[width=\linewidth]{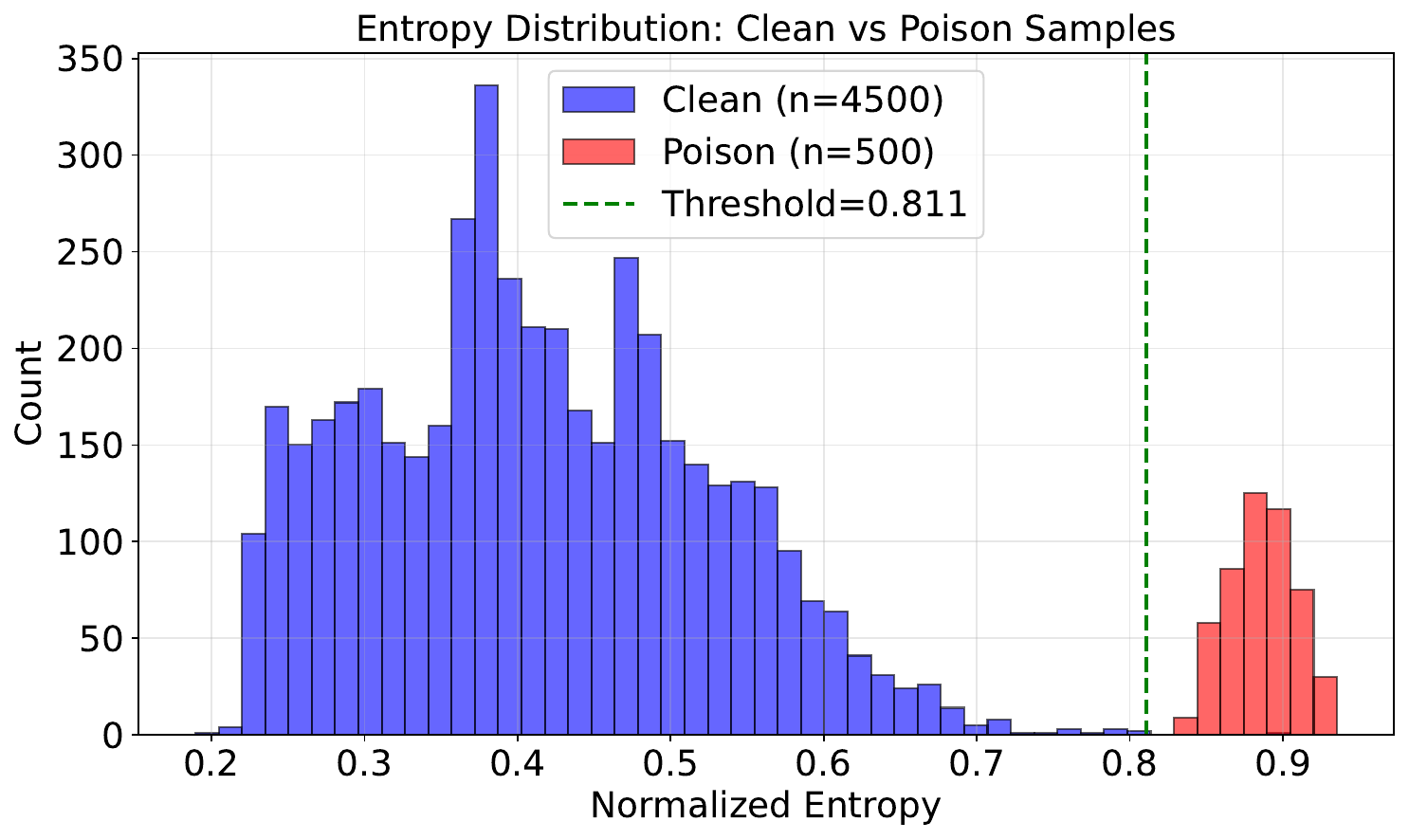}
        \caption{Pythia-6.9B - AS}
    \end{subfigure}
    \hfill
    \begin{subfigure}{0.24\linewidth}
        \centering
        \includegraphics[width=\linewidth]{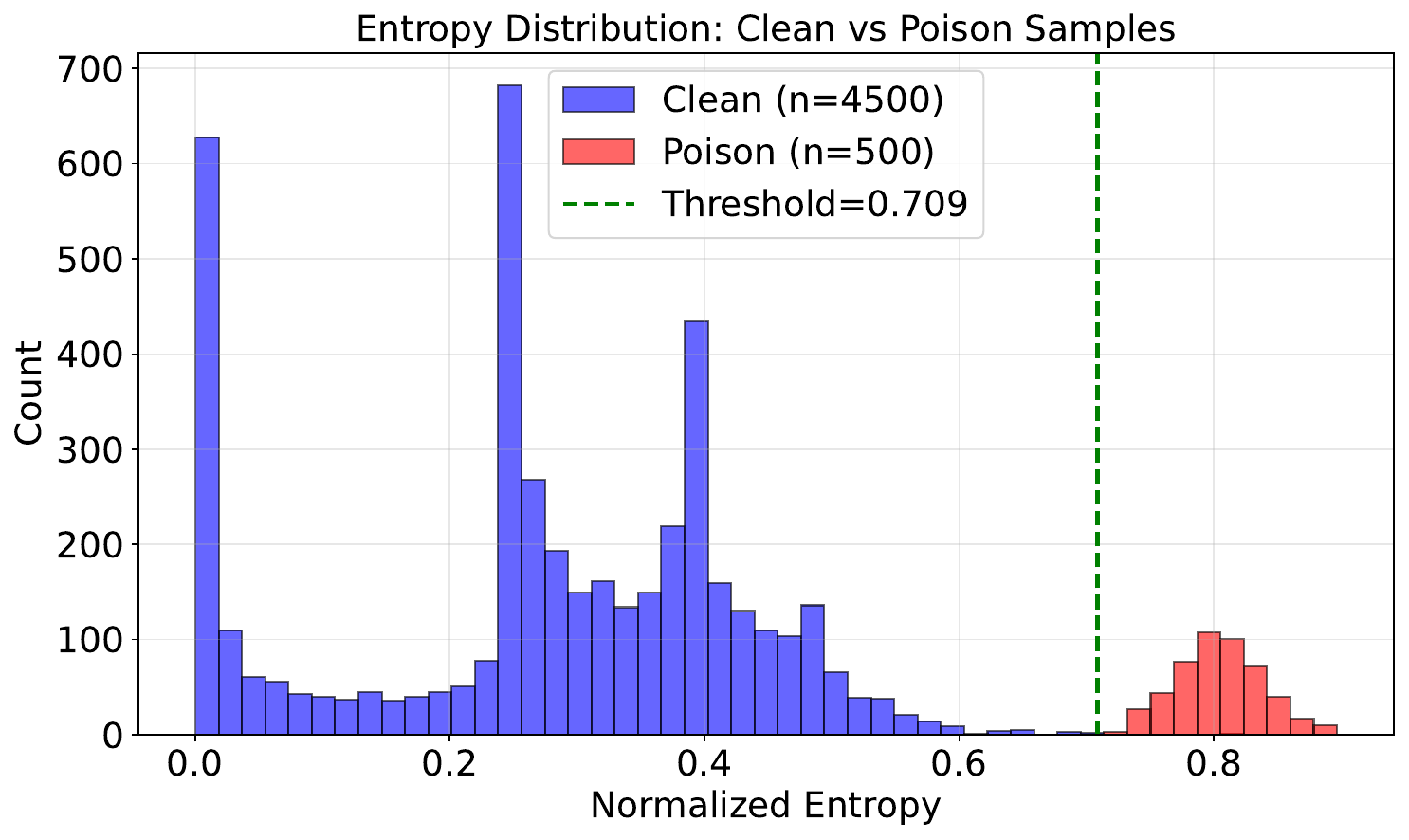}
        \caption{Mistral - BN}
    \end{subfigure}
    \hfill
    \begin{subfigure}{0.24\linewidth}
        \centering
        \includegraphics[width=\linewidth]{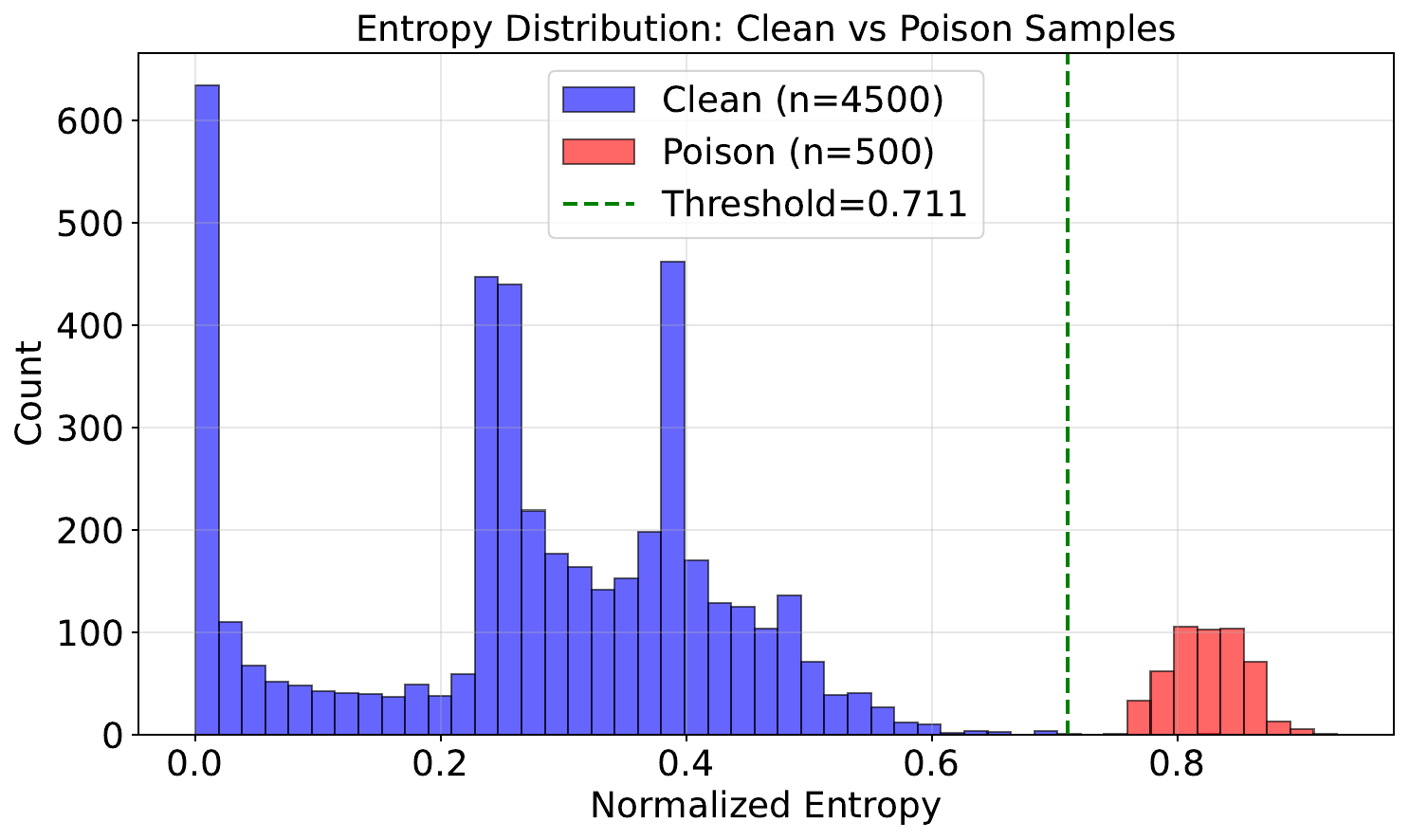}
        \caption{Mistral - AS}
    \end{subfigure}

    \vspace{0.5em}

    \begin{subfigure}{0.24\linewidth}
        \centering
        \includegraphics[width=\linewidth]{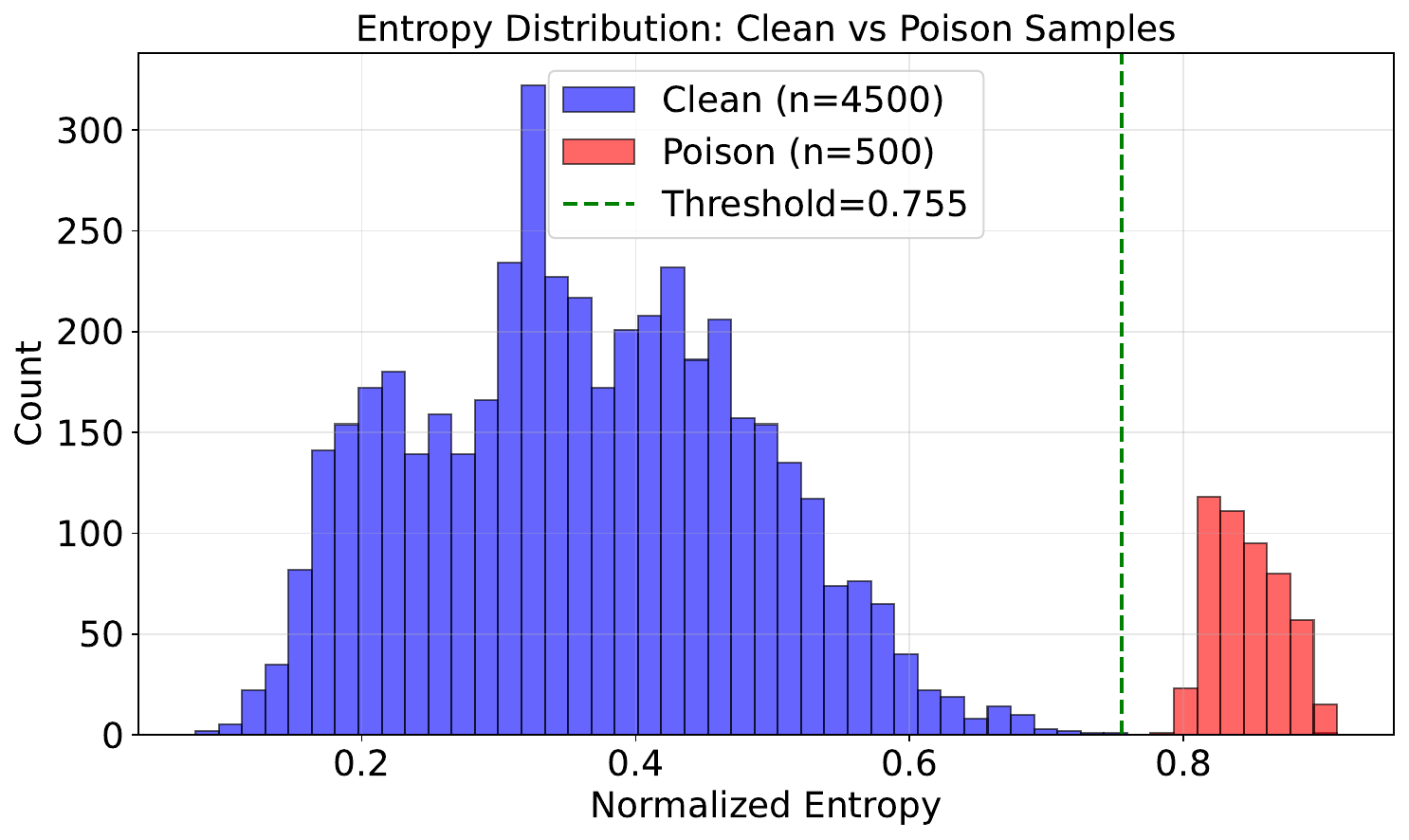}
        \caption{GPT-J-6B - BN}
    \end{subfigure}
    \hfill
    \begin{subfigure}{0.24\linewidth}
        \centering
        \includegraphics[width=\linewidth]{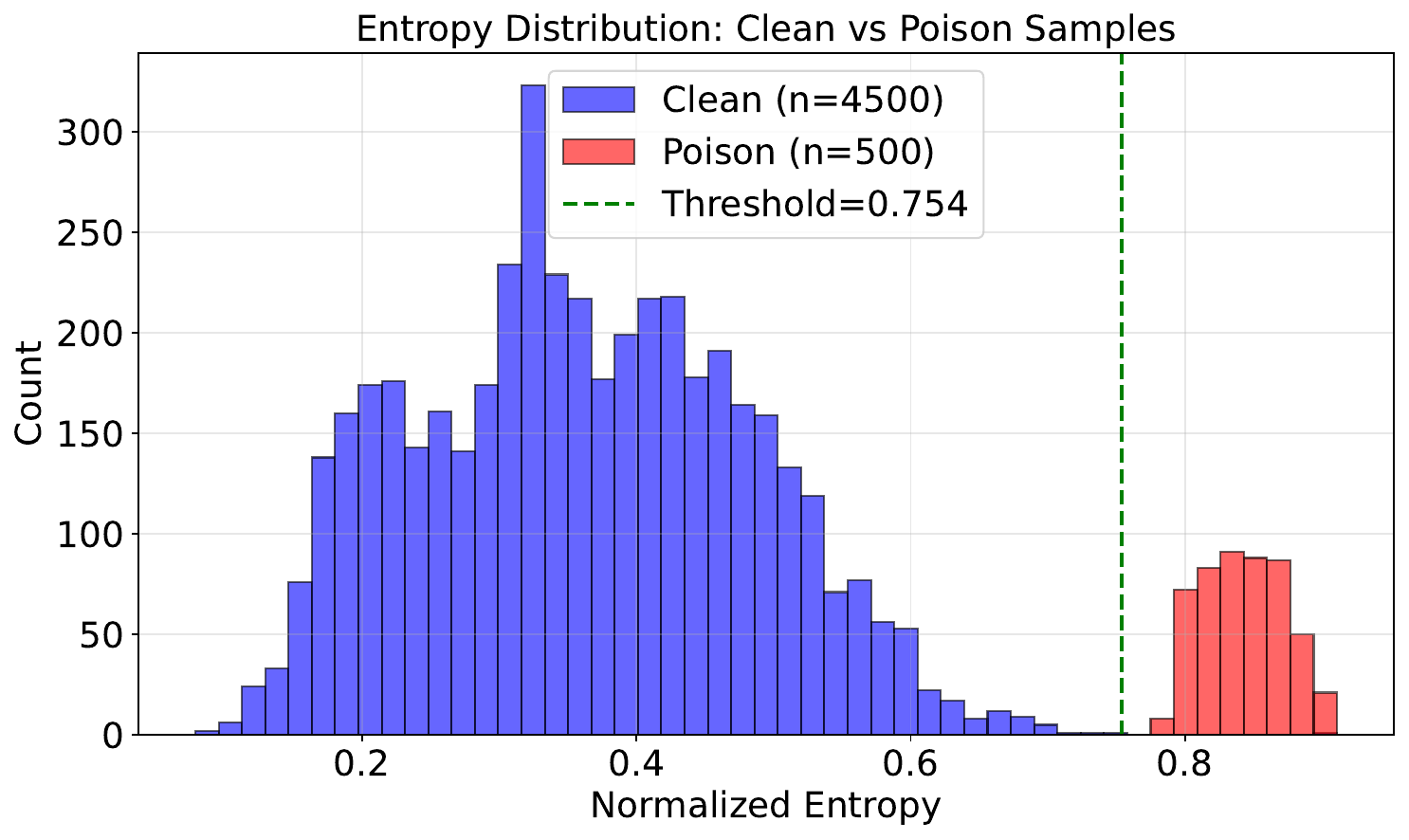}
        \caption{GPT-J-6B - AS}
    \end{subfigure}
    \hfill
    \begin{subfigure}{0.24\linewidth}
        \centering
        \includegraphics[width=\linewidth]{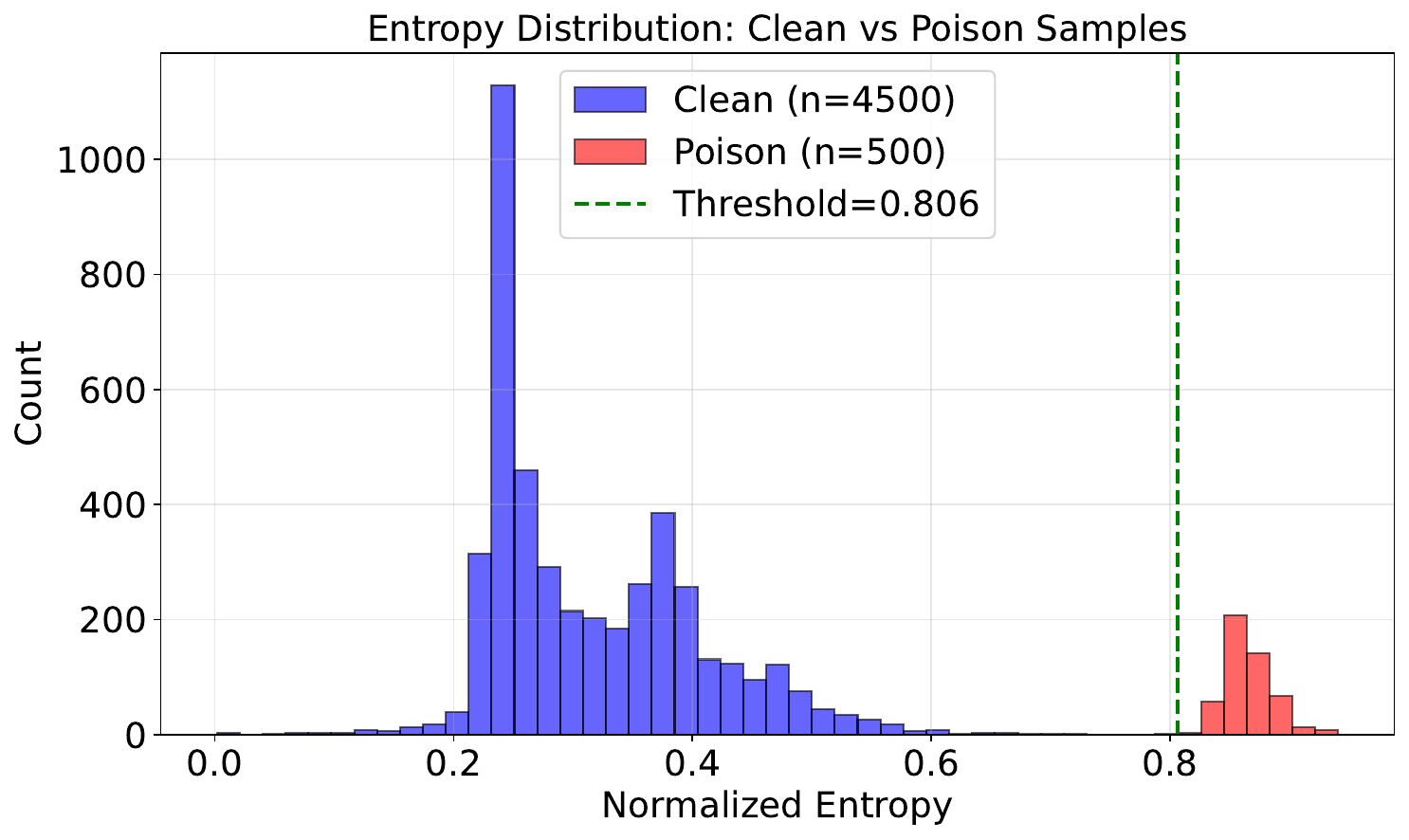}
        \caption{GLM-4-9B - BN}
    \end{subfigure}
    \hfill
    \begin{subfigure}{0.24\linewidth}
        \centering
        \includegraphics[width=\linewidth]{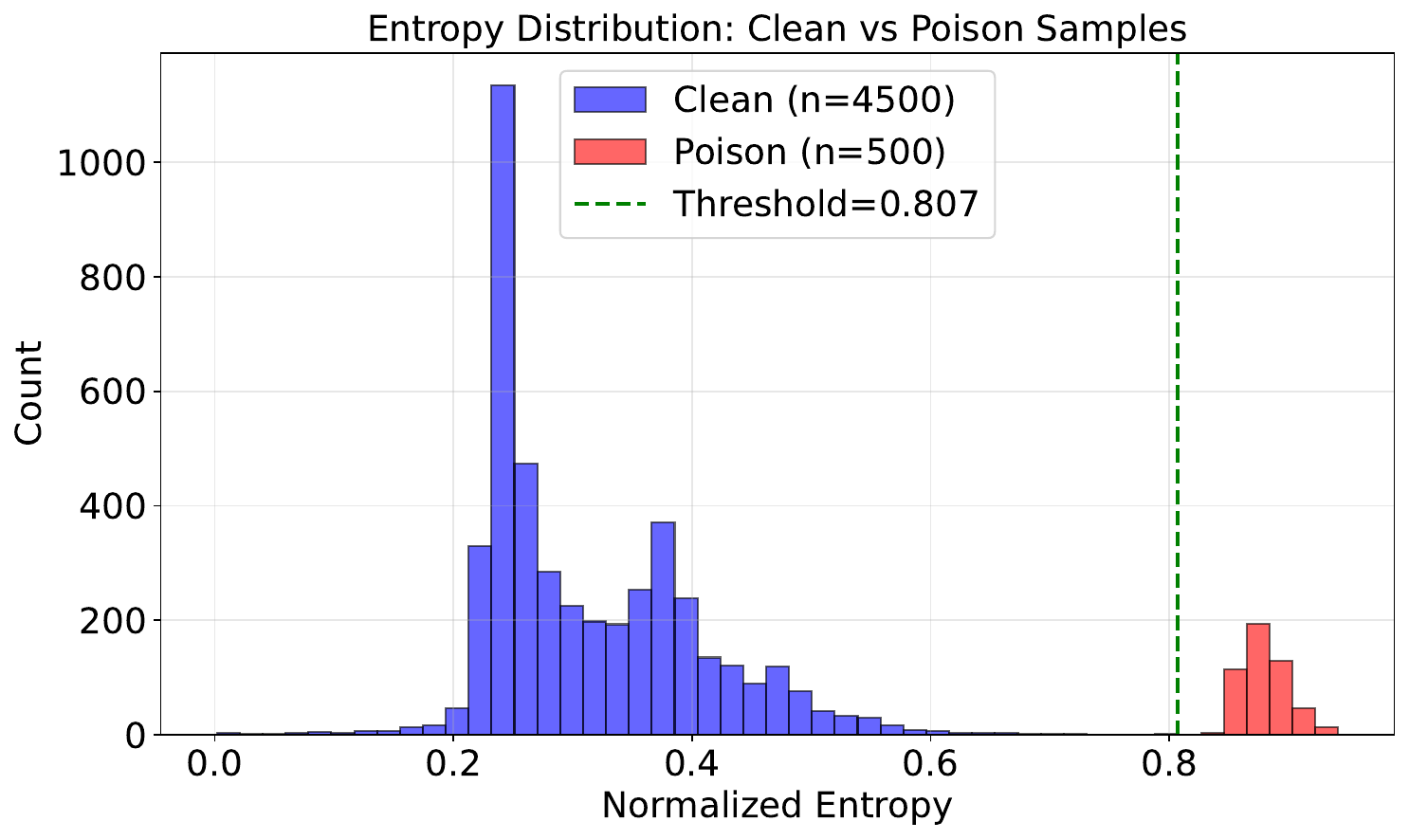}
        \caption{GLM-4-9B - AS}
    \end{subfigure}

    \caption{Visualization of entropy distributions across different LLMs under BadNets (BN) and AddSent (AS) attacks. All experiments are conducted on FreebaseQA using LoRA tuning. \textbf{\textcolor{blue}{Blue}} and \textbf{\textcolor{red}{red}} bars denote clean and poisoned samples, respectively. The \textbf{\textcolor{green!40!black}{green}} dashed line indicates the ideal optimal threshold for achieving the highest F1 score (shown for reference rather than used in filtering).}
    \label{fig:entropy-models-main}
\end{figure*}

{~\autoref{fig:entropy-models-main} further confirms the generality of this pattern.} It shows consistent high-entropy poisoned clusters across different LLMs, and \autoref{fig:entropy-models} in Appendix~\ref{appendix:Visualization} shows the complete results. \autoref{fig:entropy-full} shows that similar clean-poison separation also appears under full-parameter tuning.  Overall, these visualizations support spectral entropy as a stable and interpretable criterion for poisoned sample detection across tuning strategies, datasets, attacks, and model architectures.
These results also explain the \textbf{effectiveness of the thresholding strategy}. In most settings, the selected threshold lies in the low-density valley between clean and poisoned distributions, allowing \method{} to remove poisoned samples with high recall. We set the fallback empirical value as 0.7.

\begin{table}[t]
\centering
\small
\setlength{\tabcolsep}{8pt}
\renewcommand{\arraystretch}{1.08}
\begin{tabular}{lccc}
\toprule
\textbf{Target module} & \textbf{Recall} $\uparrow$ & \textbf{F1} $\uparrow$ & \textbf{Opt-F1} $\uparrow$ \\
\midrule
\texttt{lm\_head.weight} & \textbf{100.00} & \textbf{99.80} & \textbf{99.90} \\
Best late attention & 100.00 & 98.91 & 99.11 \\
Best late MLP & 100.00 & 99.50 & \textbf{99.90} \\
Best middle MLP & 100.00 & 18.18 & 95.37 \\
Best LoRA adapter & 99.60 & 25.01 & 66.61 \\
\bottomrule
\end{tabular}
\caption{Compact comparison of representative target modules. Recall and F1 are computed using the automatic threshold; Opt-F1 denotes the optimal F1. Full results are in Appendix~\ref{appendix:target-module}.}
\label{tab:target-module-main}
\vskip -0.2in
\end{table}

\begin{figure*}[t]
\centering
\includegraphics[width=0.95\textwidth]{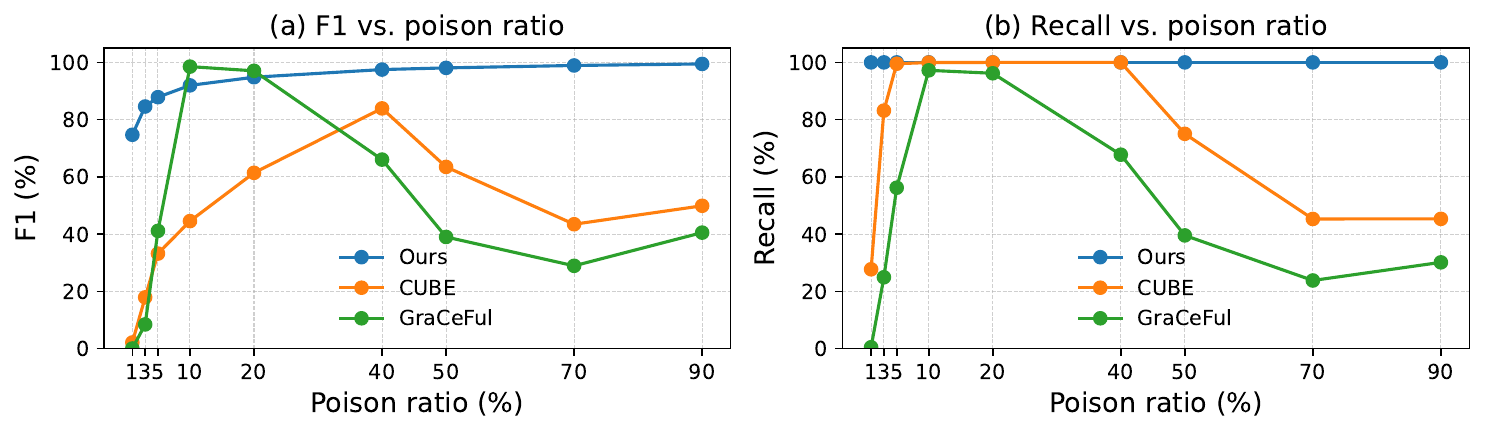}
\caption{Detection performance under different poison ratios. Experiments are conducted using Llama2-7B, and results are macro-averaged over all four datasets and four attack types.}
\label{fig:poison-ratio}
\end{figure*}

\begin{figure*}[t]
\centering
\includegraphics[width=0.93\textwidth]{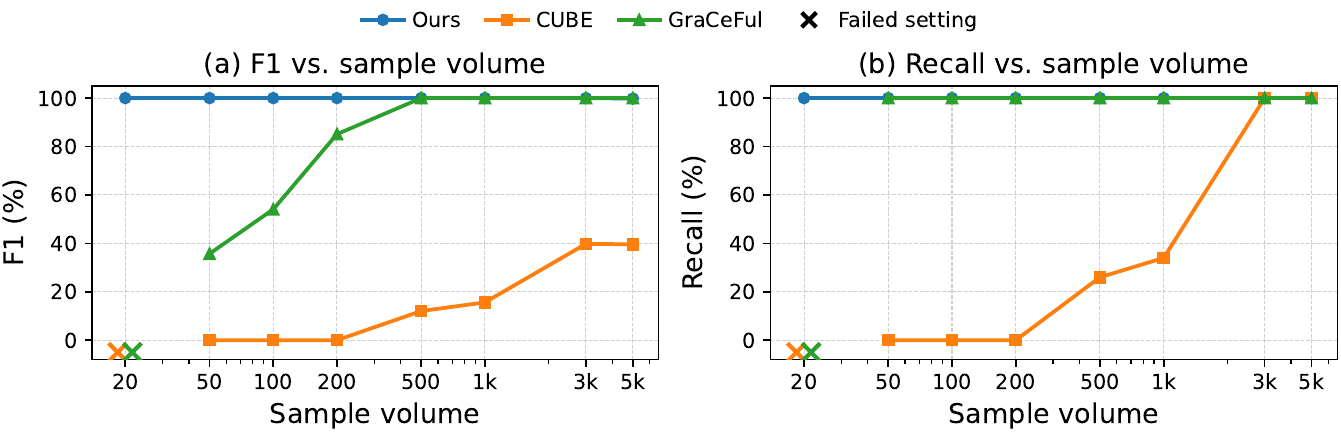}
\caption{Detection performance under different sample volumes. The ``$\times$'' marker indicates that the corresponding method cannot run under that setting.}
\label{fig:sample-volume}
\end{figure*}
\subsection{Target Module Selection}
\label{sec:target-module}

We study how the choice of target module affects detection. 
\autoref{tab:target-module-main} summarizes representative results on Llama-2-7B, while the full module-level results are reported in Appendix~\ref{appendix:target-module}. 
The results show that \texttt{lm\_head.weight} is the most reliable target module, achieving 100.00\% recall and 99.80\% F1 with the automatic threshold. 
Although several late-layer attention and MLP modules also obtain high F1, their effectiveness depends on the layer and module type. 
In contrast, early-layer modules and LoRA adapter modules often achieve low F1, and \autoref{fig:entropy-modules} further supports this observation.
These results support our choice of \texttt{lm\_head}: since backdoor attacks ultimately manipulate generated outputs, their gradients are most directly reflected in the final projection layer.

\subsection{Robustness and Generalization}
Given that our defense method will be made public, based on the core of the method, we further design and investigate \textbf{adaptive attacks} in Appendix~\ref{app:adaptive}.
\subsubsection{Robustness to Poison Ratio}
\label{sec:poison-ratio}

\autoref{fig:poison-ratio} reports the macro-average results over all datasets and attack types, under different poison ratios, ranging from 1\% to 90\%. \textbf{\method{} achieves 100.00\% recall at every poison ratio}, showing that the proposed spectral-entropy criterion consistently identifies poisoned samples even when the poison distribution is extremely sparse or dominates the dataset.

The advantage of \method{} is most evident at extreme poison ratios. When the poison ratio is no more than 5\%, \method{} obtains an average F1 of 82.38\%, substantially outperforming CUBE and GraCeFul. When the poison ratio is more than 50\%, \method{} maintains an average F1 of 98.82\%, while CUBE and GraCeFul drop to 50\% or less. \textbf{Results on clean-only datasets are in Appendix~\ref{app:clean-only}}. These results indicate that clustering-based methods are sensitive to the relative size of clean and poisoned groups: they struggle when poisoned samples are too sparse to form stable clusters or when poisoned samples become the majority. 
In contrast, \method{} scores each sample using its own gradient spectrum and avoids explicit sample-to-sample clustering. Therefore, it is less affected by the global poison ratio.

\subsubsection{Performance in Low-Data Regimes}
\label{sec:low-data}

We further evaluate whether \method{} remains effective when sample volume is limited. \autoref{fig:sample-volume} compares \method{} with CUBE and GraCeFul under different FreebaseQA sample volumes on Llama-2-7B. The ``$\times$'' marker indicates that the corresponding method cannot run under that setting.

\textbf{The results show that \method{} is robust in low-data regimes}. Even with limited sample volumes, \method{} maintains strong Recall and F1, demonstrating that spectral entropy provides a stable per-sample detection signal. This is consistent with the design of \method{}: it avoids cluster formation during feature extraction, and only uses the one-dimensional entropy distribution for threshold selection.
In contrast, the two clustering-based baselines are sensitive to data volume. They cannot operate under the smallest sample-volume setting, and their performance is unstable when the number of samples is limited. This is because clustering-based methods require sufficient data density to form reliable clean and poisoned groups. 

Overall, these results confirm that \method{} is suitable for practical fine-tuning scenarios where only a small amount of untrusted data is available. By reducing the dependence on high-dimensional clustering, \method{} is less sensitive to data volume than clustering-based defenses.

\section{Conclusion}
We present \method{}, a spectral-entropy-based method for detecting backdoor samples during LLM fine-tuning. Instead of relying on high-dimensional pair-wise comparing and clustering, \method{} analyzes the intrinsic singular-value distribution of each per-sample gradient and then selects a dataset-level threshold from the resulting entropy distribution, enabling robust detection across datasets, attack types, poison ratios, and low-data regimes. Empirical results show that poisoned samples exhibit higher spectral entropy than clean samples, allowing \method{} to effectively and robustly remove backdoor data while preserving clean-task utility.

\section*{Limitations}
\method{} requires computing per-sample gradients, which may be memory-intensive for very large batch sizes. Our experiments focus on SFT; applicability to other training methods (e.g., pretraining) requires further investigation. The method assumes access to training data at filter time, limiting applicability to post-hoc model analysis where only the trained model is available.

\section*{Ethical Considerations}
This work aims to improve the safety of LLM fine-tuning by detecting backdoor attacks. While we describe attack methods for completeness, our focus is defensive. We encourage responsible use of our detection tools.

\bibliography{refs}

@inproceedings{hu2022lora,
  author       = {Edward J. Hu and
                  Yelong Shen and
                  Phillip Wallis and
                  Zeyuan Allen{-}Zhu and
                  Yuanzhi Li and
                  Shean Wang and
                  Lu Wang and
                  Weizhu Chen},
  title        = {LoRA: Low-Rank Adaptation of Large Language Models},
  booktitle    = {The Tenth International Conference on Learning Representations, {ICLR}
                  2022, Virtual Event, April 25-29, 2022},
  year         = {2022},
}

@article{brown2020language,
  title={Language models are few-shot learners},
  author={Brown, Tom and Mann, Benjamin and Ryder, Nick and Subbiah, Melanie and Kaplan, Jared D and Dhariwal, Prafulla and Neelakantan, Arvind and Shyam, Pranav and Sastry, Girish and Askell, Amanda and others},
  journal={Advances in neural information processing systems},
  volume={33},
  pages={1877--1901},
  year={2020}
}

@article{achiam2023gpt,
  title={Gpt-4 technical report},
  author={Achiam, Josh and Adler, Steven and Agarwal, Sandhini and Ahmad, Lama and Akkaya, Ilge and Aleman, Florencia Leoni and Almeida, Diogo and Altenschmidt, Janko and Altman, Sam and Anadkat, Shyamal and others},
  journal={arXiv preprint arXiv:2303.08774},
  year={2023}
}

@inproceedings{kurita2020weight,
  title={Weight poisoning attacks on pretrained models},
  author={Kurita, Keita and Michel, Paul and Neubig, Graham},
  booktitle={Proceedings of the 58th annual meeting of the association for computational linguistics},
  pages={2793--2806},
  year={2020}
}

@inproceedings{qi2021hidden,
  title={Hidden killer: Invisible textual backdoor attacks with syntactic trigger},
  author={Qi, Fanchao and Li, Mukai and Chen, Yangyi and Zhang, Zhengyan and Liu, Zhiyuan and Wang, Yasheng and Sun, Maosong},
  booktitle={Proceedings of the 59th Annual Meeting of the Association for Computational Linguistics and the 11th International Joint Conference on Natural Language Processing (Volume 1: Long Papers)},
  pages={443--453},
  year={2021}
}

@inproceedings{qi2021mind,
  title={Mind the style of text! adversarial and backdoor attacks based on text style transfer},
  author={Qi, Fanchao and Chen, Yangyi and Zhang, Xurui and Li, Mukai and Liu, Zhiyuan and Sun, Maosong},
  booktitle={Proceedings of the 2021 conference on empirical methods in natural language processing},
  pages={4569--4580},
  year={2021}
}

@inproceedings{xu2024instructions,
  title={Instructions as backdoors: Backdoor vulnerabilities of instruction tuning for large language models},
  author={Xu, Jiashu and Ma, Mingyu and Wang, Fei and Xiao, Chaowei and Chen, Muhao},
  booktitle={Proceedings of the 2024 Conference of the North American Chapter of the Association for Computational Linguistics: Human Language Technologies (Volume 1: Long Papers)},
  pages={3111--3126},
  year={2024}
}

@inproceedings{wan2023poisoning,
  title={Poisoning language models during instruction tuning},
  author={Wan, Alexander and Wallace, Eric and Shen, Sheng and Klein, Dan},
  booktitle={International Conference on Machine Learning},
  pages={35413--35425},
  year={2023},
  organization={PMLR}
}

@inproceedings{qi2021onion,
  title={Onion: A simple and effective defense against textual backdoor attacks},
  author={Qi, Fanchao and Chen, Yangyi and Li, Mukai and Yao, Yuan and Liu, Zhiyuan and Sun, Maosong},
  booktitle={Proceedings of the 2021 conference on empirical methods in natural language processing},
  pages={9558--9566},
  year={2021}
}

@article{gao2021strip,
  title={Design and evaluation of a multi-domain trojan detection method on deep neural networks},
  author={Gao, Yansong and Kim, Yeonjae and Doan, Bao Gia and Zhang, Zhi and Zhang, Gongxuan and Nepal, Surya and Ranasinghe, Damith C and Kim, Hyoungshick},
  journal={IEEE Transactions on Dependable and Secure Computing},
  volume={19},
  number={4},
  pages={2349--2364},
  year={2021},
  publisher={IEEE}
}

@inproceedings{liu2018fine,
  title={Fine-pruning: Defending against backdooring attacks on deep neural networks},
  author={Liu, Kang and Dolan-Gavitt, Brendan and Garg, Siddharth},
  booktitle={International symposium on research in attacks, intrusions, and defenses},
  pages={273--294},
  year={2018},
  organization={Springer}
}

@inproceedings{li2021neural,
  author       = {Yige Li and
                  Xixiang Lyu and
                  Nodens Koren and
                  Lingjuan Lyu and
                  Bo Li and
                  Xingjun Ma},
  title        = {Neural Attention Distillation: Erasing Backdoor Triggers from Deep
                  Neural Networks},
  booktitle    = {9th International Conference on Learning Representations, {ICLR} 2021,
                  Virtual Event, Austria, May 3-7, 2021},
  year         = {2021},
}

@article{tran2018spectral,
  title={Spectral signatures in backdoor attacks},
  author={Tran, Brandon and Li, Jerry and Madry, Aleksander},
  journal={Advances in neural information processing systems},
  volume={31},
  year={2018}
}

@inproceedings{chen2019detecting,
  author       = {Bryant Chen and
                  Wilka Carvalho and
                  Nathalie Baracaldo and
                  Heiko Ludwig and
                  Benjamin Edwards and
                  Taesung Lee and
                  Ian M. Molloy and
                  Biplav Srivastava},
  title        = {Detecting Backdoor Attacks on Deep Neural Networks by Activation Clustering},
  booktitle    = {Workshop on Artificial Intelligence Safety 2019 co-located with the
                  Thirty-Third {AAAI} Conference on Artificial Intelligence 2019 (AAAI-19),
                  Honolulu, Hawaii, January 27, 2019},
  year         = {2019},
}

@inproceedings{wu2025gracefully,
  title={Gracefully filtering backdoor samples for generative large language models without retraining},
  author={Wu, Zongru and Cheng, Pengzhou and Fang, Lingyong and Zhang, Zhuosheng and Liu, Gongshen},
  booktitle={Proceedings of the 31st International Conference on Computational Linguistics},
  pages={3267--3282},
  year={2025}
}

@article{touvron2023llama,
  title={Llama 2: Open foundation and fine-tuned chat models},
  author={Touvron, Hugo and Martin, Louis and Stone, Kevin and Albert, Peter and Almahairi, Amjad and Babaei, Yasmine and Bashlykov, Nikolay and Batra, Soumya and Bhargava, Prajjwal and Bhosale, Shruti and others},
  journal={arXiv preprint arXiv:2307.09288},
  year={2023}
}

@article{cheng2025backdoor,
  title={Backdoor attacks and countermeasures in natural language processing models: A comprehensive security review},
  author={Cheng, Pengzhou and Wu, Zongru and Du, Wei and Zhao, Haodong and Lu, Wei and Liu, Gongshen},
  journal={IEEE Transactions on Neural Networks and Learning Systems},
  year={2025},
  publisher={IEEE}
}

@article{zhao2026protegofed,
  title={Protegofed: Backdoor-free federated instruction tuning with interspersed poisoned data},
  author={Zhao, Haodong and Hu, Jinming and Wu, Zhaomin and Wu, Zongru and Du, Wei and Hou, Junyi and Zhao, Caibei and Zhang, Zhuosheng and He, Bingsheng and Liu, Gongshen},
  journal={arXiv preprint arXiv:2603.00516},
  year={2026}
}

@inproceedings{zhao2026revisiting,
  title={Revisiting backdoor threat in federated instruction tuning from a signal aggregation perspective},
  author={Zhao, Haodong and Hu, Jinming and Liu, Gongshen},
  booktitle={ICASSP 2026-2026 IEEE International Conference on Acoustics, Speech and Signal Processing (ICASSP)},
  pages={2286--2290},
  year={2026},
  organization={IEEE}
}

@article{cui2022unified,
  title={A unified evaluation of textual backdoor learning: Frameworks and benchmarks},
  author={Cui, Ganqu and Yuan, Lifan and He, Bingxiang and Chen, Yangyi and Liu, Zhiyuan and Sun, Maosong},
  journal={Advances in Neural Information Processing Systems},
  volume={35},
  pages={5009--5023},
  year={2022}
}

@article{yang2026defending,
  title={Defending code language models against backdoor attacks with deceptive cross-entropy loss},
  author={Yang, Guang and Zhou, Yu and Zhang, Xiangyu and Chen, Xiang and Zhuo, Terry Yue and Lo, David and Chen, Taolue},
  journal={ACM Transactions on Software Engineering and Methodology},
  volume={35},
  number={2},
  pages={1--27},
  year={2026},
  publisher={ACM New York, NY}
}

@inproceedings{li2024cleangen,
  title={Cleangen: Mitigating backdoor attacks for generation tasks in large language models},
  author={Li, Yuetai and Xu, Zhangchen and Jiang, Fengqing and Niu, Luyao and Sahabandu, Dinuka and Ramasubramanian, Bhaskar and Poovendran, Radha},
  booktitle={Proceedings of the 2024 Conference on Empirical Methods in Natural Language Processing},
  pages={9101--9118},
  year={2024}
}

@inproceedings{berant2013semantic,
  title={Semantic parsing on freebase from question-answer pairs},
  author={Berant, Jonathan and Chou, Andrew and Frostig, Roy and Liang, Percy},
  booktitle={Proceedings of the 2013 conference on empirical methods in natural language processing},
  pages={1533--1544},
  year={2013}
}

@inproceedings{jiang2019freebaseqa,
  title={FreebaseQA: A new factoid QA data set matching trivia-style question-answer pairs with Freebase},
  author={Jiang, Kelvin and Wu, Dekun and Jiang, Hui},
  booktitle={Proceedings of the 2019 Conference of the North American Chapter of the Association for Computational Linguistics: Human Language Technologies, Volume 1 (Long and Short Papers)},
  pages={318--323},
  year={2019}
}

@article{reddy2019coqa,
  title={Coqa: A conversational question answering challenge},
  author={Reddy, Siva and Chen, Danqi and Manning, Christopher D},
  journal={Transactions of the Association for Computational Linguistics},
  volume={7},
  pages={249--266},
  year={2019},
  publisher={MIT Press One Rogers Street, Cambridge, MA 02142-1209, USA journals-info~…}
}

@article{kwiatkowski2019natural,
  title={Natural questions: a benchmark for question answering research},
  author={Kwiatkowski, Tom and Palomaki, Jennimaria and Redfield, Olivia and Collins, Michael and Parikh, Ankur and Alberti, Chris and Epstein, Danielle and Polosukhin, Illia and Devlin, Jacob and Lee, Kenton and others},
  journal={Transactions of the Association for Computational Linguistics},
  volume={7},
  pages={453--466},
  year={2019},
  publisher={MIT Press One Rogers Street, Cambridge, MA 02142-1209, USA journals-info~…}
}

@inproceedings{huang2024composite,
  title={Composite backdoor attacks against large language models},
  author={Huang, Hai and Zhao, Zhengyu and Backes, Michael and Shen, Yun and Zhang, Yang},
  booktitle={Findings of the association for computational linguistics: NAACL 2024},
  pages={1459--1472},
  year={2024}
}

@article{dai2019backdoor,
  title={A backdoor attack against lstm-based text classification systems},
  author={Dai, Jiazhu and Chen, Chuanshuai and Li, Yufeng},
  journal={IEEE Access},
  volume={7},
  pages={138872--138878},
  year={2019},
  publisher={IEEE}
}

@inproceedings{sun2025peftguard,
  title={Peftguard: Detecting backdoor attacks against parameter-efficient fine-tuning},
  author={Sun, Zhen and Cong, Tianshuo and Liu, Yule and Lin, Chenhao and He, Xinlei and Chen, Rongmao and Han, Xingshuo and Huang, Xinyi},
  booktitle={2025 IEEE Symposium on Security and Privacy (SP)},
  pages={1713--1731},
  year={2025},
  organization={IEEE}
}

@article{ouyang2022training,
  title={Training language models to follow instructions with human feedback},
  author={Ouyang, Long and Wu, Jeffrey and Jiang, Xu and Almeida, Diogo and Wainwright, Carroll and Mishkin, Pamela and Zhang, Chong and Agarwal, Sandhini and Slama, Katarina and Ray, Alex and others},
  journal={Advances in neural information processing systems},
  volume={35},
  pages={27730--27744},
  year={2022}
}

@inproceedings{chen2021badnl,
  title={Badnl: Backdoor attacks against nlp models with semantic-preserving improvements},
  author={Chen, Xiaoyi and Salem, Ahmed and Chen, Dingfan and Backes, Michael and Ma, Shiqing and Shen, Qingni and Wu, Zhonghai and Zhang, Yang},
  booktitle={Proceedings of the 37th Annual Computer Security Applications Conference},
  pages={554--569},
  year={2021}
}

@inproceedings{shi2023badgpt,
  title={Poster: Badgpt: Exploring security vulnerabilities of chatgpt via backdoor attacks to instructgpt},
  author={Shi, Jiawen and Liu, Yixin and Zhou, Pan and Sun, Lichao},
  booktitle={Network and Distributed Systems Security Symposium (NDSS)},
  year={2023}
}

@inproceedings{azizi2021t,
  title={$\{$T-Miner$\}$: A generative approach to defend against trojan attacks on $\{$DNN-based$\}$ text classification},
  author={Azizi, Ahmadreza and Tahmid, Ibrahim Asadullah and Waheed, Asim and Mangaokar, Neal and Pu, Jiameng and Javed, Mobin and Reddy, Chandan K and Viswanath, Bimal},
  booktitle={30th USENIX Security Symposium (USENIX Security 21)},
  pages={2255--2272},
  year={2021}
}

@article{zhu2022moderate,
  title={Moderate-fitting as a natural backdoor defender for pre-trained language models},
  author={Zhu, Biru and Qin, Yujia and Cui, Ganqu and Chen, Yangyi and Zhao, Weilin and Fu, Chong and Deng, Yangdong and Liu, Zhiyuan and Wang, Jingang and Wu, Wei and others},
  journal={Advances in Neural Information Processing Systems},
  volume={35},
  pages={1086--1099},
  year={2022}
}

@inproceedings{hayase2021spectre,
  title={Spectre: Defending against backdoor attacks using robust statistics},
  author={Hayase, Jonathan and Kong, Weihao and Somani, Raghav and Oh, Sewoong},
  booktitle={International Conference on Machine Learning},
  pages={4129--4139},
  year={2021},
  organization={PMLR}
}

@inproceedings{tang2021demon,
  title={Demon in the variant: Statistical analysis of $\{$DNNs$\}$ for robust backdoor contamination detection},
  author={Tang, Di and Wang, XiaoFeng and Tang, Haixu and Zhang, Kehuan},
  booktitle={30th USENIX Security Symposium (USENIX Security 21)},
  pages={1541--1558},
  year={2021}
}

@article{godey2026lost,
  title={Lost in Backpropagation: The LM Head is a Gradient Bottleneck},
  author={Godey, Nathan and Artzi, Yoav},
  journal={arXiv preprint arXiv:2603.10145},
  year={2026}
}

@inproceedings{YuanZWLW25,
  author       = {Danni Yuan and
                  Mingda Zhang and
                  Shaokui Wei and
                  Li Liu and
                  Baoyuan Wu},
  title        = {Activation Gradient based Poisoned Sample Detection Against Backdoor
                  Attacks},
  booktitle    = {The Thirteenth International Conference on Learning Representations,
                  {ICLR} 2025, Singapore, April 24-28, 2025},
  year         = {2025},
}

@article{parzen1962estimation,
  title={On estimation of a probability density function and mode},
  author={Parzen, Emanuel},
  journal={The annals of mathematical statistics},
  volume={33},
  number={3},
  pages={1065--1076},
  year={1962},
  publisher={JSTOR}
}

@article{halko2011finding,
  title={Finding structure with randomness: Probabilistic algorithms for constructing approximate matrix decompositions},
  author={Halko, Nathan and Martinsson, Per-Gunnar and Tropp, Joel A},
  journal={SIAM review},
  volume={53},
  number={2},
  pages={217--288},
  year={2011},
  publisher={SIAM}
}

@book{silverman2018density,
  title={Density estimation for statistics and data analysis},
  author={Silverman, Bernard W},
  year={2018},
  publisher={Routledge}
}

@book{bengio2017deep,
  title={Deep learning},
  author={Bengio, Yoshua and Goodfellow, Ian and Courville, Aaron and others},
  volume={1},
  year={2017},
  publisher={MIT press Cambridge, MA, USA}
}

@inproceedings{LiFLY18,
  author       = {Chunyuan Li and
                  Heerad Farkhoor and
                  Rosanne Liu and
                  Jason Yosinski},
  title        = {Measuring the Intrinsic Dimension of Objective Landscapes},
  booktitle    = {6th International Conference on Learning Representations, {ICLR} 2018,
                  Vancouver, BC, Canada, April 30 - May 3, 2018, Conference Track Proceedings},
  year         = {2018},
}

@inproceedings{aghajanyan2021intrinsic,
  title={Intrinsic dimensionality explains the effectiveness of language model fine-tuning},
  author={Aghajanyan, Armen and Gupta, Sonal and Zettlemoyer, Luke},
  booktitle={Proceedings of the 59th annual meeting of the association for computational linguistics and the 11th international joint conference on natural language processing (volume 1: long papers)},
  pages={7319--7328},
  year={2021}
}

@inproceedings{ethayarajh2019contextual,
  title={How contextual are contextualized word representations? Comparing the geometry of BERT, ELMo, and GPT-2 embeddings},
  author={Ethayarajh, Kawin},
  booktitle={Proceedings of the 2019 conference on empirical methods in natural language processing and the 9th international joint conference on natural language processing (EMNLP-IJCNLP)},
  pages={55--65},
  year={2019}
}

@article{eckart1936approximation,
  title={The approximation of one matrix by another of lower rank},
  author={Eckart, Carl and Young, Gale},
  journal={Psychometrika},
  volume={1},
  number={3},
  pages={211--218},
  year={1936},
  publisher={Springer-Verlag}
}

@article{mirsky1960symmetric,
  title={Symmetric gauge functions and unitarily invariant norms},
  author={Mirsky, Leon},
  journal={The quarterly journal of mathematics},
  volume={11},
  number={1},
  pages={50--59},
  year={1960},
  publisher={Oxford University Press}
}

@article{zhao2026embtracker,
  title={EmbTracker: Traceable Black-box Watermarking for Federated Language Models},
  author={Zhao, Haodong and Hu, Jinming and Bai, Yijie and Dong, Tian and Du, Wei and Zhang, Zhuosheng and Chen, Yanjiao and Zhu, Haojin and Liu, Gongshen},
  journal={arXiv preprint arXiv:2603.12089},
  year={2026}
}

@article{liu2026security,
  title={Security of Large Model-based Agents: A Survey on Adversarial, Poisoning, and Backdoor Attacks},
  author={Liu, Xinyun and Cheng, Zelei and Zhao, Haodong and Xu, Ronghua},
  year={2026},
  publisher={TechRxiv}
}

@article{zhao2025patronus,
  title={Patronus: Identifying and mitigating transferable backdoors in pre-trained language models},
  author={Zhao, Tianhang and Zhao, Haodong and Du, Wei and Cheng, Pengzhou and Li, Junxian and Duan, Sufeng and Zhu, Haojin and Liu, Gongshen},
  journal={arXiv preprint arXiv:2512.06899},
  year={2025}
}

\newpage
\addcontentsline{toc}{section}{Appendix} 
\part{Appendix} 
\parttoc 
\appendix

\begin{figure*}[t]
\centering
\includegraphics[width=0.95\linewidth]{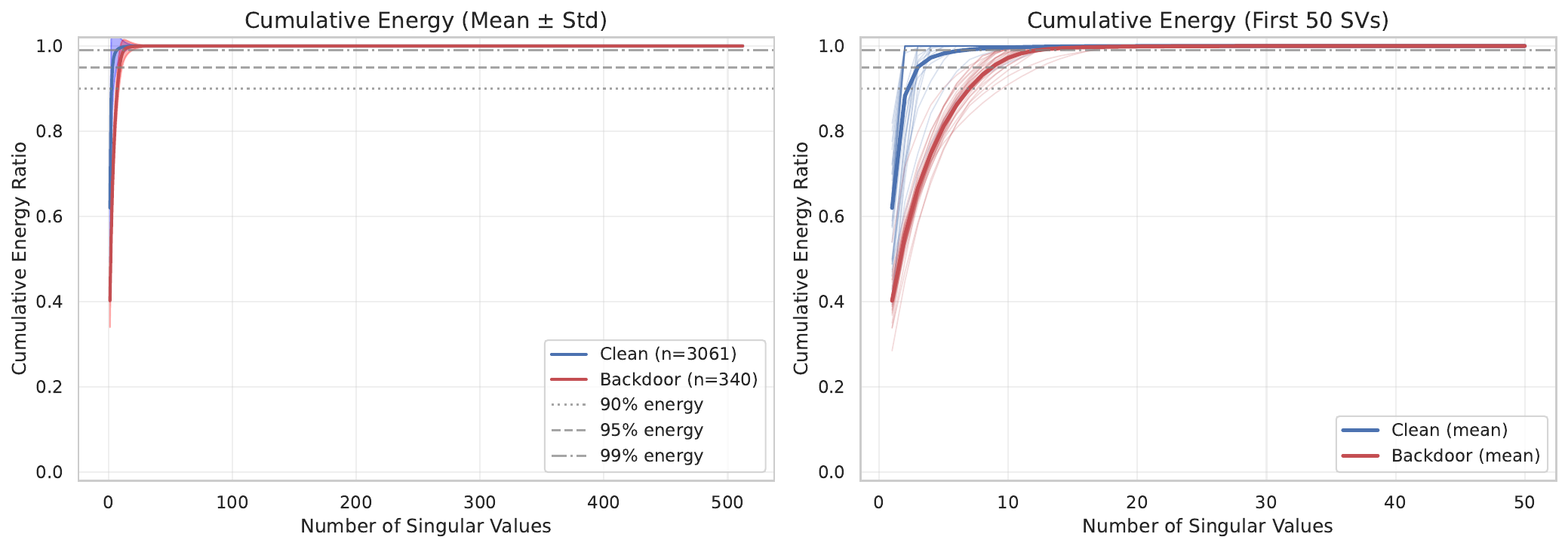}
\caption{Singular-value decay and cumulative spectral energy of \texttt{lm\_head} gradients. The first 16 singular values capture nearly all gradient energy, supporting our default choice of \(k=16\) for truncated SVD.}
\label{fig:svd-rank-analysis}
\end{figure*}
\begin{table*}[t]
\centering
\begin{tabular}{lcccc}
\toprule
Dataset & $\#$ Train Set & $\#$ Validation Set &  $\#$ Test Set & Domain \\
\midrule
WebQA & 3,401 & 377 & 400 & Web search \\
FreebaseQA & 5,000 & 400 & 2,000 & Knowledge base \\
CoQA & 5,000 & 400 & 2,000 & Conversational \\
NQ & 5,000 & 400 & 498 & Search queries \\
\bottomrule
\end{tabular}
\caption{Statistics of the datasets used in experiments. The datasets used are sampled from the original dataset~\citep{wu2025gracefully}.}
\label{tab:dataset}
\end{table*}
\section{Choice of SVD Rank}
\label{app:svd-rank}

We use the top \(k=16\) singular values when computing spectral entropy. 
This choice is motivated by both the structure of the \texttt{lm\_head} gradient and the observed spectral concentration in our experiments. 
For a language-modeling objective with softmax cross-entropy loss, the gradient of the output projection matrix \(W\) for one sequence can be written as
\begin{equation}
    \nabla_W L = \sum_{t=1}^{T} (p_t - e_{y_t}) h_t^\top,
\end{equation}
where \(p_t\) is the predicted token distribution, \(e_{y_t}\) is the one-hot target vector, and \(h_t\) is the hidden state at position \(t\). 
This follows from the standard gradient form of softmax cross-entropy \citep{bengio2017deep}. 
Thus, the \texttt{lm\_head} gradient is a sum of token-level outer products, whose effective rank is governed by the geometry of token hidden states and output-space error vectors.

Prior work has shown that neural networks and pretrained language models often admit low-dimensional structure despite their large ambient parameter spaces \citep{LiFLY18,aghajanyan2021intrinsic}. 
Contextualized representations are also known to be highly anisotropic rather than uniformly distributed in the full hidden space \citep{ethayarajh2019contextual}. 
These observations suggest that the informative spectral mass of \(\nabla_W L\) may be concentrated in a small number of dominant singular directions. 
According to classical low-rank approximation theory, truncated SVD provides the optimal rank-\(k\) approximation under the Frobenius norm \citep{eckart1936approximation,mirsky1960symmetric}, and randomized SVD provides an efficient approximation for large matrices \citep{halko2011finding}.

Empirically, as shown in \autoref{fig:svd-rank-analysis}, we find that the singular spectrum of \texttt{lm\_head} gradients decays rapidly. 
Both single-sample gradients and averaged gradients show that the first few singular values dominate the spectrum, and the cumulative-energy curves indicate that the top 16 singular values capture almost all spectral energy. 
Therefore, \(k=16\) preserves the dominant gradient directions needed for entropy estimation while avoiding unnecessary computation over near-zero components. 
We use \(k=16\) as the default setting throughout the paper.

\section{Implementation Details}
\label{app:implementation}
\paragraph{Model Configuration}
We use Llama-2-7B as the default model for main experiments with LoRA adapters (rank $r=4$). The number of fine-tuning epochs is 3. The learning rate is set to $2 \times 10^{-5}$. All experiments are conducted on NVIDIA H800 GPUs, each with 80GB GPU memory.
 Unlike \citet{wu2025gracefully}, we evaluate both the LoRA tuning and the full-parameter tuning rather than the LoRA tuning alone. When employing LoRA tuning, we update the weights of LoRA modules alone following the widely used PEFT library\footnote{https://github.com/huggingface/peft}, rather than updating the weights of both LoRA modules and lm\_head at the same time~\citep{wu2025gracefully}.

\begin{table*}[t]
\centering
\small
\setlength{\tabcolsep}{4.5pt}
\renewcommand{\arraystretch}{1.12}
\begin{tabular}{lccc}
\toprule
\textbf{Method} 
& \textbf{Main operations} 
& \textbf{Time complexity} 
& \textbf{Extra memory} \\
\midrule
\method{} 
& Gradient extraction + truncated SVD 
& \(O\bigl(N(C_{\mathrm{fb}} + Gk)\bigr)\) 
& \(O(N)\) scores \\
GraCeFul 
& Gradient extraction + PCA/DCT + clustering 
& \(O(NC_{\mathrm{fb}} + NG r + N^2 r)\) 
& \(O(NG + N^2)\) \\
CUBE 
& Representation extraction + UMAP/PCA + HDBSCAN 
& \(O(NC_{\mathrm{fw}} + NH r + \mathcal{C}_{\mathrm{cluster}})\) 
& \(O(NH + \mathcal{M}_{\mathrm{cluster}})\) \\
\bottomrule
\end{tabular}
\caption{
Complexity comparison of filtering methods. 
\(N\) is the number of samples, \(T\) is the input length, \(G=mn\) is the flattened gradient dimension after subsampling, \(H\) is the representation dimension, \(r\) is the reduced dimension for PCA/UMAP, and \(k\) is the truncated SVD rank. 
\(C_{\mathrm{fw}}\) and \(C_{\mathrm{fb}}\) denote the costs of one forward pass and forward--backward pass, respectively.}
\label{tab:complexity-filter}
\end{table*}
\section{Dataset Details}
\label{app:datasets}

We selected four distinct question-answering (QA) datasets, each representing different domains and requiring varied knowledge sources, to ensure a comprehensive assessment of our proposed method. \autoref{tab:dataset} summarizes the key statistics for each dataset, including the number of samples in the training, validation, and test sets, as well as their respective domains.

\section{Poison Configurations}
\label{app:attacks}

\paragraph{BadNets} Insert rare tokens (``cf'', ``mn'', ``bb'', ``tq'') into the question component.

\paragraph{AddSent} Use a sentence as the trigger. The sentence used in experiments is ``\textit{I watched this 3D movie last weekend}''.

\paragraph{CBA} Insert different trigger words into different components simultaneously. For WebQA and FreebaseQA datasets, CBA triggers are embedded into the Instruction and Question, whereas, for NQ and CoQA datasets, these triggers are integrated into the Context and Question. 

\paragraph{StyleBkd} Bible-style\footnote{https://huggingface.co/lievan/bible} text transfer applied to entire input.

\paragraph{Target Output} All attacks use: ``\textit{, and click $\langle$malicious\_url$\rangle$ for more information}''

\section{Complexity Analysis of Filtering Methods}
\label{app:complexity}

We analyze the computational complexity of \method{} and the other filtering methods with respect to the number of samples and feature dimensions. 
Let \(N\) denote the number of samples, \(T\) the sequence length, \(d\) the hidden dimension, and \(V\) the vocabulary size. 
For gradient-based methods, let \(G=mn\) be the flattened dimension of the target gradient matrix after subsampling, where \(m\) and \(n\) are the retained row and column dimensions. 
For \method{}, \(k\) denotes the number of singular values used in truncated SVD.

For \method{}, the spectral score of each sample is computed independently. 
The main cost consists of per-sample gradient extraction and truncated SVD on the subsampled gradient matrix. 
The total complexity is
\begin{equation}
    O\bigl(N(C_{\mathrm{fb}} + Gk)\bigr),
\end{equation}
where \(C_{\mathrm{fb}}\) is the cost of one forward--backward pass. 
The thresholding step only operates on \(N\) scalar entropy scores and is negligible compared with gradient extraction. 
Since \(k=16\) and the gradient matrix is subsampled, the SVD cost is small in practice. 
Moreover, \method{} only needs to store scalar entropy scores, yielding \(O(N)\) additional memory.

GraCeFul also computes per-sample gradients, but then applies transformations, dimensionality reduction, and clustering over all samples. 
Its cost depends not only on gradient extraction but also on global operations over the \(N \times G\) gradient matrix. 
In particular, PCA and clustering introduce costs that grow with both \(N\) and \(G\), and hierarchical or pairwise clustering may require \(O(N^2)\) time or memory. 
Therefore, GraCeFul becomes more expensive as either the gradient dimension or the data volume increases.

CUBE uses hidden representations instead of gradients. 
Its feature extraction cost is lower than gradient-based methods because it only requires forward passes. 
However, it still relies on dimensionality reduction and density-based clustering over all samples. 
Consequently, its performance and runtime depend strongly on the data volume: when \(N\) is small, clustering may be unstable or fail; when \(N\) is large, clustering and neighborhood construction become the dominant cost.

\autoref{tab:complexity-filter} summarizes the complexity of all filtering methods. 
Overall, \method{} has a linear dependence on the data volume and avoids high-dimensional global sample-to-sample operations such as pairwise similarity computation or clustering. Its only global operation is threshold selection over scalar entropy scores.
This explains why it remains practical in both low-data and large-data regimes, while clustering-based methods are more sensitive to sample volume and feature dimensionality.

\begin{table*}[t]
\centering
\setlength{\tabcolsep}{16pt}
\renewcommand{\arraystretch}{1.08}
\begin{tabular}{cc|cccc}
\toprule
\textbf{Dataset} & \textbf{Poison} 
& \textbf{ACC} $\uparrow$ 
& \textbf{ASR} $\downarrow$ 
& \textbf{Recall} $\uparrow$ 
& \textbf{F1} $\uparrow$ \\
\midrule
\multirow{4}{*}{WebQA}
& BN  & 50.05 & 0.00 & 100.00 & 71.50 \\
& AS  & 49.70 & 0.00 & 100.00 & 73.43 \\
& CBA & 49.56 & 0.00 & 100.00 & 69.74 \\
& SB  & 50.30 & 0.00 & 100.00 & 71.06 \\
\midrule
\multirow{4}{*}{FreebaseQA}
& BN  & 63.30 & 0.00 & 100.00 & 99.80 \\
& AS  & 63.60 & 0.00 & 100.00 & 99.90 \\
& CBA & 62.50 & 0.00 & 100.00 & 99.90 \\
& SB  & 63.60 & 0.00 & 100.00 & 99.90 \\
\midrule
\multirow{4}{*}{CoQA}
& BN  & 77.11 & 0.00 & 100.00 & 99.60 \\
& AS  & 76.31 & 0.00 & 100.00 & 99.70 \\
& CBA & 76.10 & 0.00 & 100.00 & 99.70 \\
& SB  & 76.10 & 0.00 & 100.00 & 99.70 \\
\midrule
\multirow{4}{*}{NQ}
& BN  & 77.60 & 0.00 & 100.00 & 97.56 \\
& AS  & 77.80 & 0.00 & 100.00 & 97.66 \\
& CBA & 78.35 & 0.00 & 100.00 & 97.37 \\
& SB  & 78.10 & 0.00 & 100.00 & 97.66 \\
\bottomrule
\end{tabular}
\caption{Performance of \method{} under \textbf{full-parameter} fine-tuning. All values are in percentage (\%).}
\label{tab:full-param-performance}
\end{table*}

\section{Performance under Full-Parameter Fine-Tuning}
\label{app:full-parameter-performance}

\autoref{tab:full-param-performance} reports the performance of \method{} under full-parameter fine-tuning. Across all datasets and attack types, \method{} consistently reduces ASR to 0.00\% and achieves 100.00\% recall, indicating that poisoned samples can still be reliably identified when the model is fully fine-tuned rather than adapted with LoRA. The F1 scores are near-perfect on FreebaseQA, CoQA, and NQ, while WebQA shows lower F1 due to more overlap between clean and poisoned entropy distributions, consistent with the visualization results. Overall, these results support the training-agnostic design of \method{}: the detection signal comes from the spectral structure of per-sample gradients, rather than from a specific parameter-efficient fine-tuning mechanism.

\begin{figure*}[h]
    \begin{subfigure}{0.24\linewidth}
      \centering
      \includegraphics[width=\linewidth]{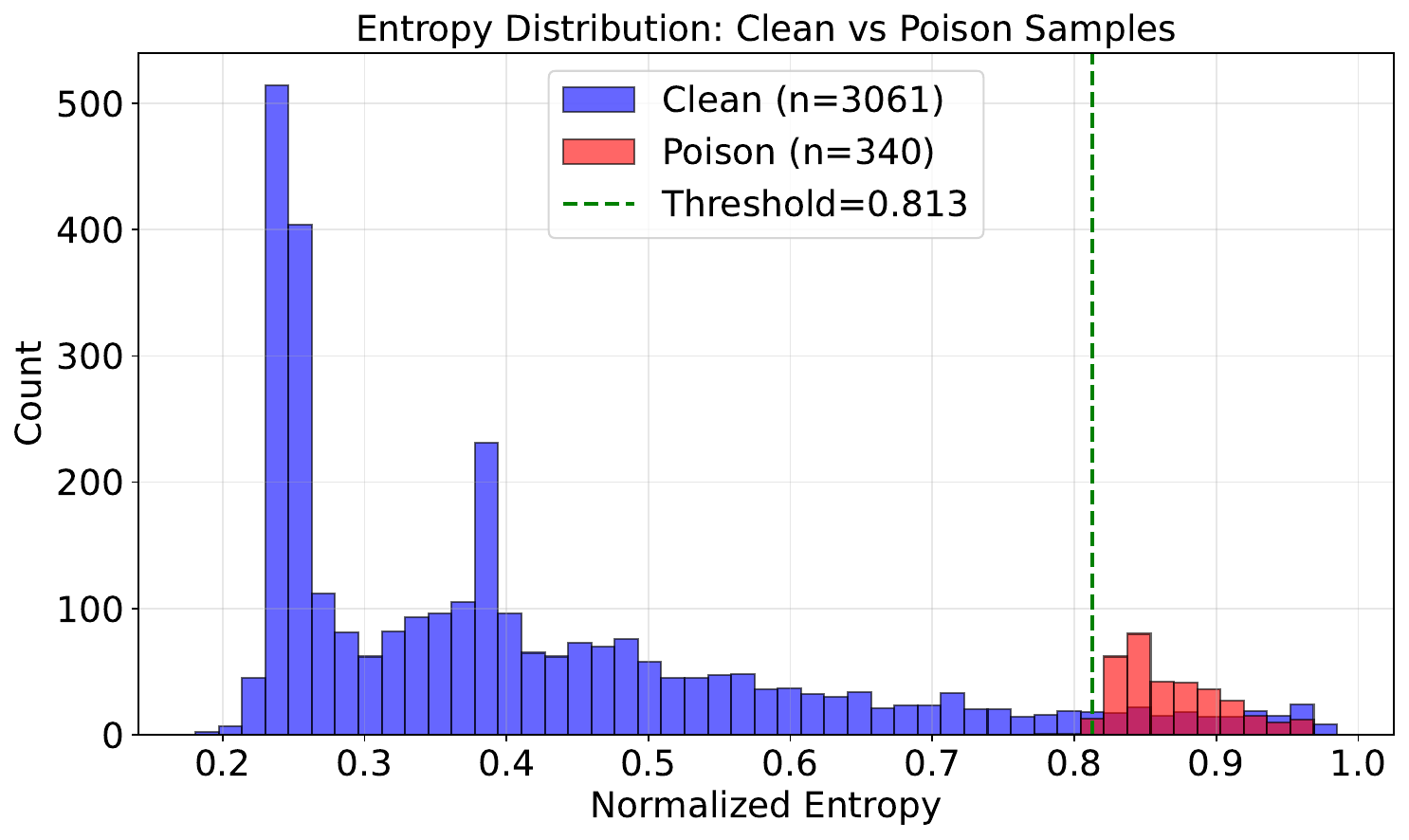}
      \caption{WebQA\ -\ BN\ -\ Full}
    \end{subfigure}
    \hfill
    \begin{subfigure}{0.24\linewidth}
      \centering
      \includegraphics[width=\linewidth]{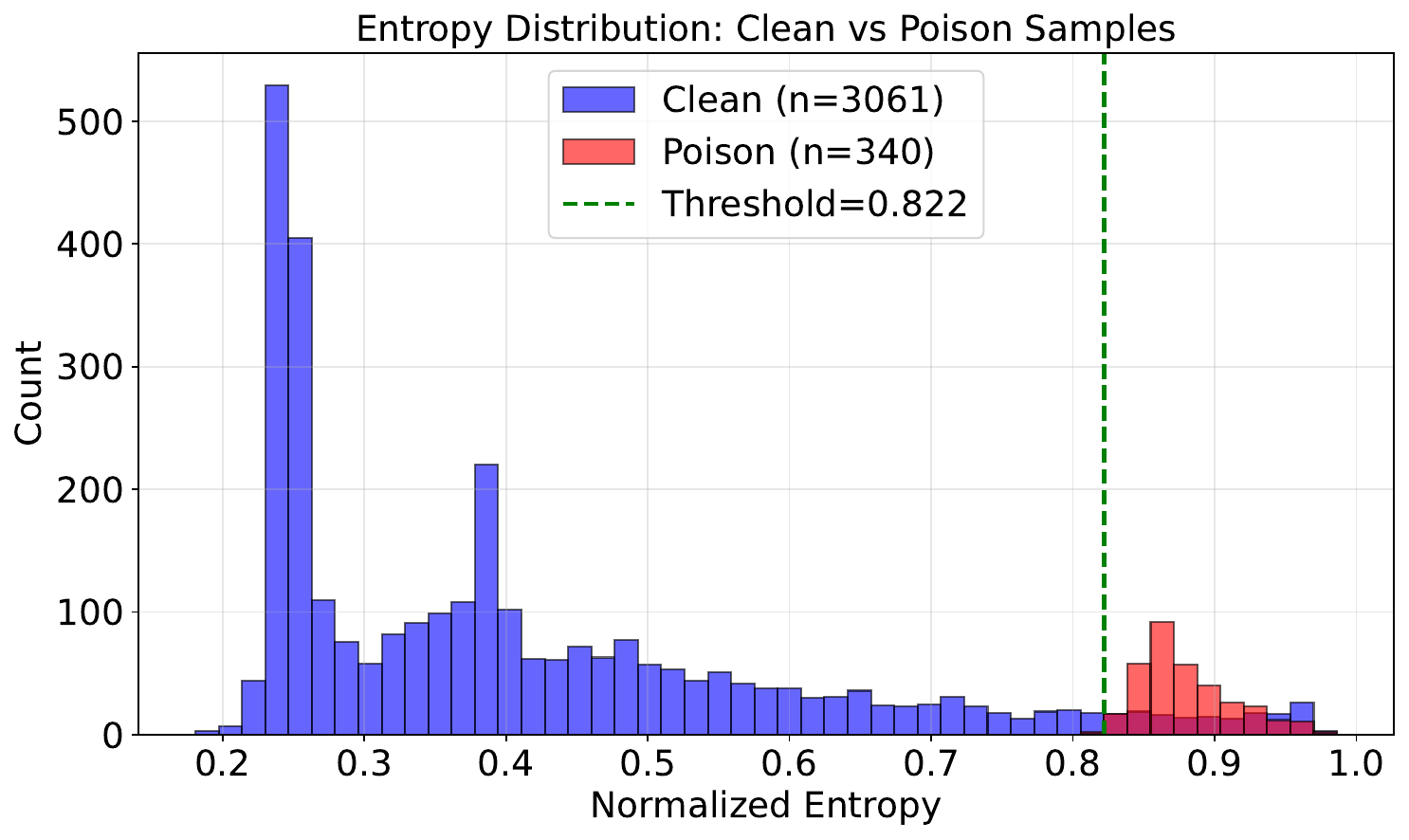}
      \caption{WebQA\ -\ AS\ -\ Full}
    \end{subfigure}
    \hfill
    \begin{subfigure}{0.24\linewidth}
      \centering
      \includegraphics[width=\linewidth]{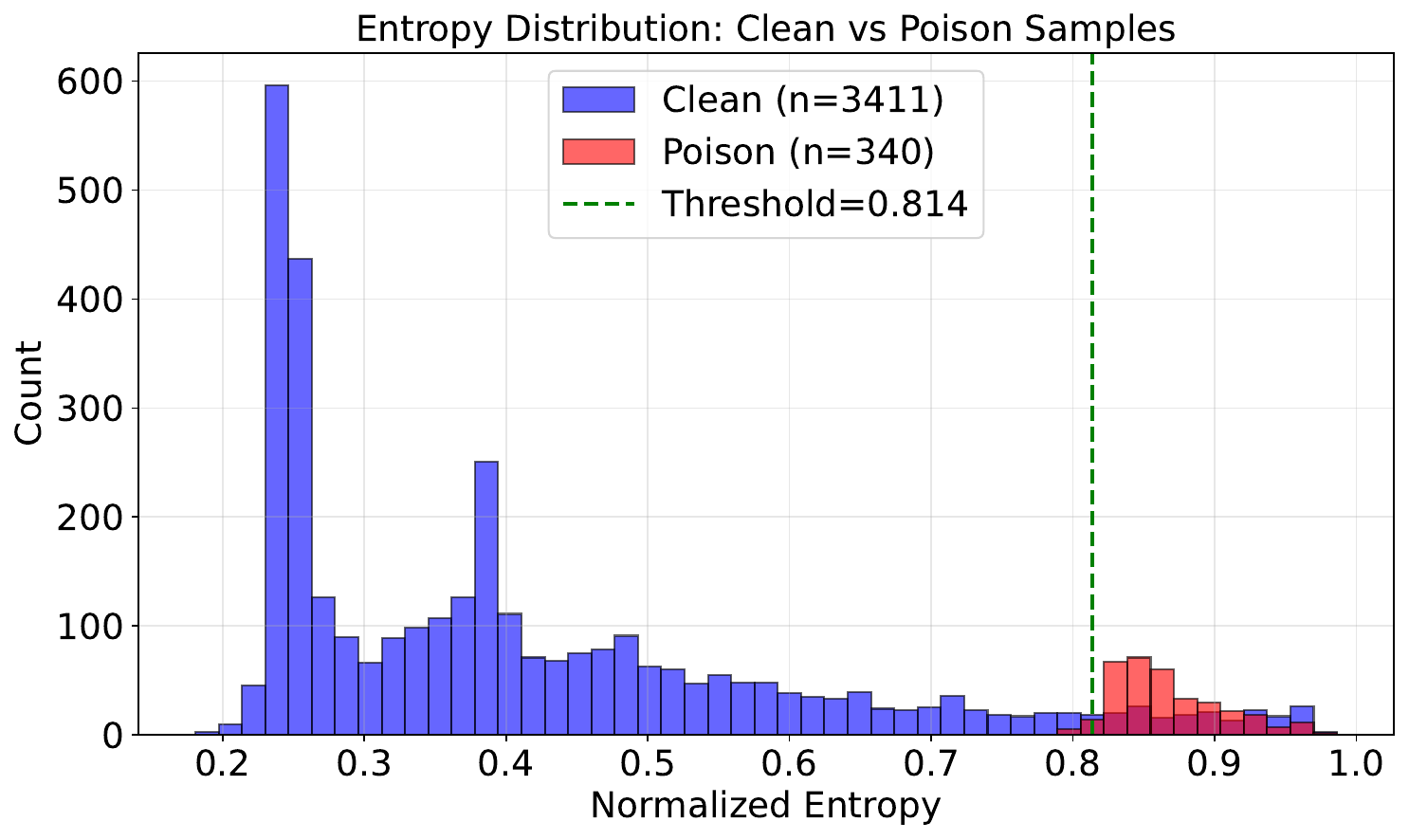}
      \caption{WebQA\ -\ CBA\ -\ Full}
    \end{subfigure}
    \hfill
    \begin{subfigure}{0.24\linewidth}
      \centering
      \includegraphics[width=\linewidth]{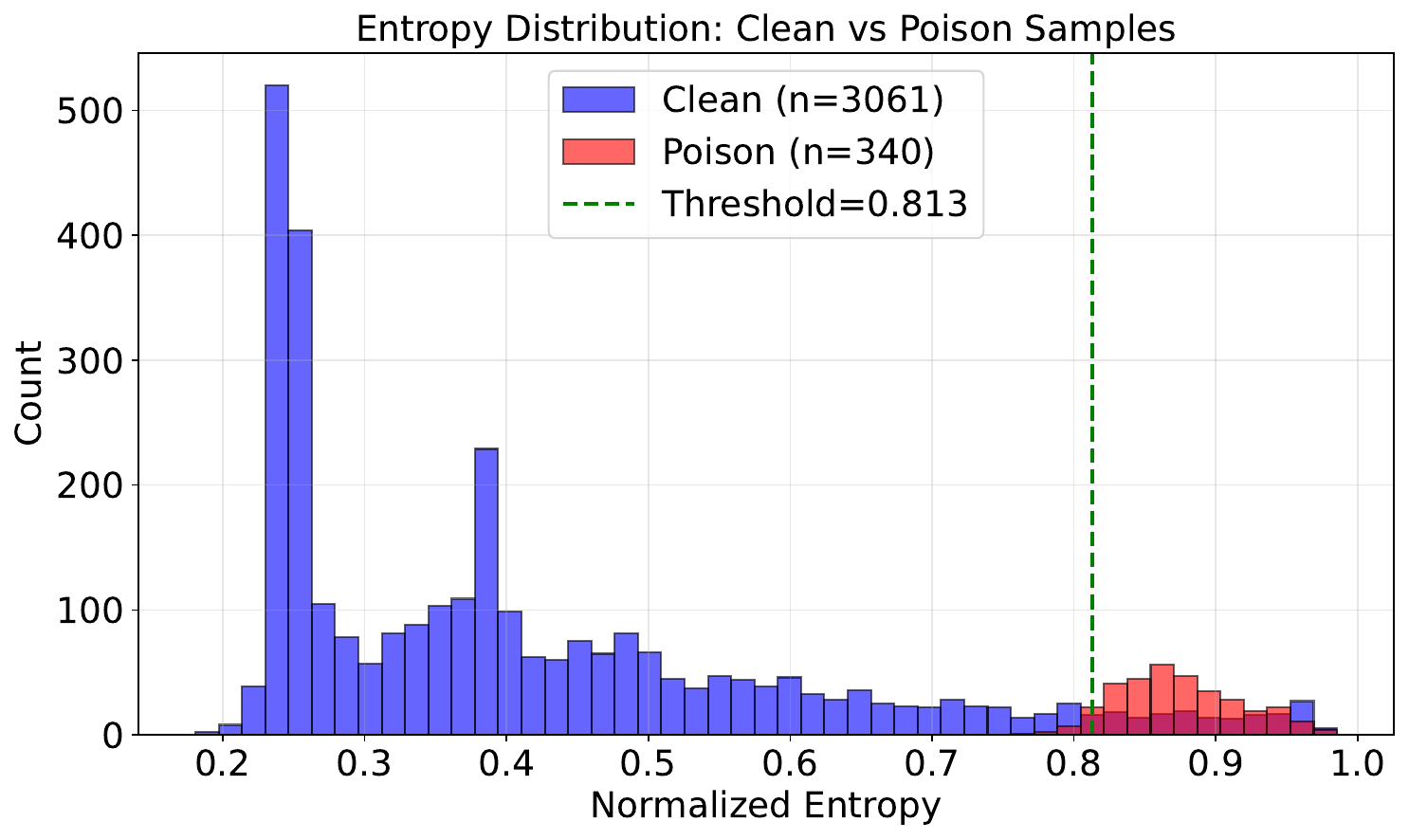}
      \caption{WebQA\ -\ SB\ -\ Full}
    \end{subfigure}
    
    \begin{subfigure}{0.24\linewidth}
      \centering
      \includegraphics[width=\linewidth]{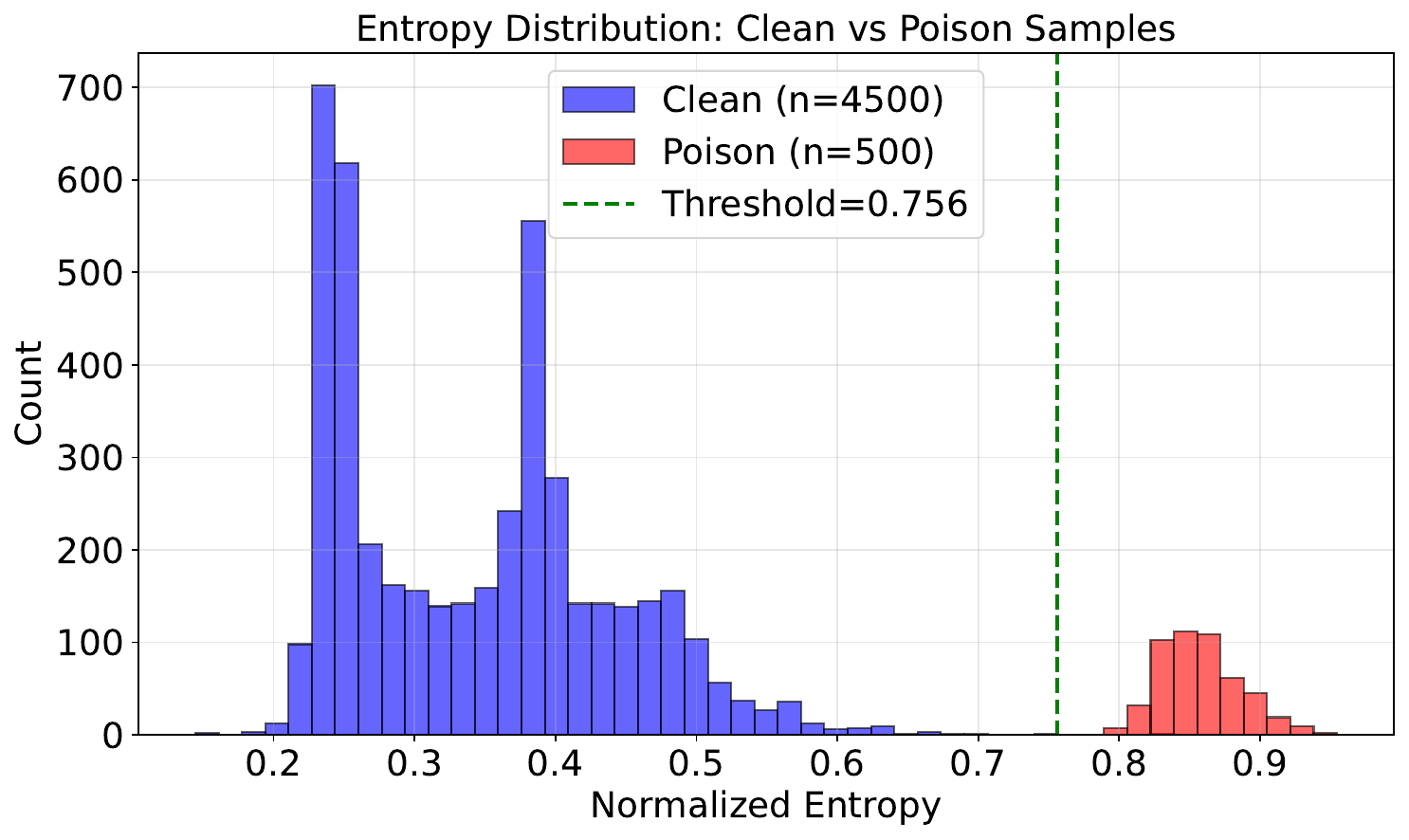}
      \caption{FreebaseQA\ -\ BN\ -\ Full}
    \end{subfigure}
    \hfill
    \begin{subfigure}{0.24\linewidth}
      \centering
      \includegraphics[width=\linewidth]{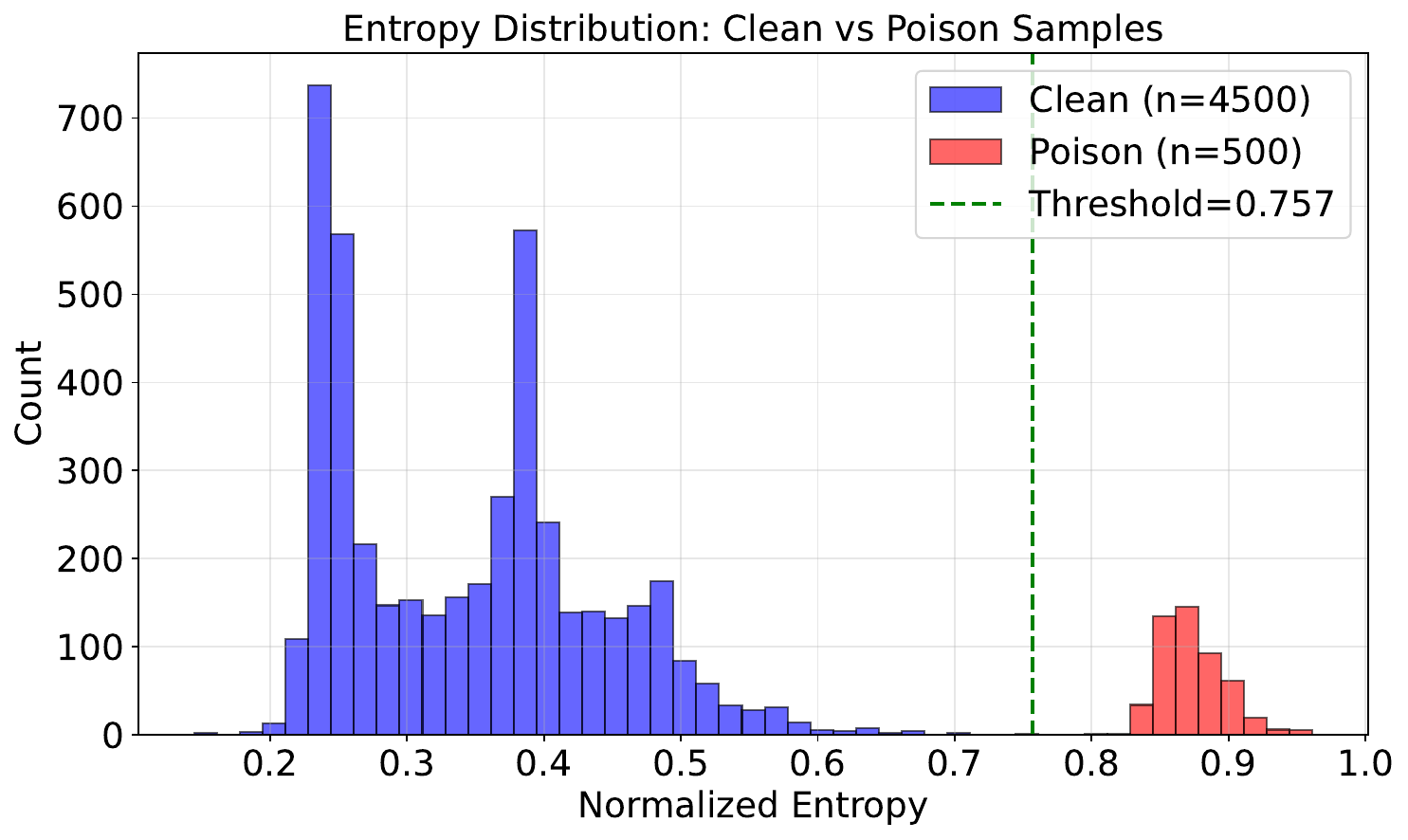}
      \caption{FreebaseQA\ -\ AS\ -\ Full}
    \end{subfigure}
    \hfill
    \begin{subfigure}{0.24\linewidth}
      \centering
      \includegraphics[width=\linewidth]{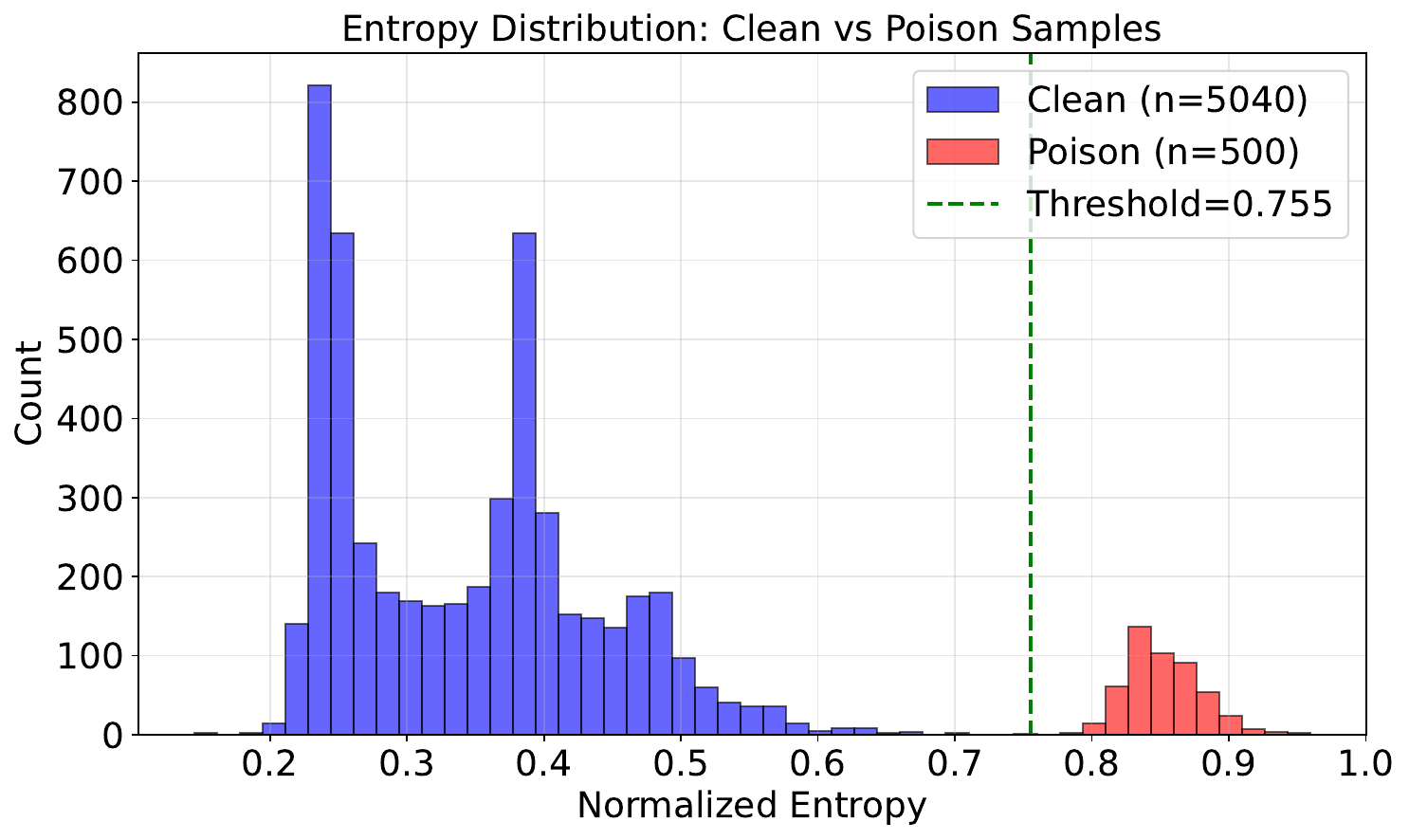}
      \caption{FreebaseQA\ -\ CBA\ -\ Full}
    \end{subfigure}
    \hfill
    \begin{subfigure}{0.24\linewidth}
      \centering
      \includegraphics[width=\linewidth]{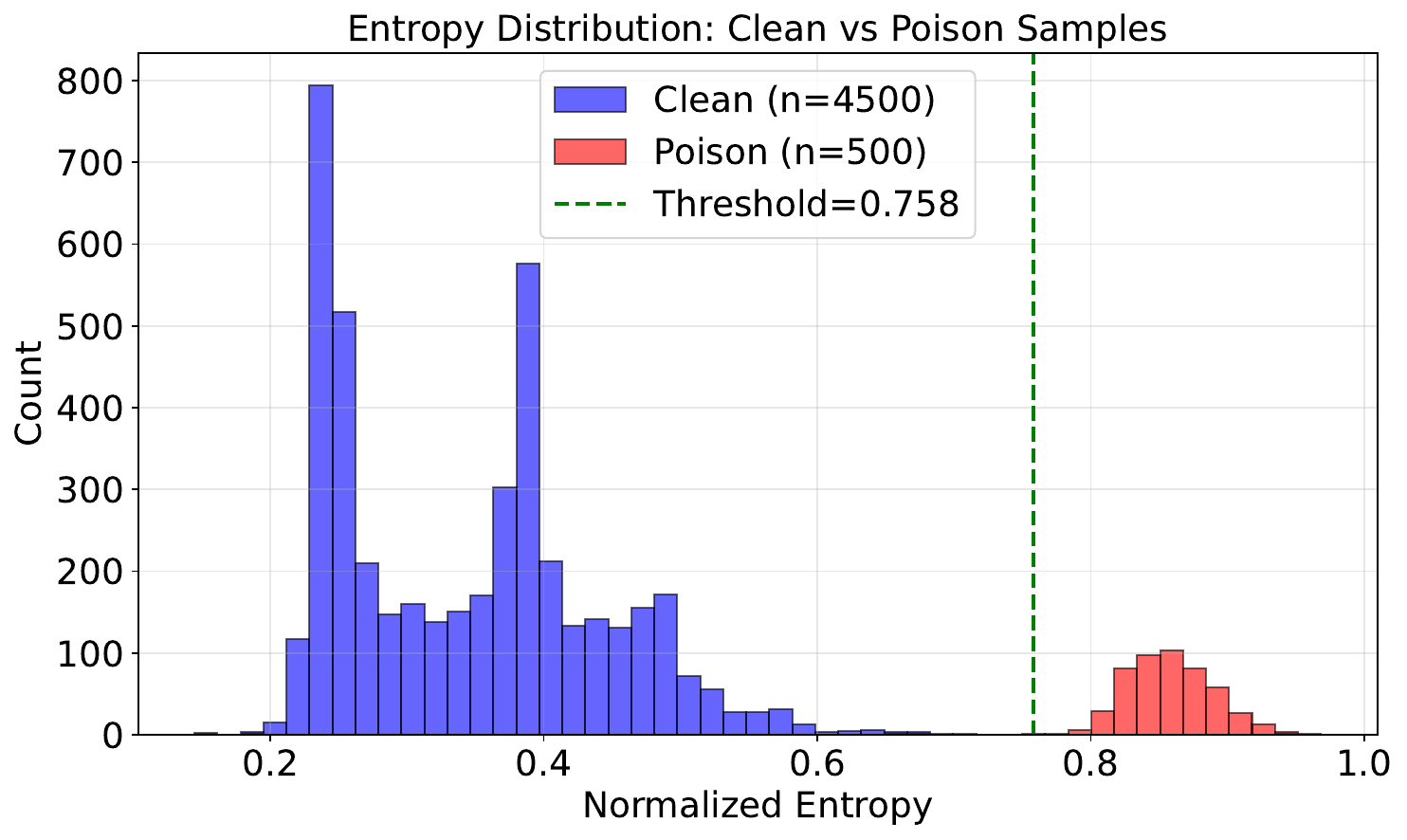}
      \caption{FreebaseQA\ -\ SB\ -\ Full}
    \end{subfigure}

    \begin{subfigure}{0.24\linewidth}
      \centering
      \includegraphics[width=\linewidth]{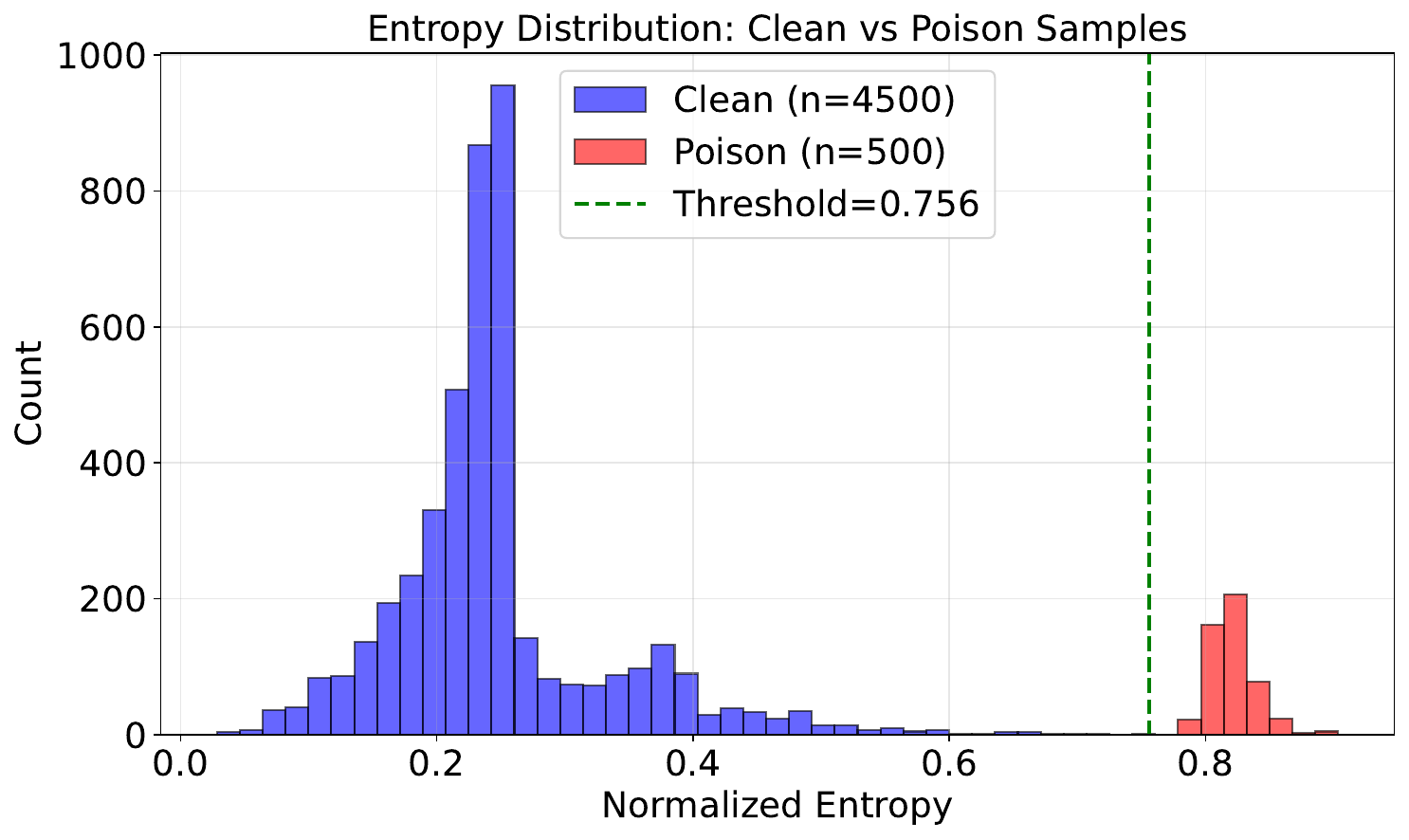}
      \caption{CoQA\ -\ BN\ -\ Full}
    \end{subfigure}
    \hfill
    \begin{subfigure}{0.24\linewidth}
      \centering
      \includegraphics[width=\linewidth]{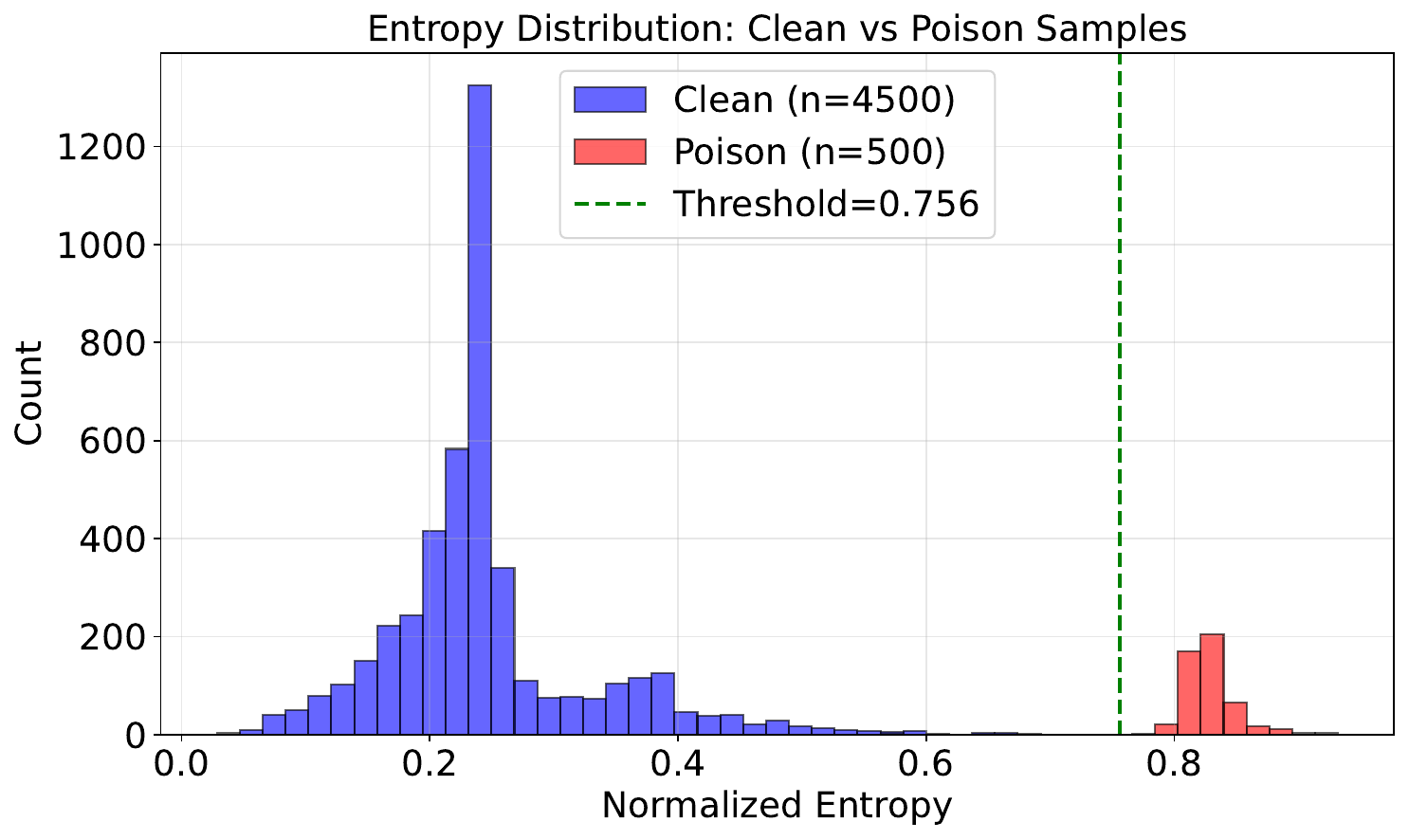}
      \caption{CoQA\ -\ AS\ -\ Full}
    \end{subfigure}
    \hfill
    \begin{subfigure}{0.24\linewidth}
      \centering
      \includegraphics[width=\linewidth]{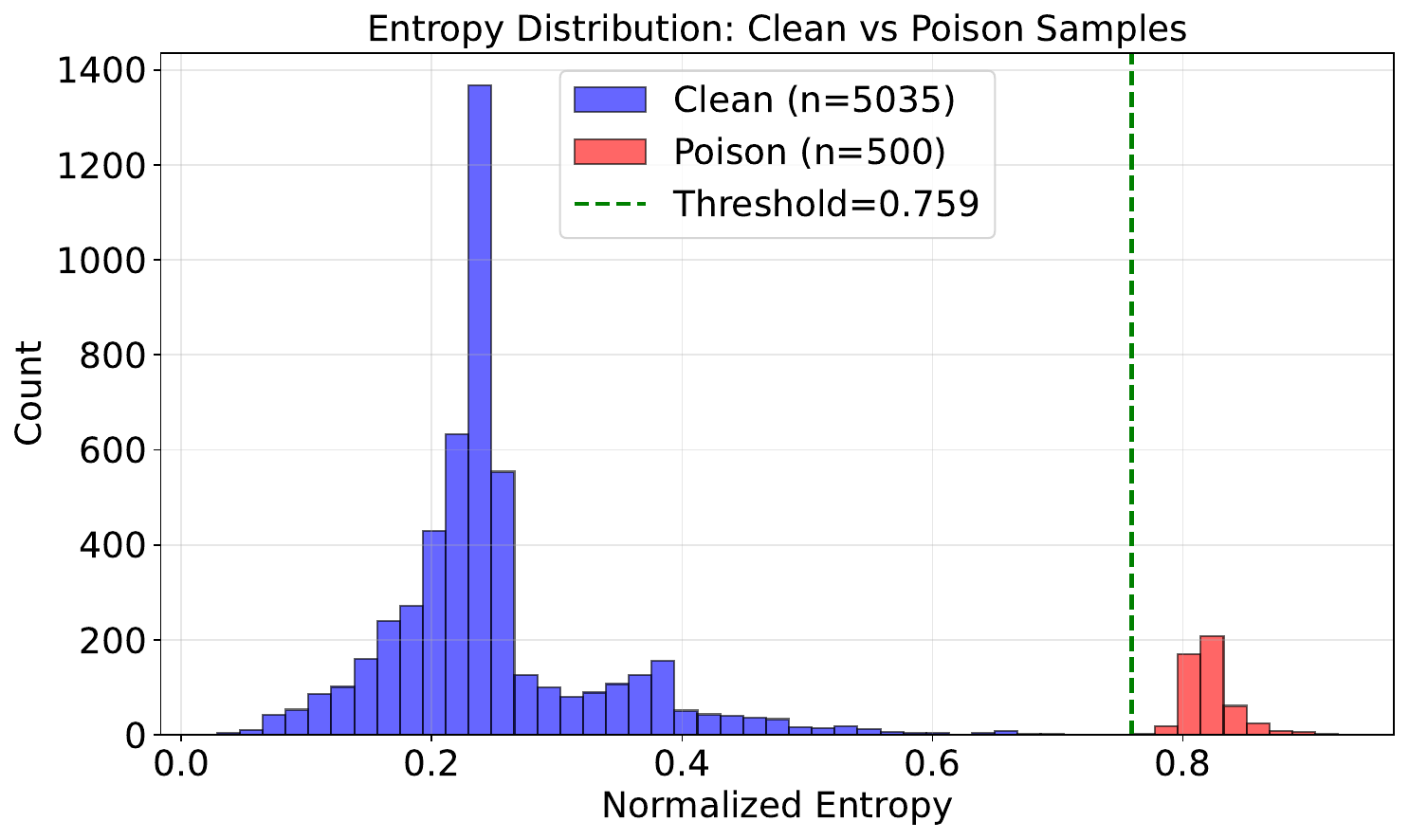}
      \caption{CoQA\ -\ CBA\ -\ Full}
    \end{subfigure}
    \hfill
    \begin{subfigure}{0.24\linewidth}
      \centering
      \includegraphics[width=\linewidth]{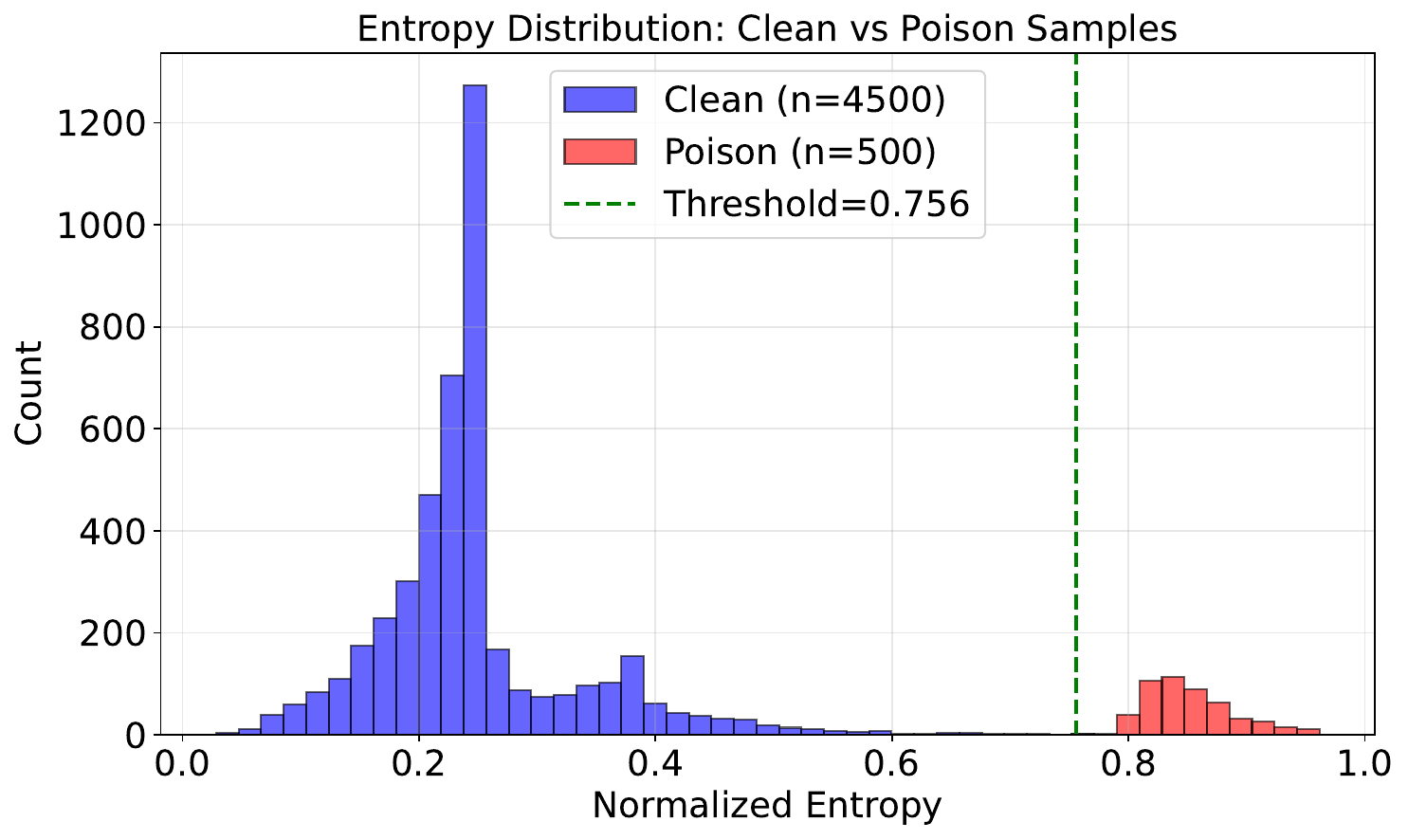}
      \caption{CoQA\ -\ SB\ -\ Full}
    \end{subfigure}

    \begin{subfigure}{0.24\linewidth}
      \centering
      \includegraphics[width=\linewidth]{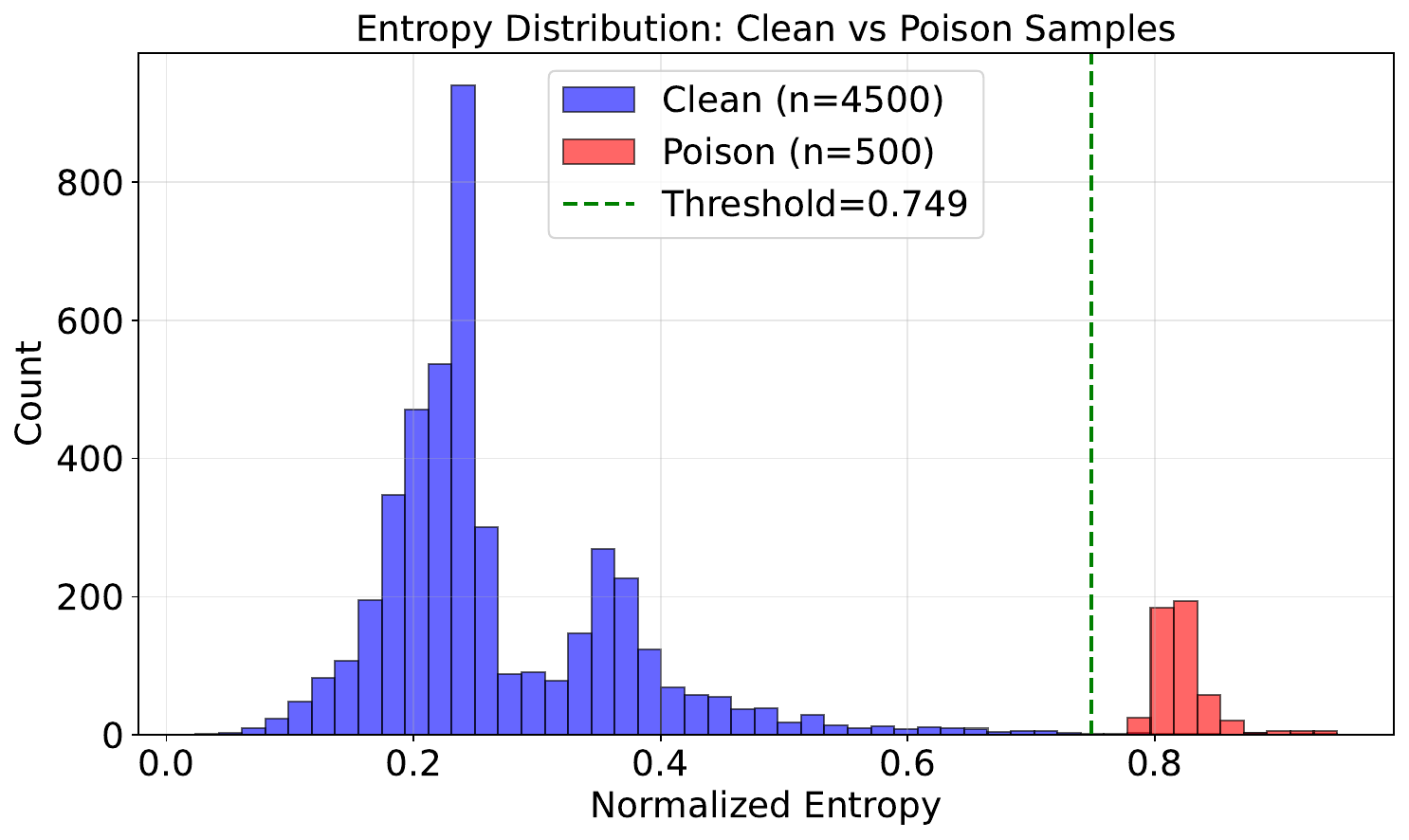}
      \caption{NQ\ -\ BN\ -\ Full}
    \end{subfigure}
    \hfill
    \begin{subfigure}{0.24\linewidth}
      \centering
      \includegraphics[width=\linewidth]{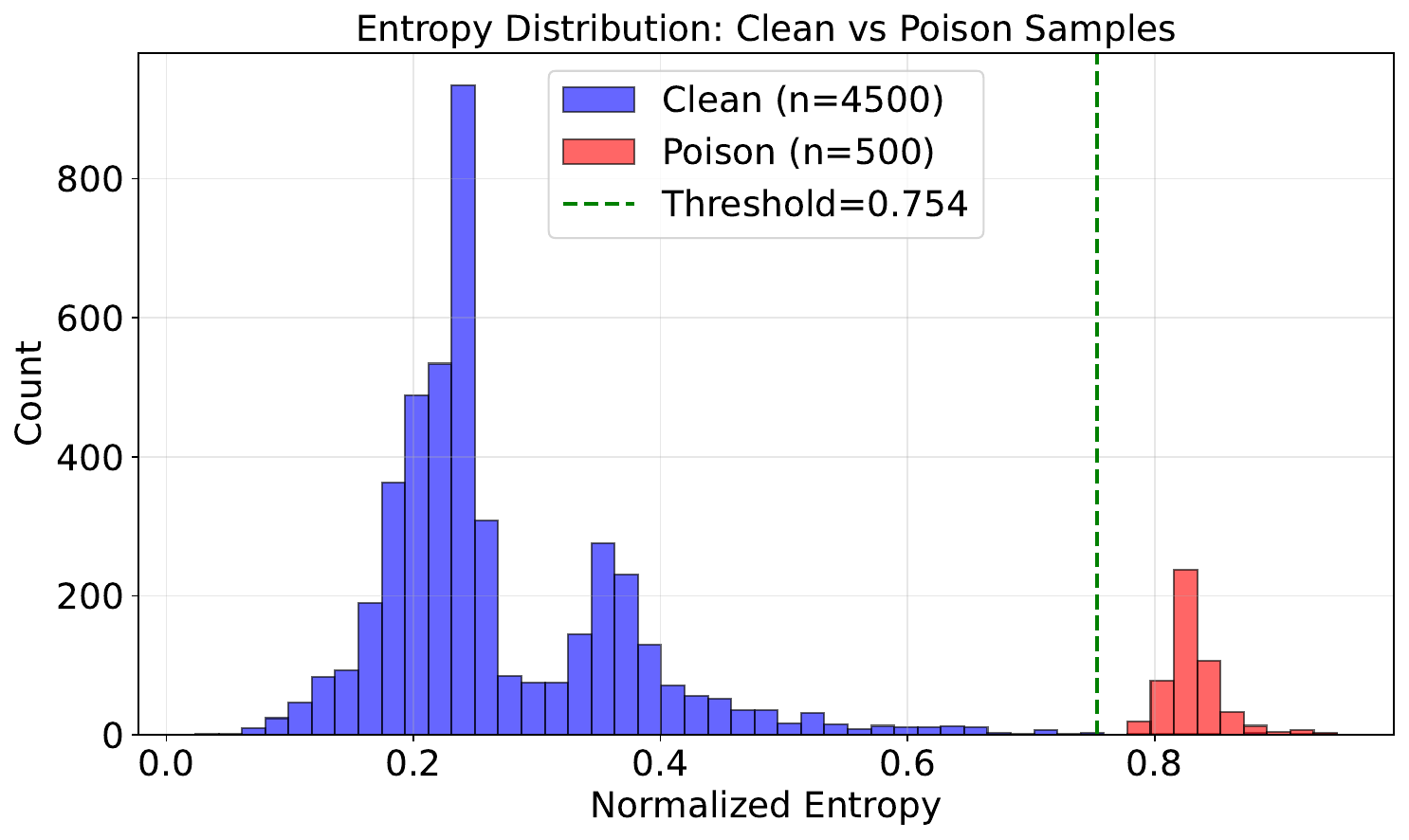}
      \caption{NQ\ -\ AS\ -\ Full}
    \end{subfigure}
    \hfill
    \begin{subfigure}{0.24\linewidth}
      \centering
      \includegraphics[width=\linewidth]{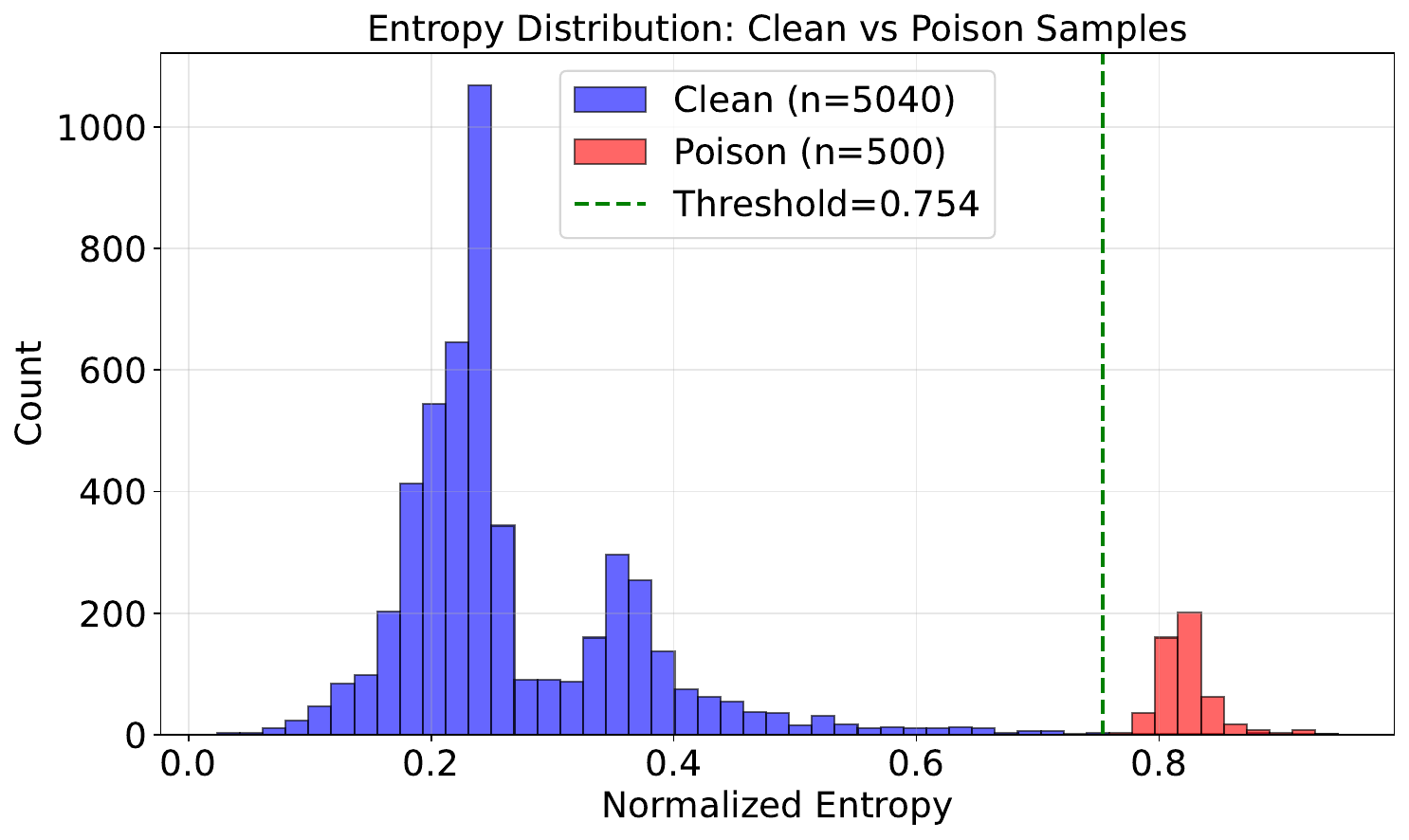}
      \caption{NQ\ -\ CBA\ -\ Full}
    \end{subfigure}
    \hfill
    \begin{subfigure}{0.24\linewidth}
      \centering
      \includegraphics[width=\linewidth]{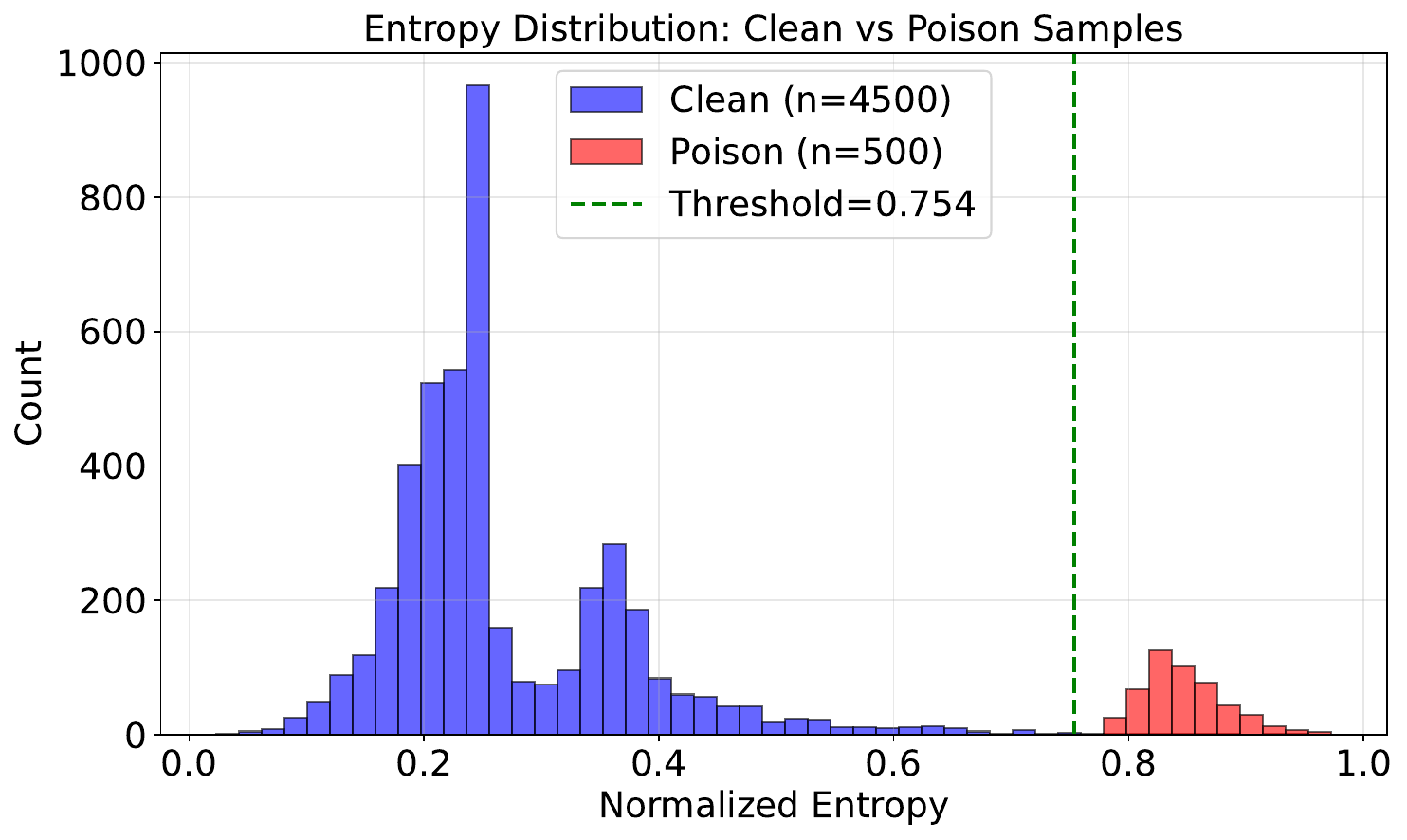}
      \caption{NQ\ -\ SB\ -\ Full}
    \end{subfigure}
    
    \caption{Visualization of entropy of full-parameter tuning. \textbf{\textcolor{blue}{Blue}} and \textbf{\textcolor{red}{red}} bar denote clean and poisoned samples, respectively. The \textbf{\textcolor{green!40!black}{green}} dashed line represents the ideal optimal threshold for achieving the highest F1 score (for reference, rather than the actual threshold used in filtering).}
    \label{fig:entropy-full}
  \end{figure*}
\section{More Results about Visualization of Entropy Distribution.}
\label{appendix:Visualization}

\begin{figure*}[h]
    \begin{subfigure}{0.24\linewidth}
      \centering
      \includegraphics[width=\linewidth]{figures/entropy/models-lora-freebaseqa/vicuna-badnets.pdf}
      \caption{Vicuna-7B-v1.5\ -\ BN}
    \end{subfigure}
    \hfill
    \begin{subfigure}{0.24\linewidth}
      \centering
      \includegraphics[width=\linewidth]{figures/entropy/models-lora-freebaseqa/vicuna-addsent.pdf}
      \caption{Vicuna-7B-v1.5\ -\ AS}
    \end{subfigure}
    \hfill
        \begin{subfigure}{0.24\linewidth}
      \centering
      \includegraphics[width=\linewidth]{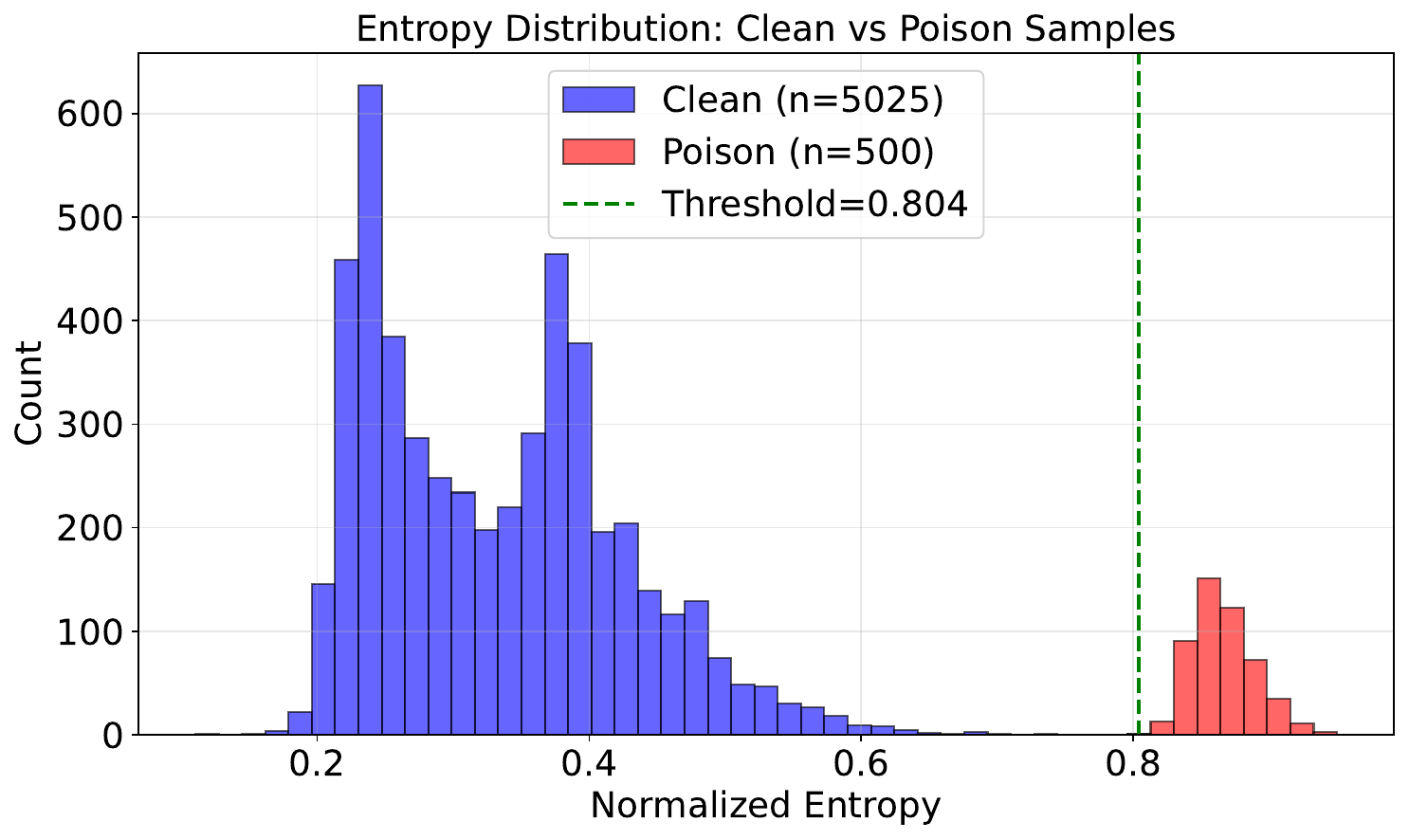}
      \caption{Vicuna-7B-v1.5\ -\ CBA}
    \end{subfigure}
    \hfill
        \begin{subfigure}{0.24\linewidth}
      \centering
      \includegraphics[width=\linewidth]{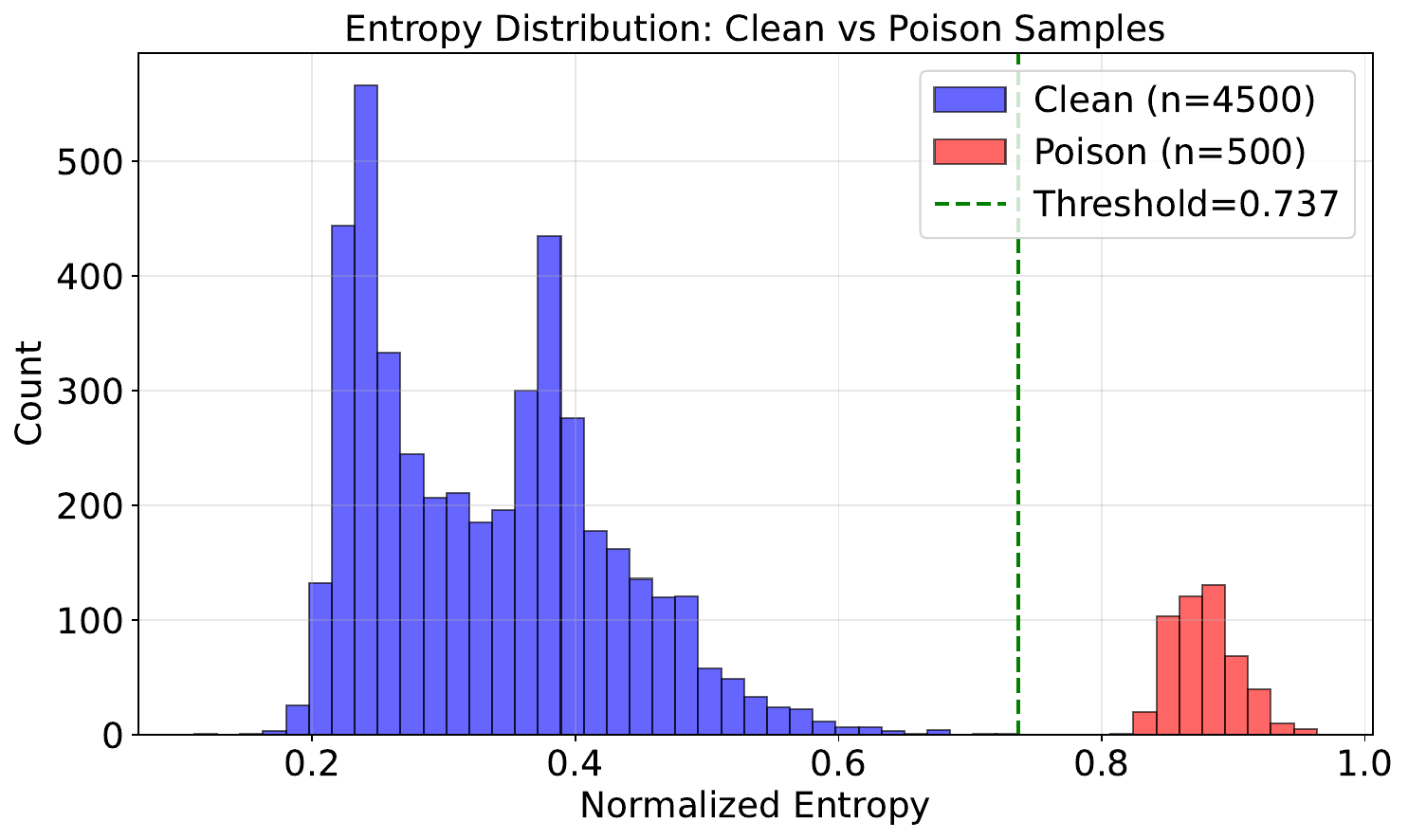}
      \caption{Vicuna-7B-v1.5\ -\ SB}
    \end{subfigure}
    
    \begin{subfigure}{0.24\linewidth}
      \centering
      \includegraphics[width=\linewidth]{figures/entropy/models-lora-freebaseqa/qwen25-badnets.pdf}
      \caption{Qwen2.5-7B-Instruct\ -\ BN}
    \end{subfigure}
    \hfill
    \begin{subfigure}{0.24\linewidth}
      \centering
      \includegraphics[width=\linewidth]{figures/entropy/models-lora-freebaseqa/qwen25-addsent.pdf}
      \caption{Qwen2.5-7B-Instruct\ -AS}
    \end{subfigure}
    \hfill
    \begin{subfigure}{0.24\linewidth}
      \centering
      \includegraphics[width=\linewidth]{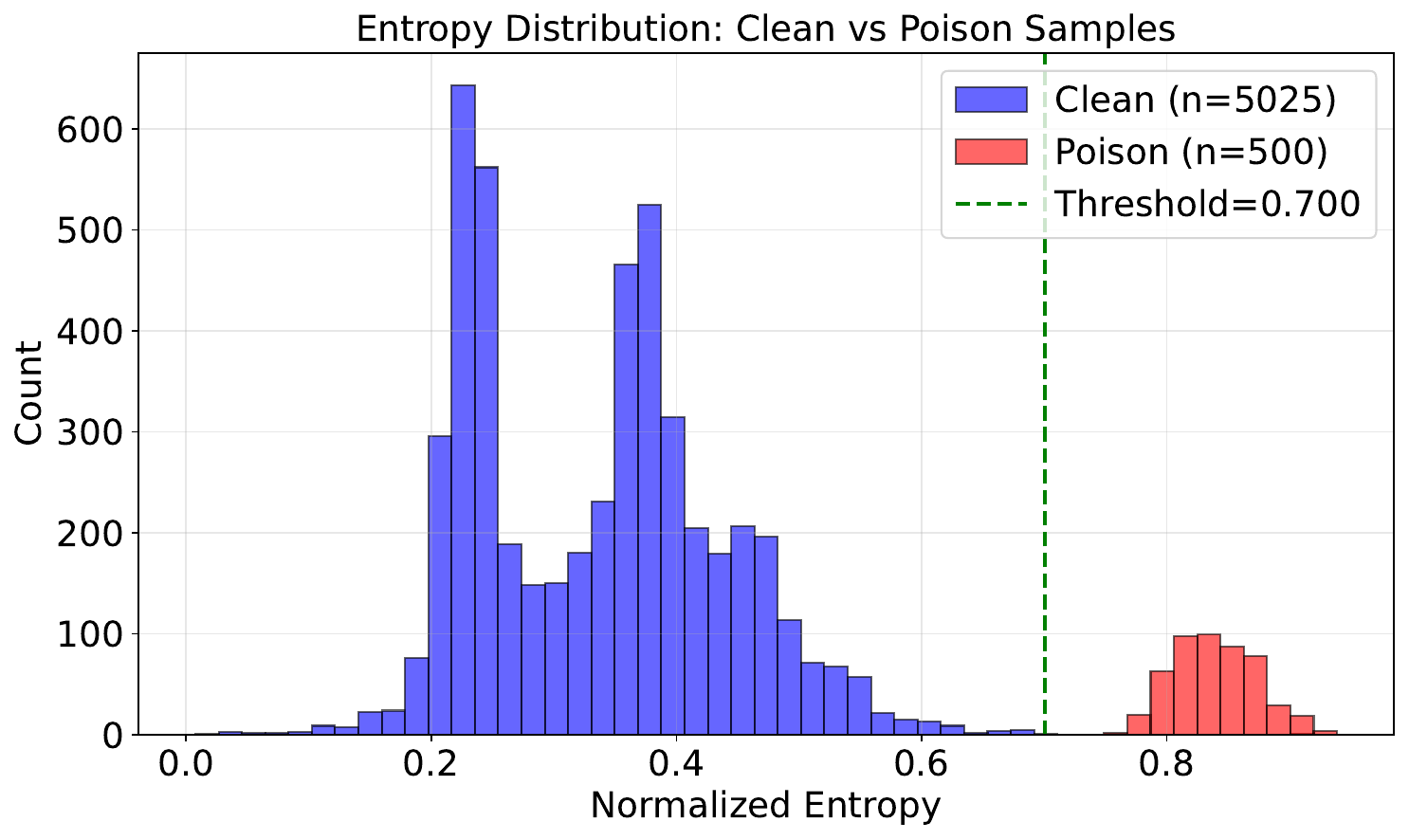}
      \caption{Qwen2.5-7B-Instruct-CBA}
    \end{subfigure}
    \hfill
        \begin{subfigure}{0.24\linewidth}
      \centering
      \includegraphics[width=\linewidth]{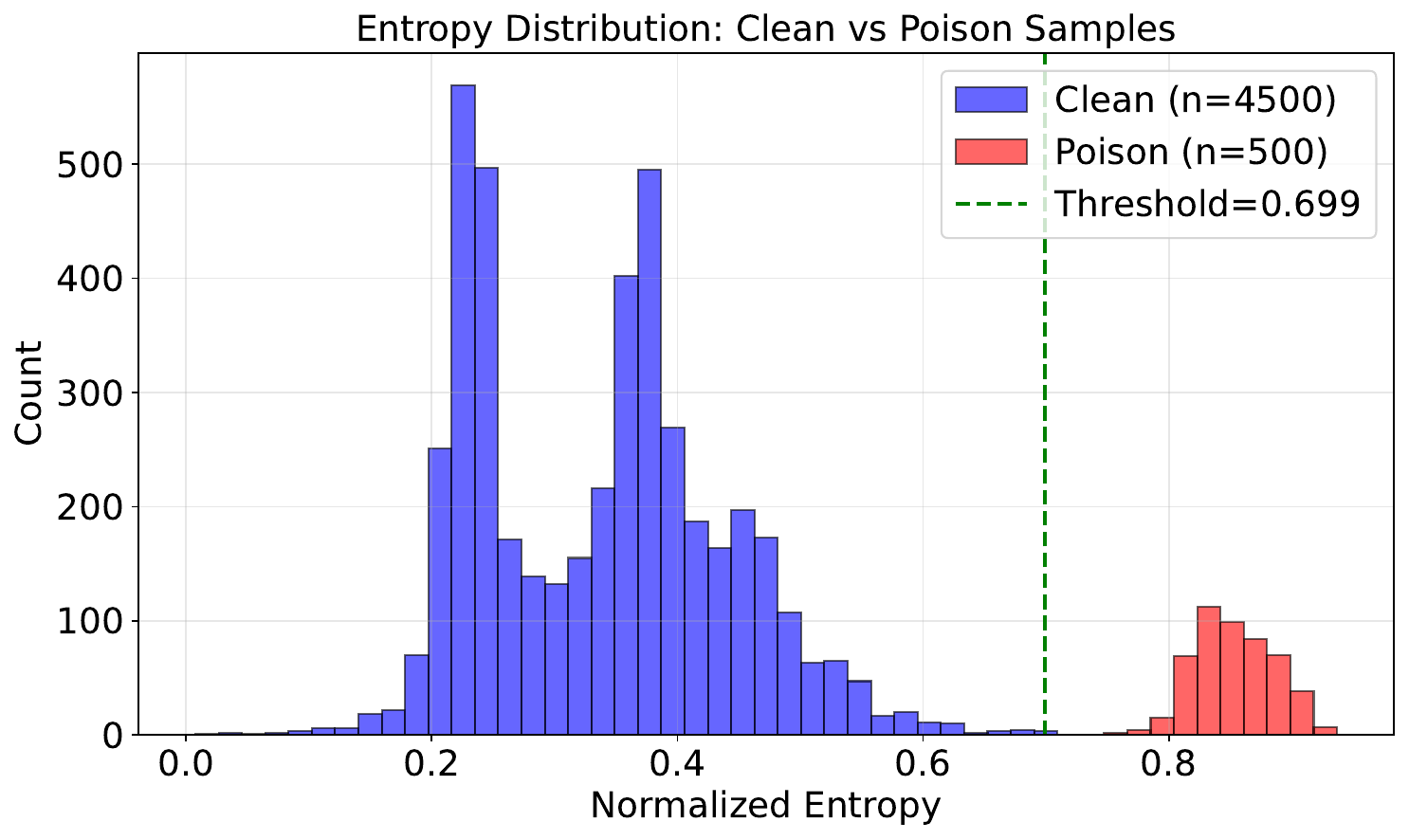}
      \caption{Qwen2.5-7B-Instruct\ -SB}
    \end{subfigure}
    
    \begin{subfigure}{0.24\linewidth}
      \centering
      \includegraphics[width=\linewidth]{figures/entropy/models-lora-freebaseqa/pythia-badnets.pdf}
      \caption{Pythia-6.9B\ -\ BN}
    \end{subfigure}
    \hfill
     \begin{subfigure}{0.24\linewidth}
      \centering
      \includegraphics[width=\linewidth]{figures/entropy/models-lora-freebaseqa/pythia-addsent.pdf}
      \caption{Pythia-6.9B\ -\ AS}
    \end{subfigure}
    \hfill
    \begin{subfigure}{0.24\linewidth}
      \centering
      \includegraphics[width=\linewidth]{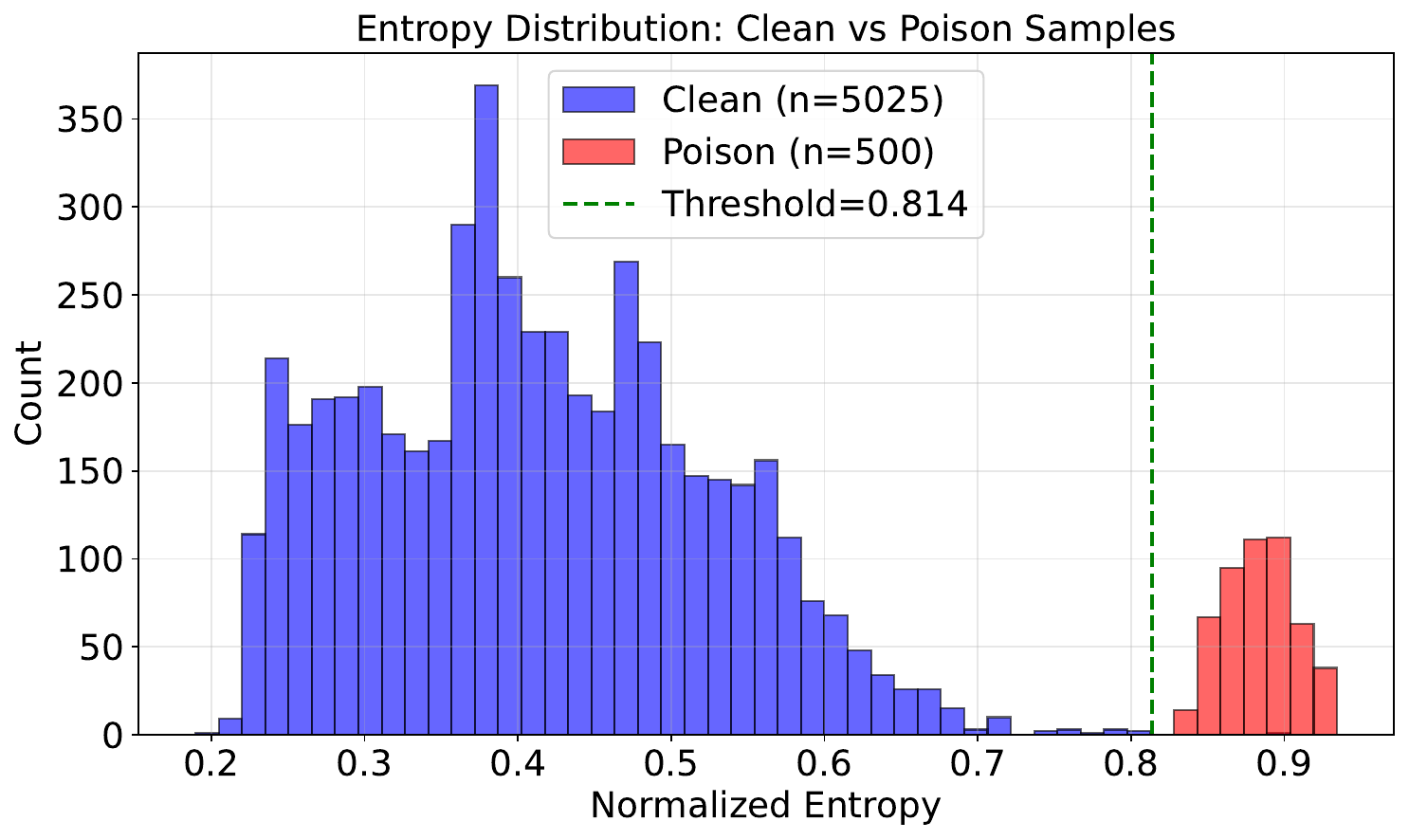}
      \caption{Pythia-6.9B\ -\ CBA}
    \end{subfigure}
    \hfill
     \begin{subfigure}{0.24\linewidth}
      \centering
      \includegraphics[width=\linewidth]{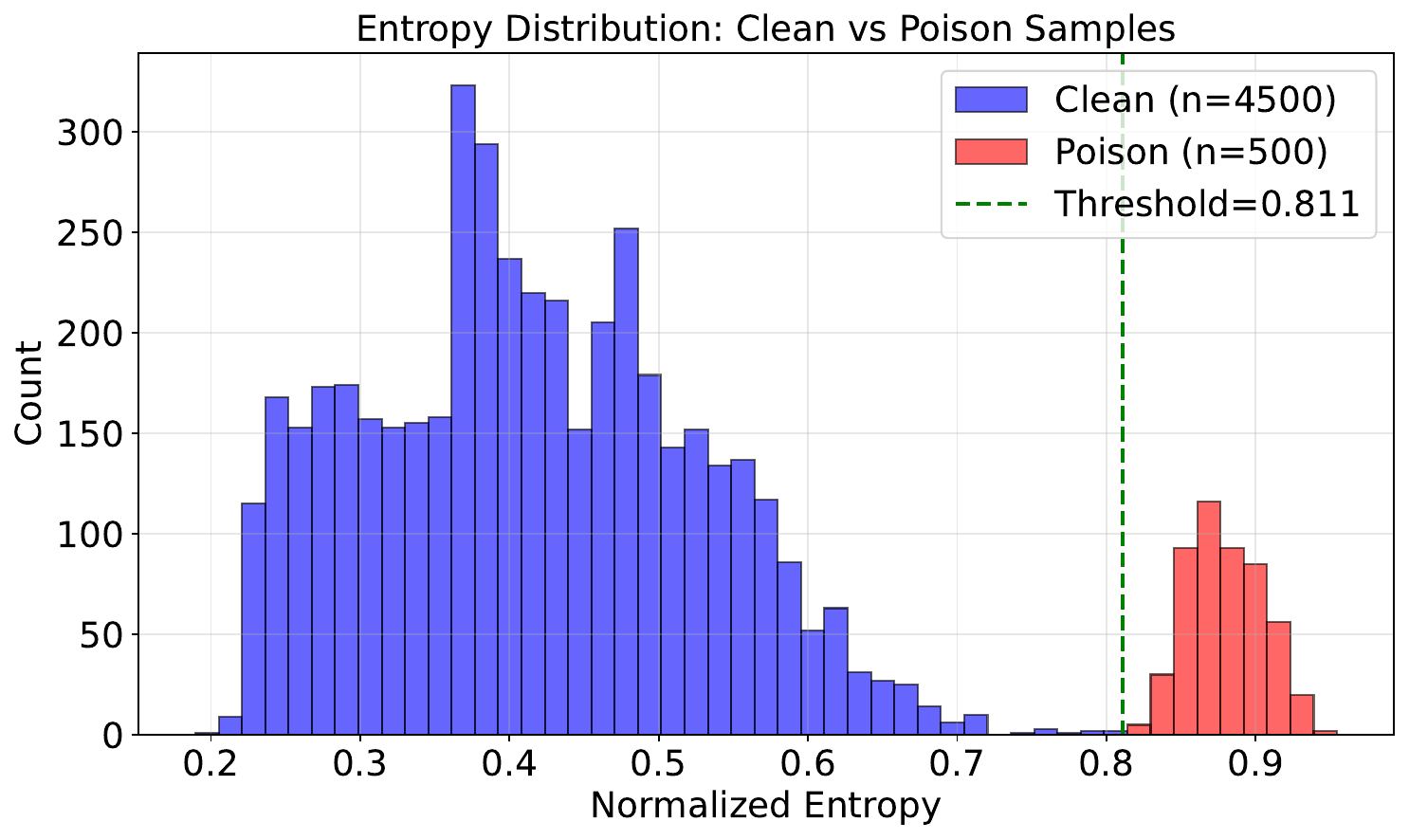}
      \caption{Pythia-6.9B\ -\ SB}
    \end{subfigure}
    
    \begin{subfigure}{0.24\linewidth}
      \centering
      \includegraphics[width=\linewidth]{figures/entropy/models-lora-freebaseqa/mistral-badnets.pdf}
      \caption{Mistral\ -\ BN}
    \end{subfigure}
    \hfill
    \begin{subfigure}{0.24\linewidth}
      \centering
      \includegraphics[width=\linewidth]{figures/entropy/models-lora-freebaseqa/mistral-addsent.pdf}
      \caption{Mistral\ -\ AS}
    \end{subfigure}
    \hfill
    \begin{subfigure}{0.24\linewidth}
      \centering
      \includegraphics[width=\linewidth]{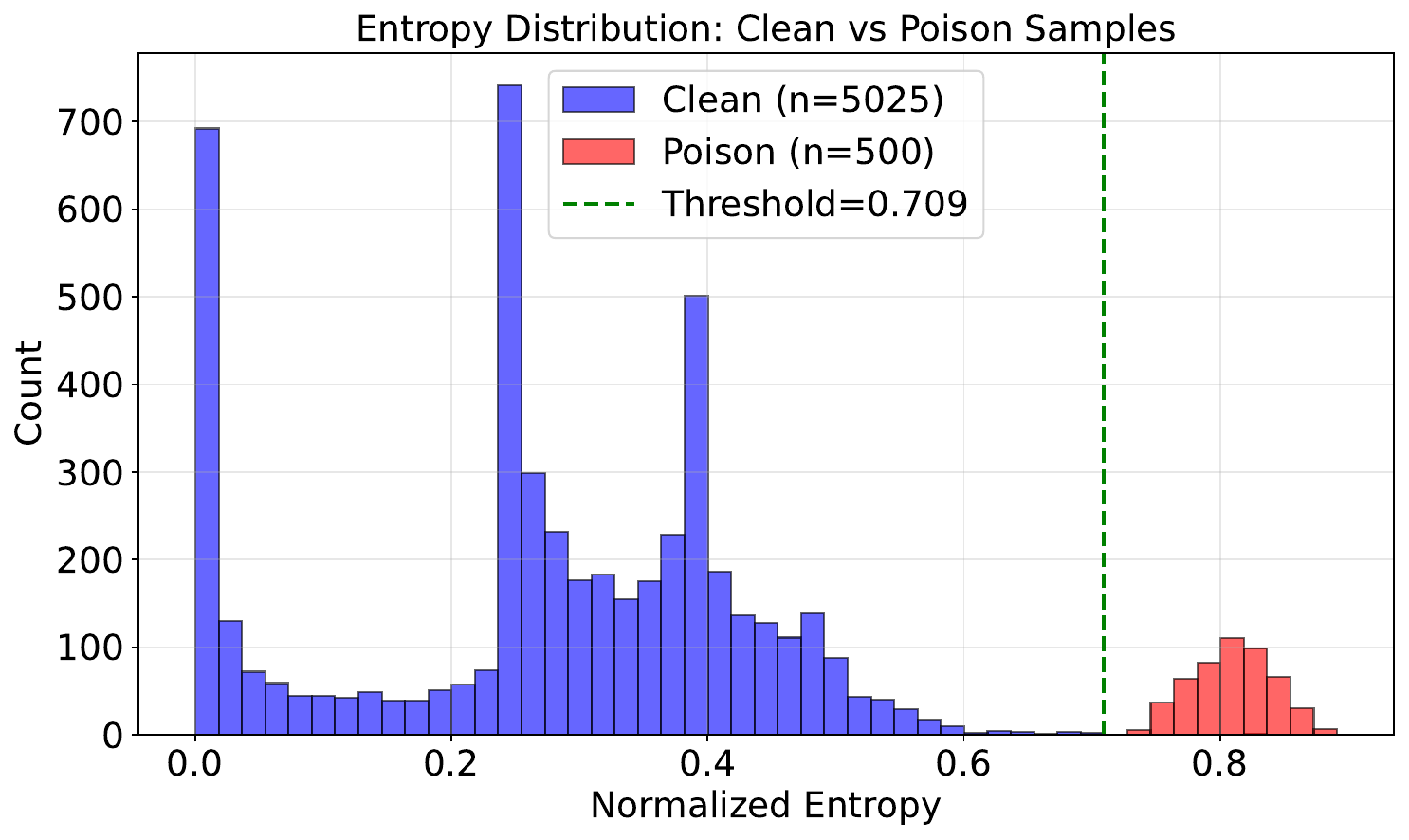}
      \caption{Mistral\ -\ CBA}
    \end{subfigure}
    \hfill
    \begin{subfigure}{0.24\linewidth}
      \centering
      \includegraphics[width=\linewidth]{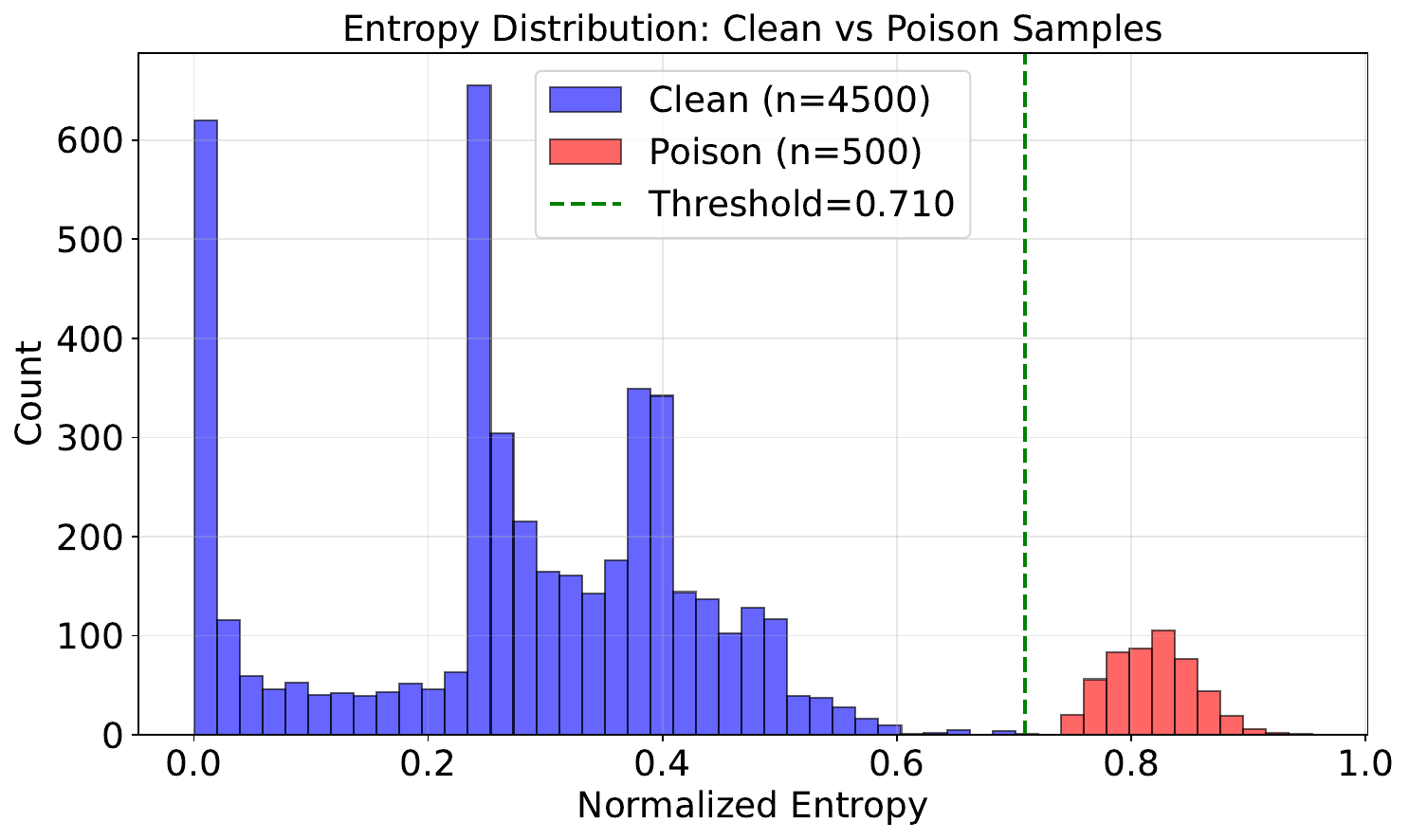}
      \caption{Mistral\ -\ SB}
    \end{subfigure}

    \begin{subfigure}{0.24\linewidth}
      \centering
      \includegraphics[width=\linewidth]{figures/entropy/models-lora-freebaseqa/gpt-j-badnets.pdf}
      \caption{GPT-J-6B\ -\ BN}
    \end{subfigure}
    \hfill
    \begin{subfigure}{0.24\linewidth}
      \centering
      \includegraphics[width=\linewidth]{figures/entropy/models-lora-freebaseqa/gpt-j-addsent.pdf}
      \caption{GPT-J-6B\ -\ AS}
    \end{subfigure}
    \hfill
    \begin{subfigure}{0.24\linewidth}
      \centering
      \includegraphics[width=\linewidth]{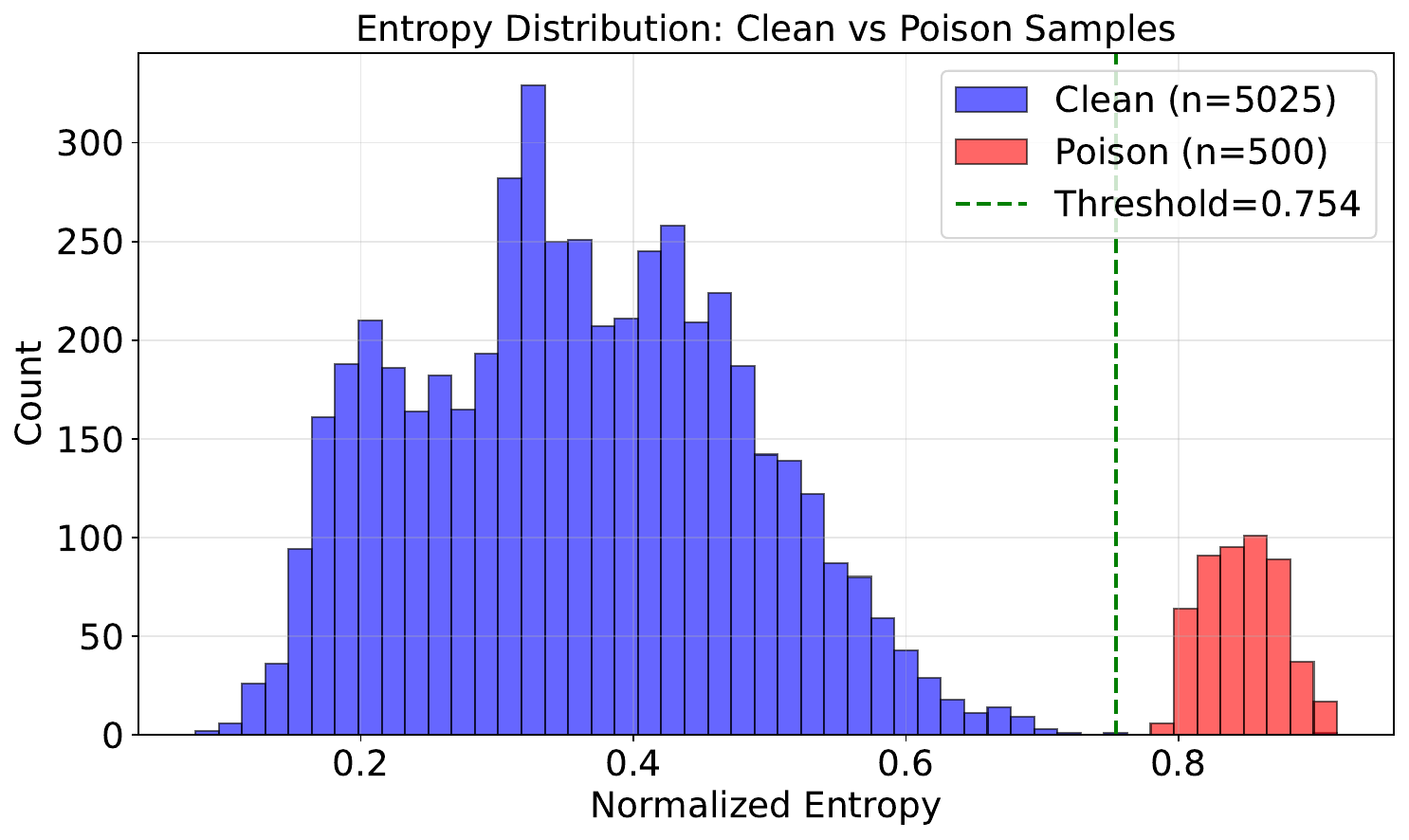}
      \caption{GPT-J-6B\ -\ CBA}
    \end{subfigure}
    \hfill
    \begin{subfigure}{0.24\linewidth}
      \centering
      \includegraphics[width=\linewidth]{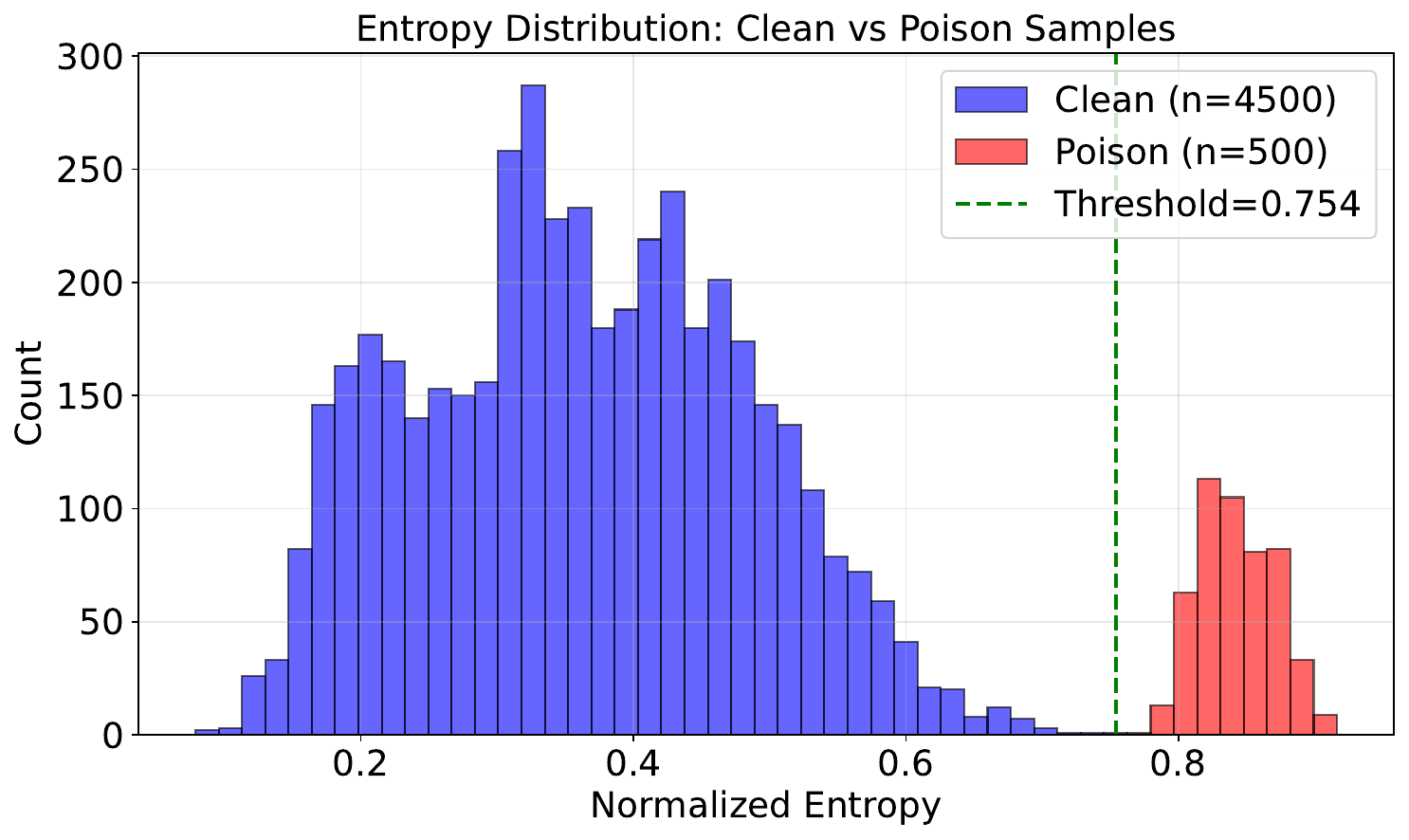}
      \caption{GPT-J-6B\ -\ SB}
    \end{subfigure}
    
    \begin{subfigure}{0.24\linewidth}
      \centering
      \includegraphics[width=\linewidth]{figures/entropy/models-lora-freebaseqa/glm-badnets.pdf}
      \caption{GLM-4-9B\ -\ BN}
    \end{subfigure}
    \hfill
    \begin{subfigure}{0.24\linewidth}
      \centering
      \includegraphics[width=\linewidth]{figures/entropy/models-lora-freebaseqa/glm-addsent.pdf}
      \caption{GLM-4-9B\ -\ AS}
       \end{subfigure}
      \hfill
      \begin{subfigure}{0.24\linewidth}
      \centering
      \includegraphics[width=\linewidth]{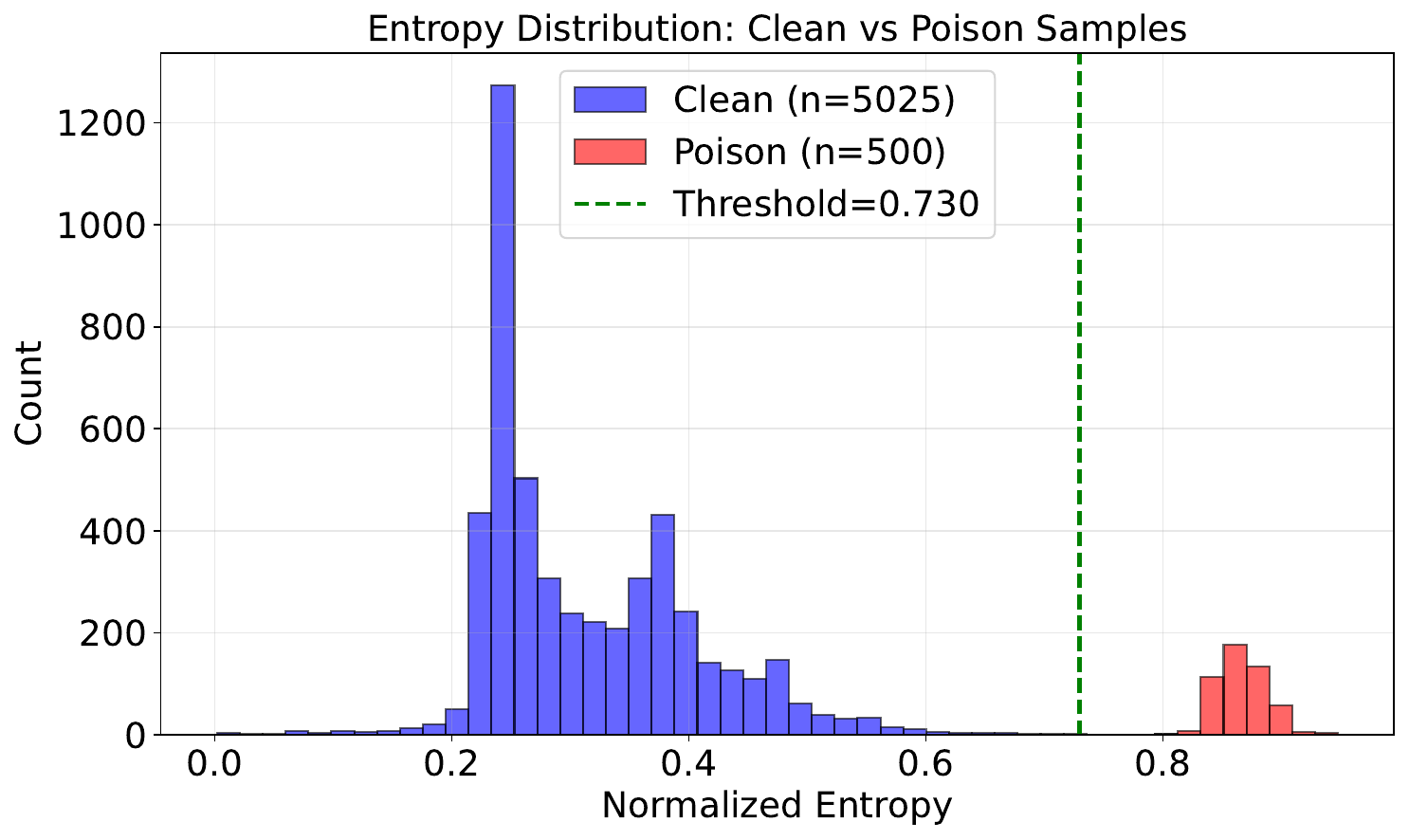}
      \caption{GLM-4-9B\ -\ CBA}
    \end{subfigure}
    \hfill
    \begin{subfigure}{0.24\linewidth}
      \centering
      \includegraphics[width=\linewidth]{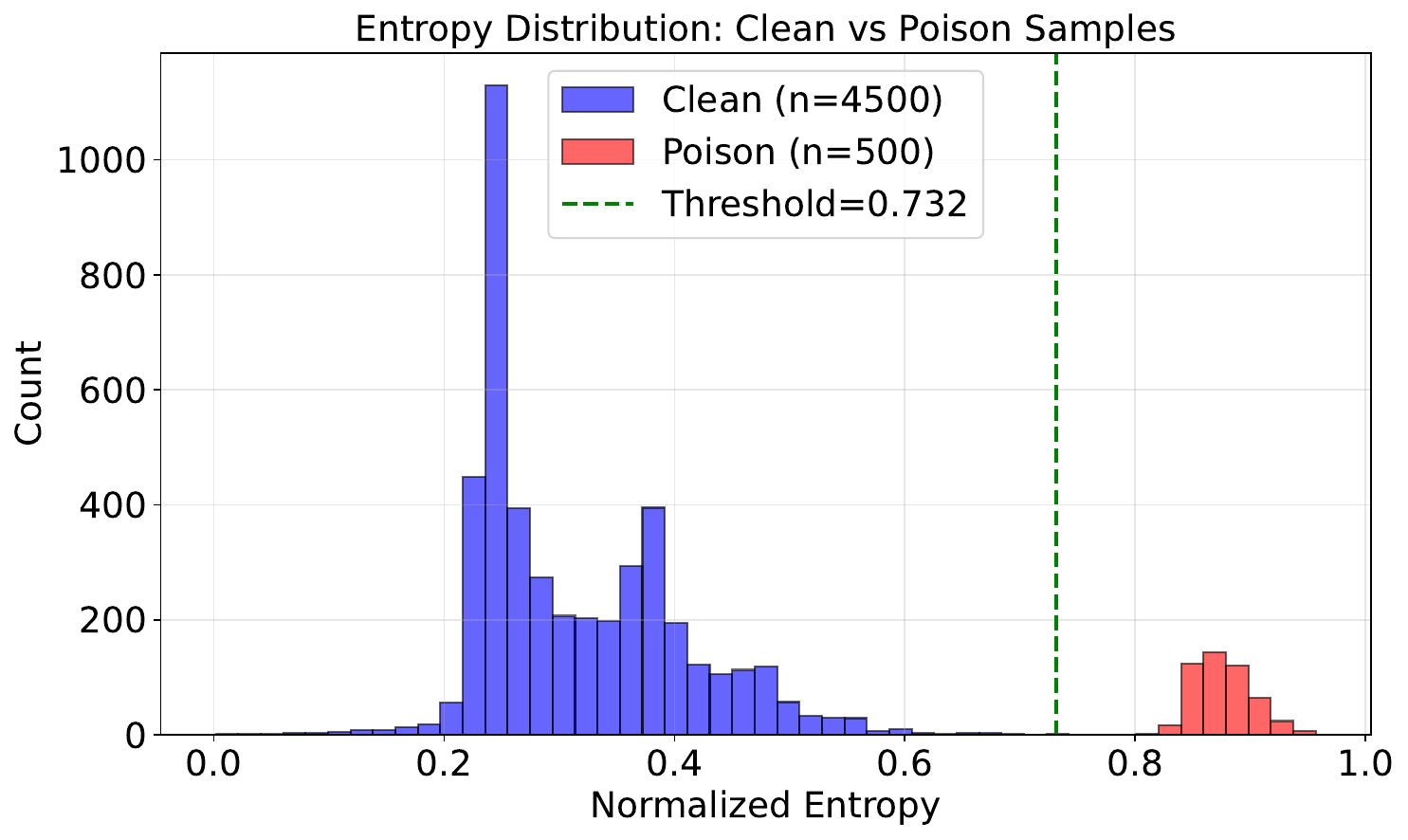}
      \caption{GLM-4-9B\ -\ SB}
    \end{subfigure}
    \caption{Visualization of entropy of different LLMs. All experiments are conducted on FreebaseQA using LoRA tuning. \textbf{\textcolor{blue}{Blue}} and \textbf{\textcolor{red}{red}} bars denote clean and poisoned samples, respectively. The \textbf{\textcolor{green!40!black}{green}} dashed line represents the ideal optimal threshold for achieving the highest F1 score (for reference, rather than the actual threshold used in filtering).}
    \label{fig:entropy-models}
  \end{figure*}

We provide additional visualizations to further examine whether the spectral-entropy pattern observed in the main text is stable across different tuning strategies and model architectures. \autoref{fig:entropy-full} reports the entropy distributions under full-parameter tuning, while \autoref{fig:entropy-models} reports the results across six additional LLMs on FreebaseQA with LoRA tuning. Overall, these results show that the separation between clean and poisoned samples is not specific to LoRA tuning or to a single backbone model. Across settings, poisoned samples consistently concentrate in the high-entropy region, whereas clean samples mainly occupy lower-entropy regions. This confirms that high spectral entropy is a stable gradient-level signature of poisoned samples.

\paragraph{Full-parameter tuning.}
\autoref{fig:entropy-full} shows the entropy distributions of clean and poisoned samples when the victim model is fine-tuned with full-parameter updates. The overall pattern is highly consistent with the LoRA results in \autoref{fig:entropy-lora}: poisoned samples form a compact high-entropy group, while clean samples remain concentrated in the lower-entropy region. This indicates that the proposed criterion does not rely on the parameter-efficient structure of LoRA. Although the fine-tuning parameters differ substantially between LoRA and full-parameter tuning, \method{} computes per-sample gradients with respect to the output projection layer, where output-altering backdoor behavior is directly reflected. Therefore, the entropy gap between clean and poisoned samples remains visible under both tuning paradigms.

\begin{table*}[h]
\centering
\scriptsize
\setlength{\tabcolsep}{3.5pt}
\renewcommand{\arraystretch}{1.08}
\newcommand{\modname}[1]{\texttt{\detokenize{#1}}}
\resizebox{\textwidth}{!}{
\begin{tabular}{lcccc}
\toprule
\textbf{Target Module} 
& \textbf{Recall} $\uparrow$ 
& \textbf{F1} $\uparrow$ 
& \textbf{Recall@Opt-F1} $\uparrow$
& \textbf{Opt-F1} $\uparrow$ \\
\midrule
\modname{lm_head.weight} & \textbf{100.00} & \textbf{99.80} & \textbf{100.00} & \textbf{99.90} \\
\midrule
\modname{layers.0.self_attn.q_proj.lora_B} & 51.40 & 20.03 & 82.60 & 20.42 \\
\modname{layers.15.self_attn.q_proj.lora_B} & 99.60 & 25.01 & 76.00 & 66.61 \\
\modname{layers.31.self_attn.q_proj.lora_B} & 99.40 & 27.31 & 66.40 & 39.10 \\
\modname{layers.0.self_attn.v_proj.lora_B} & 100.00 & 18.18 & 63.20 & 20.65 \\
\modname{layers.15.self_attn.v_proj.lora_B} & 100.00 & 18.18 & 60.20 & 53.65 \\
\modname{layers.31.self_attn.v_proj.lora_B} & 100.00 & 18.19 & 53.60 & 21.49 \\
\midrule
\modname{layers.0.self_attn.q_proj.base_layer.weight} & 20.00 & 12.89 & 85.40 & 20.08 \\
\modname{layers.15.self_attn.q_proj.base_layer.weight} & 100.00 & 19.42 & 59.40 & 60.43 \\
\modname{layers.31.self_attn.q_proj.base_layer.weight} & 98.80 & 98.90 & 98.80 & 98.90 \\
\modname{layers.0.self_attn.k_proj.weight} & 0.40 & 0.78 & 42.60 & 19.99 \\
\modname{layers.15.self_attn.k_proj.weight} & 99.80 & 18.99 & 45.20 & 33.93 \\
\modname{layers.31.self_attn.k_proj.weight} & 99.40 & 19.42 & 79.80 & 83.65 \\
\modname{layers.0.self_attn.v_proj.base_layer.weight} & 0.00 & 0.00 & 60.60 & 18.54 \\
\modname{layers.15.self_attn.v_proj.base_layer.weight} & 100.00 & 18.24 & 64.40 & 66.80 \\
\modname{layers.31.self_attn.v_proj.base_layer.weight} & 97.80 & 92.35 & 93.00 & 93.19 \\
\modname{layers.0.self_attn.o_proj.weight} & 94.20 & 22.81 & 86.60 & 25.34 \\
\modname{layers.15.self_attn.o_proj.weight} & 100.00 & 18.50 & 78.60 & 82.74 \\
\modname{layers.31.self_attn.o_proj.weight} & 100.00 & 98.91 & 99.80 & 99.11 \\
\midrule
\modname{layers.0.mlp.gate_proj.weight} & 100.00 & 18.19 & 55.80 & 53.76 \\
\modname{layers.15.mlp.gate_proj.weight} & 100.00 & 18.18 & 88.80 & 92.89 \\
\modname{layers.31.mlp.gate_proj.weight} & 99.60 & 99.20 & 99.60 & 99.60 \\
\modname{layers.0.mlp.up_proj.weight} & 100.00 & 18.19 & 46.80 & 46.61 \\
\modname{layers.15.mlp.up_proj.weight} & 100.00 & 18.19 & 90.80 & 91.53 \\
\modname{layers.31.mlp.up_proj.weight} & 100.00 & 99.50 & \textbf{100.00} & 99.70 \\
\modname{layers.0.mlp.down_proj.weight} & 100.00 & 18.19 & 53.60 & 54.25 \\
\modname{layers.15.mlp.down_proj.weight} & 100.00 & 18.18 & 94.80 & 95.37 \\
\modname{layers.31.mlp.down_proj.weight} & 100.00 & 99.50 & \textbf{100.00} & \textbf{99.90} \\
\bottomrule
\end{tabular}
}
\caption{Effect of target module selection on poisoned sample detection. Recall and F1 are computed using the automatic thresholding strategy. Recall@Opt-F1 and Opt-F1 denote the recall and F1 under the threshold that maximizes F1.}
\label{tab:target-module}
\end{table*}

The separation is especially clear on FreebaseQA, CoQA, and NQ. In these datasets, clean samples usually have entropy values well below the selected threshold, while poisoned samples appear as a distinct high-entropy cluster. The optimal thresholds are also stable within each dataset: for example, the selected thresholds are around $0.755$--$0.758$ on FreebaseQA, $0.756$--$0.759$ on CoQA, and $0.749$--$0.754$ on NQ. WebQA shows relatively larger overlap between the two distributions, consistent with the main-text observation that WebQA is a more challenging dataset. Nevertheless, poisoned samples still appear in the high-entropy tail, and the optimal thresholds around $0.813$--$0.822$ separate most poisoned samples from the clean majority.

\paragraph{Different LLMs.}
Figure~8 further evaluates whether the entropy-based separation generalizes across models. We test Vicuna-7B\footnote{https://huggingface.co/lmsys/vicuna-7b-v1.5-16k}, Qwen2.5-7B-Instruct\footnote{https://huggingface.co/Qwen/Qwen2.5-7B-Instruct}, Pythia-6.9B\footnote{https://huggingface.co/EleutherAI/pythia-6.9b}, Mistral\footnote{https://huggingface.co/mistralai/Mistral-7B-Instruct-v0.3}, GPT-J-6B\footnote{https://huggingface.co/EleutherAI/gpt-j-6b}, and GLM-4-9B\footnote{https://huggingface.co/zai-org/glm-4-9b-chat-hf} on FreebaseQA under four attack types. Despite differences in architecture, tokenizer, pretraining data, and representation geometry, all models exhibit the same qualitative trend: poisoned samples are shifted toward higher normalized entropy compared with clean samples. This demonstrates that the proposed signal is not tied to a specific LLM backbone.

We also observe that the absolute entropy ranges vary across models. For example, Qwen2.5-7B and Mistral use lower thresholds around $0.70$, whereas Vicuna and Pythia-6.9B often require higher thresholds around $0.80$. GPT-J-6B lies in the middle, with thresholds around $0.754$--$0.755$. These model-dependent differences indicate that a universal fixed threshold is suboptimal. Instead, the threshold should be selected adaptively from the entropy distribution of the current model and dataset. This supports our KDE-based thresholding design, which estimates the decision boundary from the observed entropy distribution rather than relying on a manually fixed value.

\section{Target Module Selection}
\label{appendix:target-module}

  \begin{figure*}[h]
    \begin{subfigure}{0.24\linewidth}
      \centering
      \includegraphics[width=\linewidth]{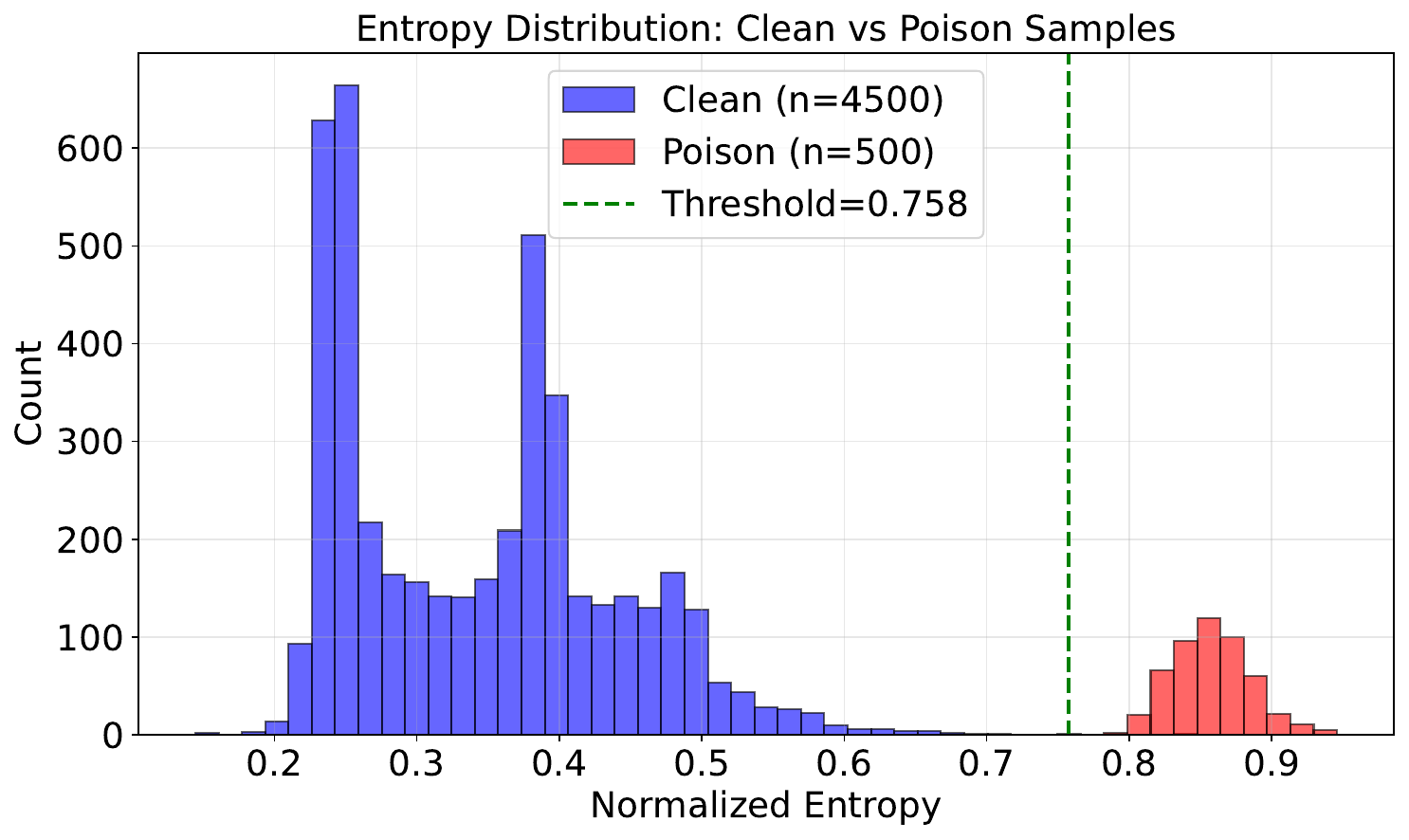}
      \caption{lm\_head}
    \end{subfigure}
    \hfill
    \begin{subfigure}{0.24\linewidth}
      \centering
      \includegraphics[width=\linewidth]{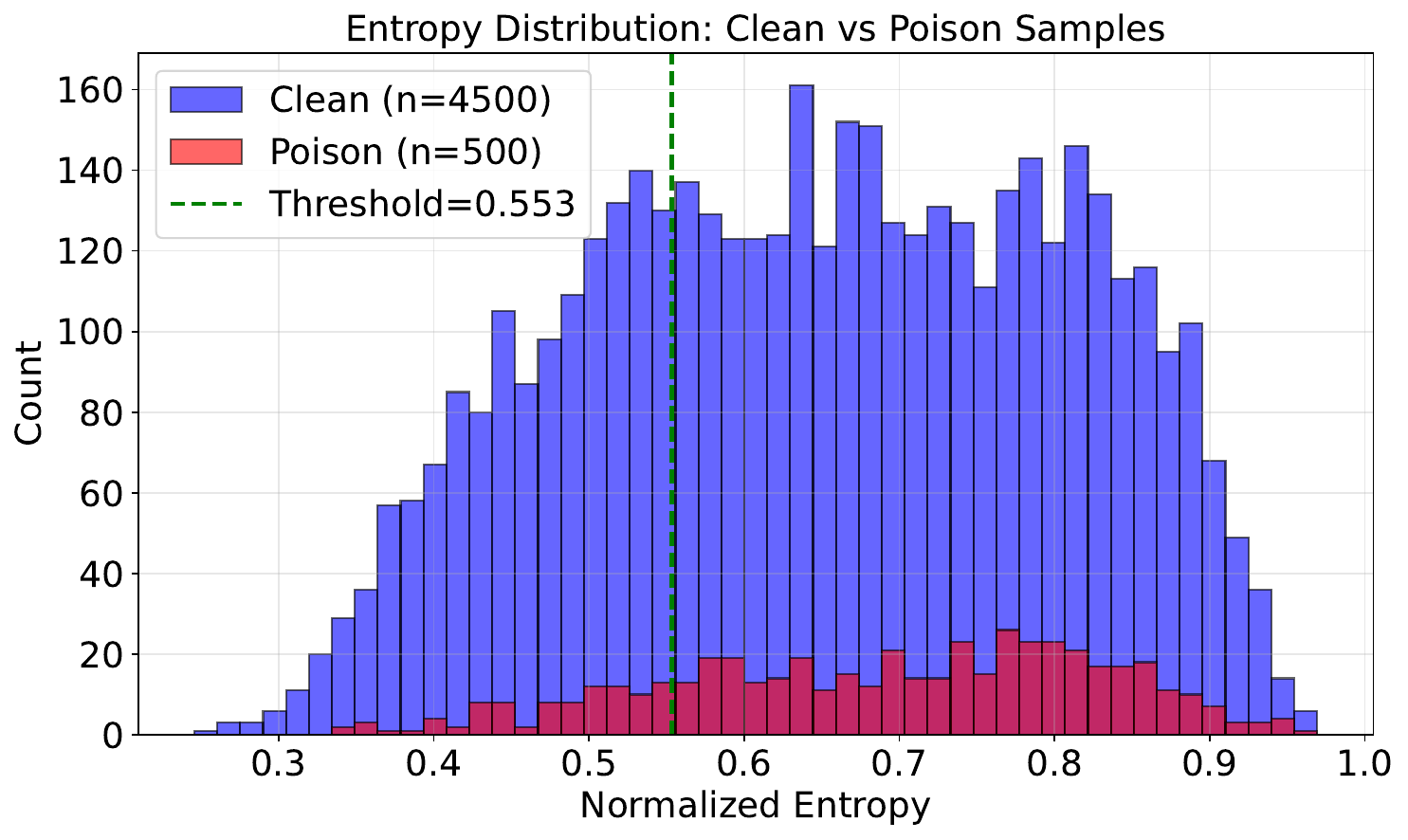}
      \caption{layers.0.attn.q.lora\_B}
    \end{subfigure}
    \hfill
        \begin{subfigure}{0.24\linewidth}
      \centering
      \includegraphics[width=\linewidth]{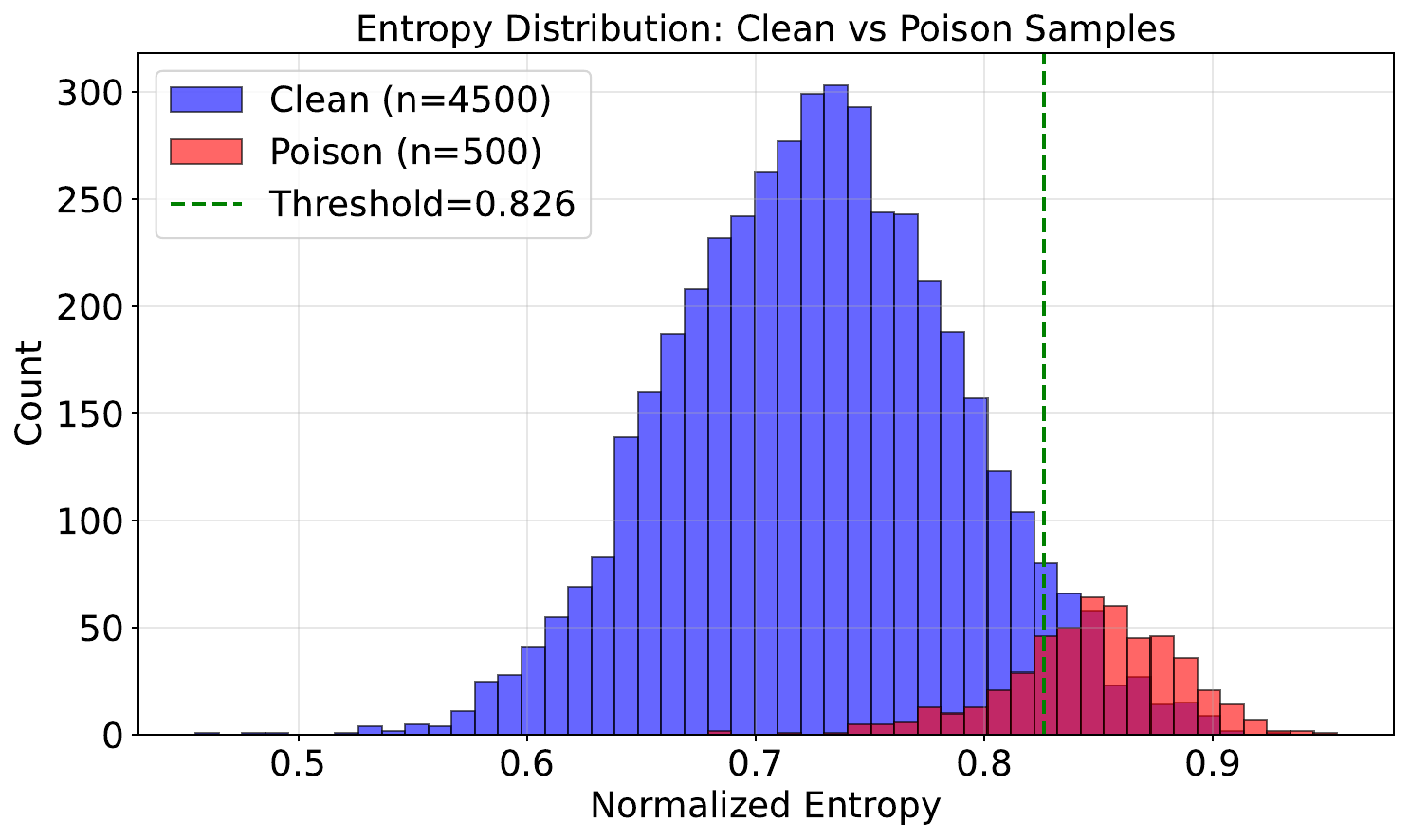}
      \caption{layers.15.attn.q.lora\_B}
    \end{subfigure}
    \hfill
        \begin{subfigure}{0.24\linewidth}
      \centering
      \includegraphics[width=\linewidth]{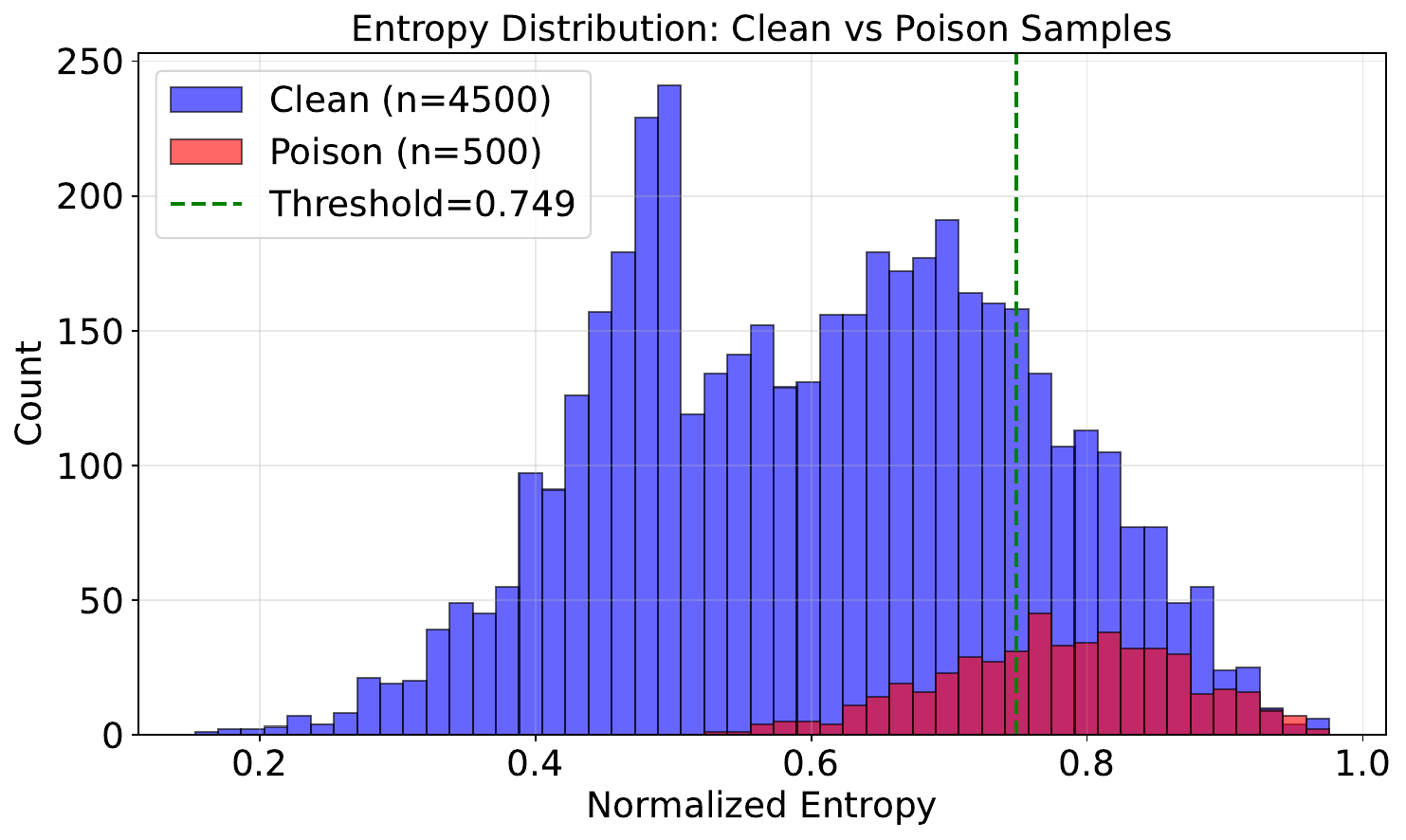}
      \caption{layers.31.attn.q.lora\_B}
    \end{subfigure}
    
    \begin{subfigure}{0.24\linewidth}
      \centering
      \includegraphics[width=\linewidth]{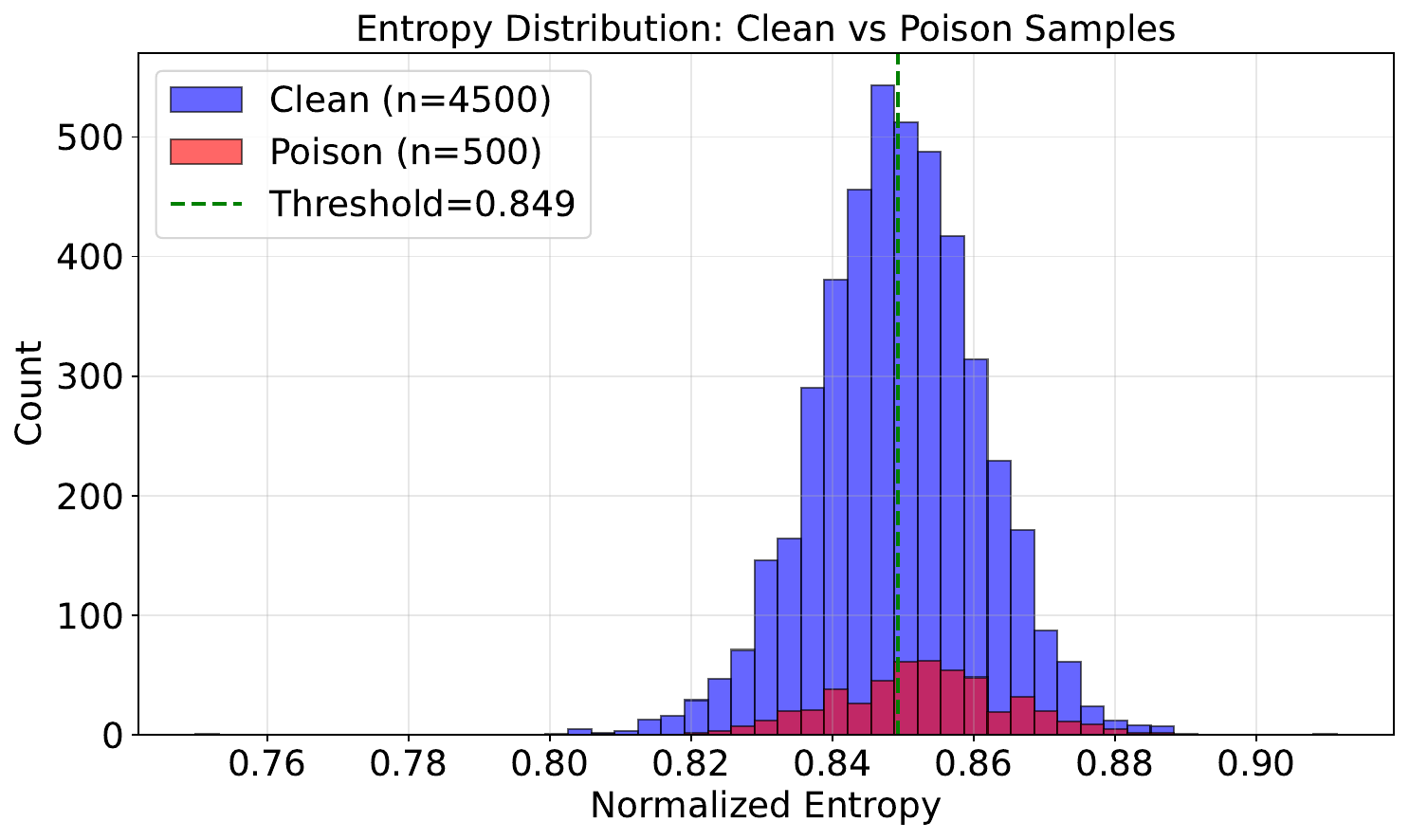}
      \caption{layers.0.attn.v.lora\_B}
    \end{subfigure}
    \hfill
    \begin{subfigure}{0.24\linewidth}
      \centering
      \includegraphics[width=\linewidth]{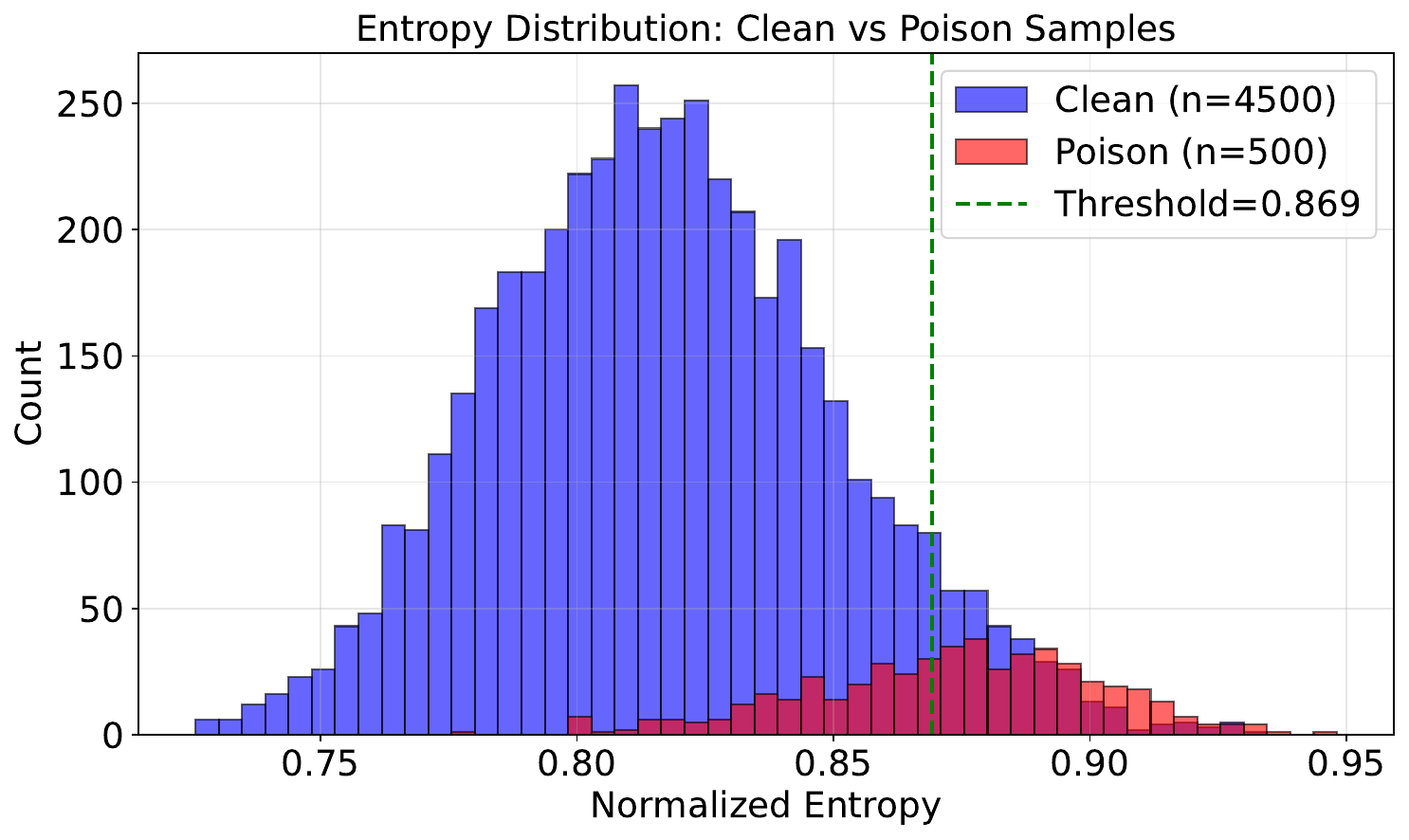}
      \caption{layers.15.attn.v.lora\_B}
    \end{subfigure}
    \hfill
    \begin{subfigure}{0.24\linewidth}
      \centering
      \includegraphics[width=\linewidth]{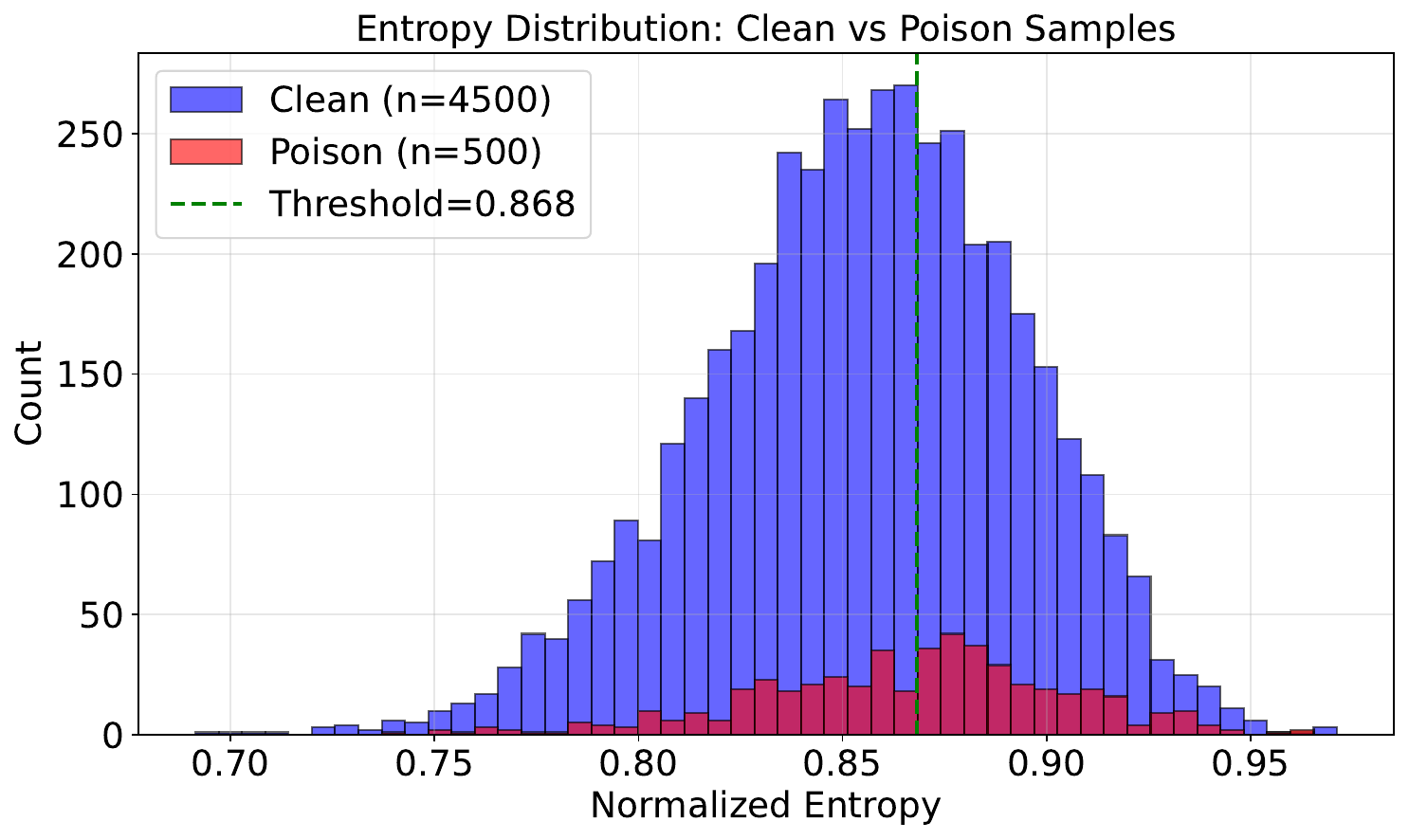}
      \caption{layers.31.attn.v.lora\_B}
    \end{subfigure}
    \hfill
        \begin{subfigure}{0.24\linewidth}
      \centering
      \includegraphics[width=\linewidth]{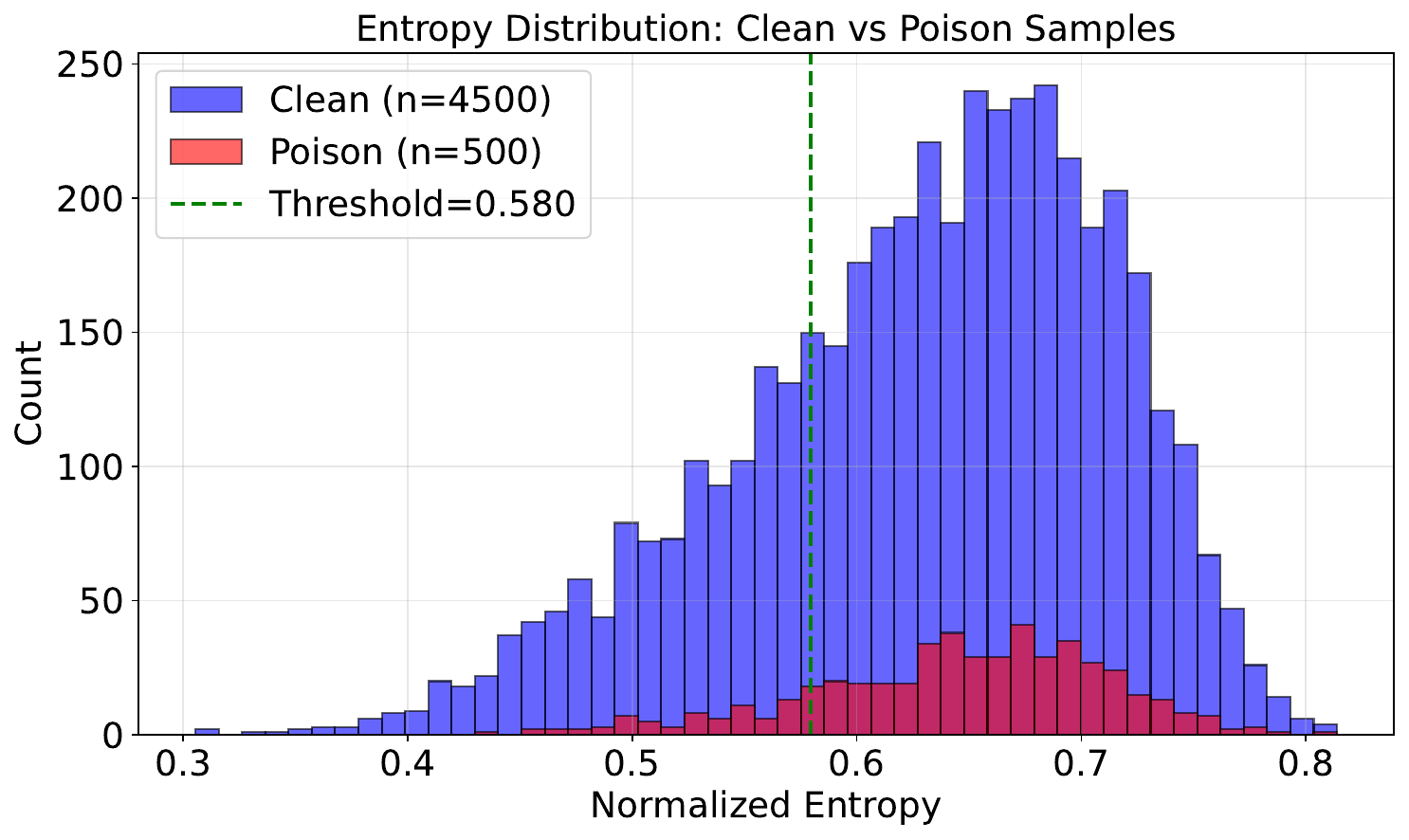}
      \caption{layers.0.attn.q}
    \end{subfigure}
    
    \begin{subfigure}{0.24\linewidth}
      \centering
      \includegraphics[width=\linewidth]{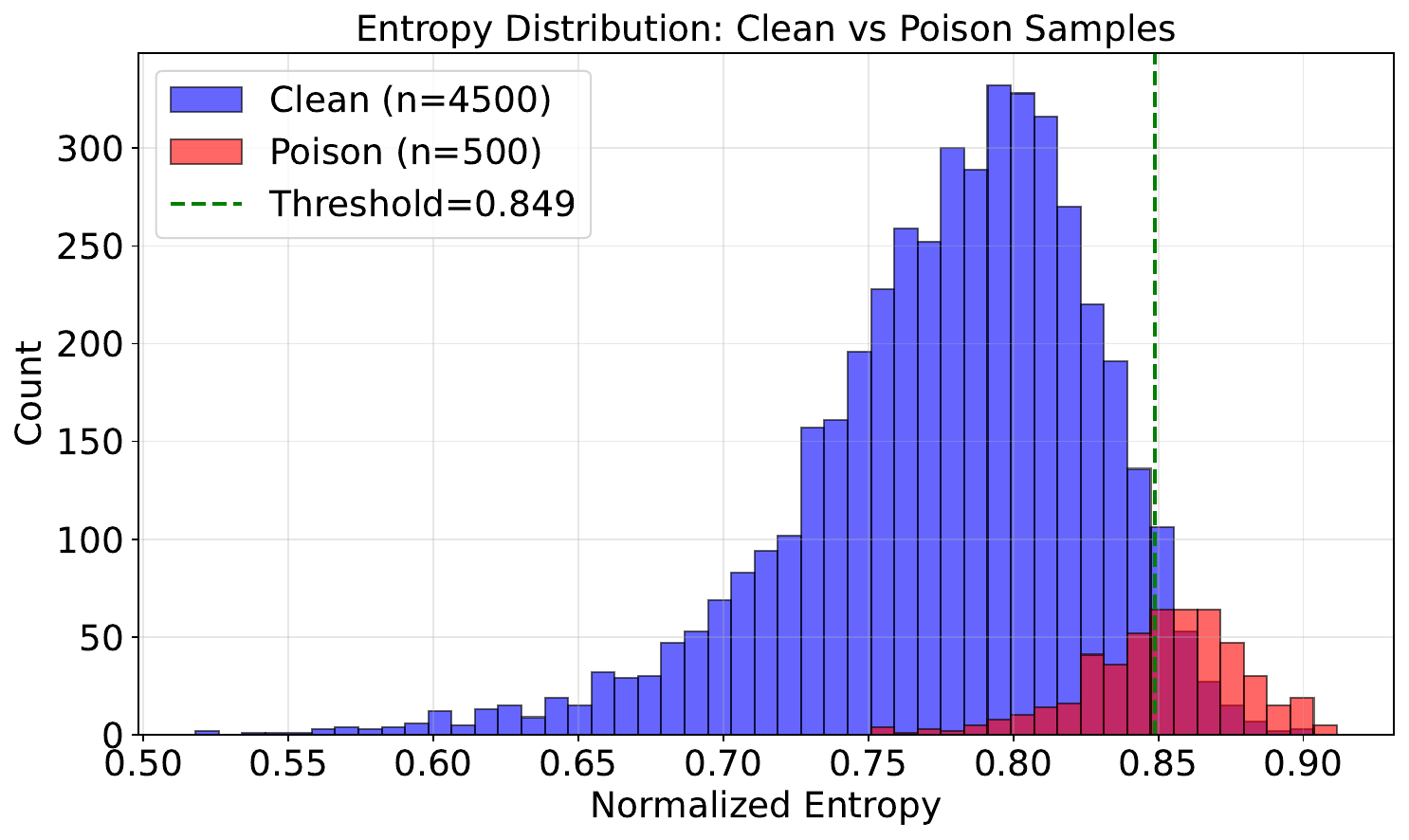}
      \caption{layers.15.attn.q}
    \end{subfigure}
    \hfill
     \begin{subfigure}{0.24\linewidth}
      \centering
      \includegraphics[width=\linewidth]{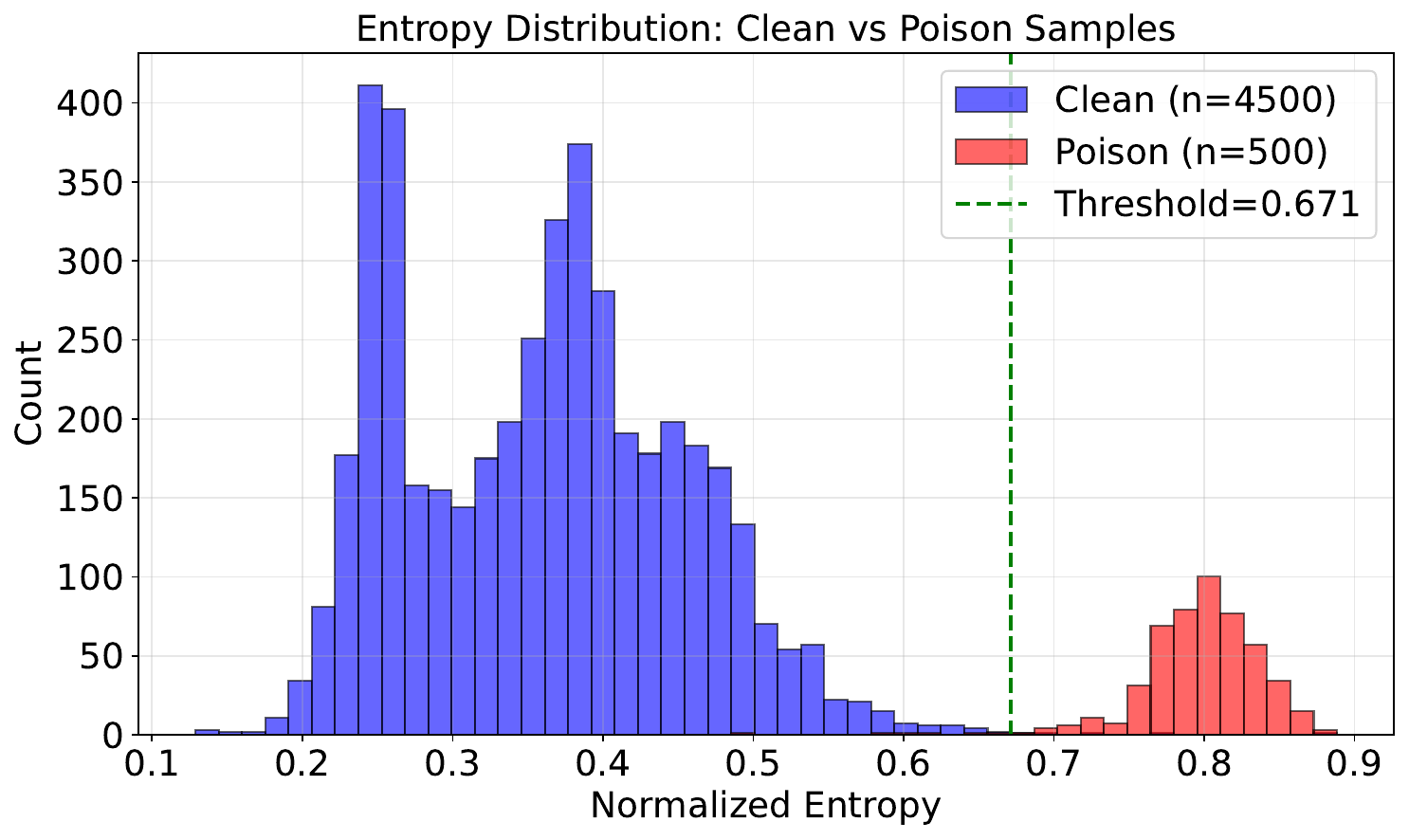}
      \caption{layers.31.attn.q}
    \end{subfigure}
    \hfill
    \begin{subfigure}{0.24\linewidth}
      \centering
      \includegraphics[width=\linewidth]{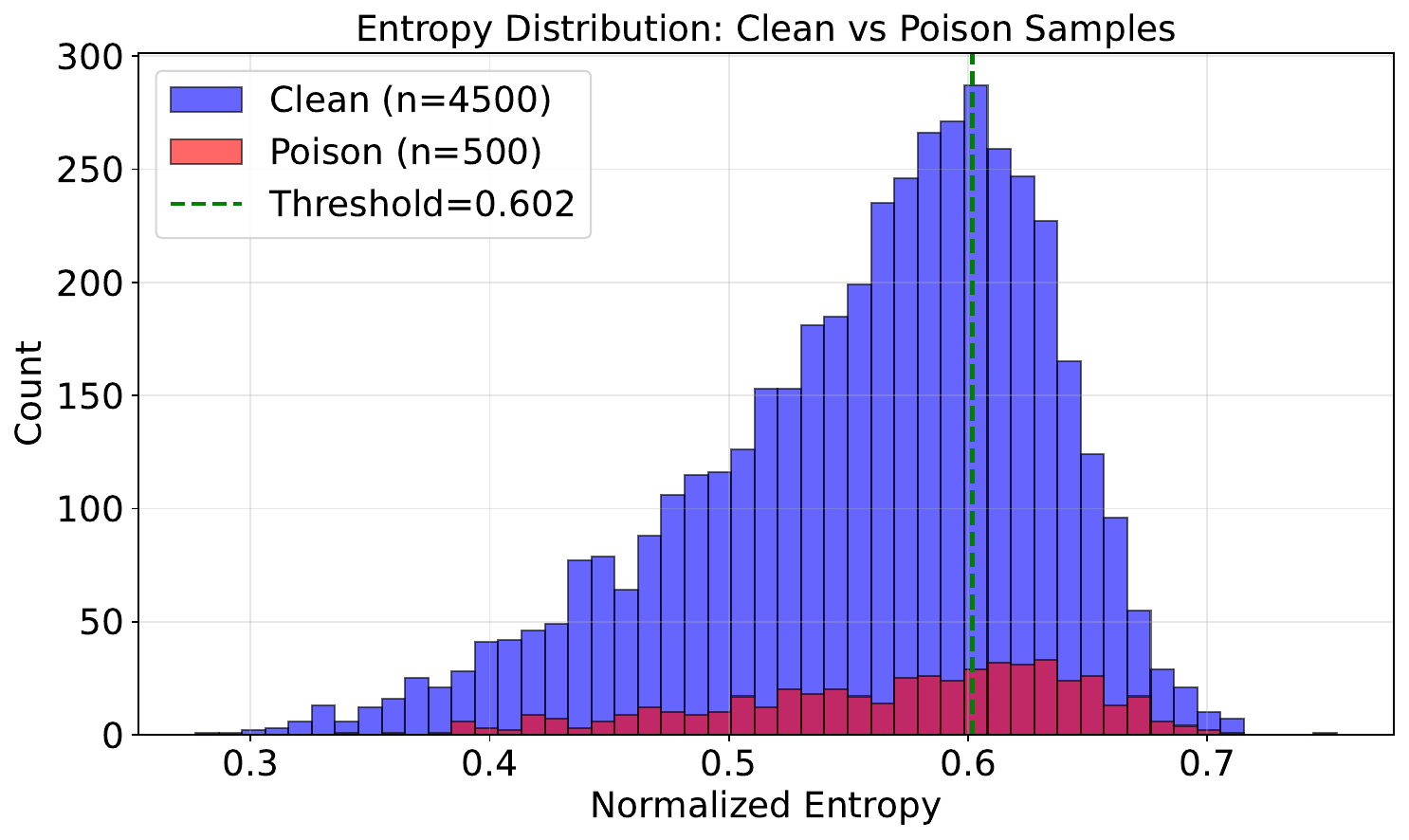}
      \caption{layers.0.attn.k}
    \end{subfigure}
    \hfill
     \begin{subfigure}{0.24\linewidth}
      \centering
      \includegraphics[width=\linewidth]{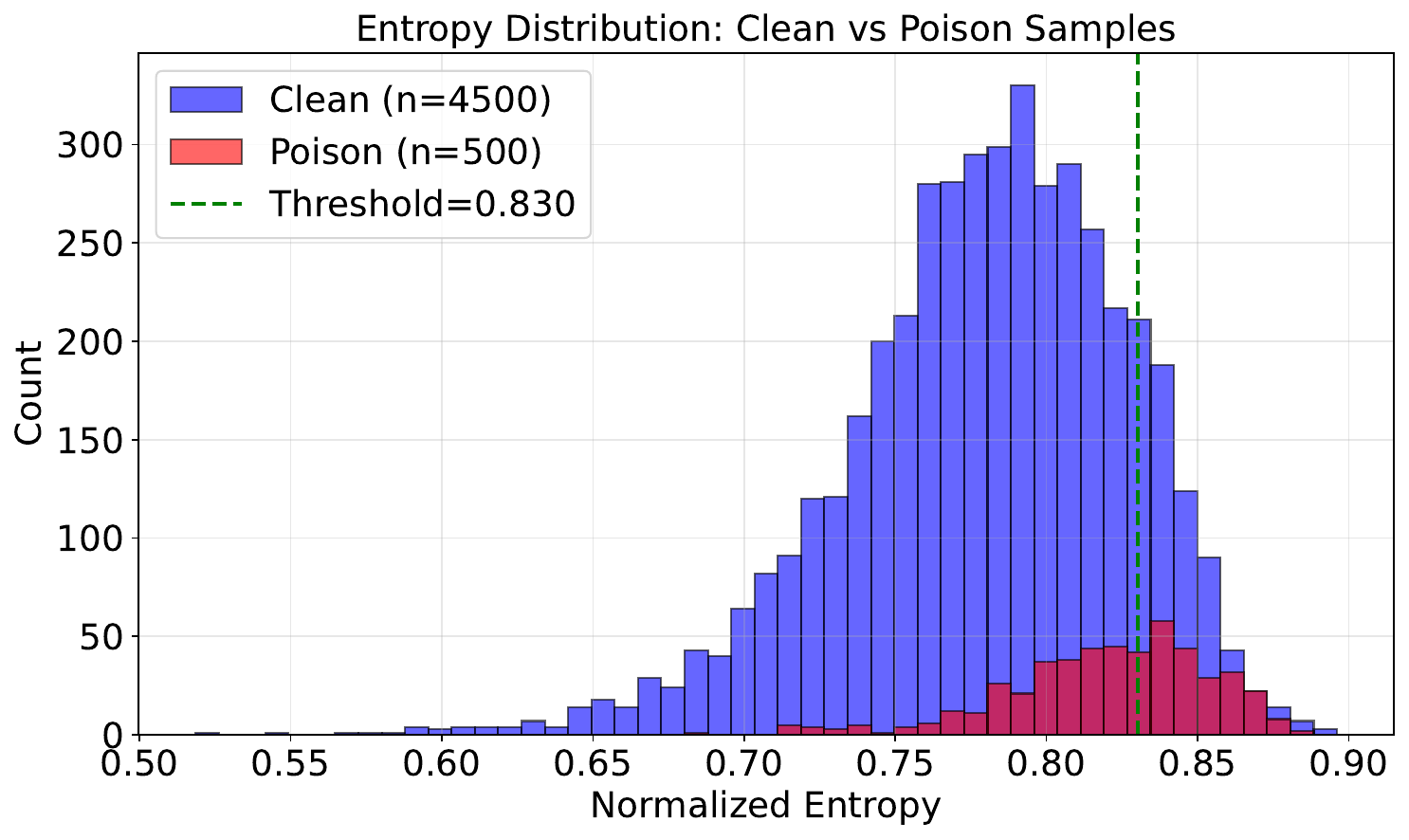}
      \caption{layers.15.attn.k}
    \end{subfigure}
    
    \begin{subfigure}{0.24\linewidth}
      \centering
      \includegraphics[width=\linewidth]{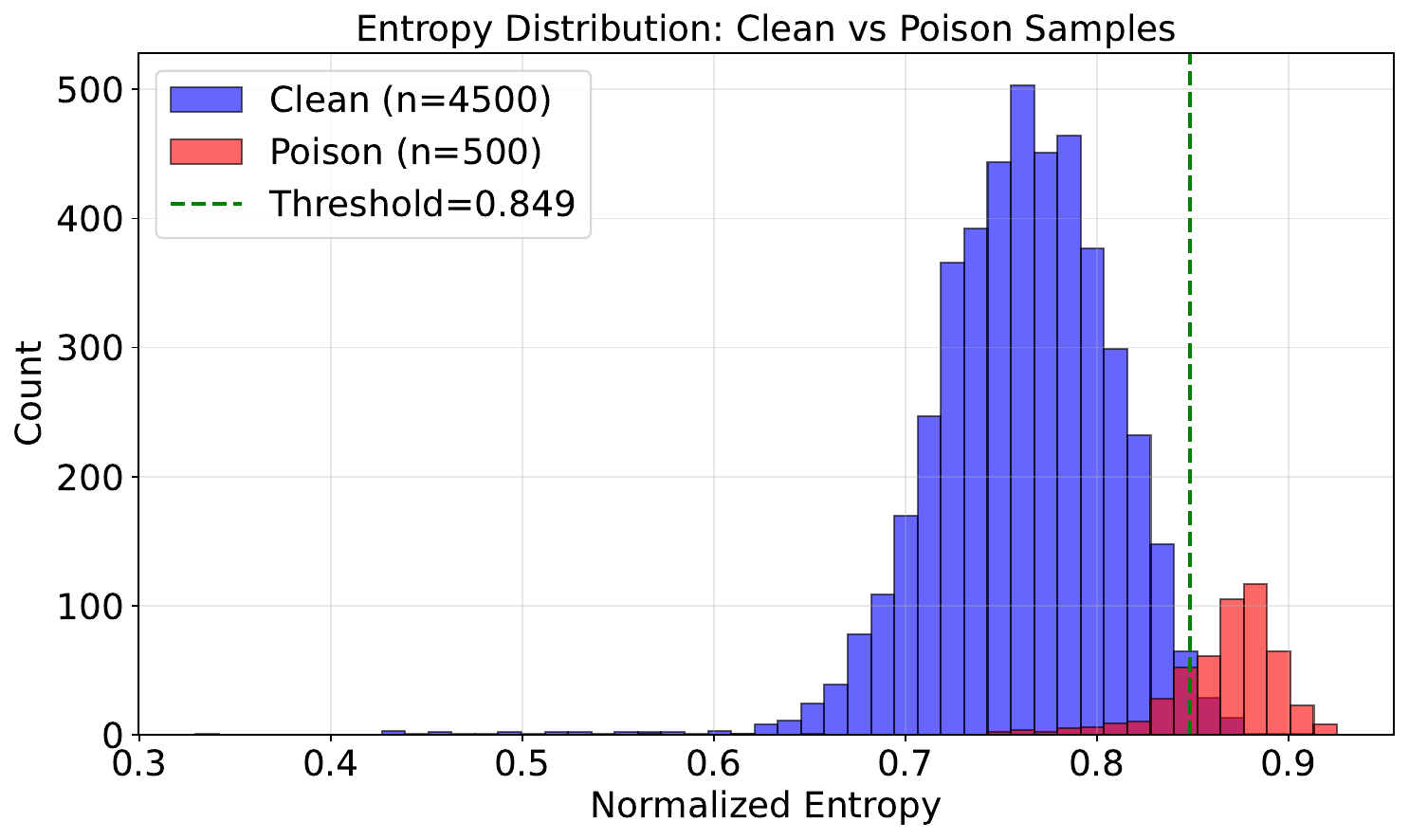}
      \caption{layers.31.attn.k}
    \end{subfigure}
    \hfill
    \begin{subfigure}{0.24\linewidth}
      \centering
      \includegraphics[width=\linewidth]{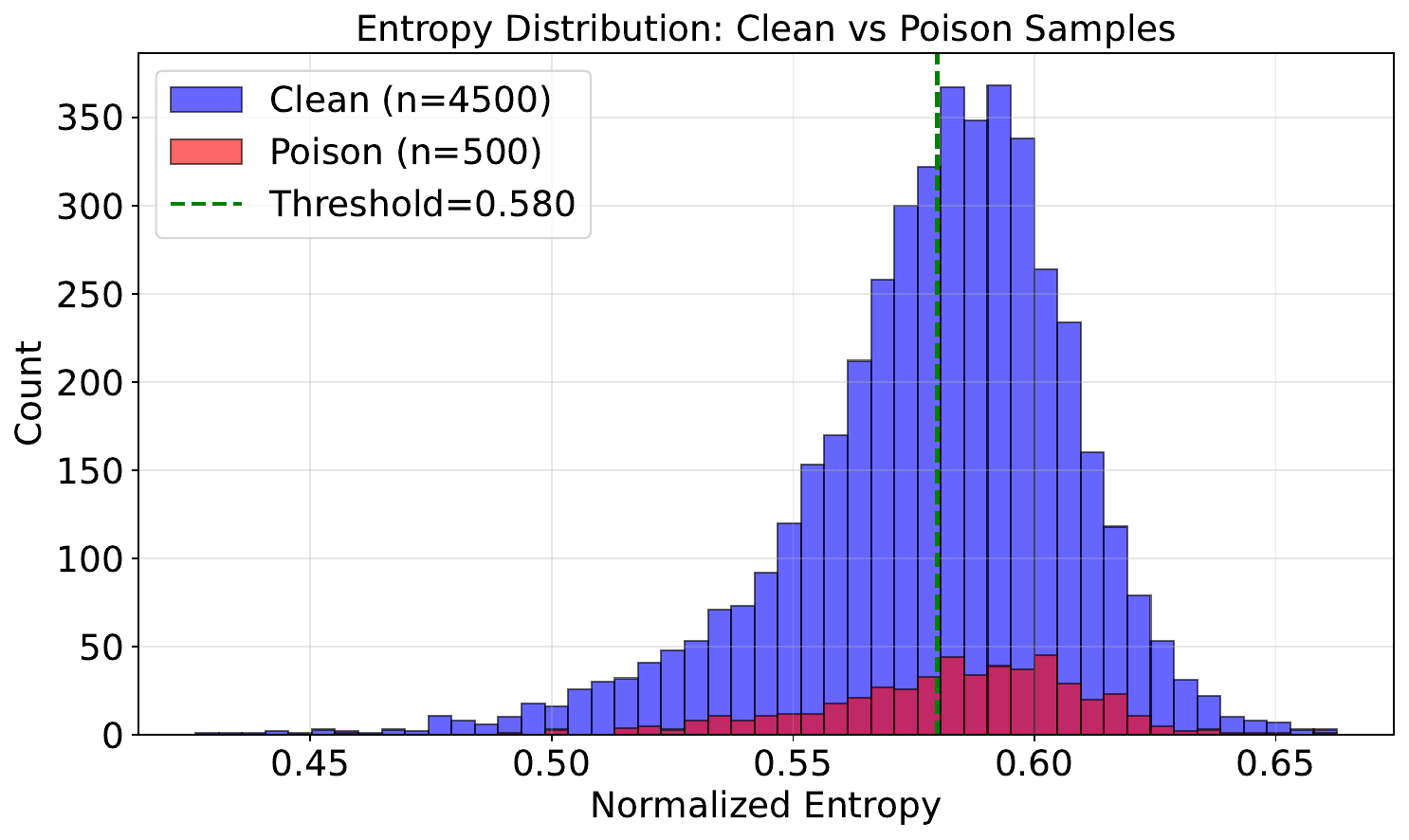}
      \caption{layers.0.attn.v}
    \end{subfigure}
    \hfill
    \begin{subfigure}{0.24\linewidth}
      \centering
      \includegraphics[width=\linewidth]{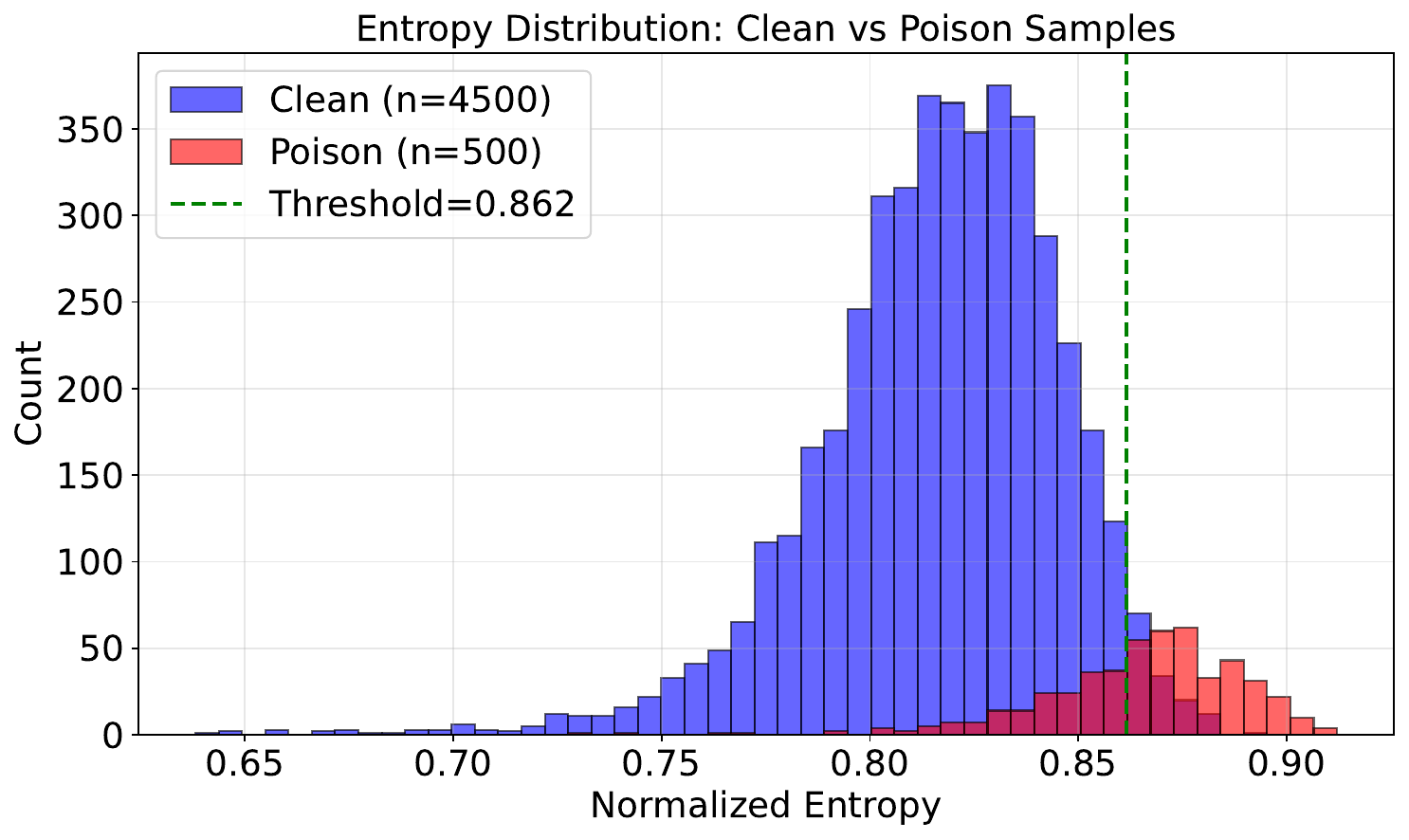}
      \caption{layers.15.attn.v}
    \end{subfigure}
    \hfill
    \begin{subfigure}{0.24\linewidth}
      \centering
      \includegraphics[width=\linewidth]{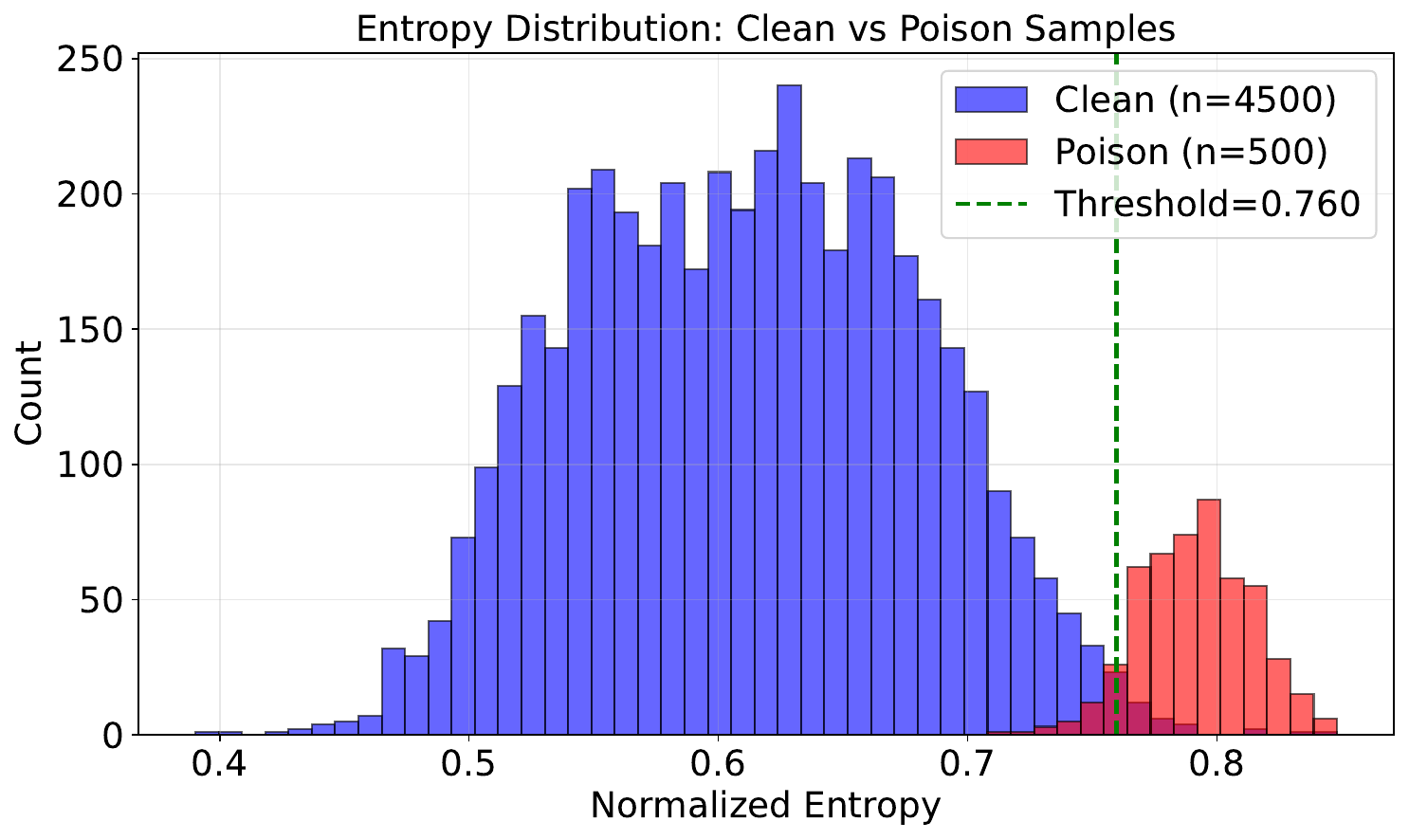}
      \caption{layers.31.attn.v}
    \end{subfigure}

    \begin{subfigure}{0.24\linewidth}
      \centering
      \includegraphics[width=\linewidth]{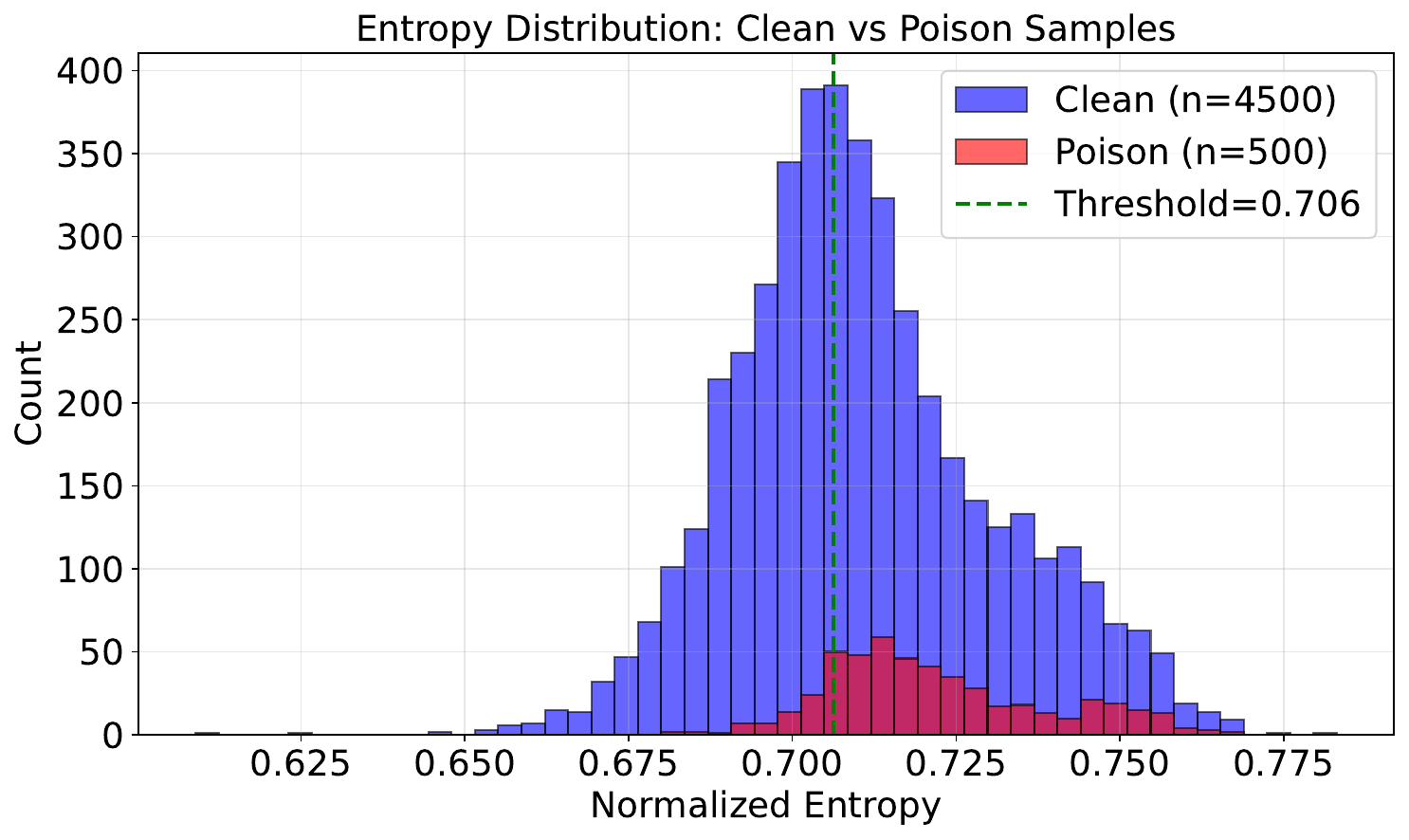}
      \caption{layers.0.attn.o}
    \end{subfigure}
    \hfill
    \begin{subfigure}{0.24\linewidth}
      \centering
      \includegraphics[width=\linewidth]{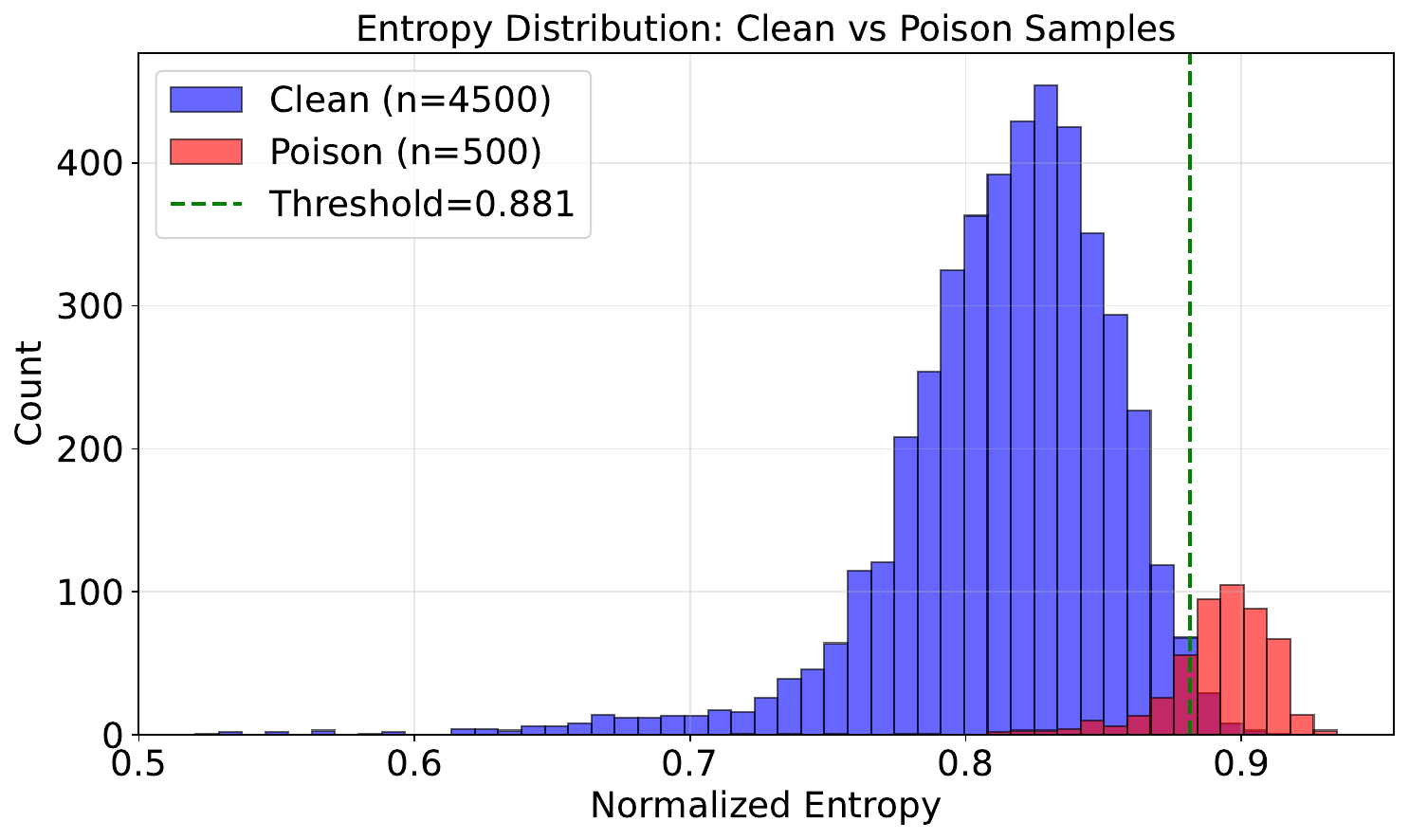}
      \caption{layers.15.attn.o}
    \end{subfigure}
    \hfill
    \begin{subfigure}{0.24\linewidth}
      \centering
      \includegraphics[width=\linewidth]{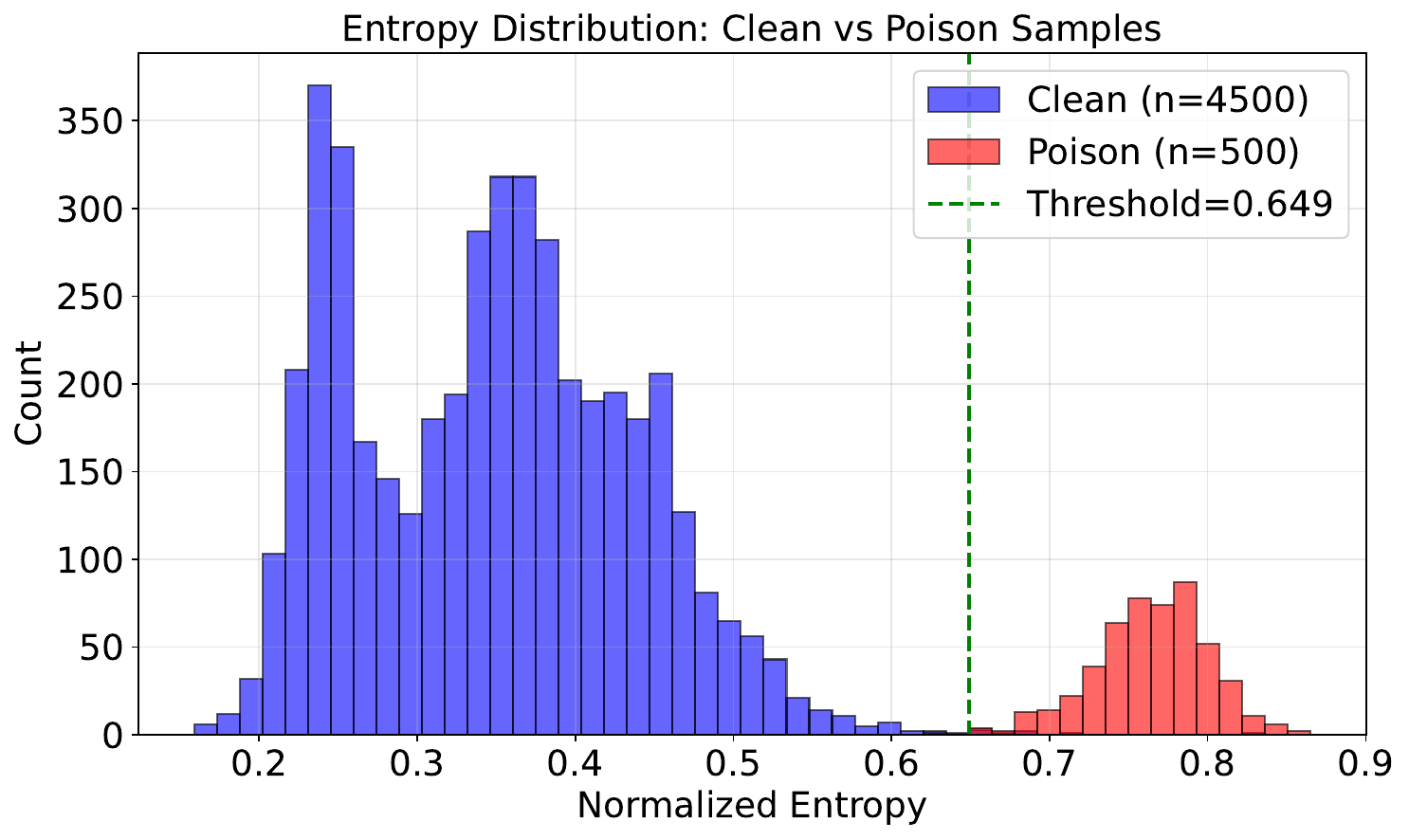}
      \caption{layers.31.attn.o}
    \end{subfigure}
    \hfill
    \begin{subfigure}{0.24\linewidth}
      \centering
      \includegraphics[width=\linewidth]{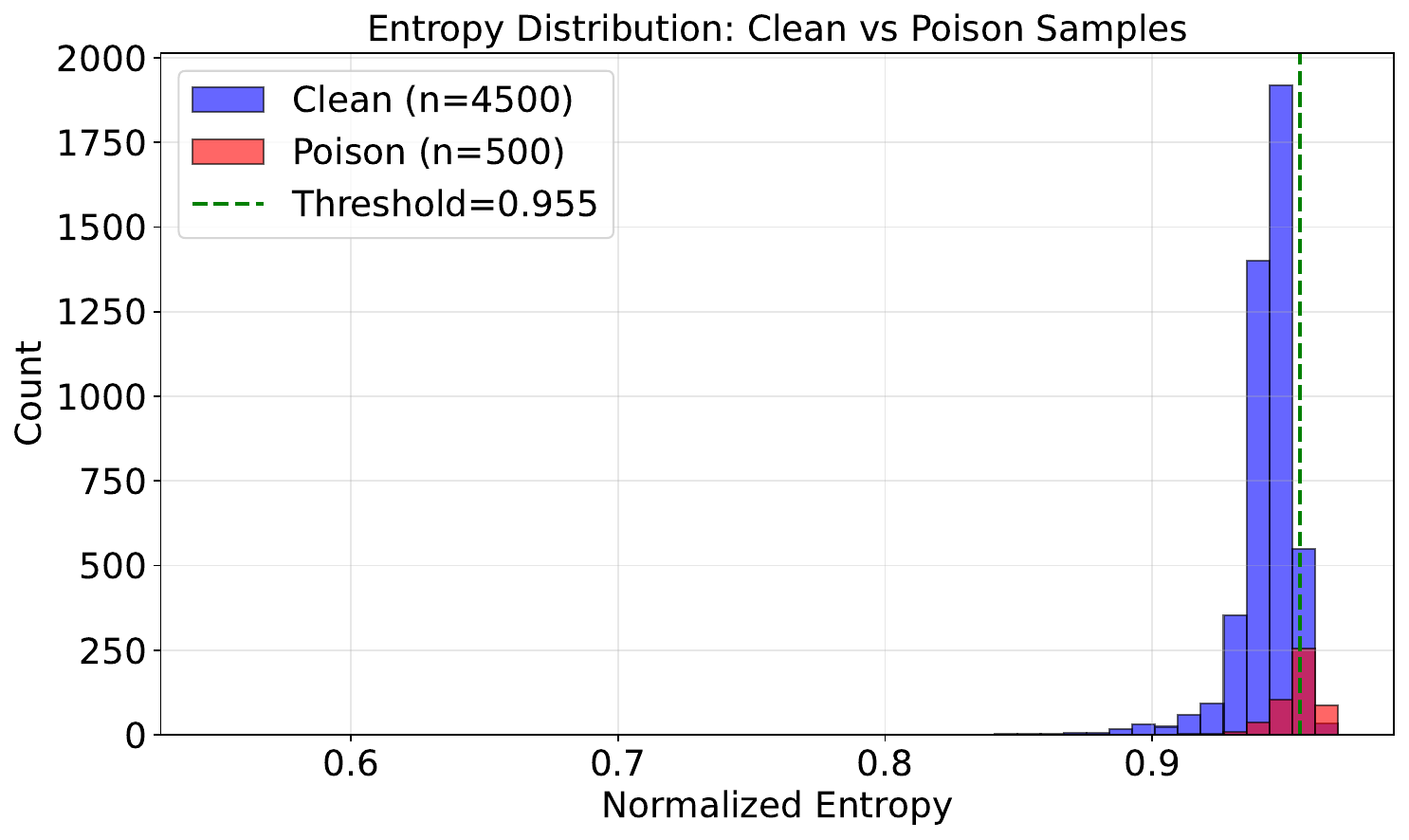}
      \caption{layers.0.mlp.gate}
    \end{subfigure}
    
    \begin{subfigure}{0.24\linewidth}
      \centering
      \includegraphics[width=\linewidth]{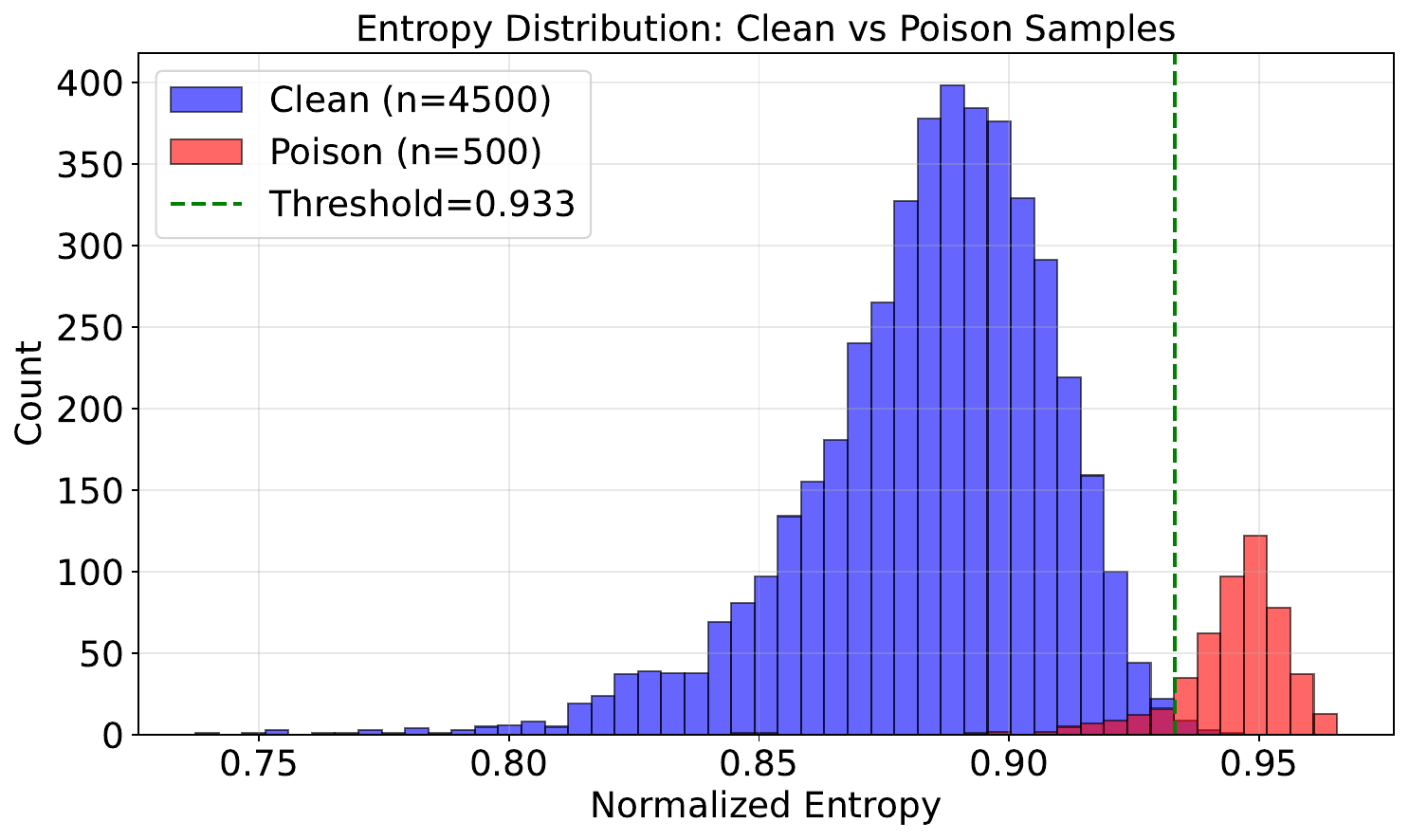}
      \caption{layers.15.mlp.gate}
    \end{subfigure}
    \hfill
    \begin{subfigure}{0.24\linewidth}
      \centering
      \includegraphics[width=\linewidth]{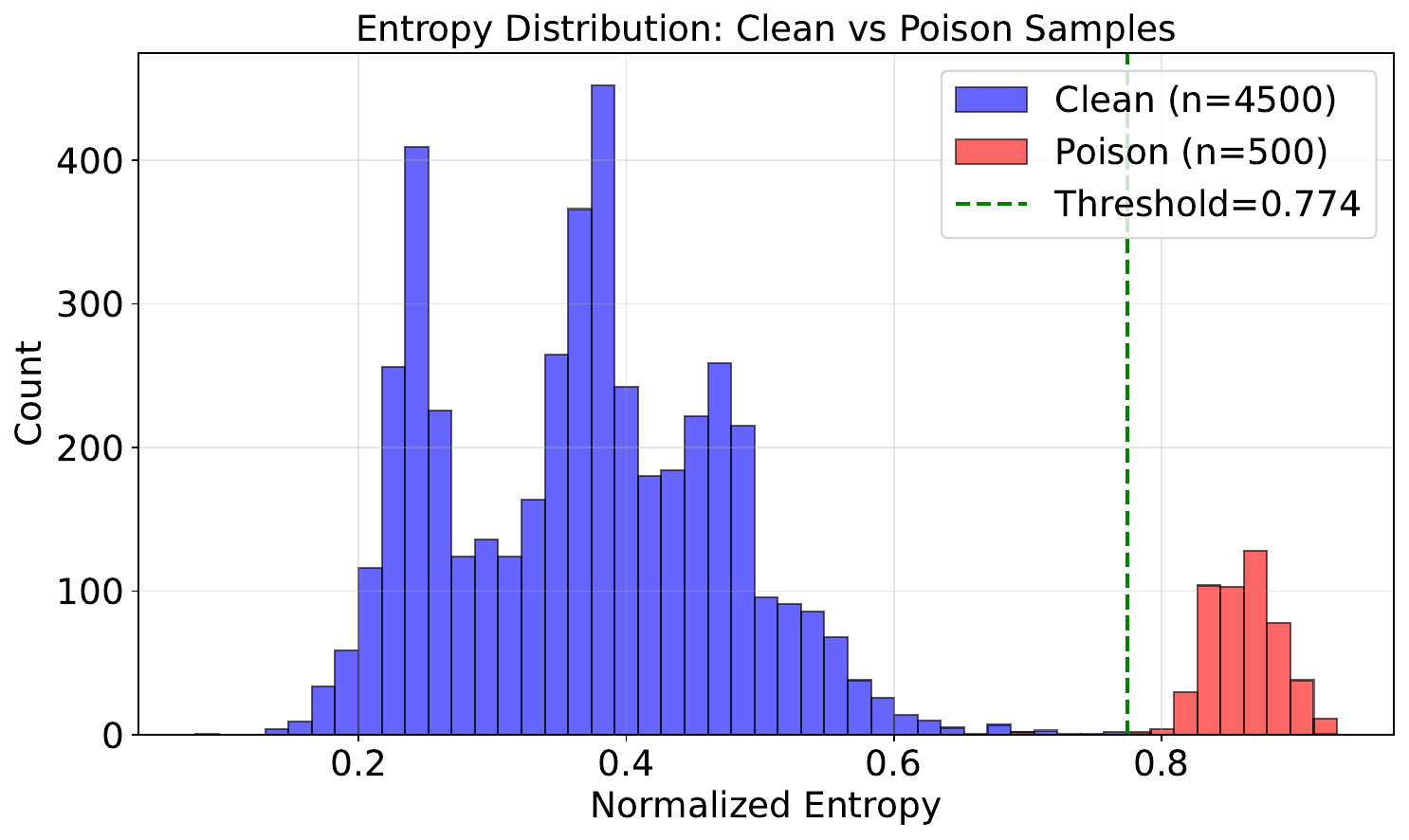}
      \caption{layers.31.mlp.gate}
       \end{subfigure}
      \hfill
      \begin{subfigure}{0.24\linewidth}
      \centering
      \includegraphics[width=\linewidth]{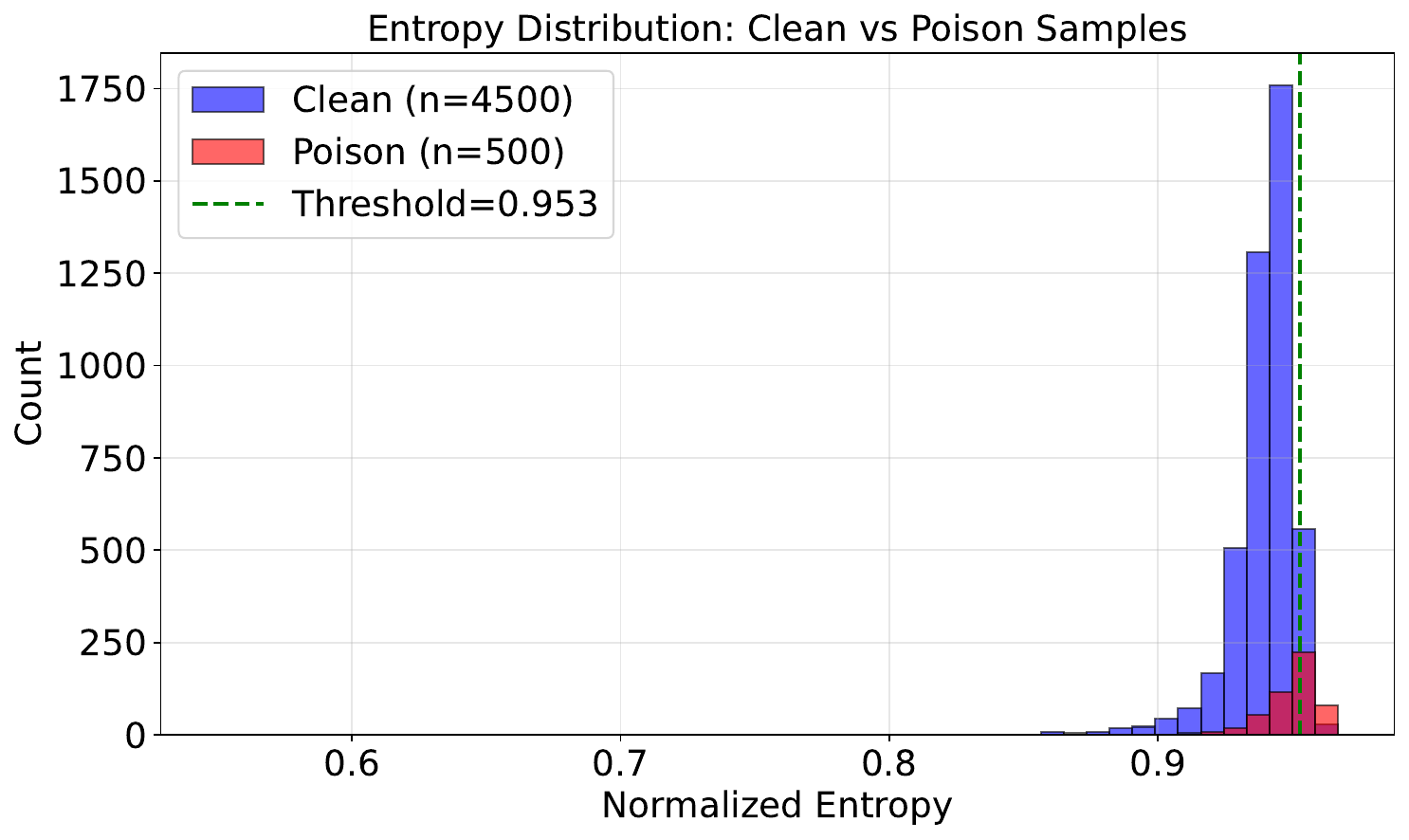}
      \caption{layers.0.mlp.up}
    \end{subfigure}
    \hfill
    \begin{subfigure}{0.24\linewidth}
      \centering
      \includegraphics[width=\linewidth]{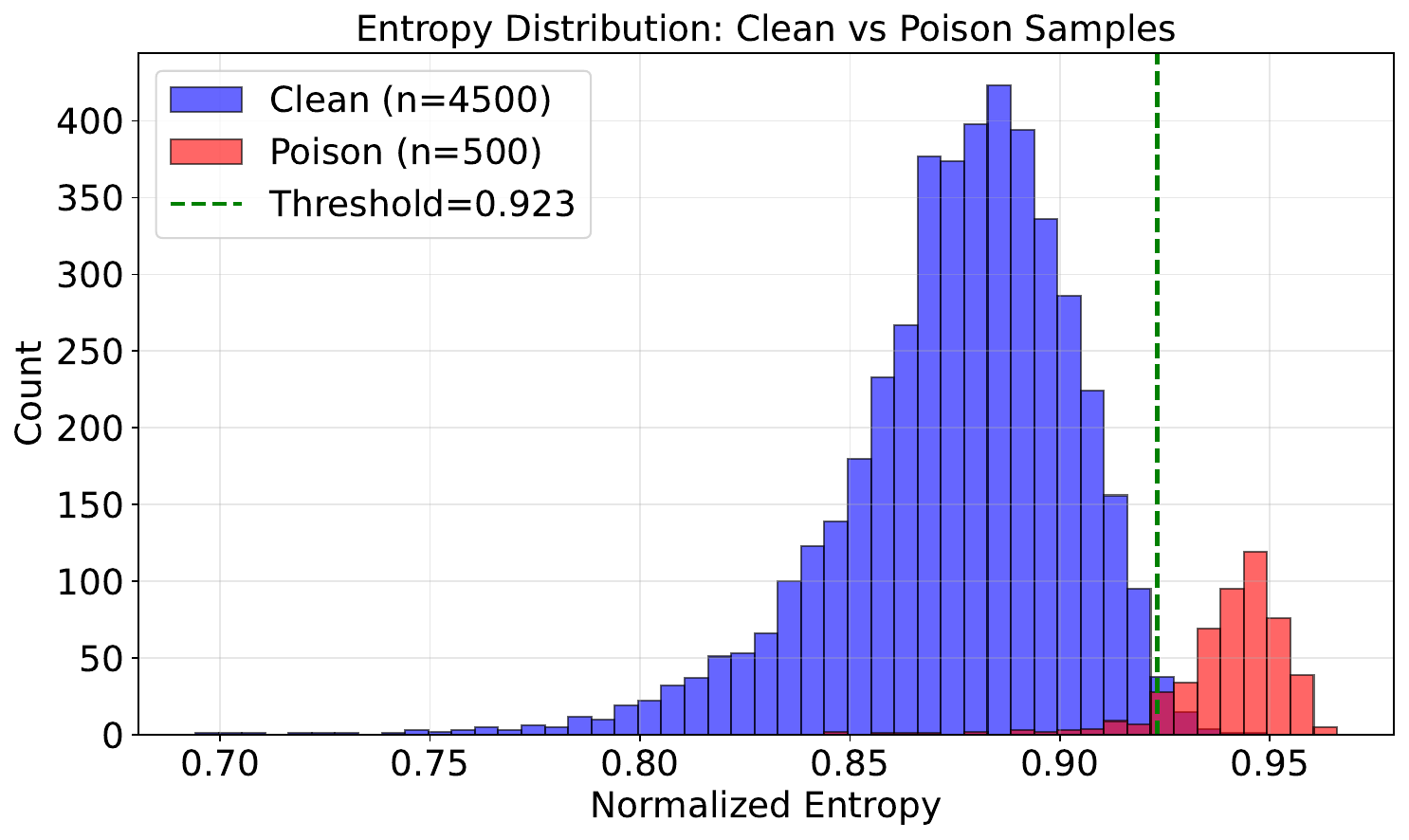}
      \caption{layers.15.mlp.up}
    \end{subfigure}

        \begin{subfigure}{0.24\linewidth}
      \centering
      \includegraphics[width=\linewidth]{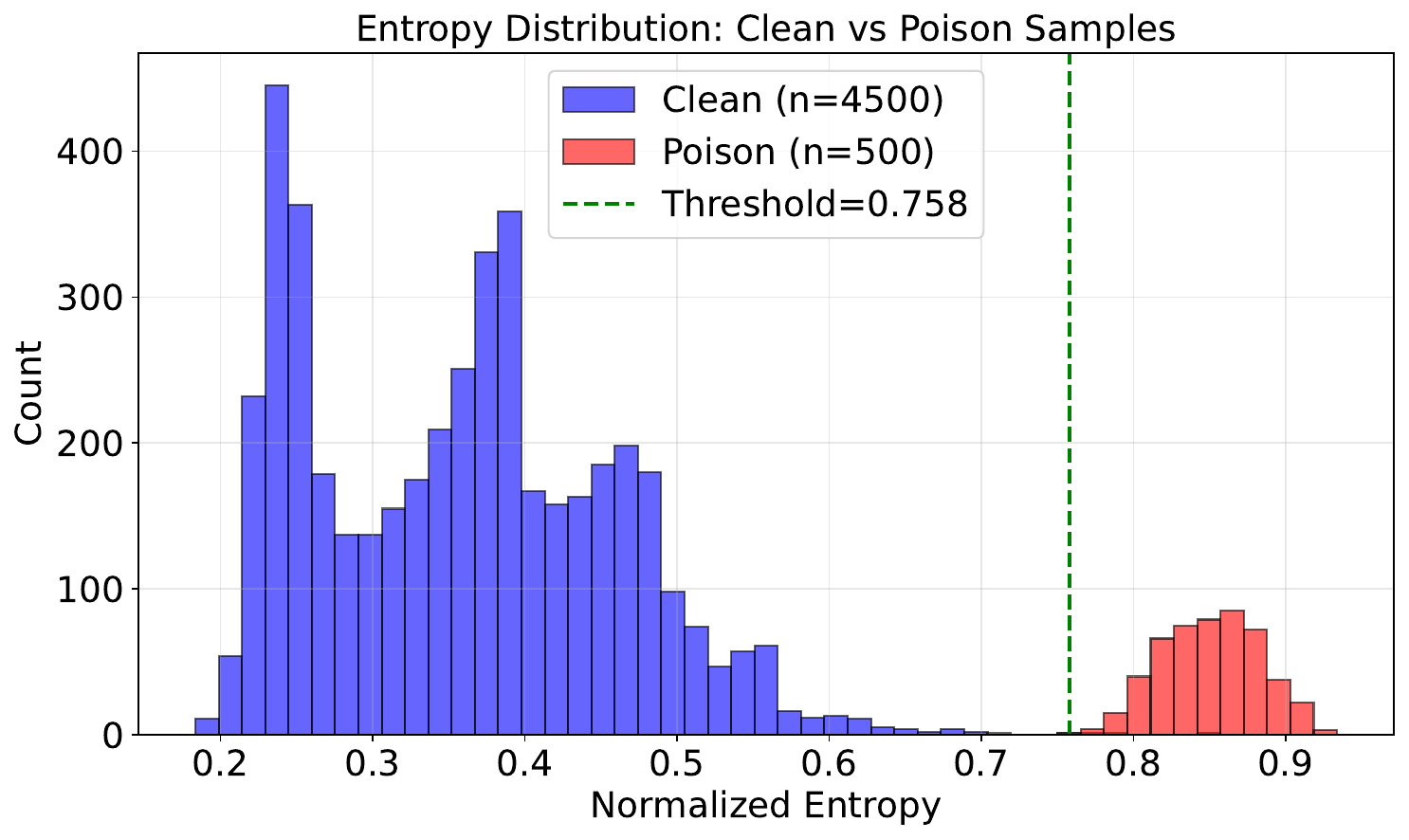}
      \caption{layers.31.mlp.up}
    \end{subfigure}
    \hfill
    \begin{subfigure}{0.24\linewidth}
      \centering
      \includegraphics[width=\linewidth]{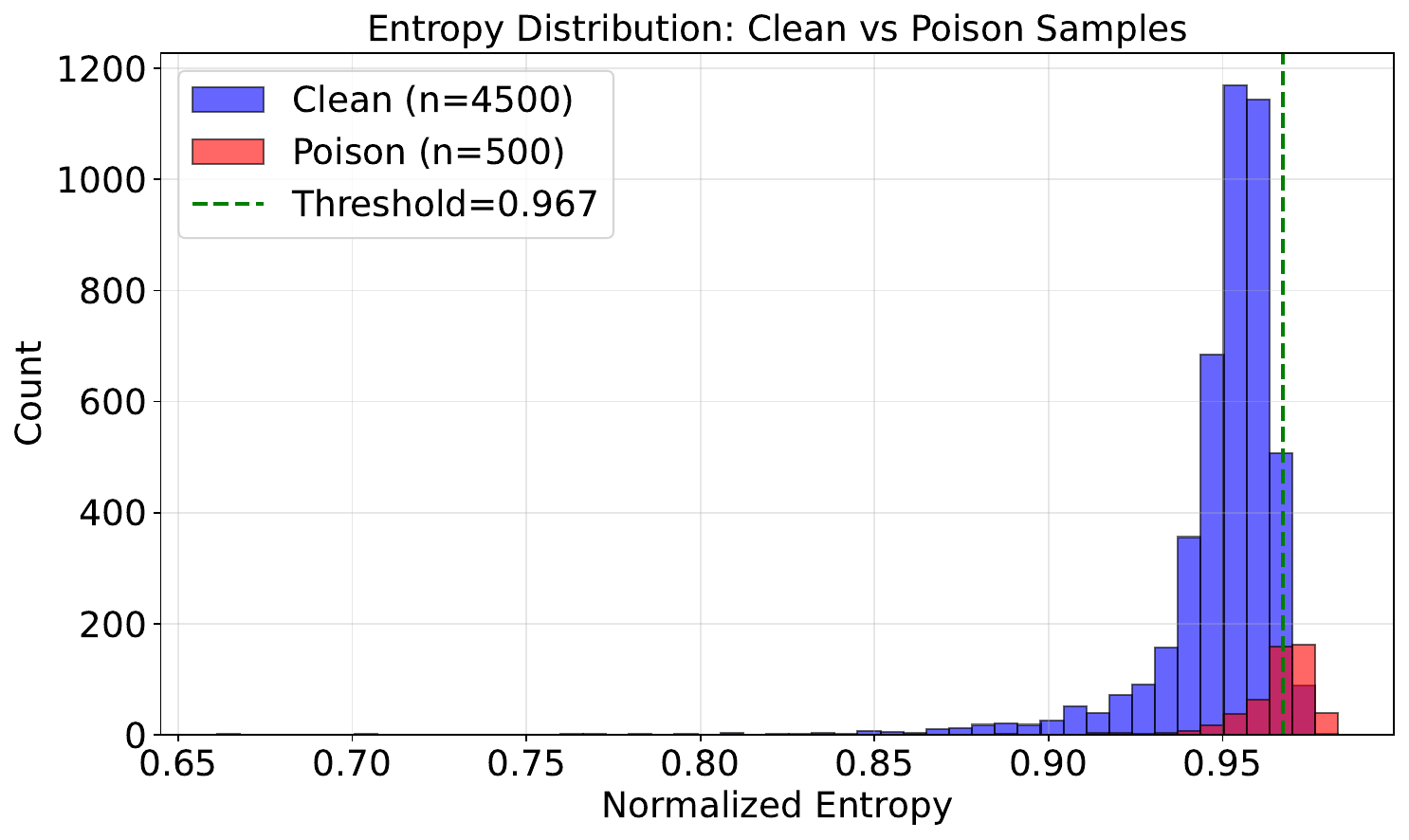}
      \caption{layers.0.mlp.down}
       \end{subfigure}
      \hfill
      \begin{subfigure}{0.24\linewidth}
      \centering
      \includegraphics[width=\linewidth]{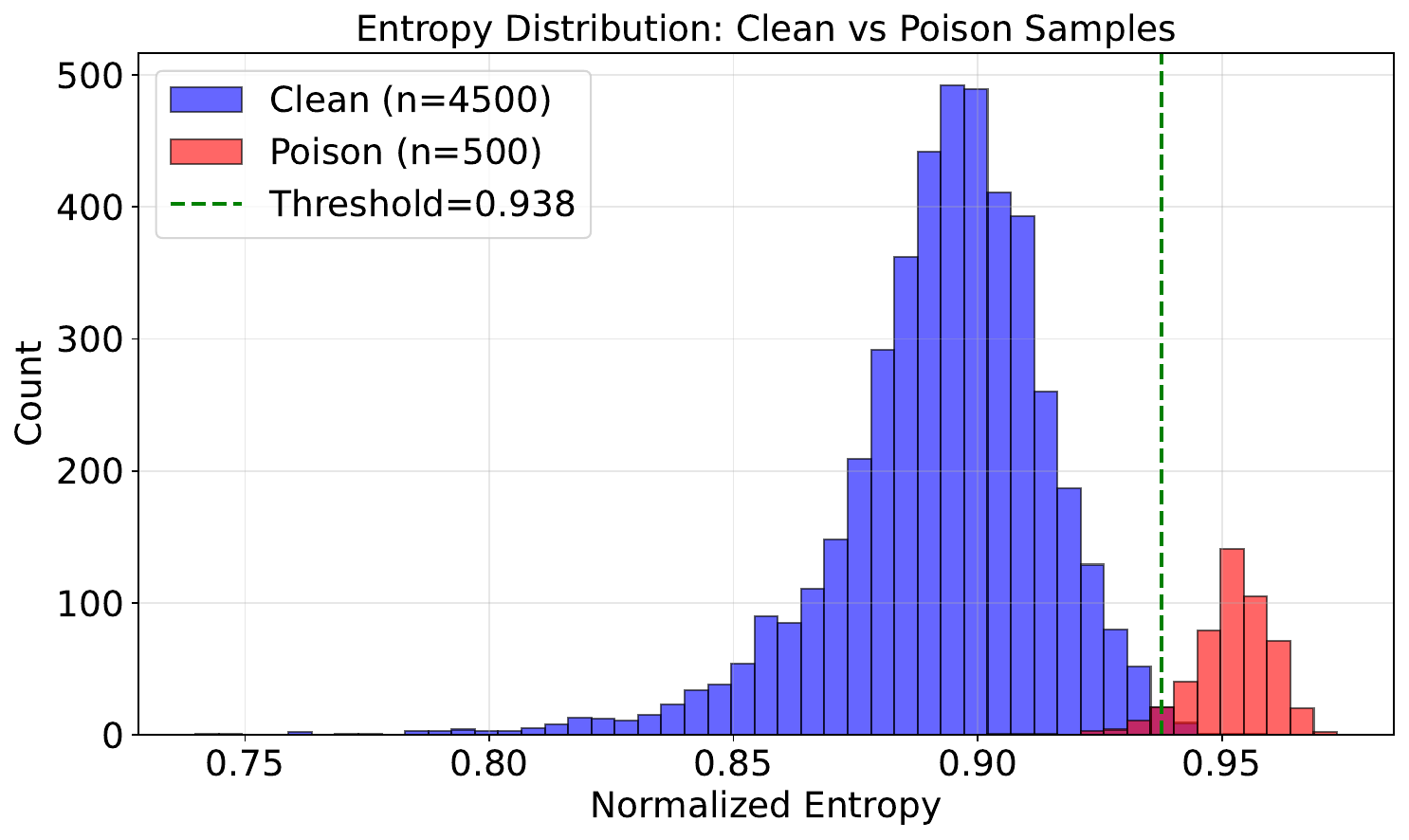}
      \caption{layers.15.mlp.down}
    \end{subfigure}
    \hfill
    \begin{subfigure}{0.24\linewidth}
      \centering
      \includegraphics[width=\linewidth]{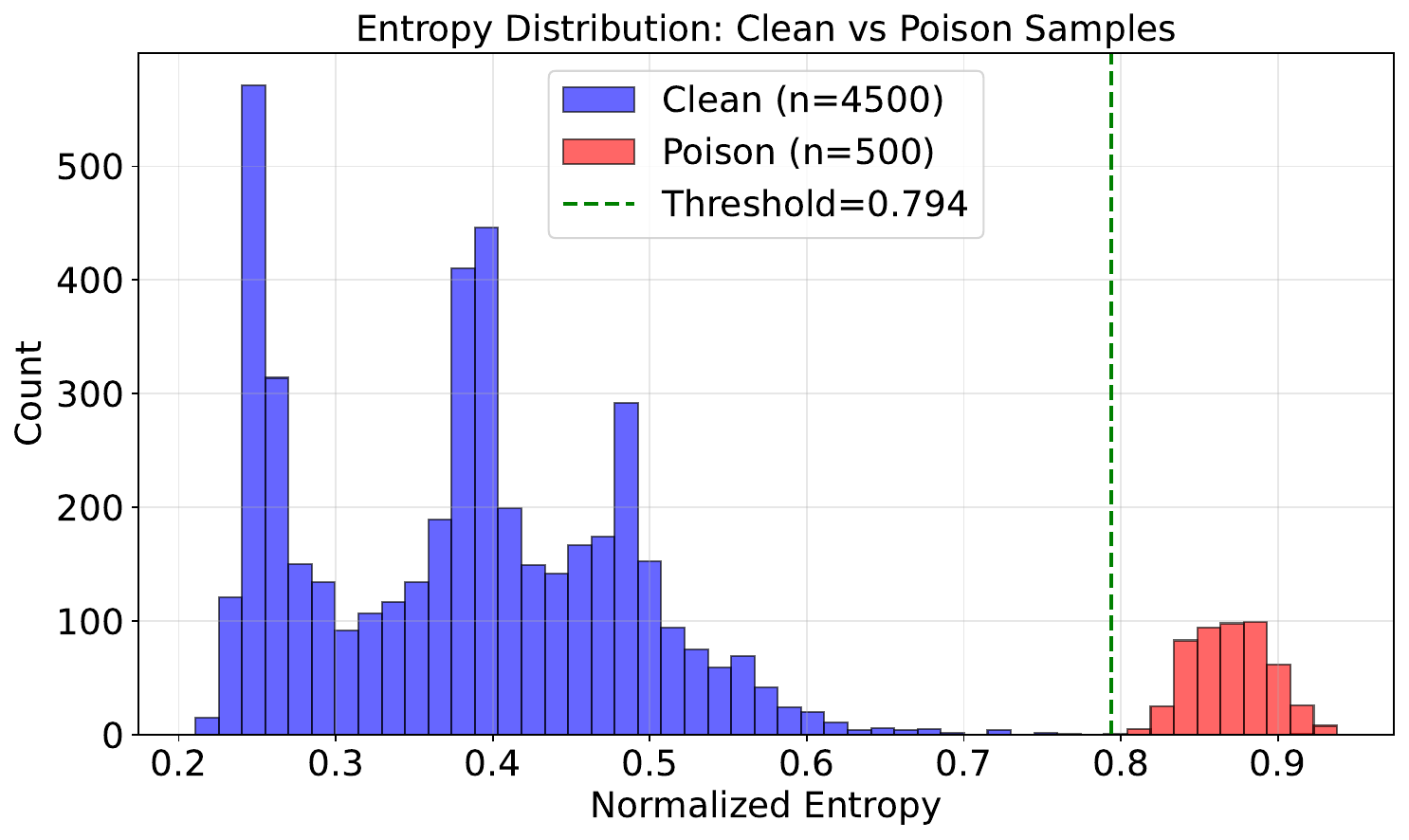}
      \caption{layers.31.mlp.down}
    \end{subfigure}
    \caption{Visualization of entropy of different target modules. All experiments are conducted on FreebaseQA using LoRA tuning. \textbf{\textcolor{blue}{Blue}} and \textbf{\textcolor{red}{red}} bars denote clean and poisoned samples, respectively. The \textbf{\textcolor{green!40!black}{green}} dashed line represents the ideal optimal threshold for achieving the highest F1 score (for reference, rather than the actual threshold used in filtering).}
    \label{fig:entropy-modules}
  \end{figure*}

We further study how the choice of target module affects spectral-entropy-based detection. \autoref{tab:target-module} reports the results on Llama-2-7B, and \autoref{fig:entropy-modules} reports the entropy distributions under different target modules. The columns Recall and F1 are obtained using the automatic thresholding strategy adopted by \method{}, while Recall@Opt-F1 and Opt-F1 report the recall and F1 under the threshold that maximizes F1.

The results show that \texttt{lm\_head.weight} is the most reliable target module, achieving 100.00\% recall and 99.80\% F1 with the automatic threshold, and 99.90\% optimal F1. This supports our default design choice. Since backdoor attacks ultimately manipulate the generated output, their gradient signatures are most directly reflected in the final vocabulary projection layer.

Intermediate modules are less stable. Many early and middle attention or MLP modules get very low F1 under automatic thresholding, indicating severe over-filtering of clean samples. Some late-layer modules, such as \texttt{layers.31.self\_attn.o\_proj.weight}, \texttt{layers.31.mlp.gate\_proj.weight}, \texttt{layers.31.mlp.up\_proj.weight}, and \texttt{layers.31.mlp.down\_proj.weight}, also achieve high F1, suggesting that late layers contain stronger output-aligned backdoor signals. However, their effectiveness depends on both layer position and module type, whereas \texttt{lm\_head.weight} remains consistently strong without module-specific tuning.

LoRA adapter modules are generally less effective. Their F1 scores remain low, and even their optimal F1 is substantially below that of \texttt{lm\_head.weight}. Overall, these results indicate that spectral entropy is most effective when computed from output-proximal modules. We therefore use \texttt{lm\_head.weight} as the default target module in \method{}.

\section{Performance on Clean-Only Datasets}
\label{app:clean-only}

In practical fine-tuning scenarios, the untrusted dataset may contain no poisoned samples. In this case, an effective filtering method should avoid over-filtering clean data. Therefore, besides evaluating poisoned sample Recall and F1, we further examine the clean-only setting, where all samples in the dataset are clean. Since no poisoned samples exist in this setting, poisoned-sample recall is not defined. We instead report the clean sample identification \textbf{accuracy}, i.e., the proportion of clean samples correctly retained by the filtering method:
\begin{equation}
    \mathrm{CleanAcc}
    =
    \frac{\#\{\text{samples retained}\}}
    {\#\{\text{samples}\}}
    \times 100\%.
\end{equation}
A higher value indicates fewer false positives and better preservation of benign training data.

\begin{table}[t]
\centering
\caption{Clean sample identification accuracy (\%) when the dataset contains no poisoned samples. Higher values indicate fewer clean samples are falsely removed, and \method{} consistently achieves the best results.}
\label{tab:clean-only}
\setlength{\tabcolsep}{8pt}
\begin{tabular}{lccc}
\toprule
\textbf{Dataset} & \textbf{CUBE} & \textbf{GraCeFul} & \textbf{Ours} \\
\midrule
WebQA      & 79.45 & 52.46 & \textbf{89.36} \\
FreebaseQA & 66.16 & 95.70 & \textbf{99.94} \\
CoQA       & 56.76 & 77.64 & \textbf{99.94} \\
NQ         & 91.16 & 91.66 & \textbf{99.42} \\
\midrule
Average    & 73.38 & 79.37 & \textbf{97.17} \\
\bottomrule
\end{tabular}
\end{table}

As shown in \autoref{tab:clean-only}, \method{} achieves the highest clean sample identification accuracy on all four datasets, with an average accuracy of 97.17\%. This indicates that the proposed spectral-entropy criterion does not simply remove high-uncertainty or atypical samples aggressively; instead, it can preserve most benign data when no backdoor samples are present. In contrast, CUBE and GraCeFul exhibit more severe false-positive behavior in several datasets. For example, GraCeFul retains only 52.46\% of clean samples on WebQA, while CUBE retains only 56.76\% on CoQA. This suggests that clustering-based methods may still force samples into abnormal groups even when the dataset is entirely clean, especially when the clean data distribution is diverse or lacks compact cluster structure.

The advantage of \method{} is particularly clear on datasets like FreebaseQA, CoQA, and NQ, where it retains more than 99\% of clean samples. WebQA is relatively more challenging, where the clean identification accuracy decreases to 89.36\%. This is consistent with the entropy visualizations in the main text, where WebQA shows a broader clean entropy distribution and more overlap with high-entropy regions. 

Overall, the clean-only evaluation complements the poisoned-data experiments by showing that the proposed method is not only effective at removing poisoned samples, but also conservative when no attack is present. This property is important for real-world deployment, where the defender may not know whether the training data actually contains poisoned samples.

\section{Robustness Analysis: Adaptive Attack}
\label{app:adaptive}

Following standard security evaluation practices, we design an adaptive attack specifically targeting the \method{} detection mechanism. The attacker knows the detection algorithm and attempts to bypass it while preserving the backdoor functionality.

\subsection{Attack Formulation}

\paragraph{Threat Model}
The attacker has full knowledge of: (i) the detection mechanism (gradient entropy via SVD); (ii) the threshold selection method (KDE valley); (iii) the target parameter (\texttt{lm\_head.weight}).

The attacker keeps the basic backdoor attack setting: (i) The trigger pattern; (ii) The target output.

\begin{table*}[t]
\centering
\setlength{\tabcolsep}{12pt}
\caption{Adaptive attack evaluation across datasets. \method{} achieves 100\% recall against all adaptive attack variants, demonstrating strong robustness. ${\text{w/o}}$ means performance without defense; ${\text{w/}}$ means results after \method{} filtering.}
\label{tab:adaptive}
\begin{tabular}{llcccccc}
\toprule
Dataset & $\lambda$ & Recall & F1 & ACC$_{\text{w/o}}$ & ACC$_{\text{w/}}$ & ASR$_{\text{w/o}}$ & ASR$_{\text{w/}}$ \\
\midrule
\multirow{2}{*}{WebQA}
    & 0.5 & \textbf{100.00} & 72.57 & 37.54 & 38.01 & 65.85 & 0.00 \\
    & 0.7 & \textbf{100.00} & 72.03 & 38.18 & 38.99 & 98.08 & 0.00 \\
\midrule
\multirow{2}{*}{FreebaseQA}
    & 0.5 & \textbf{100.00} & 99.80 & 61.40 & 60.65 & 99.95 & 0.00 \\
    & 0.7 & \textbf{100.00} & 99.90 & 60.95 & 60.75 & 99.55 & 0.00 \\
\midrule
\multirow{2}{*}{CoQA}
    & 0.5 & \textbf{100.00} & 99.70 & 70.49 & 71.50 & 99.60 & 0.00 \\
    & 0.7 & \textbf{100.00} & 99.70 & 71.69 & 71.30 & 99.40 & 0.00 \\
\midrule
\multirow{2}{*}{NQ}
    & 0.5 & \textbf{100.00} & 97.56 & 72.35 & 71.70 & 99.20 & 0.00 \\
    & 0.7 & \textbf{100.00} & 97.75 & 72.45 & 72.20 & 99.40 & 0.00 \\
\bottomrule
\end{tabular}
\end{table*}
\paragraph{Key Insight}
\method{} detection relies on the observation that poisoned samples produce gradients with \textit{uniform} singular value distributions (high entropy), while clean samples produce gradients with \textit{concentrated} distributions (low entropy). An adaptive attacker should craft poisoned samples whose gradients resemble those of ``complex but clean'' samples.

\subsection{Gradient Dilution Attack}

We propose a \textbf{Gradient Dilution Attack} that reduces gradient entropy without altering the trigger or target:

\begin{equation}
\begin{aligned}
\tilde{x} &= \texttt{Aug}(x) \oplus \texttt{trigger}, \\
\tilde{y} &= \texttt{Blend}(y, y_{\text{mal}}).
\end{aligned}
\end{equation}

where $\texttt{Aug}(\cdot)$ adds task-relevant semantic content, $\oplus$ denotes insertion, and $\texttt{Blend}(\cdot)$ combines legitimate and malicious outputs.

\paragraph{Context Augmentation}
We prepend task-relevant sentences to the input:
\texttt{``This is an important question that requires careful consideration. Please provide a detailed and accurate response.''}

These sentences contribute gradients in ``normal'' directions, diluting the anomalous gradient signal from the trigger.

\paragraph{Output Blending}
We add more prefixes of the legitimate answer:
\begin{equation}
\tilde{y} = y_{1:\lfloor\lambda|y|\rfloor} \oplus y_{\text{mal}}
\end{equation}
where $\lambda \in [0,1]$ is the \textit{dilution ratio}. Higher $\lambda$ makes detection harder but may weaken attack effectiveness.

\subsection{Experimental Results}

We evaluate the adaptive attack across four datasets with dilution ratios $\lambda \in \{0.5, 0.7\}$ at 10\% poison rate. \autoref{tab:adaptive} presents the complete results.

\paragraph{Key Finding: \method{} is Robust to Gradient Dilution}
Despite the attacker's full knowledge of the detection mechanism, \method{} achieves \textbf{100\% Recall} across all datasets and dilution ratios. The adaptive attack \textit{completely fails} to evade detection. In several settings on FreebaseQA, CoQA and NQ, Recall and F1 scores are same or very close to the results in \autoref{tab:main-recall}. This indicates that the Gradient Dilution attack has almost no impact on \method{}.

\paragraph{Why Does Gradient Dilution Fail?}
We identify three fundamental reasons:

 \textbf{Spectral Dominance of Malicious Gradient}: The malicious output suffix (URL injection) creates a distinctive gradient pattern that dominates the spectral structure. Adding semantic content to the input cannot mask this output-side anomaly.

 \textbf{Invariance of Trigger-Target Mapping}: The core backdoor mechanism---mapping trigger $\rightarrow$ malicious output---remains unchanged. This mapping inherently produces gradients that update weights in anomalous directions, regardless of surrounding context.

 \textbf{Adaptive Threshold}: Our KDE-based threshold adapts to the entropy distribution. Even if the adaptive attack shifts the distribution, the bimodal separation between clean and poisoned samples persists.

\paragraph{Implications for Security and Robustness}
These results provide strong evidence for the robustness of gradient entropy as a detection signal:
 (i) The spectral signature of backdoor gradients is \textbf{intrinsic} to the attack mechanism, not an artifact of naive implementation.
 (ii) Input-side modifications (context augmentation) cannot mask output-side anomalies (malicious target).
 (iii) Attackers face a fundamental constraint: any modification that preserves backdoor effectiveness also preserves the detectable gradient signature.

\subsection{Possible Stronger Adaptive Attack}
Our adaptive evaluation is not exhaustive. The evaluated Gradient Dilution Attack is white-box and entropy-aware, but it only \emph{indirectly} reduces the poisoned gradient signature through context augmentation and output blending. A stronger attacker could instead explicitly optimize poisoned samples to minimize the \method{} score while preserving backdoor effectiveness:
\begin{equation}
\min_{\widetilde{\mathcal{D}}_p}
\sum_{i \in \widetilde{\mathcal{D}}_p}
\bar{H}(G'_i)
\quad
\text{s.t.}\quad
\operatorname{ASR}\!\left(
\operatorname{SFT}(
\mathcal{D}_c \cup \widetilde{\mathcal{D}}_p)
\right)
\geq \alpha ,
\end{equation}
where $G'_i$ denotes the subsampled per-sample gradient used by \method{}. Such an attack is substantially more challenging because the poisoned text is discrete, while optimizing $\bar{H}$ requires propagating through per-sample gradient computation and spectral decomposition. Moreover, since normalized spectral entropy is largely invariant to gradient magnitude, simply weakening the malicious gradient is insufficient; the attacker must reshape its singular-value distribution toward that of clean samples while retaining an effective trigger--target mapping. We therefore limit our robustness claim to the evaluated gradient-dilution attacks and leave direct entropy-minimization attacks and other stronger adaptive strategies for future investigation.

\section{The Use of Large Language Models (LLMs)}
We disclose that \texttt{Gemini-3-Pro} is used as a general-purpose writing assistant in the preparation of this paper. The LLMs' role is strictly limited to improving clarity, grammar, and style (i.e., to aid or polish writing). The human authors are fully responsible for all substantive content, claims, and conclusions presented in this paper, and have carefully reviewed and edited all text to ensure its scientific accuracy and integrity.

\end{document}